\documentclass[
aps,
prb,
amsmath,amssymb,
reprint,
groupedaddress,
longbibliography
]{revtex4-2}

\usepackage{graphicx}
\usepackage[percent]{overpic}

\usepackage[justification=raggedright,singlelinecheck=false]{caption}

\usepackage{hyperref}

\usepackage{mathrsfs} 
\usepackage{physics}

\begin{document}

\title{Dynamic helimagnetism via charged spin current}

\author{Masaru Onoda}
\affiliation{%
Mathematical Science Course,
Graduate School of Engineering Science,
Akita University, Akita 010-8502, Japan
}%

\date{October 26, 2026}

\begin{abstract}
We present a theoretical and numerical study of classical, localized spins coupled to conduction electrons that carry both charge and spin currents, deriving the effective equations of motion from the $s$-$d$ exchange Hamiltonian. Applied to a periodically arranged one-dimensional spin chain, these equations yield two current-driven helical phases: a static helix stabilized by a pure spin current alone, and a coherently precessing helix that appears once a small charge-current component is added. For a given helical wavenumber, the precession angular frequency $\omega$ and the canting ratio $\Delta$ toward the hard axis follow analytically from the equations of motion, while the wavenumber $q_{\mathrm{helix}}$ of the precessing helix is selected by a heuristically constructed effective Lagrangian. Direct numerical integration of the same equations of motion reproduces all three observables wherever it yields a coherent helix. The helix chirality is set by a long-range, current-generated Dzyaloshinskii-Moriya (DM)-type coupling rather than an intrinsic DM interaction, so materials without bulk or interfacial DM interactions could also host current-controlled helical order. Because the helical pitch, precession frequency, and canting ratio are all continuously tunable by the current amplitudes, these helices could serve as probes of spin currents.
\end{abstract}

\maketitle

\section{Introduction}
\label{sec:introduction}

Controlling magnetization with electric currents is a central theme of spintronics~\cite{Wolf2001, Manchon2019, Fert2024}. In 1996, Slonczewski~\cite{Slonczewski1996} and Berger~\cite{Berger1996} predicted spin-transfer torque (STT): a spin-polarized current flowing perpendicular to magnetic multilayers can exert enough torque to reorient the magnetization. Zhang \textit{et al.}~\cite{Zhang2002} subsequently showed that the transverse spin accumulation of conduction electrons produces both a spin torque and an effective field of comparable magnitude. 
These studies raised the prospect of magnetic memory and logic devices operated purely electrically.
Together with the earlier discoveries of the giant magnetoresistive effect~\cite{Baibich1988} and the spin-valve effect~\cite{Dieny1991}, which established that magnetic configurations control electrical transport, this prospect prompted extensive studies of the reverse action of currents on magnetic textures.
A prominent example is current-driven domain-wall motion.
On the theoretical side, Bazaliy \textit{et al.}~\cite{Bazaliy1998} incorporated a spin-polarized current into the Landau-Lifshitz equation for a continuously varying magnetization; Tatara and Kohno~\cite{Tatara2004} distinguished the spin-transfer and momentum-transfer contributions to wall motion; Zhang and Li~\cite{Zhang2004} identified an additional torque exerted by nonequilibrium conduction electrons that is essential for interpreting wall-motion experiments; and Barnes and Maekawa~\cite{Barnes2005} showed that, without pinning, angular momentum is transferred completely from the current to the wall.
Experimentally, current-induced wall displacement was inferred from the nonlinear transport of Fe$_3$O$_4$ nanocontacts~\cite{Versluijs2001}, achieved in a ferromagnetic semiconductor at current densities below $10^{5}$~A\,cm$^{-2}$~\cite{Yamanouchi2004}, and imaged directly by magnetic force microscopy in submicron Ni$_{81}$Fe$_{19}$ wires~\cite{Yamaguchi2004}.

In parallel, two main methods for generating pure spin currents without accompanying charge flow have been proposed and confirmed experimentally.
One is the spin battery method, which relies on spin pumping: a precessing magnetization pumps a spin current into an adjacent normal metal~\cite{Tserkovnyak2002}, so that a ferromagnet driven at ferromagnetic resonance acts as a source of pure spin current~\cite{Brataas2002} (for a review, see Ref.~\cite{Tserkovnyak2005}). Experimentally, spin pumping accounts for the enhanced Gilbert damping observed earlier in permalloy films sandwiched between Pt or Pd layers~\cite{Mizukami2001JJAP, Mizukami2001JMMM}; a dc voltage accompanying ferromagnetic resonance was later reported as possible evidence of spin pumping~\cite{Azevedo2005}, and the pumped spin current was detected electrically in a Pt film~\cite{Saitoh2006}. 
Another is the spin Hall effect (SHE), predicted by Dyakonov and Perel~\cite{Dyakonov1971} in 1971 and revived by Hirsch~\cite{Hirsch1999} in 1999.
The SHE arises from spin-orbit coupling in nonmagnetic materials and converts a charge current into a transverse spin current.
In the early 2000s, theories of the intrinsic SHE based on Berry's geometric phase~\cite{Berry1984} in momentum space predicted a dissipationless spin current in hole-doped semiconductors~\cite{Murakami2003} and a universal spin Hall conductivity in two-dimensional Rashba systems~\cite{Sinova2004}.
The SHE was subsequently observed in semiconductors~\cite{Kato2004, Wunderlich2005} and metals~\cite{Valenzuela2006}.
It is now widely used to inject pure spin currents into adjacent magnetic layers as an alternative to spin-polarized charge currents.

In recent years, spin-orbit torque (SOT)~\cite{Manchon2019} switching has advanced rapidly: it now reaches picosecond time scales~\cite{Jhuria2020, Polley2023} and, through symmetry engineering, achieves deterministic switching without an external field~\cite{Yu2014, Fukami2016}.
Van der Waals (vdW) materials have also entered SOT devices, both as switched ferromagnets and as spin sources: a Pt spin Hall layer switches the perpendicularly magnetized vdW ferromagnet Fe$_3$GaTe$_2$ at room temperature~\cite{Kajale2024}; out-of-plane antidamping torques from low-symmetry WTe$_2$ switch a vdW ferromagnet without an external field~\cite{Kao2022}; the topological insulator Bi$_2$Te$_3$ enables room-temperature switching in an all-vdW Fe$_3$GeTe$_2$ heterostructure~\cite{Wang2023}; and wafer-scale PtTe$_2$/WTe$_2$ bilayers combine high spin Hall conductivity with out-of-plane spins for field-free, low-power switching~\cite{WangNat2024}. In these devices, a spin current injected from an adjacent layer exerts the torque on the local moments; the present work analyzes a related situation, in which the spin current flows along a chain of localized moments. These developments have extended current-driven magnetization dynamics from conventional metals to quantum materials. Electric currents have also been used to control the spin helicity of the itinerant helimagnet MnP~\cite{Jiang2020} and, through spin-orbit torque, to rotate the helical antiferromagnetic order of the vdW magnet Ni$_{1/3}$NbS$_{2}$ collectively~\cite{Zhang2025}; such control may lead to magnetic memories based on chiral degrees of freedom.

Despite extensive theoretical and experimental progress on STT~\cite{Tsoi1998, Katine2000, Ralph2008, Brataas2006} and SOT, previous studies have focused primarily on dynamics driven either by spin-polarized charge currents or by pure spin currents, whereas realistic spintronic devices often involve an interplay between charge and spin currents. For continuous ferromagnets, spin-transfer theories have shown that a sufficiently strong spin-polarized current destabilizes uniform order against spin waves~\cite{Bazaliy1998, Fernandez-Rossier2004} and nucleates domains above a critical spin current~\cite{Shibata2005}. Spin currents are also closely tied to the Dzyaloshinskii-Moriya (DM) interaction~\cite{Dzyaloshinsky1958, Moriya1960} in equilibrium~\cite{Katsura2005, Kikuchi2016}. Out of equilibrium, a current can in turn modify the DM interaction itself or even generate a DM-type coupling: the interfacial DM interaction has been modulated experimentally by an in-plane current~\cite{Kato2019}, a current-induced DM interaction has been formulated for spin-orbit-coupled metallic magnets~\cite{Freimuth2020}, and DM-type couplings without spin-orbit coupling have been derived for localized spins in tunnel junctions and open quantum systems carrying spin-polarized currents~\cite{Fransson2014, Liu2018}. By contrast, much less attention has been paid to the ordered states that emerge when localized spins are driven by both charge and pure spin currents, and to how their chirality is selected, particularly in extended spin systems. This combined regime is also of practical interest: a pure spin current exerts torques without net charge flow through the magnetic region, whereas a small superimposed charge-current component---forming what we call a charged spin current---provides a second, independent control parameter that sets the magnetic texture into coherent motion.

Precisely patterned arrays of single-domain nanomagnets on Si substrates are now routinely fabricated, as exemplified by artificial spin ices~\cite{Wang2006, Skjaervo2020}, and self-organized semiconductor surfaces offer large-area templates for such arrays~\cite{Teichert2003}.
Such platforms may offer a route toward periodic chains of localized moments, although a quantitative description of extended nanomagnets would require the point-like exchange coupling assumed here to be replaced by one distributed over the magnet volume, and the shape anisotropy and dipolar coupling of the magnets, absent from the present model, to be weak compared with the current-induced couplings.

In a previous letter~\cite{Onoda2006}, we studied the dynamics of localized spins coupled to conduction electrons carrying both charge and spin currents and obtained, for a two-spin system, a phase diagram with distinct dynamic regimes---steady states, periodic oscillations, and chaotic motion---depending on the current ratio. Owing to the letter format, the microscopic derivation of the current-induced torques could not be presented in detail, and the treatment of extended spin systems was only sketched: a nearest-neighbor argument suggested a static helical order in a one-dimensional array under a pure spin current and its possible use as a spin-current detector. Here we derive the effective equations of motion systematically from the $s$-$d$ exchange Hamiltonian for arbitrary combinations of charge and spin currents and apply them to extended spin chains.
In these chains, the long-range Ruderman-Kittel-Kasuya-Yosida (RKKY)-type couplings~\cite{Ruderman1954, Kasuya1956, Yosida1957} reverse the chirality expected from the nearest-neighbor argument, and both the static and the precessing helices can be treated quantitatively.

For a periodically arranged one-dimensional spin chain, we analytically predict two distinct phenomena: (i) static helical magnetic order, in which a pure spin current alone stabilizes a chiral spin texture without any net charge flow, and (ii) dynamic helical magnetic order, in which a charged spin current makes the helix precess coherently at a current-controlled angular frequency. Given a wavenumber, the equations of motion determine the precession angular frequency $\omega$ and the canting ratio $\Delta$ toward the hard axis; the helical wavenumber $q_{\mathrm{helix}}$ itself is selected variationally, for the precessing helix by a heuristically constructed effective Lagrangian. Numerical simulations of the same spin chain agree quantitatively with the predicted $q_{\mathrm{helix}}$, $\omega$, and $\Delta$ wherever they yield a coherent helix. Because the current amplitudes and polarization tune the helical pitch and precession and set the chirality, these results lend quantitative support to the spin-current sensing suggested in Ref.~\cite{Onoda2006} and point to electrically switchable chirality. They also complement theoretical work on the current-driven dynamics of helical spin-density waves~\cite{Wessely2006} and spin spirals~\cite{Hals2013} and on pure spin currents injected into helimagnets~\cite{Ustinov2024}: there the helix is intrinsic to the material, whereas here the spin current itself stabilizes the helix.

The remainder of this paper is organized as follows. Section~\ref{sec:Model} introduces the model and outlines the derivation in the closed-time-path Green's function (CTPGF) formalism, Sec.~\ref{sec:Approximations} describes the approximations, and Sec.~\ref{sec:EOM} presents the resulting effective equations of motion, including Gilbert-type and nonlocal damping, spin torques, and current-induced DM-type torques; Sec.~\ref{sec:F-transformed_RKKY} summarizes the Fourier-transformed RKKY range functions entering them. Section~\ref{sec:helix-theory} determines $q_{\mathrm{helix}}$, $\omega$, and $\Delta$ analytically for a periodic spin chain, Sec.~\ref{sec:simulation} tests these predictions numerically, Sec.~\ref{sec:discussion} discusses the physical implications and possible extensions, and Sec.~\ref{sec:conclusion} concludes. The Appendixes derive the equations of motion and the RKKY range functions in detail.

\section{Theoretical model}
\label{sec:Model}
This section introduces a model for classical localized spins coupled to quantum conduction electrons with charge and spin currents and outlines the derivation of the equations of motion for the localized spins. 
Sections~\ref{sec:Approximations} and \ref{sec:EOM} complete this derivation.
These equations of motion enable us to analyze the effects of conduction electrons' charge and spin currents on the dynamics of localized spins and to discuss the possibility of helimagnetism in spin chains induced by several types of spin currents, including its extension to the dynamic regime induced by charged spin currents.
We use units with $\hbar = c = 1$ throughout the paper 
and denote a localized spin by $\vb*{S}_{\vb*{r}}(t)$ and
the annihilation (creation) spinor operator of a conduction electron in the Heisenberg picture by $\hat{\psi}^{(\dagger)}(t, \vb*{r})$,
where $t$ and $\vb*{r}$ stand for time and position, respectively.
With the standard Pauli matrices $\vb*{\sigma} = (\sigma^{1}, \sigma^{2}, \sigma^{3})$, the spin density operator of the conduction electrons is $\hat{\vb*{\sigma}}(t, \vb*{r})/2$, where $\hat{\vb*{\sigma}}(t, \vb*{r})=\hat{\psi}^{\dagger}(t, \vb*{r})\vb*{\sigma}\hat{\psi}(t, \vb*{r})$.
We also introduce the symbol $\sigma^{0}$ for the 2-by-2 identity matrix in spin space.
The symbols $\mu_{\mathrm{B}}$, $g$, and $g_{\mathrm{e}}$ stand for the Bohr magneton and the $g$ factors of $\vb*{S}_{\vb*{r}}(t)$ and $\hat{\vb*{\sigma}}(t, \vb*{r})/2$, respectively; the signs of $g$ and $g_{\mathrm{e}}$ are chosen so that $g\mu_{\mathrm{B}}\vb*{S}_{\vb*{r}}(t)$ and $g_{\mathrm{e}}\mu_{\mathrm{B}}\hat{\vb*{\sigma}}(t, \vb*{r})/2$ are the magnetic moment of a localized spin and the magnetic moment density of the conduction electrons.
We consider the case in which $\vb*{S}_{\vb*{r}}(t)$ and $\hat{\vb*{\sigma}}(t, \vb*{r})/2$ couple to each other through the $s$-$d$ (Kondo-type) exchange coupling $-J_{\mathrm{ex}}\vb*{S}_{\vb*{r}}(t)\vdot \qty{v_{\mathrm{uc}}\hat{\vb*{\sigma}}(t, \vb*{r})/2}$, where $J_{\mathrm{ex}}$ and $v_{\mathrm{uc}}$ represent the coupling energy and the volume of a unit cell of the background lattice structure.
For simplicity, the magnitude of every localized spin is assumed to be saturated as $\abs{\vb*{S}_{\vb*{r}}(t)}=S$, where $S$ is a half-integer- or integer-valued constant.

A dynamic spin texture $\qty{\vb*{S}_{\vb*{r}}(t)}$ (a set of localized spin variables) influences the quantum state of the conduction-electron system, which in turn acts back on the spin texture.
The localized spins thus interact with each other through the conduction electrons, as in the RKKY mechanism~\cite{Ruderman1954, Kasuya1956, Yosida1957}.  
The CTPGF formalism~\cite{Schwinger1961, Keldysh1965, Chou1985}, introduced below, is well suited to analyzing such systems.
For example, integrating out the electrons on the Keldysh contour yields a time-nonlocal effective action for a single spin embedded in a Josephson junction~\cite{Zhu2004}, and the RKKY interaction between two spins across a biased tunneling junction has been analyzed in the Keldysh formalism~\cite{Schwabe1996}. The exchange interaction between localized spins on a surface driven out of equilibrium by a tunneling current, including a DM-type term, has also been derived from such an effective action~\cite{Fransson2010}; the present work follows the same route for many localized spins coupled to current-carrying conduction electrons.

Before presenting the model in this formalism, we introduce some notation.
For an arbitrary function $f(t)$ defined on the closed time path, we represent the closed-time-path integral as follows:
\begin{align}
\oint\dd{t}  f(t)
&= 
 \int_{-\infty}^{\infty}\dd{t_{+}}f(t_{+})+\int_{\infty}^{-\infty}\dd{t_{-}}f(t_{-})
\nonumber\\
& =  \int_{-\infty}^{\infty}\dd{t_{+}}f(t_{+})-\int_{-\infty}^{\infty}\dd{t_{-}}f(t_{-})
.
\end{align}
Note that, in the CTPGF formalism, $f(t_{+})$ on the forward path and $f(t_{-})$ on the backward path can differ from each other even when $t_{+}$ and $t_{-}$ take the same value.
We also apply the above closed-time-path integral concept to operator-valued functions.

Here, we introduce the partition function $\mathcal{Z}[\vb*{S}]$ of our model as follows:
\begin{widetext}
\begin{align}
&\mathcal{Z}[\vb*{S}] 
= e^{i\qty(\mathcal{A}_{\mathrm{WZWN}}[\vb*{S}]+\mathcal{A}_{\vb*{B}}[\vb*{S}])} \mathcal{Z}_{\mathrm{Q}}[\vb*{S}] 
,\quad
\mathcal{Z}_{\mathrm{Q}}[\vb*{S}] 
= \Tr[\hat{\rho}_{0}\mathcal{T}_{\mathrm{ctp}}e^{-i\oint\dd{t}\hat{H}_{\mathrm{S}}(t)}]
= \Tr[\hat{\rho}_{0}\mathcal{T}_{\mathrm{ctp}}e^{-i\oint\dd{t}\,\hat{V}_{\mathrm{I}}(t)}]
\label{eq:partition-function}
,\\
&\mathcal{A}_{\mathrm{WZWN}}[\vb*{S}] 
=
-\frac{1}{S^{2}}\int_{0}^{1} \dd{\tau}\oint \dd{t}\sum_{i}
 \vb*{S}_{\vb*{r}_{i}}(\tau, t)\vdot \qty{{\pdv{ \vb*{S}_{\vb*{r}_{i}}}{\tau}} (\tau, t) \cross {\pdv{\vb*{S}_{\vb*{r}_{i}}}{t}}(\tau, t)}
\label{eq:WZWN}
, \quad
\mathcal{A}_{\vb*{B}}[\vb*{S}] = 
\oint\dd{t} \sum_{i}g\mu_{\mathrm{B}}\vb*{B}(t, \vb*{r}_{i})
\vdot \vb*{S}_{\vb*{r}_{i}}(t)
,\\
&\hat{H}_{\mathrm{S}}(t)  
= 
\int \dd{\vb*{r}}
\qty[\frac{1}{2m^{*}}\qty{\vb*{\mathcal{D}}\hat{\psi}_{\mathrm{S}}(\vb*{r})}^{\dagger}\vdot\vb*{\mathcal{D}}\hat{\psi}_{\mathrm{S}}(\vb*{r})
+\hat{\psi}_{\mathrm{S}}^{\dagger}(\vb*{r})
\qty{
e\phi(t, \vb*{r})\sigma^{0}+e_{\mathrm{s}}\vb*{\phi}_{\mathrm{s}}(t, \vb*{r})\vdot\vb*{\sigma}
-\frac{g_{\mathrm{e}}\mu_{\mathrm{B}}}{2}\vb*{B}(t, \vb*{r})\vdot\vb*{\sigma}
}
\hat{\psi}_{\mathrm{S}}(\vb*{r})]
 \nonumber\\
&\qquad\qquad
-\frac{J_{\mathrm{ex}}v_{\mathrm{uc}}}{2}\sum_{i}\vb*{S}_{\vb*{r}_{i}}(t) \vdot \hat{\psi}_{\mathrm{S}}^{\dagger}(\vb*{r}_{i})\vb*{\sigma}\hat{\psi}_{\mathrm{S}}(\vb*{r}_{i})
,\qquad
\vb*{\mathcal{D}} = \qty{\grad-ie\vb*{A}(t, \vb*{r})}\sigma^{0}-\sum_{\mu=1}^{3}ie_{\mathrm{s}}\vb*{A}^{\mu}_{\mathrm{s}}(t, \vb*{r})\sigma^{\mu}
, \\
&
\hat{H}_{0} =
\int \dd{\vb*{r}}
\frac{1}{2m^{*}}\grad\hat{\psi}_{\mathrm{S}}^{\dagger}(\vb*{r})\vdot\grad\hat{\psi}_{\mathrm{S}}(\vb*{r})
, \quad 
\hat{V}_{\mathrm{S}}(t)  
=  \hat{H}_{\mathrm{S}}(t) -\hat{H}_{0}
, 
\end{align}
\end{widetext}
where
$\mathcal{A}_{\mathrm{WZWN}}[\vb*{S}]$ is the Wess-Zumino-Witten-Novikov term~\cite{Wess1971, Novikov1982, Witten1983} describing Berry's geometric phase~\cite{Berry1984} of the localized spins in the spin coherent-state path integral~\cite{Klauder1979, Kuratsuji1980},
$\mathcal{A}_{\vb*{B}}[\vb*{S}]$  represents the coupling of the localized spins
to the external magnetic flux density $\vb*{B}(t,\vb*{r})$,
$\hat{\rho}_{0}$ is the initial density matrix,
$\mathcal{T}_{\mathrm{ctp}}$ orders operators from right to left along the closed time path,
$\hat{H}_{\mathrm{S}}(t)$ is the Hamiltonian of the conduction-electron system,
and $\hat{H}_{0}$ and $\hat{V}_{\mathrm{S}}(t)$ are its free and interaction parts, respectively.
The subscript ``$\mathrm{S}$'' of $\hat{\psi}_{\mathrm{S}}^{(\dagger)}(\vb*{r})$, $\hat{H}_{\mathrm{S}}(t)$, and $\hat{V}_{\mathrm{S}}(t)$ stands for the Schr\"{o}dinger picture, and the subscript ``$\mathrm{I}$'' of $\hat{V}_{\mathrm{I}}(t)$ stands for the interaction picture.
Appendix~\ref{sec:Three-pictures_in_quantum-theory} summarizes the relations among the Schr\"{o}dinger, Heisenberg, and interaction pictures on the closed time path.
We assume that the initial density matrix $\hat{\rho}_{0}$ describes an equilibrium state of $\hat{H}_{0}$.

To represent $\mathcal{A}_{\mathrm{WZWN}}[\vb*{S}]$ explicitly, we extend the set of localized spin variables $\qty{\vb*{S}_{\vb*{r}}(t)}$ to $\qty{\vb*{S}_{\vb*{r}}(\tau, t)}$ by introducing an auxiliary parameter $\tau\in[0,1]$ such that 
\begin{align}
&\vb*{S}_{\vb*{r}}(0, t) = \vb*{S}_{0}, \quad \vb*{S}_{\vb*{r}}(1, t) = \vb*{S}_{\vb*{r}}(t),
\label {eq:extended_variables}
\end{align}
where $\vb*{S}_{0}$ is a reference constant vector.
This extension is not unique, but the choice does not affect the equations of motion, which involve only the physical values $\vb*{S}_{\vb*{r}}(1, t)=\vb*{S}_{\vb*{r}}(t)$ (Appendix~\ref{sec:Variational-procedure_for_the-action}).

As a matter of practical convenience, we model the conduction electrons as a three-dimensional electron gas with quadratic dispersion and an isotropic effective mass $m^{*}$, in which the localized spins are embedded. Its Fermi wavenumber $k_{\mathrm{F}}$ determines the Fermi energy $\epsilon_{\mathrm{F}}=k_{\mathrm{F}}^{2}/(2m^{*})$ and the number density $n_{\mathrm{e}}=k_{\mathrm{F}}^{3}/(3\pi^{2})$. We represent the effect of driving forces through sets of potentials $(\phi, \vb*{A})$ and $(\phi^{\mu}_{\mathrm{s}}, \vb*{A}^{\mu}_{\mathrm{s}})$ for the charge and spin sectors, respectively.
The spin sector $(\phi^{\mu}_{\mathrm{s}}, \vb*{A}^{\mu}_{\mathrm{s}})$ is virtual and serves only as a technical device, and the parameter $e_{\mathrm{s}} = 1/2$ makes the charge and spin sectors formally symmetric.
Although we can interpret this sector as a non-Abelian gauge field, we do not consider such aspects (e.g., nonlinearity) and, for simplicity, define its driving force to linear order.
In any case, we do not treat the potentials directly and consider only the case in which the driving forces, $\vb*{E} = -\grad\phi-\partial_{t}\vb*{A}$ and $\vb*{E}^{\mu}_{\mathrm{s}} = -\grad\phi^{\mu}_{\mathrm{s}}-\partial_{t}\vb*{A}^{\mu}_{\mathrm{s}}$, are static and homogeneous.

The CTPGF formalism can treat unphysical situations such as $\vb*{S}_{\vb*{r}}(\tau, t_{+})$ on the forward path and  $\vb*{S}_{\vb*{r}}(\tau, t_{-})$ on the backward path as different from each other even when $t_{+}$ and $t_{-}$ take the same value.
This feature helps us derive the equations of motion for the localized spins under the physically appropriate condition in which $\vb*{S}_{\vb*{r}}(\tau, t_{+})$ and $\vb*{S}_{\vb*{r}}(\tau, t_{-})$ coincide with each other when $t_{+}$ and $t_{-}$ take the same value.
To organize the derivation, we introduce variables with the subscripts ``$\pm$'' and the superscripts ``$\mathrm{c}$'' and ``$\Delta$''.
Let $f(t)$ be a variable on the closed time path, i.e., a function of $t$ parameterizing the closed time path. Then, we define $f_{\pm}(t)$, $f^{\mathrm{c}}(t)$, and $f^{\Delta}(t)$ as follows:
\begin{align}
f_{\pm}(t) 
&= f(t_{\pm})\vert_{t_{\pm}=t}
\label{eq:superscripts_pm}
,\\
f^{\mathrm{c}}(t)
&= \frac{1}{2}\qty{f_{+}(t) + f_{-}(t)},
\label{eq:superscripts_c&Delta}
\\
f^{\Delta}(t) 
&= f_{+}(t) - f_{-}(t),\quad f^{\Delta}(\pm\infty) = 0,
\label{eq:superscript_Delta}
\end{align}
where the boundary conditions of $f^{\Delta}(t)$ at $\pm\infty$ come from the closedness of the time path $f_{+}(\pm\infty) = f_{-}(\pm\infty)$.

The physically appropriate condition for $\qty{\vb*{S}_{\vb*{r}_{i}}(\tau, t)}$ is then $\vb*{S}^{\Delta}_{\vb*{r}_{i}}(\tau, t)=0$ for all $(i, \tau, t)$, which we abbreviate as $\vb*{S}^{\Delta}=0$.
For $(\phi, \vb*{A})$ and $(\phi^{\mu}_{\mathrm{s}}, \vb*{A}^{\mu}_{\mathrm{s}})$, we impose the physically appropriate condition from the outset, because the driving forces $\vb*{E}$ and $\vb*{E}^{\mu}_{\mathrm{s}}$ are assumed to be static and homogeneous and their equations of motion are not needed.

Next, we define the expectation value of an arbitrary Heisenberg operator $\hat{\mathcal{O}}(t)$ under the condition $\vb*{S}^{\Delta}=0$ as
\begin{align}
\ev{\hat{\mathcal{O}}(t)}_{\vb*{S}^{\Delta}=0}
&=\Tr[\hat{\rho}_{0}\hat{\mathcal{O}}(t)]_{\vb*{S}^{\Delta}=0}
\nonumber\\
&=\Tr[\hat{\rho}_{0}\mathcal{T}_{\mathrm{ctp}}\qty{\hat{\mathcal{O}}_{\mathrm{I}}(t)e^{-i\oint\dd{s}\,\hat{V}_{\mathrm{I}}(s)}}]_{\vb*{S}^{\Delta}=0}
.
\end{align}

The effective action $\mathcal{A}_{\mathrm{eff}}[\vb*{S}]$ for the localized spins is defined by $\mathcal{A}_{\mathrm{eff}}[\vb*{S}]=-i\ln\mathcal{Z}[\vb*{S}]$ and
 is formally evaluated up to the linear order in $\qty{\vb*{S}^{\Delta}_{\vb*{r}}(\tau, t)}$ as follows:
 \begin{align}
\mathcal{A}_{\mathrm{eff}}[\vb*{S}]
&\cong
\mathcal{A}_{\mathrm{WZWN}}[\vb*{S}]
+g\mu_{\mathrm{B}}\int_{-\infty}^{\infty} \dd{t} \sum_{i}
\vb*{S}^{\Delta}_{\vb*{r}_{i}}(t)\vdot\vb*{B}(t, \vb*{r}_{i})
\nonumber\\
&\quad+\frac{J_{\mathrm{ex}}v_{\mathrm{uc}}}{2}
\int_{-\infty}^{\infty} \dd{t} \sum_{i}
\vb*{S}^{\Delta}_{\vb*{r}_{i}}(t)\vdot\ev{\hat{\vb*{\sigma}}(t, \vb*{r}_{i})}_{\vb*{S}^{\Delta}=0}
.
\label{eq:effective-action}
\end{align}

The variational principle on $\mathcal{A}_{\mathrm{eff}}[\vb*{S}]$ with respect to $\qty{\vb*{S}^{\Delta}_{\vb*{r}}(\tau, t)}$ at $\vb*{S}^{\Delta}=0$
gives the equations of motion for $\qty{\vb*{S}^{\mathrm{c}}_{\vb*{r}}(\tau, t)}_{\vb*{S}^{\Delta}=0}$ as shown in Appendix~\ref{sec:Variational-procedure_for_the-action}. More precisely, it gives nontrivial equations of motion only for the physically meaningful parts $\qty{\vb*{S}^{\mathrm{c}}_{\vb*{r}}(1, t)}_{\vb*{S}^{\Delta}=0} = \qty{\vb*{S}^{\mathrm{c}}_{\vb*{r}}(t)}_{\vb*{S}^{\Delta}=0}$.
Hereafter, we drop the superscript ``c'' of $\qty{\vb*{S}^{\mathrm{c}}_{\vb*{r}}(t)}$, with the condition $\vb*{S}^{\Delta}=0$ understood, because $\vb*{S}^{\mathrm{c}}_{\vb*{r}}(t)=\vb*{S}_{\vb*{r}}(t_{+})\vert_{t_{+}=t}=\vb*{S}_{\vb*{r}}(t_{-})\vert_{t_{-}=t}$ under this condition.
We thus obtain the equations of motion for $\qty{\vb*{S}_{\vb*{r}}(t)}$ in the formal form 
\begin{align}
{\pdv{\vb*{S}_{\vb*{r}}}{t}}(t)
&=
\vb*{S}_{\vb*{r}}(t)\times
\biggl\{g\mu_{\mathrm{B}}\vb*{B}(t, \vb*{r})
\nonumber\\
&\qquad
+\frac{J_{\mathrm{ex}}v_{\mathrm{uc}}}{2}\ev*{\hat{\vb*{\sigma}}(t, \vb*{r})}_{\vb*{S}^{\Delta}=0}
\biggr\}.
\label{eq:EOM-formal}
\end{align}
The remaining task to reach the equations of motion in a specific form is the evaluation of $\ev*{\hat{\vb*{\sigma}}(t, \vb*{r})}_{\vb*{S}^{\Delta}=0}$
as a functional of $\qty{\vb*{S}_{\vb*{r}}(t)}$.

This paper focuses on the weak-coupling regime $\abs{J_{\mathrm{ex}}}\ll \epsilon_{\mathrm{F}}$ and retains terms up to the linear order of $\qty{\vb*{S}_{\vb*{r}}(t)}$ in the evaluation of $\ev*{\hat{\vb*{\sigma}}(t, \vb*{r})}_{\vb*{S}^{\Delta}=0}$. 
Because the localized spins are classical, the conduction electrons feel them only through the exchange field, which is proportional to $J_{\mathrm{ex}}S$.
The expansion in the localized spins is therefore controlled more precisely by the Born parameter $v_{\mathrm{uc}}n_{\mathrm{e}}\abs{J_{\mathrm{ex}}}S/\epsilon_{\mathrm{F}}$, which must also be small; in terms of the parameters introduced in Sec.~\ref{sec:EOM}, it is of the order of $\gamma_{\mathrm{ex}}S=\tilde{\gamma}\sqrt{S}$.
In addition, for simplicity, we consider the case without external magnetic flux density, $\vb*{B}(t, \vb*{r})=0$, from this point forward.
Within these approximations, we obtain
\begin{widetext}
\begin{align}
\ev*{\hat{\sigma}^{\mu}(t, \vb*{r})}_{\vb*{S}^{\Delta}=0}
&\cong \frac{J_{\mathrm{ex}}v_{\mathrm{uc}}}{2}i\int_{-\infty}^{\infty}\dd{t'}\theta(t-t')\sum_{i}
\ev{\qty[\hat{\sigma}^{\mu}(t, \vb*{r}), \vb*{S}_{\vb*{r}_{i}}(t')\vdot \hat{\vb*{\sigma}}(t', \vb*{r}_{i})]}_{J_{\mathrm{ex}}=0}
\nonumber\\
&= -\frac{J_{\mathrm{ex}}v_{\mathrm{uc}}}{2}\int_{-\infty}^{\infty} \dd{t'} \sum_{i}\sum_{\nu}
\Pi^{\mathrm{R}\mu\nu}(t, \vb*{r};t',\vb*{r}_{i})S^{\nu}_{\vb*{r}_{i}}(t')
\label{eq:linear-approximation}
,\\
\Pi^{\mathrm{R}\mu\nu}(t,\vb*{r}; t', \vb*{r}') 
&=
-i\theta(t-t')
\ev{\qty[\hat{\sigma}^{\mu}(t, \vb*{r}),\hat{\sigma}^{\nu}(t',\vb*{r}')]}_{J_{\mathrm{ex}}=0}
.
\label{eq:spin-spin_RGF}
\end{align}
\end{widetext}
Here, $\Pi^{\mathrm{R}\mu\nu}(t,\vb*{r}; t', \vb*{r}')$ is the spin-spin retarded Green function of the conduction-electron system in the situation without coupling to the localized spins.
Note, however, that it contains the influence of external driving forces for the charge and spin currents of conduction electrons.
In other words, Eq.~(\ref{eq:linear-approximation}) is linear in the localized spins but does not yet restrict its dependence on the driving forces.

This paper assumes that the external driving forces are sufficiently weak and considers their responses up to linear order.
To capture these responses, the next section will introduce a Drude-type phenomenological approximation rather than a regular diagrammatic expansion, owing to its technical simplicity.

\section{Approximate evaluation of the spin density\texorpdfstring{ $\ev*{\hat{\vb*{\sigma}}(t, \vb*{r})}_{\vb*{S}^{\Delta}=0}$}{}}
\label{sec:Approximations}
This section discusses prescriptions to evaluate the conduction-electron spin density $\ev*{\hat{\vb*{\sigma}}(t, \vb*{r})}_{\vb*{S}^{\Delta}=0}$ as a functional of $\qty{\vb*{S}_{\vb*{r}}(t)}$ to derive a specific form of the equations of motion for $\qty{\vb*{S}_{\vb*{r}}(t)}$ from their formal expression, Eq.~(\ref{eq:EOM-formal}), in the case without external magnetic flux density, $\vb*{B}(t, \vb*{r})=0$.
As mentioned in the previous section, we focus on the weak-coupling regime, $\abs{J_{\mathrm{ex}}}\ll \epsilon_{\mathrm{F}}$, and retain the terms up to linear order in $\qty{\vb*{S}_{\vb*{r}}(t)}$, as in Eq.~(\ref{eq:linear-approximation}).
The remaining task is thus to evaluate $\Pi^{\mathrm{R}\mu\nu}(t,\vb*{r}; t', \vb*{r}')$ defined by Eq.~(\ref{eq:spin-spin_RGF}).

Here we adopt three major and two minor prescriptions for the approximate evaluation of $\Pi^{\mathrm{R}\mu\nu}(t,\vb*{r}; t', \vb*{r}')$.
The first major prescription is to approximate the anticommutation relations such as $\qty{\hat{\psi}_{\xi}(t, \vb*{r}),\hat{\psi}^{\dagger}_{\xi'}(t', \vb*{r}')}_{J_{\mathrm{ex}}=0}$ by their c-number functions in equilibrium without any external driving forces or coupling to the localized spins.
The second major prescription is to take into account the nonequilibrium nature caused by the external driving forces through 
the nonequilibrium distribution function $\overleftrightarrow{g}(\vb*{k})$ in the lesser Green function,
\begin{align}
&-iG^{<}_{\xi'\xi}(t',\vb*{r}'; t, \vb*{r}) =
\ev{\hat{\psi}^{\dagger}_{\xi}(t,\vb*{r})\hat{\psi}_{\xi'}(t',\vb*{r}')}_{J_{\mathrm{ex}}=0}.
\end{align}
The third major prescription is to phenomenologically incorporate the effect of energy decoherence by damping as $e^{-\eta(t-t')}$  ($\eta>0$), which will also turn out to be necessary for the derivation of the effective equations of motion in a local form with respect to the time variable.
Two minor prescriptions are related to the second major prescription.
The first minor prescription is the low-temperature approximation for the nonequilibrium distribution function $\overleftrightarrow{g}(\vb*{k})$ and
the second minor prescription is a Drude-type approximation to relate the driving forces to the expectation values of charge and spin currents.

The specific procedure of the first major prescription is to set the anticommutation relation among annihilation (creation) operators to zero and to adopt the following form for the anticommutation relation of annihilation and creation operators:
\begin{align}
 &\qty{\hat{\psi}_{\xi}(t, \vb*{r}),\hat{\psi}^{\dagger}_{\xi'}(t', \vb*{r}')}_{J_{\mathrm{ex}}=0}
\nonumber\\
&\qquad
\cong \delta_{\xi\xi'}\int\frac{\dd{\vb*{k}}}{(2\pi)^{3}}e^{i\vb*{k}\vdot(\vb*{r}-\vb*{r}')-i\epsilon_{\vb*{k}}(t-t')}
,
\end{align}
where $\epsilon_{\vb*{k}}$ is the energy of a free electron with momentum $\vb*{k}$, i.e.,  $\epsilon_{\vb*{k}} = k^{2}/\qty(2m^{*})$ ($k=\abs{\vb*{k}}$) in the present model. This is equivalent to the following approximation:
\begin{align}
&\hat{\psi}_{\xi}(t,\vb*{r})
\cong \int\frac{\dd{\vb*{k}}}{(2\pi)^{3}}\hat{\psi}_{\xi}(\vb*{k}) e^{i\vb*{k}\vdot\vb*{r}-i\epsilon_{\vb*{k}}t}
, \\
&\qty{\hat{\psi}_{\xi}(\vb*{k}), \hat{\psi}^{\dagger}_{\xi'}(\vb*{k}')}_{J_{\mathrm{ex}}=0}
\cong \delta_{\xi\xi'}(2\pi)^{3}\delta(\vb*{k}-\vb*{k}')
.
\end{align}

The specific procedure of the second major prescription is to incorporate the influence of static and homogeneous driving forces on the lesser Green function in the following form:
\begin{align}
&-iG^{<}_{\xi'\xi}(t',\vb*{r}'; t, \vb*{r})
\cong
\int\!\frac{\dd{\vb*{k}}}{(2\pi)^{3}}
e^{-i\vb*{k}\vdot(\vb*{r}-\vb*{r}')+i\epsilon_{\vb*{k}}(t-t')}
g_{\xi'\xi}(\vb*{k})
,
\end{align}
where $g_{\xi'\xi}(\vb*{k})$ is the $\xi'\xi$-component of the tensor function $\overleftrightarrow{g}(\vb*{k})$ defined by
\begin{align}
&\overleftrightarrow{g}(\vb*{k})
=f_{\mathrm{F}}(\epsilon_{\vb*{k}})\sigma^{0}
\nonumber\\
&\qquad\quad-
\tau_{\mathrm{e}}\vb*{v}_{\vb*{k}}\vdot\qty(e\vb*{E}\sigma^{0}+\sum_{\mu=1}^{3}e_{\mathrm{s}}\vb*{E}_{\mathrm{s}}^{\mu}\sigma^{\mu})
{\dv{f_{\mathrm{F}}}{\epsilon}}(\epsilon_{\vb*{k}})
.
\label{eq:distribution_function}
\end{align}
In the above, $f_{\mathrm{F}}(\epsilon)$ is the Fermi distribution function, $\tau_{\mathrm{e}}$ is the relaxation time of conduction electrons, and 
$\vb*{v}_{\vb*{k}}$ is the group velocity of an electronic state with momentum $\vb*{k}$, i.e., $\vb*{v}_{\vb*{k}} = \vb*{k}/m^{*}$ in the present model.
Equation~(\ref{eq:distribution_function}) approximates the nonequilibrium deviation from $f_{\mathrm{F}}(\epsilon_{\vb*{k}})$ to linear order in the driving forces.
The antisymmetric, DM-type structure of the spin-current response derived below follows from $\Tr[\sigma^{\mu}\sigma^{\nu}\sigma^{\lambda}]=2i\epsilon_{\mu\nu\lambda}$ for any nonequilibrium distribution with a spin-polarized component odd in $\vb*{k}$, whereas its range function depends on the displaced-Fermi-surface form of Eq.~(\ref{eq:distribution_function}); spin currents generated by other means, such as spin Hall injection or spin pumping, may therefore modify the magnitude and range of the induced coupling. Such spin currents may also be accompanied by a spin accumulation, which Eq.~(\ref{eq:distribution_function}) does not contain because its spin-polarized part is odd in $\vb*{k}$; a spin accumulation would exert an additional torque on each localized spin already at first order in $J_{\mathrm{ex}}$.

For later convenience, we rewrite $\Pi^{\mathrm{R}\mu\nu}(t,\vb*{r}; t', \vb*{r}')$ as follows:
\begin{align}
&\Pi^{\mathrm{R}\mu\nu}(t,\vb*{r}; t', \vb*{r}')
= -i\theta(t-t') \sum_{\xi_{1},\xi_{2},\xi_{3},\xi_{4}}\sigma^{\mu}_{\xi_{1}\xi_{2}}\sigma^{\nu}_{\xi_{3}\xi_{4}}
\nonumber\\
&\times
\ev{\qty[\hat{\psi}^{\dagger}_{\xi_{1}}(t, \vb*{r})\hat{\psi}_{\xi_{2}}(t, \vb*{r}),\hat{\psi}^{\dagger}_{\xi_{3}}(t',\vb*{r}')\hat{\psi}_{\xi_{4}}(t',\vb*{r}')]}_{J_{\mathrm{ex}}=0}
.
\end{align}
The first and second major prescriptions enable us to evaluate the expectation value of the following commutator in $\Pi^{\mathrm{R}\mu\nu}(t,\vb*{r}; t', \vb*{r}')$:
\begin{align}
& \qty[\hat{\psi}^{\dagger}_{\xi_{1}}(t, \vb*{r})\hat{\psi}_{\xi_{2}}(t, \vb*{r}),\hat{\psi}^{\dagger}_{\xi_{3}}(t',\vb*{r}')\hat{\psi}_{\xi_{4}}(t',\vb*{r}')]
\nonumber\\
&= \hat{\psi}^{\dagger}_{\xi_{1}}(t, \vb*{r}){\qty{\hat{\psi}_{\xi_{2}}(t, \vb*{r}),\hat{\psi}^{\dagger}_{\xi_{3}}(t',\vb*{r}')}}\hat{\psi}_{\xi_{4}}(t',\vb*{r}')
\nonumber\\
&\quad
-\hat{\psi}^{\dagger}_{\xi_{1}}(t, \vb*{r})\hat{\psi}^{\dagger}_{\xi_{3}}(t',\vb*{r}'){\qty{\hat{\psi}_{\xi_{2}}(t, \vb*{r}),\hat{\psi}_{\xi_{4}}(t',\vb*{r}')}}
\nonumber\\
&\quad
-\hat{\psi}^{\dagger}_{\xi_{3}}(t',\vb*{r}'){\qty{\hat{\psi}_{\xi_{4}}(t',\vb*{r}'),\hat{\psi}^{\dagger}_{\xi_{1}}(t, \vb*{r})}}\hat{\psi}_{\xi_{2}}(t, \vb*{r})
\nonumber\\
&\quad
+{\qty{\hat{\psi}^{\dagger}_{\xi_{1}}(t, \vb*{r}),\hat{\psi}^{\dagger}_{\xi_{3}}(t',\vb*{r}')}}\hat{\psi}_{\xi_{4}}(t',\vb*{r}')\hat{\psi}_{\xi_{2}}(t, \vb*{r}).
\end{align}
Incorporating the third major prescription about the damping $e^{-\eta (t-t')}$ ($\eta>0$), we reach the following result:
\begin{align}
&\Pi^{\mathrm{R}\mu\nu}(t, \vb*{r}; t', \vb*{r}')
\cong \Pi^{\mathrm{R}\mu\nu}_{\mathrm{eff}}(t- t', \vb*{r}-\vb*{r}'),
\label{eq:approximation_of_spin-spin_RGF}
\\
&\Pi^{\mathrm{R}\mu\nu}_{\mathrm{eff}}(t, \vb*{r})
=
-i \theta(t) \int\frac{\dd{\vb*{k}}}{(2\pi)^{3}}\int\frac{\dd{\vb*{k}'}}{(2\pi)^{3}}
\nonumber\\
&\hspace{1.5cm}\times
\Biggl\{
e^{-i(\vb*{k}-\vb*{k}')\vdot\vb*{r}+i(\epsilon_{\vb*{k}}-\epsilon_{\vb*{k}'})t-\eta t}\Tr\qty[\sigma^{\mu}\sigma^{\nu}\overleftrightarrow{g}(\vb*{k})]
\Biggr.
\nonumber\\
&\hspace{2cm}
\Biggl.
-e^{i(\vb*{k}-\vb*{k}')\vdot\vb*{r}-i(\epsilon_{\vb*{k}}-\epsilon_{\vb*{k}'})t-\eta t}\Tr\qty[\sigma^{\nu}\sigma^{\mu}\overleftrightarrow{g}(\vb*{k})]
\Biggr\}
.
\label{eq:spin-spin_RGF_eff}
\end{align}

Taking advantage of the homogeneity in time of $\Pi^{\mathrm{R}\mu\nu}_{\mathrm{eff}}(t- t', \vb*{r}-\vb*{r}')$, we consider the time-derivative expansion of $\qty{S^{\mu}_{\vb*{r}}(t)}$ in $\ev*{\hat{\sigma}^{\mu}(t, \vb*{r})}_{\vb*{S}^{\Delta}=0}$ as follows:
\begin{align}
&\ev*{\hat{\sigma}^{\mu}(t, \vb*{r})}_{\vb*{S}^{\Delta}=0}
\nonumber\\
&\cong -\frac{J_{\mathrm{ex}}v_{\mathrm{uc}}}{2}\int_{-\infty}^{\infty} \dd{t'} \sum_{i}\sum_{\nu}
\Pi^{\mathrm{R}\mu\nu}_{\mathrm{eff}}\qty(t-t',\vb*{r}-\vb*{r}_{i})S^{\nu}_{\vb*{r}_{i}}(t')
\nonumber\\
&= -\frac{J_{\mathrm{ex}}v_{\mathrm{uc}}}{2}\int_{-\infty}^{\infty} \dd{t'} \sum_{i}\sum_{\nu}
\nonumber\\
&\qquad\times
\Pi^{\mathrm{R}\mu\nu}_{\mathrm{eff}}\qty(t-t',\vb*{r}-\vb*{r}_{i})\sum_{n=0}^{\infty}\frac{(t'-t)^{n}}{n!}{\pdv[n]{S^{\nu}_{\vb*{r}_{i}}}{t}}(t)
\nonumber\\
&= -\frac{J_{\mathrm{ex}}v_{\mathrm{uc}}}{2}\sum_{i}\sum_{\nu}\sum_{n=0}^{\infty}
\Pi^{\mathrm{R}\mu\nu(n)}_{\mathrm{eff}}(\vb*{r}-\vb*{r}_{i}){\pdv[n]{S^{\nu}_{\vb*{r}_{i}}}{t}}(t)
,
\end{align}
where
\begin{align}
&\Pi^{\mathrm{R}\mu\nu(n)}_{\mathrm{eff}}(\vb*{r})
 = \int_{-\infty}^{\infty} \dd{t}\frac{(-t)^n}{n!}\Pi^{\mathrm{R}\mu\nu}_{\mathrm{eff}}\qty(t,\vb*{r})
.
\label{eq:Pi_eff}
\end{align}

It remains to evaluate $\Pi^{\mathrm{R}\mu\nu(n)}_{\mathrm{eff}}(\vb*{r})$. 
Later in this section, we present a specific form of $\ev*{\hat{\sigma}^{\mu}(t, \vb*{r})}_{\vb*{S}^{\Delta}=0}$ with a discussion about the first and second minor prescriptions to approximately evaluate $\Pi^{\mathrm{R}\mu\nu(n)}_{\mathrm{eff}}(\vb*{r})$.
Appendix~\ref{sec:Spin-spin_retarded-Green-function} gives the details of the evaluation.

We now return to the damping.
A finite $\eta>0$ will turn out to be necessary for the convergence of $\Pi^{\mathrm{R}\mu\nu(n)}_{\mathrm{eff}}(\vb*{r})$ in the long-distance limit $\abs{\vb*{r}}\to\infty$ and thus for the validity of the above expansion.
In the weak-coupling regime, $\abs{J_{\mathrm{ex}}}\ll\epsilon_{\mathrm{F}}$, it is reasonable to assume that the localized spins move much more slowly than the conduction electrons.
When $\omega_{\mathrm{S}}$ is a typical frequency of the localized-spin motion, the $n$th-order derivative term will turn out to be of the order of $(\omega_{\mathrm{S}}/\epsilon_{\mathrm{F}})^{n}$.
Because the spin density of the conduction electrons cannot exceed their local number density, which equals $n_{\mathrm{e}}$ to leading order in the weak-coupling regime, Eq.~(\ref{eq:EOM-formal}) bounds the angular velocity of any localized-spin motion, steady or not, by about $v_{\mathrm{uc}}n_{\mathrm{e}}\abs{J_{\mathrm{ex}}}/2$, so that $\omega_{\mathrm{S}}/\epsilon_{\mathrm{F}}\ll1$ in the weak-coupling regime; a magnetic field, set to zero here, would add its Larmor frequency $\abs{g\mu_{\mathrm{B}}\vb*{B}}$.
Consequently, owing to a finite damping factor $\eta>0$, higher-derivative terms can be less important than lower-derivative terms.
This paper focuses on such a situation, which ensures the validity of the effective equations of motion in a form local in time.

As mentioned above, we interpret the damping as a phenomenological representation of energy decoherence, which inevitably exists in real systems.
If $\tau_{\varepsilon}$ is the typical decoherence time, then $\eta \sim 1/\tau_{\varepsilon}$.
We treat $\eta$ as an independent parameter: the relaxation time $\tau_{\mathrm{e}}$ in Eq.~(\ref{eq:distribution_function}) enters the results only through the currents [Eq.~(\ref{eq:Drude-type_approximation})], so that $\eta$ and the currents can be chosen independently.
Here we will consider only systems with relatively good coherence, $\epsilon_{\mathrm{F}}\tau_{\varepsilon} \gg 1$, and set $\eta$ finite but sufficiently small, $0<\eta/\epsilon_{\mathrm{F}}\ll 1$.
This setting justifies neglecting the second- and higher-order derivative terms and setting $\eta \to 0$ in the remaining terms up to the first order.
More precisely, because the $n$th-order kernels contain the factor $(2k_{\mathrm{F}}r)^{m}$ ($m\le n$), the relevant expansion parameter for a pair of spins separated by $r$ is $(\omega_{\mathrm{S}}/\epsilon_{\mathrm{F}})k_{\mathrm{F}}r$ rather than $\omega_{\mathrm{S}}/\epsilon_{\mathrm{F}}$. Since the decoherence cuts off the kernels beyond the decay length $\ell\simeq v_{\mathrm{F}}/\eta$, i.e., beyond $k_{\mathrm{F}}r\sim\epsilon_{\mathrm{F}}/\eta$, where $v_{\mathrm{F}}=k_{\mathrm{F}}/m^{*}$ is the Fermi velocity, the truncation at $n=1$ requires $\omega_{\mathrm{S}}\ll\eta$, i.e., the localized spins should hardly move during the decoherence time, while setting $\eta\to0$ in the remaining terms requires $\ell$ to exceed the range relevant to the spin texture. For the steadily precessing representative helix of Sec.~\ref{sec:simulation}, $\omega_{\mathrm{S}}$ is the precession frequency $\omega\approx4\times10^{-3}\epsilon_{\mathrm{F}}$, and the helical pitch is about $26a_{0}$, where $a_{0}$ is the spacing of the localized spins and $k_{\mathrm{F}}a_{0}=1$; both conditions are thus met, roughly, for $10^{-2}\lesssim\eta/\epsilon_{\mathrm{F}}\lesssim3\times10^{-2}$.

The remaining parts of this paper focus on the low-temperature regime $k_{\mathrm{B}}T \ll \epsilon_{\mathrm{F}}$, where $k_{\mathrm{B}}$ is the Boltzmann constant and $T$ is the temperature, and adopt the approximation $f_{\mathrm{F}}(\epsilon) \cong \theta(\epsilon_{\mathrm{F}}-\epsilon)$ in addition to the conditions $\abs{J_{\mathrm{ex}}}\ll \epsilon_{\mathrm{F}}$ and $\vb*{B}(t, \vb*{r})=0$.
Furthermore, we replace the virtual driving force $\vb*{E}_{\mathrm{s}}^{\mu}$ for the spin current with the expectation value of the spin current, because this virtual force serves only as a technical device.
We also make a similar replacement for the charge sector to treat the charge and spin sectors on an equal footing, while the driving force $\vb*{E}$ for the charge current is not virtual.
The following Drude-type approximation is the simplest tool to relate the driving forces to the expectation values of charge and spin currents, $\ev*{\hat{\vb*{j}}_{\mathrm{cc}}}_{J_{\mathrm{ex}}=0}$ and $\ev*{\hat{\vb*{j}}_{\mathrm{sc}}^{\mu}}_{J_{\mathrm{ex}}=0}$,
\begin{align}
\ev*{\hat{\vb*{j}}_{\mathrm{cc}}}_{J_{\mathrm{ex}}=0}
= \frac{e^{2}n_{\mathrm{e}}\tau_{\mathrm{e}}}{m^{*}}\vb*{E},
\quad
\ev*{\hat{\vb*{j}}_{\mathrm{sc}}^{\mu}}_{J_{\mathrm{ex}}=0} 
= \frac{e_{\mathrm{s}}^{2}n_{\mathrm{e}}\tau_{\mathrm{e}}}{m^{*}}\vb*{E}_{\mathrm{s}}^{\mu}
,
\label{eq:Drude-type_approximation}
\end{align}
where $n_{\mathrm{e}}$ is the number density of conduction electrons (Sec.~\ref{sec:Model}).
For later convenience in numerical calculations, we transform the above currents to the following dimensionless form:
\begin{align}
\vb*{J}_{\mathrm{cc}}
&= 
\frac{3\lambda_{\mathrm{F}}^{2}}{4e\epsilon_{\mathrm{F}}}
\ev*{\hat{\vb*{j}}_{\mathrm{cc}}}_{J_{\mathrm{ex}}=0},
\quad
\vb*{J}_{\mathrm{sc}}^{\mu} 
= 
\frac{3\lambda_{\mathrm{F}}^{2}}{4e_{\mathrm{s}}\epsilon_{\mathrm{F}}}
\ev*{\hat{\vb*{j}}_{\mathrm{sc}}^{\mu}}_{J_{\mathrm{ex}}=0},
\label{eq:nondimensionalization}
\end{align}
where $\lambda_{\mathrm{F}} = 2\pi/k_{\mathrm{F}}$ is the Fermi wavelength.
We set the units of charge and spin currents as 
$4e\epsilon_{\mathrm{F}}/(3\lambda_{\mathrm{F}}^{2})$
and
$4e_{\mathrm{s}}\epsilon_{\mathrm{F}}/(3\lambda_{\mathrm{F}}^{2})$,
respectively.
They are two-thirds of the units used in Ref.~\cite{Onoda2006}. 
This rescaling is merely a device to simplify the coefficients of the functions that appear in the dimensionless form of the effective equations of motion, which we present below.
Physically, the order estimate of the present units is almost the same as before.
The unit of charge current is estimated 
as $\sim 10^{5}$~A\,cm$^{-2}$ for a semiconductor with $n_{\mathrm{e}}\sim 10^{18}$~cm$^{-3}$ and $\epsilon_{\mathrm{F}} \sim 1$~meV,
and as $\sim 10^{10}$~A\,cm$^{-2}$ for a metal with $n_{\mathrm{e}}\sim 10^{21}$~cm$^{-3}$ and $\epsilon_{\mathrm{F}} \sim 1$~eV.

 Appendix~\ref{sec:Spin-spin_retarded-Green-function} describes in detail the evaluation of $\Pi^{\mathrm{R}\mu\nu(n)}_{\mathrm{eff}}(\vb*{r})$ under the above conditions and replacements.
 It contains the explanation of the following decomposition:
\begin{align}
\Pi^{\mathrm{R}\mu\nu(n)}_{\mathrm{eff}}(\vb*{r})
& =
\delta_{\mu\nu} \qty{\Pi^{\mathrm{R}(n)}_{\mathrm{eq}}(\vb*{r})+\Pi^{\mathrm{R}(n)}_{\mathrm{cc}}(\vb*{r})}
\nonumber\\
& \quad
+\sum_{\lambda}\epsilon_{\mu\nu\lambda}\Pi^{\mathrm{R}\lambda(n)}_{\mathrm{sc}}(\vb*{r})
,
\end{align}
where $\Pi^{\mathrm{R}(n)}_{\mathrm{eq}}(\vb*{r})$, $\Pi^{\mathrm{R}(n)}_{\mathrm{cc}}(\vb*{r})$, and $\Pi^{\mathrm{R}\lambda(n)}_{\mathrm{sc}}(\vb*{r})$ are the part independent of both currents, the part proportional to the charge current, and the part proportional to the spin current, respectively.
Note that $\Pi^{\mathrm{R}(0)}_{\mathrm{eq}}(\vb*{r})$ is the RKKY response function.
Here, we introduce the dimensionless functions $F^{(n)}_{\mathrm{eq}}(\vb*{r})$, $F^{(n)}_{\mathrm{cc}}(\vb*{r})$, and $\vb*{F}^{(n)}_{\mathrm{sc}}(\vb*{r})$ for later convenience in numerical calculations, 
\begin{align}
\Pi^{\mathrm{R}(n)}_{\mathrm{eq}}(\vb*{r}) & \cong -\frac{1}{\epsilon_{\mathrm{F}}^{n+1}}\qty(\frac{3n_{\mathrm{e}}}{2})^{2} F^{(n)}_{\mathrm{eq}}(\vb*{r})
\label{eq:Pi(n)eq_F(n)eq}
,
\\
\Pi^{\mathrm{R}(n)}_{\mathrm{cc}}(\vb*{r}) & \cong -\frac{1}{\epsilon_{\mathrm{F}}^{n+1}}\qty(\frac{3n_{\mathrm{e}}}{2})^{2} F^{(n)}_{\mathrm{cc}}(\vb*{r})
\label{eq:Pi(n)cc_F(n)cc}
,
\\
\vb*{\Pi}^{\mathrm{R}(n)}_{\mathrm{sc}}(\vb*{r}) & \cong -\frac{1}{\epsilon_{\mathrm{F}}^{n+1}}\qty(\frac{3n_{\mathrm{e}}}{2})^{2} \vb*{F}^{(n)}_{\mathrm{sc}}(\vb*{r})
\label{eq:Pi(n)sc_F(n)sc}
.
\end{align}
These functions, $F^{(n)}_{\mathrm{eq}}(\vb*{r})$, $F^{(n)}_{\mathrm{cc}}(\vb*{r})$, and $\vb*{F}^{(n)}_{\mathrm{sc}}(\vb*{r})$, correspond to $F^{(n)}_{0}(\vb*{r})$, $F^{(n)}_{\mathrm{cc}}(\vb*{r})$, and $\vb*{F}^{(n)}_{\mathrm{sc}}(\vb*{r})$ in Ref.~\cite{Onoda2006}, respectively.
However, we extracted the factor $(3/2)^{2}$ from the old functions to define the new ones.
Their explicit expressions are given below.
We can see that $\Pi^{\mathrm{R}(0)}_{\mathrm{eq}}(\vb*{r})$ is ill-defined at $\vb*{r}=0$ because $F^{(0)}_{\mathrm{eq}}(\vb*{r})$ diverges in this limit.
The function $\Pi^{\mathrm{R}\mu(0)}_{\mathrm{sc}}(\vb*{r})$ is also ill-defined in the limit $\vb*{r}\to0$ because $F^{\mu(0)}_{\mathrm{sc}}(\vb*{r})$ approaches $(\pi/3)\vb*{J}_{\mathrm{sc}}^{\mu}\vdot \vb*{r}/r$, which depends on the direction of $\vb*{r}$.
We will regularize the limit $\vb*{r}\to0$ of terms including $F^{(0)}_{\mathrm{eq}}(\vb*{r})$ by excluding the ferromagnetic self-interaction caused by $\Pi^{\mathrm{R}(0)}_{\mathrm{eq}}(\vb*{r})$.
On the other hand, since the self-interaction caused by $\Pi^{\mathrm{R}\mu(n)}_{\mathrm{sc}}(\vb*{r})$ vanishes due to the symmetry of this type of interaction, we will implement no special measures for this.
We will take into account the limit $\vb*{r}\to0$ for all other interactions.

Using the results on the above dimensionless functions in Appendix~\ref{sec:Spin-spin_retarded-Green-function}, we obtain the following evaluation of  $\ev*{\hat{\vb*{\sigma}}(t, \vb*{r})}_{\vb*{S}^{\Delta}=0}$:
\begin{widetext}
\begin{align}
\ev*{\hat{\vb*{\sigma}}(t, \vb*{r})}_{\vb*{S}^{\Delta}=0}
&\cong
\frac{J_{\mathrm{ex}}v_{\mathrm{uc}}}{2} 
\sum_{i}\sum_{n=0}^{\infty}\frac{1}{\epsilon_{\mathrm{F}}^{n+1}}\qty(\frac{3n_{\mathrm{e}}}{2})^{2}
\qty[
\qty{F^{(n)}_{\mathrm{eq}}(\vb*{r}-\vb*{r}_{i})+F^{(n)}_{\mathrm{cc}}(\vb*{r}-\vb*{r}_{i})}
{\pdv[n]{\vb*{S}_{\vb*{r}_{i}}}{t}}(t)
-\vb*{F}^{(n)}_{\mathrm{sc}}(\vb*{r}-\vb*{r}_{i})\times
{\pdv[n]{\vb*{S}_{\vb*{r}_{i}}}{t}}(t)
],
\label{eq:evaluation-of-sigma}
\\
F^{(n)}_{\mathrm{eq}}(\vb*{r})
&= 
\frac{\pi}{2^{2n}n!}
\sum_{m=0}^{n} C^{(n)}_{m} \qty(2k_{\mathrm{F}}r)^{m} I^{(n)}_{m}\qty(\frac{\eta}{\epsilon_{\mathrm{F}}};k_{\mathrm{F}}r)
\label{eq:F(n)eq_I(n)m-rep}
,\\
F^{(n)}_{\mathrm{cc}}(\vb*{r})
&=
\qty(\frac{\vb*{r}}{r}\vdot\vb*{J}_{\mathrm{cc}})
\frac{\pi}{2^{2n}n!}
\sum_{m=0}^{n} 
C^{(n)}_{m} \qty(2k_{\mathrm{F}}r)^{m} 
J^{(n)}_{m}\qty(\frac{\eta}{\epsilon_{\mathrm{F}}}; k_{\mathrm{F}}r)
\sin\qty(\Theta^{(n)}_{m}\qty(\frac{\eta}{\epsilon_{\mathrm{F}}}; k_{\mathrm{F}}r))
,\\
F^{\mu(n)}_{\mathrm{sc}}(\vb*{r})
&=
\qty(\frac{\vb*{r}}{r}\vdot\vb*{J}_{\mathrm{sc}}^{\mu})
\frac{\pi}{2^{2n}n!}
\sum_{m=0}^{n} 
C^{(n)}_{m} \qty(2k_{\mathrm{F}}r)^{m}
J^{(n)}_{m}\qty(\frac{\eta}{\epsilon_{\mathrm{F}}}; k_{\mathrm{F}}r)
\cos\qty(\Theta^{(n)}_{m}\qty(\frac{\eta}{\epsilon_{\mathrm{F}}}; k_{\mathrm{F}}r))
,\\
I^{(n)}_{m}\qty(\frac{\eta}{\epsilon_{\mathrm{F}}};k_{\mathrm{F}}r)
&=
\frac{1}{\qty(k_{\mathrm{F}}r)^{2}}\int_{0}^{1}\dd{u}\,u\sin\qty(k_{\mathrm{F}}ru)
\qty{
\frac{(-i)^{n+m}e^{i k_{\mathrm{F}}r\sqrt{u^{2}+i\frac{\eta}{\epsilon_{\mathrm{F}}}}}}{\qty(u^{2}+i\frac{\eta}{\epsilon_{\mathrm{F}}})^{\frac{2n-m}{2}}}
+\frac{i^{n+m}e^{-i k_{\mathrm{F}}r\sqrt{u^{2}-i\frac{\eta}{\epsilon_{\mathrm{F}}}}}}{\qty(u^{2}-i\frac{\eta}{\epsilon_{\mathrm{F}}})^{\frac{2n-m}{2}}}
}
\nonumber\\
&= \frac{1}{\qty(k_{\mathrm{F}}r)^{2}}\sum_{p,q=0}^{\infty} \frac{(-1)^{p}\qty(k_{\mathrm{F}}r)^{2p+q+1}}{(2p+1)! q!} 
\nonumber\\
&\quad\times\frac{1}{2p+3}
\Biggl\{
(-i)^{n+m-q}\qty(1+i\frac{\eta}{\epsilon_{\mathrm{F}}})^{\frac{q-2n+m}{2}} {}_{2}F_{1}\qty(\frac{2n-m-q}{2}, 1; \frac{2p+5}{2}; \frac{1}{1+i\frac{\eta}{\epsilon_{\mathrm{F}}}})
\Biggr.
\nonumber\\
&\hspace{24mm}
\Biggl.
+
i^{n+m-q}\qty(1-i\frac{\eta}{\epsilon_{\mathrm{F}}})^{\frac{q-2n+m}{2}} {}_{2}F_{1}\qty(\frac{2n-m-q}{2}, 1; \frac{2p+5}{2}; \frac{1}{1-i\frac{\eta}{\epsilon_{\mathrm{F}}}})
\Biggr\}
\label{eq:I(n)m}
,\\
J^{(n)}_{m}\qty(\frac{\eta}{\epsilon_{\mathrm{F}}}; k_{\mathrm{F}}r)
&=
\frac{\sin\qty(k_{\mathrm{F}}r)-\qty(k_{\mathrm{F}}r)\cos\qty(k_{\mathrm{F}}r)}{\qty(k_{\mathrm{F}}r)^{3}}
\cdot
\frac{e^{-k_{\mathrm{F}}r\qty(1+\frac{\eta^{2}}{\epsilon_{\mathrm{F}}^{2}})^{\frac{1}{4}}\sin\qty(\frac{1}{2}\arctan\frac{\eta}{\epsilon_{\mathrm{F}}})}
}{\qty(1+\frac{\eta^{2}}{\epsilon_{\mathrm{F}}^{2}})^{\frac{2n-m}{4}}}
,\\
\Theta^{(n)}_{m}\qty(\frac{\eta}{\epsilon_{\mathrm{F}}}; k_{\mathrm{F}}r)
&=
-\frac{\pi}{2}(n+m)
+k_{\mathrm{F}}r\qty(1+\frac{\eta^{2}}{\epsilon_{\mathrm{F}}^{2}})^{\frac{1}{4}}\cos\qty(\frac{1}{2}\arctan\frac{\eta}{\epsilon_{\mathrm{F}}})
-\frac{2n-m}{2}\arctan\frac{\eta}{\epsilon_{\mathrm{F}}}
\label{eq:Theta(n)m}
,
\end{align}
\end{widetext}
where 
${}_{2}F_{1}(\alpha,\beta;\gamma;z)$ ($\abs{z}<1$) is the hypergeometric function defined by
\begin{align}
{}_{2}F_{1}(\alpha,\beta;\gamma;z) 
&= \sum_{j=0}^{\infty}\frac{(\alpha)_{j}(\beta)_{j}}{(\gamma)_{j}} \cdot \frac{z^{j}}{j!}
,
\end{align}
with the Pochhammer symbol
$(\delta)_{j} 
=\Gamma\qty(\delta+j)/\Gamma\qty(\delta)$,
and $C^{(n)}_{m}$ is the coefficient defined by
\begin{align}
C^{(n)}_{m}
=
\frac{m}{2n-m}
\cdot \frac{\qty(2n-m)!}{m!\qty(n-m)!}
.\label{eq:C(n)m}
\end{align}
We keep the factor $m/(2n-m)$ in $C^{(n)}_{m}$ so that the same expression also covers $m=0$, with the value at $n=m=0$ defined by continuation from $C^{(n)}_{n}=1$: 
\begin{align}
C^{(n)}_{0}& =
\begin{cases}
 \left.C^{(n)}_{n}\right\vert_{n=0} = 1 & \qty(n=0)
 \\
0 & \qty(n \ge 1)
\end{cases}
.\label{eq:C(n)0}
\end{align}

The above expressions give the time-derivative expansion of $\ev*{\hat{\vb*{\sigma}}(t, \vb*{r})}_{\vb*{S}^{\Delta}=0}$ in closed form to every order $n$ at finite $\eta$, with $I^{(n)}_{m}\qty(\eta/\epsilon_{\mathrm{F}};k_{\mathrm{F}}r)$ expressed through the hypergeometric function.

\section{Effective equations of motion}
\label{sec:EOM}
Section~\ref{sec:Approximations} introduced energy decoherence phenomenologically through the damping factor $e^{-\eta t}$.
Its effect is explicit in the above results: the functions $J^{(n)}_{m}\qty(\eta/\epsilon_{\mathrm{F}};k_{\mathrm{F}}r)$ in $F^{(n)}_{\mathrm{cc}}(\vb*{r})$ and $F^{\mu(n)}_{\mathrm{sc}}(\vb*{r})$ contain the factor $e^{-r/\ell}$ with the decay length $\ell\simeq v_{\mathrm{F}}/\eta$, while the damping of $F^{(n)}_{\mathrm{eq}}(\vb*{r})$ appears implicitly in the integrand of $I^{(n)}_{m}\qty(\eta/\epsilon_{\mathrm{F}};k_{\mathrm{F}}r)$.
If we set $\eta \to 0$ from the outset, the functions $F^{(n)}_{\mathrm{eq}}(\vb*{r})$, $F^{(n)}_{\mathrm{cc}}(\vb*{r})$, and $F^{\mu(n)}_{\mathrm{sc}}(\vb*{r})$ with $n \ge 2$ decay only slowly, if at all, and some of them do not even converge in the long-distance limit $k_{\mathrm{F}}r\to\infty$; the time-derivative expansion then breaks down.
As discussed in Sec.~\ref{sec:Approximations}, we therefore retain only the terms with $n\le1$.

As preparation to present the effective equations of motion for $\qty{\vb*{S}_{\vb*{r}}(t)}$, we introduce an abbreviation and a dimensionless parameter as follows:
\begin{align}
&\vb*{S}_{\vb*{r}_{i}}(t) \to \vb*{S}_{i},
\quad
\gamma_{\mathrm{ex}} = \frac{3v_{\mathrm{uc}}n_{\mathrm{e}}\abs{J_{\mathrm{ex}}}}{4\epsilon_{\mathrm{F}}}
.
\end{align}
Inserting the spin density of Eq.~(\ref{eq:evaluation-of-sigma}) into the formal equations of motion, Eq.~(\ref{eq:EOM-formal}), we obtain the following effective equations of motion:
\begin{widetext}
\begin{align}
\pdv{\vb*{S}_{i}}{t}
&\cong
\gamma_{\mathrm{ex}}^{2}
F^{(1)}_{\mathrm{eq}}(0)\vb*{S}_{i}
\times\pdv{\vb*{S}_{i}}{t}
\nonumber\\
&\quad
+
\gamma_{\mathrm{ex}}^{2}
\epsilon_{\mathrm{F}}
\sum_{j\neq i}
\qty[
\qty{F^{(0)}_{\mathrm{eq}}(\vb*{r}_{i}-\vb*{r}_{j})+F^{(0)}_{\mathrm{cc}}(\vb*{r}_{i}-\vb*{r}_{j})}
\vb*{S}_{i}\times\vb*{S}_{j}
-\vb*{S}_{i}\times\qty{\vb*{F}^{(0)}_{\mathrm{sc}}(\vb*{r}_{i}-\vb*{r}_{j})\times\vb*{S}_{j}}
]
\nonumber\\
&\quad
+\gamma_{\mathrm{ex}}^{2}
\sum_{j\neq i}
\qty[
\qty{F^{(1)}_{\mathrm{eq}}(\vb*{r}_{i}-\vb*{r}_{j})+F^{(1)}_{\mathrm{cc}}(\vb*{r}_{i}-\vb*{r}_{j})}
\vb*{S}_{i}\times\pdv{\vb*{S}_{j}}{t}
-
\vb*{S}_{i}\times
\qty{\vb*{F}^{(1)}_{\mathrm{sc}}(\vb*{r}_{i}-\vb*{r}_{j})\times\pdv{\vb*{S}_{j}}{t}}
].
\label{eq:EOM-S}
\end{align}
\end{widetext}
The first term with $F^{(1)}_{\mathrm{eq}}(0)$ is the Gilbert damping~\cite{Gilbert2004} due to the Fermi bath of conduction electrons, and the terms with $F^{(0)}_{\mathrm{eq}}(\vb*{r})$ come from the RKKY interaction.
The terms with $F^{(1)}_{\mathrm{eq}}(\vb*{r})$ come from the decoherence effect involving two different spins and work as nonlocal damping or pumping, which is qualitatively consistent with the results of related works on spin systems with dynamic RKKY interaction~\cite{Barnes1974, Rikitake2005} and with more recent quantum-classical and Schwinger-Keldysh studies of nonlocal damping in metallic magnets~\cite{Bajpai2020, Petrovic2021, ReyesOsorio2024, Lenzing2025}.
The Gilbert damping is the on-site ($j=i$) counterpart of the same decoherence effect.
In Refs.~\cite{Bajpai2020, Petrovic2021, ReyesOsorio2024, Lenzing2025}, the conduction electrons are driven out of equilibrium only by the localized-spin dynamics itself; here they additionally carry charge and spin currents, which supply the kernels $F^{(n)}_{\mathrm{cc}}(\vb*{r})$ and $\vb*{F}^{(n)}_{\mathrm{sc}}(\vb*{r})$. The former give spin torques induced by the charge current, of which the one with $F^{(0)}_{\mathrm{cc}}(\vb*{r})$ drives the precession of the dynamic helix [Eq.~(\ref{eq:Gcc})], and the latter generate the DM-type coupling that underlies the helical order in Sec.~\ref{sec:helix-theory}.

For the numerical simulations, we also present a normalized version of the above equations, using the fixed spin magnitude $\abs{\vb*{S}_{\vb*{r}}(t)}=S$ (Sec.~\ref{sec:Model}).
To this end, we introduce an abbreviation and dimensionless parameters as follows:
\begin{align}
&\vb*{\Omega}_{\vb*{r}_{i}}(\tilde{t})=\frac{\vb*{S}_{\vb*{r}_{i}}(t)}{S} \to \vb*{\Omega}_{i},
\quad
\tilde{t} = \tilde{\gamma}^{2} \epsilon_{\mathrm{F}} t,
\quad
\tilde{\gamma} = \sqrt{S}\gamma_{\mathrm{ex}}
.
\end{align}
This definition of $\tilde{\gamma}$ is three-halves of $\gamma$ in Ref.~\cite{Onoda2006}.
Accordingly, the unit of time in this work is four-ninths of that in Ref.~\cite{Onoda2006}.
As with the units of charge and spin currents, this rescaling simplifies the coefficients of $F^{(n)}_{\mathrm{eq}}(\vb*{r})$, $F^{(n)}_{\mathrm{cc}}(\vb*{r})$, and $F^{\mu(n)}_{\mathrm{sc}}(\vb*{r})$ ($n\le 1$) in the limit $\eta \to 0$. The dimensionless equations of motion for the normalized spin variables are then as follows:
\begin{widetext}
\begin{align}
\pdv{\vb*{\Omega}_{i}}{\tilde{t}}
&\cong
\tilde{\gamma}^{2}
F^{(1)}_{\mathrm{eq}}(0)\vb*{\Omega}_{i}
\times\pdv{\vb*{\Omega}_{i}}{\tilde{t}}
+\sum_{j\neq i}
\Biggl[
\qty{F^{(0)}_{\mathrm{eq}}(\vb*{r}_{i}-\vb*{r}_{j})+F^{(0)}_{\mathrm{cc}}(\vb*{r}_{i}-\vb*{r}_{j})}
\vb*{\Omega}_{i}\times\vb*{\Omega}_{j}
-\vb*{\Omega}_{i}\times\qty{\vb*{F}^{(0)}_{\mathrm{sc}}(\vb*{r}_{i}-\vb*{r}_{j})\times\vb*{\Omega}_{j}}
\nonumber\\
&\quad
+\tilde{\gamma}^{2}
\qty{F^{(1)}_{\mathrm{eq}}(\vb*{r}_{i}-\vb*{r}_{j})+F^{(1)}_{\mathrm{cc}}(\vb*{r}_{i}-\vb*{r}_{j})}
\vb*{\Omega}_{i}\times\pdv{\vb*{\Omega}_{j}}{\tilde{t}}
-\tilde{\gamma}^{2}
\vb*{\Omega}_{i}\times
\qty{\vb*{F}^{(1)}_{\mathrm{sc}}(\vb*{r}_{i}-\vb*{r}_{j})\times\pdv{\vb*{\Omega}_{j}}{\tilde{t}}}
\Biggr].
\label{eq:EOM}
\end{align}
\end{widetext}
In these normalized equations, the strength of the spin-spin coupling enters only through $\tilde{\gamma}^{2}= S\gamma_{\mathrm{ex}}^{2}$, not through $\gamma_{\mathrm{ex}}^{2}$ alone.
We therefore measure the coupling strength by $\tilde{\gamma}^{2}$ hereafter. Because $\tilde{\gamma}^{2}=(\gamma_{\mathrm{ex}}S)^{2}/S$, a large spin $S$, often invoked to justify the classical treatment of the localized spins, does not by itself violate the weak-coupling condition of Sec.~\ref{sec:Model}: at a fixed small Born parameter $\gamma_{\mathrm{ex}}S$, i.e., for $\abs{J_{\mathrm{ex}}}\propto1/S$, it corresponds to a small $\tilde{\gamma}^{2}\propto1/S$.

For $0 < \eta/\epsilon_{\mathrm{F}} \ll 1$, we approximate the functions $F^{(n)}_{\mathrm{eq}}(\vb*{r})$, $F^{(n)}_{\mathrm{cc}}(\vb*{r})$, and $F^{\mu(n)}_{\mathrm{sc}}(\vb*{r})$ ($n\le 1$) by those in the limit $\eta \to 0$:
\begin{align}
\left.F^{(n)}_{\mathrm{eq}}(\vb*{r})\right\vert_{\eta\to0}
&= f^{(n)}_{\mathrm{eq}}(2k_{\mathrm{F}}r)
\label{eq:Feq_limit_f}
,\\
\left.F^{(n)}_{\mathrm{cc}}(\vb*{r})\right\vert_{\eta\to0}
&= \qty(\frac{\vb*{r}}{r}\vdot\vb*{J}_{\mathrm{cc}}) f^{(n)}_{\mathrm{cc}}(2k_{\mathrm{F}}r)
,\\
\left.F^{\mu(n)}_{\mathrm{sc}}(\vb*{r})\right\vert_{\eta\to0}
&= \qty(\frac{\vb*{r}}{r}\vdot\vb*{J}_{\mathrm{sc}}^{\mu})  f^{(n)}_{\mathrm{sc}}(2k_{\mathrm{F}}r)
,
\end{align}
where
\begin{align}
f^{(0)}_{\mathrm{eq}}\qty(x)
&=
4\pi
\cdot
\frac{\sin x-x\cos x}{x^{4}}
\label{eq:f0eq}
,\\
f^{(1)}_{\mathrm{eq}}\qty(x)
&=
-
\pi
\cdot
\frac{1-\cos x}{x^{2}}
\label{eq:f1eq}
,\\
f^{(n)}_{\mathrm{cc}}\qty(x)
&=
\qty(-1)^{n}
\frac{2\pi}{2^{2n}}
\cdot
\frac{2\qty(1-\cos x)-x\sin x}{x^{3-n}}
,\\
f^{(n)}_{\mathrm{sc}}\qty(x)
&=
\qty(-1)^{n}
\frac{2\pi}{2^{2n}}
\cdot
\frac{2\sin x-x\qty(1+\cos x)}{x^{3-n}}
.
\end{align}

\begin{figure}[htb]
  \centering
  \includegraphics[width=\linewidth]{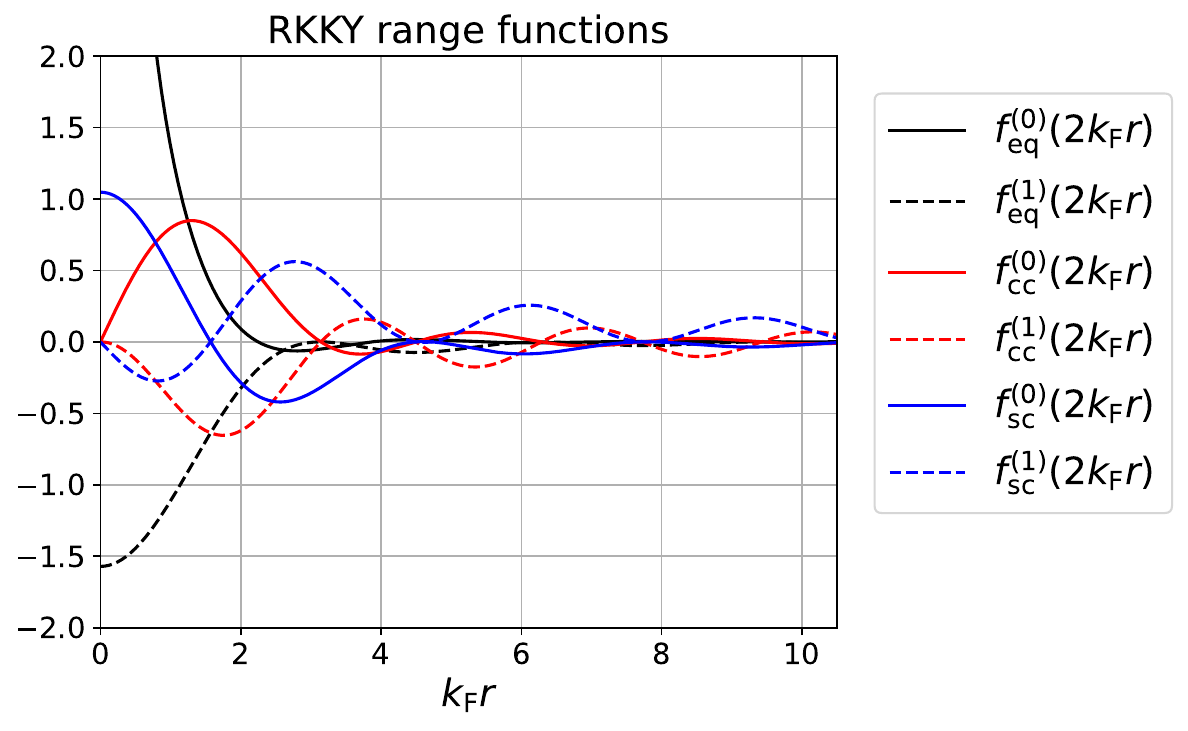}
  \caption{
  RKKY range functions $f_{\mathrm{eq}}^{(0)}(2k_{\mathrm{F}}r)$,
  $f_{\mathrm{cc}}^{(0)}(2k_{\mathrm{F}}r)$, and
  $f_{\mathrm{sc}}^{(0)}(2k_{\mathrm{F}}r)$ (solid lines) 
  and $f_{\mathrm{eq}}^{(1)}(2k_{\mathrm{F}}r)$, $f_{\mathrm{cc}}^{(1)}(2k_{\mathrm{F}}r)$, and $f_{\mathrm{sc}}^{(1)}(2k_{\mathrm{F}}r)$ (dashed lines), plotted as functions of $k_{\mathrm{F}}r$.
  All of them decay with increasing distance. The function $f_{\mathrm{eq}}^{(0)}(2k_{\mathrm{F}}r)$ of the RKKY interaction and the functions $f_{\mathrm{cc}}^{(n)}(2k_{\mathrm{F}}r)$ change sign repeatedly, whereas $f_{\mathrm{eq}}^{(1)}(2k_{\mathrm{F}}r)$ and $f_{\mathrm{sc}}^{(n)}(2k_{\mathrm{F}}r)$ keep a nearly fixed sign at large $k_{\mathrm{F}}r$.
  }
  \label{fig:RKKY}
\end{figure}

We refer to the functions $f^{(n)}_{\mathrm{eq}}(x)$, $f^{(n)}_{\mathrm{cc}}(x)$, and $f^{(n)}_{\mathrm{sc}}(x)$ ($n\le1$) as RKKY range functions. Figure~\ref{fig:RKKY} shows them. Because $f^{(1)}_{\mathrm{eq}}(x)\to-\pi/2$ as $x\to0$, the first term on the right-hand side of Eq.~(\ref{eq:EOM}) is the Gilbert damping with the damping constant $(\pi/2)\tilde{\gamma}^{2}$.

\section{Fourier-transformed RKKY range functions}
\label{sec:F-transformed_RKKY}
This section provides some mathematical preparation for the discussion on the dynamics of a one-dimensional periodic array of localized spins in the next section.
Here we introduce the following Fourier transformation of the RKKY range functions $f^{(n)}_{\mathrm{eq}}(x)$, $f^{(n)}_{\mathrm{cc}}(x)$, and $f^{(n)}_{\mathrm{sc}}(x)$ ($n\le1$):
\begin{widetext}
\begin{alignat}{2}
\tilde{f}^{(0)}_{\mathrm{eq}}\qty(qa_{0})
&=
\quad
\frac{1}{2}
\sum_{\substack{m=-\infty\\ m\neq0}}^{\infty} f^{(0)}_{\mathrm{eq}}\qty(2\abs{m}k_{\mathrm{F}}a_{0}) \qty{1-\cos\qty(mqa_{0})}
&&=
\sum_{m=1}^{\infty} f^{(0)}_{\mathrm{eq}}\qty(2mk_{\mathrm{F}}a_{0}) \qty{1-\cos\qty(mqa_{0})}
,\\
\tilde{f}^{(1)}_{\mathrm{eq}}\qty(qa_{0})
&=
-\frac{1}{2}
\sum_{m=-\infty}^{\infty} 
f^{(1)}_{\mathrm{eq}}\qty(2\abs{m}k_{\mathrm{F}}a_{0}) \cos\qty(mqa_{0})
&&=
\frac{\pi}{4}-
\sum_{m=1}^{\infty} f^{(1)}_{\mathrm{eq}}\qty(2mk_{\mathrm{F}}a_{0}) \cos\qty(mqa_{0})
,\\
\tilde{f}^{(n)}_{\mathrm{cc}}\qty(qa_{0})
&=
\quad
\frac{1}{2}
\sum_{m=-\infty}^{\infty} 
\mathrm{sgn}(m) 
f^{(n)}_{\mathrm{cc}}\qty(2\abs{m}k_{\mathrm{F}}a_{0})\sin\qty(mqa_{0})
&&=
\sum_{m=1}^{\infty}
f^{(n)}_{\mathrm{cc}}\qty(2mk_{\mathrm{F}}a_{0})\sin\qty(mqa_{0})
,\\
\tilde{f}^{(n)}_{\mathrm{sc}}\qty(qa_{0})
&=
\quad
\frac{1}{2}
\sum_{m=-\infty}^{\infty} 
\mathrm{sgn}(m) 
f^{(n)}_{\mathrm{sc}}\qty(2\abs{m}k_{\mathrm{F}}a_{0})\sin\qty(mqa_{0})
&&=
\sum_{m=1}^{\infty}
f^{(n)}_{\mathrm{sc}}\qty(2mk_{\mathrm{F}}a_{0})\sin\qty(mqa_{0})
,
\end{alignat}
\end{widetext}
where $a_{0}$ is a constant with the dimension of length and $q$ is a real wavenumber; for the chain of Sec.~\ref{sec:helix-theory}, $a_{0}$ is the spacing of the localized spins, the index $m$ labels the spins $j=i+m$ relative to a given spin $i$, and $q$ is the signed wavenumber of the helical configurations of Eqs.~(\ref{eq:static-S_helix}) and (\ref{eq:dynamic-S_helix}).

The factor $1-\cos\qty(mqa_{0})$ in $\tilde{f}^{(0)}_{\mathrm{eq}}\qty(qa_{0})$ corresponds to $S^{2}-\vb*{S}_{i}\vdot\vb*{S}_{j}$ in Eq.~(\ref{eq:Heff}): it measures the RKKY energy of a helix relative to that of the ferromagnetic state, so that $\tilde{f}^{(0)}_{\mathrm{eq}}(0)=0$.
The factor $\mathrm{sgn}(m)$ in $\tilde{f}^{(n)}_{\mathrm{cc}}\qty(qa_{0})$ and $\tilde{f}^{(n)}_{\mathrm{sc}}\qty(qa_{0})$ reflects that $F^{(n)}_{\mathrm{cc}}(\vb*{r})$ and $\vb*{F}^{(n)}_{\mathrm{sc}}(\vb*{r})$ are odd in $\vb*{r}$.
The sum in $\tilde{f}^{(1)}_{\mathrm{eq}}\qty(qa_{0})$ includes the term with $m=0$, i.e., the Gilbert damping [$f^{(1)}_{\mathrm{eq}}(x)\to-\pi/2$ as $x\to0$], which gives $\pi/4$; with the minus sign, $\tilde{f}^{(1)}_{\mathrm{eq}}\qty(qa_{0})$ is nonnegative for every $q$.
Section~\ref{sec:helix-theory} uses $\tilde{f}^{(0)}_{\mathrm{eq}}\qty(qa_{0})$ and $\tilde{f}^{(0)}_{\mathrm{sc}}\qty(qa_{0})$ in the variational energy of the static helix [Eqs.~(\ref{eq:variational-energy}) and (\ref{eq:F_H})] and all six functions for the dynamic helix [Eqs.~(\ref{eq:Gcc}) and (\ref{eq:Delta})].

These functions have analytic expressions in terms of the generalized Clausen functions $\mathrm{S}_{z}\qty(\theta)$ and $\mathrm{C}_{z}\qty(\theta)$ defined by
\begin{align}
\mathrm{S}_{z}\qty(\theta) &= \sum_{m=1}^{\infty}\frac{\sin\qty(m\theta)}{m^{z}}
,&
\mathrm{C}_{z}\qty(\theta) &= \sum_{m=1}^{\infty}\frac{\cos\qty(m\theta)}{m^{z}}
,
\end{align}
where $z$ is a complex number satisfying $\Re z > 0$.
For $\Re z\le0$, which arises when the derivatives of the functions with $n=1$ are taken, the series are understood in the Abel sense, e.g., $\mathrm{S}_{0}(\theta)=(1/2)\cot(\theta/2)$ and $\mathrm{C}_{0}(\theta)=-1/2$ for $0<\theta<2\pi$.

In terms of $\mathrm{S}_{z}\qty(\theta)$ and $\mathrm{C}_{z}\qty(\theta)$, the Fourier-transformed RKKY range functions are
\begin{widetext}
\begin{align}
\tilde{f}^{(0)}_{\mathrm{eq}}\qty(qa_{0})
&=
\frac{2\pi}{\qty(2k_{\mathrm{F}}a_{0})^{4}}\qty{2\mathrm{S}_{4}\qty(2k_{\mathrm{F}}a_{0})-\mathrm{S}_{4}\qty(2k_{\mathrm{F}}a_{0}+qa_{0})-\mathrm{S}_{4}\qty(2k_{\mathrm{F}}a_{0}-qa_{0})}
\nonumber\\
&\quad
-\frac{2\pi}{\qty(2k_{\mathrm{F}}a_{0})^{3}}\qty{2\mathrm{C}_{3}\qty(2k_{\mathrm{F}}a_{0})-\mathrm{C}_{3}\qty(2k_{\mathrm{F}}a_{0}+qa_{0})-\mathrm{C}_{3}\qty(2k_{\mathrm{F}}a_{0}-qa_{0})}
,\\
\tilde{f}^{(1)}_{\mathrm{eq}}\qty(qa_{0})
&=
\frac{\pi}{4}+\frac{\pi}{2}\cdot\frac{1}{\qty(2k_{\mathrm{F}}a_{0})^{2}}\qty{2\mathrm{C}_{2}\qty(qa_{0})-\mathrm{C}_{2}\qty(2k_{\mathrm{F}}a_{0}+qa_{0})-\mathrm{C}_{2}\qty(2k_{\mathrm{F}}a_{0}-qa_{0})}
,\\
\tilde{f}^{(n)}_{\mathrm{cc}}\qty(qa_{0})
&=
\quad
\qty(-1)^{n}
\frac{2\pi}{2^{2n}}\cdot\frac{1}{\qty(2k_{\mathrm{F}}a_{0})^{3-n}}
\qty{2\mathrm{S}_{3-n}\qty(qa_{0})-\mathrm{S}_{3-n}\qty(2k_{\mathrm{F}}a_{0}+qa_{0})+\mathrm{S}_{3-n}\qty(2k_{\mathrm{F}}a_{0}-qa_{0})}
\nonumber\\
&\quad
+\qty(-1)^{n}\frac{\pi}{2^{2n}}\cdot\frac{1}{\qty(2k_{\mathrm{F}}a_{0})^{2-n}}
\qty{\mathrm{C}_{2-n}\qty(2k_{\mathrm{F}}a_{0}+qa_{0})-\mathrm{C}_{2-n}\qty(2k_{\mathrm{F}}a_{0}-qa_{0})}
,\\
\tilde{f}^{(n)}_{\mathrm{sc}}\qty(qa_{0})
&=
-\qty(-1)^{n}
\frac{2\pi}{2^{2n}}\cdot\frac{1}{\qty(2k_{\mathrm{F}}a_{0})^{3-n}}
\qty{\mathrm{C}_{3-n}\qty(2k_{\mathrm{F}}a_{0}+qa_{0})-\mathrm{C}_{3-n}\qty(2k_{\mathrm{F}}a_{0}-qa_{0})}
\nonumber\\
&\quad
-\qty(-1)^{n}\frac{\pi}{2^{2n}}\cdot\frac{1}{\qty(2k_{\mathrm{F}}a_{0})^{2-n}}
\qty{2\mathrm{S}_{2-n}\qty(qa_{0})+\mathrm{S}_{2-n}\qty(2k_{\mathrm{F}}a_{0}+qa_{0})-\mathrm{S}_{2-n}\qty(2k_{\mathrm{F}}a_{0}-qa_{0})}
.
\end{align}
\end{widetext}
Their derivatives with respect to $q$ follow from the relations
\begin{align}
\dv{\mathrm{S}_{z}}{\theta}\qty(\theta) &= \mathrm{C}_{z-1}\qty(\theta)
,&
\dv{\mathrm{C}_{z}}{\theta}\qty(\theta) &= -\mathrm{S}_{z-1}\qty(\theta) 
.
\end{align}

For a nonnegative integer $k$, $\mathrm{S}_{n}\qty(\theta)$ and $\mathrm{C}_{n}\qty(\theta)$ with $n=2k$ and $2k+1$ relate to the standard Clausen functions $\mathrm{Sl}_{n}\qty(\theta)$ and $\mathrm{Cl}_{n}\qty(\theta)$ as
\begin{align}
\mathrm{S}_{2k}\qty(\theta) &= \mathrm{Cl}_{2k}\qty(\theta)
,&
\mathrm{C}_{2k}\qty(\theta) &= \mathrm{Sl}_{2k}\qty(\theta)
,\\
\mathrm{S}_{2k+1}\qty(\theta) &= \mathrm{Sl}_{2k+1}\qty(\theta)
,&
\mathrm{C}_{2k+1}\qty(\theta) &= \mathrm{Cl}_{2k+1}\qty(\theta) 
.
\end{align}
Further, the standard Clausen functions $\mathrm{Sl}_{n}\qty(\theta)$ relate to the Bernoulli polynomials $B_{n}\qty(\theta/(2\pi))$ for $0\le\theta\le2\pi$ ($0<\theta<2\pi$ for $\mathrm{Sl}_{0}$ and $\mathrm{Sl}_{1}$; outside this interval, $\theta$ is to be reduced modulo $2\pi$) as 
\begin{align}
 \mathrm{Sl}_{2k}\qty(\theta) &= \qty(-1)^{k-1}\frac{\qty(2\pi)^{2k}}{2\qty(2k)!}B_{2k}\qty(\frac{\theta}{2\pi})
 ,\\
 \mathrm{Sl}_{2k+1}\qty(\theta) &= \qty(-1)^{k+1}\frac{\qty(2\pi)^{2k+1}}{2\qty(2k+1)!}B_{2k+1}\qty(\frac{\theta}{2\pi})
.
\end{align}

\section{Helical magnetism in spin chains}
\label{sec:helix-theory}
This section investigates the dynamics of a one-dimensional periodic array of localized spins.
Section~\ref{sec:four-type} classifies the electron flow into four types and recalls the phase diagram of the two-spin system; Secs.~\ref{sec:helix-static} and \ref{sec:helix-dynamic} construct the effective Hamiltonian for static helimagnetism and the effective Lagrangian for dynamic helimagnetism, respectively.

\subsection{Four types of electron flow}
\label{sec:four-type}
For simplicity, we focus on charge and spin currents of the form
\begin{align}
&\vb*{J}_{\mathrm{cc}}
= J_{\mathrm{cc}} \vb*{n}_{\mathrm{c}},
\quad
\vb*{J}_{\mathrm{sc}}^{\mu} 
= J_{\mathrm{sc}}n_{\mathrm{s}}^{\mu}\vb*{n}_{\mathrm{c}},
\end{align}
where $J_{\mathrm{cc}}$ and $J_{\mathrm{sc}}$ are the magnitudes of the charge-current and spin-current components, $\vb*{n}_{\mathrm{c}}$ is the direction of the electron flow (for a pure spin current, that of the electrons with spin parallel to $\vb*{n}_{\mathrm{s}}$), and $\vb*{n}_{\mathrm{s}}$ is the spin polarization carried along $\vb*{n}_{\mathrm{c}}$.
Here the two components flow along a common axis; in general, they can flow along different axes.
By Eqs.~(\ref{eq:distribution_function}), (\ref{eq:Drude-type_approximation}), and (\ref{eq:nondimensionalization}), the electrons with spin parallel and antiparallel to $\vb*{n}_{\mathrm{s}}$ drift with velocities $(J_{\mathrm{cc}}+J_{\mathrm{sc}})(v_{\mathrm{F}}/2)\vb*{n}_{\mathrm{c}}$ and $(J_{\mathrm{cc}}-J_{\mathrm{sc}})(v_{\mathrm{F}}/2)\vb*{n}_{\mathrm{c}}$, respectively.
For nonzero currents, we accordingly classify the electron flow into four types: pure charge current ($J_{\mathrm{sc}}=0$); spin-polarized charge current ($0<J_{\mathrm{sc}}\le J_{\mathrm{cc}}$), in which no electrons drift against $\vb*{n}_{\mathrm{c}}$; charged spin current ($0<J_{\mathrm{cc}}<J_{\mathrm{sc}}$), in which the two spin species drift in opposite directions; and pure spin current ($J_{\mathrm{cc}}=0$).
At the boundary $J_{\mathrm{sc}}=J_{\mathrm{cc}}$, the electrons with spin antiparallel to $\vb*{n}_{\mathrm{s}}$ are at rest, and the charge current is fully spin polarized.
The expressions below remain valid when $J_{\mathrm{cc}}$ and $J_{\mathrm{sc}}$ are given signs; reversing $\vb*{n}_{\mathrm{c}}$ is then equivalent to reversing the signs of both $J_{\mathrm{cc}}$ and $J_{\mathrm{sc}}$, and reversing $\vb*{n}_{\mathrm{s}}$ to reversing the sign of $J_{\mathrm{sc}}$ alone.

\begin{figure}[htb]
  \centering
  \includegraphics[width=\linewidth]{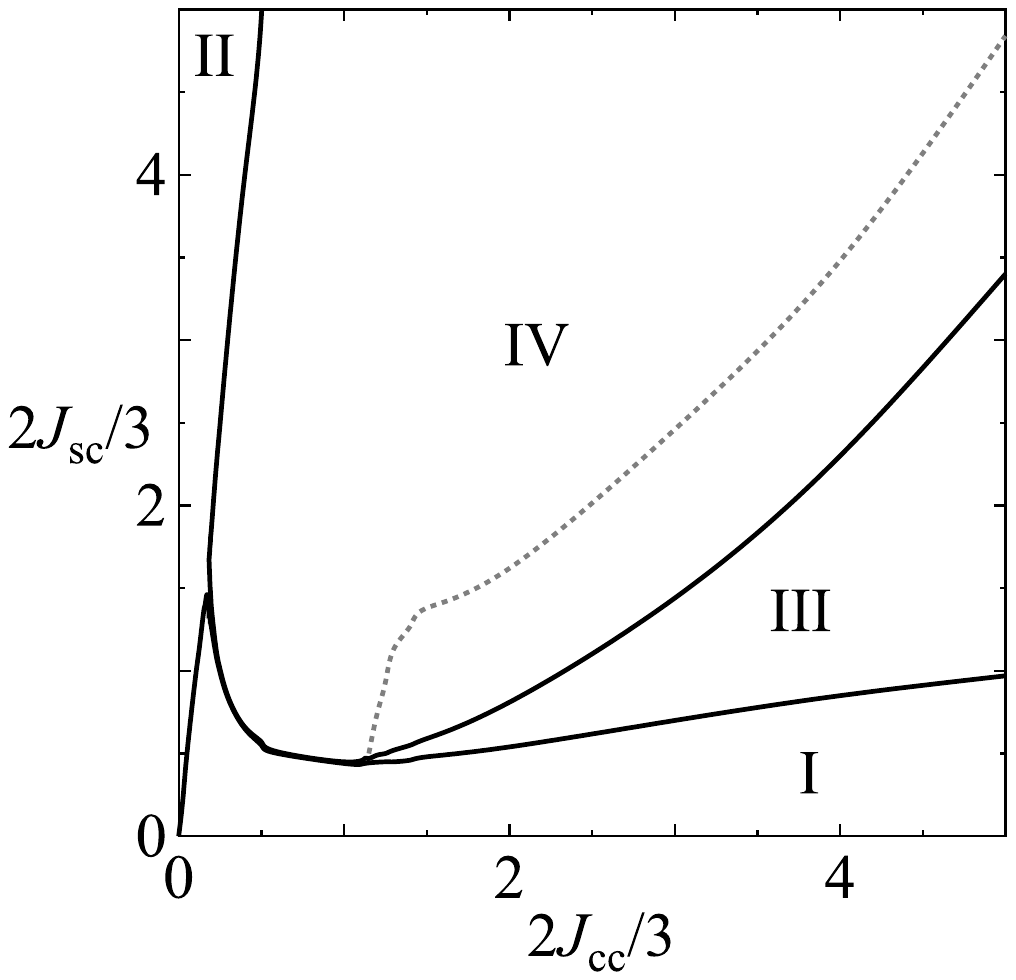}
  \caption{
  Phase diagram of the dynamics of the two-spin system, adapted from Ref.~\cite{Onoda2006} and plotted in the dimensionless currents $2J_{\mathrm{cc}}/3$ and $2J_{\mathrm{sc}}/3$ of the present work (equal to the normalized currents $J^{n}$ and $J^{s}$ of Ref.~\cite{Onoda2006}), for two spins separated by $r=a_{0}$ with $k_{\mathrm{F}}a_{0}=1$ and $\tilde{\gamma}^{2}=(3/2)^{2}\times0.05=0.1125$.
  Regions I--IV correspond to a static collinear configuration, two types of periodic motion, and chaotic or complicated motion, respectively; the dotted line in region IV marks a relatively clear change in the type of motion (Sec.~\ref{sec:helix-dynamic}). See Ref.~\cite{Onoda2006} for details.
  }
  \label{fig:phase-diagram}
\end{figure}

In the $(J_{\mathrm{cc}}, J_{\mathrm{sc}})$ plane, the pure currents lie on the axes, the spin-polarized charge currents on and below the diagonal $J_{\mathrm{cc}}=J_{\mathrm{sc}}$, and the charged spin currents above it. Figure~\ref{fig:phase-diagram} shows, in this plane, the phase diagram obtained in our previous letter~\cite{Onoda2006} for the simplest nontrivial realization of Eq.~(\ref{eq:EOM-S}): two localized spins separated by $r=a_{0}$ with $k_{\mathrm{F}}a_{0}=1$, the same nearest-neighbor separation adopted below for the periodic spin chain. Its regions I--IV, which are distinct from the four types of electron flow, correspond to a static collinear configuration, two types of periodic motion, and chaotic or complicated motion, respectively, and we refer to them repeatedly later in this section.

The remainder of this paper focuses on chains of localized spins evenly spaced along $\vb*{n}_{\mathrm{c}}$, $\vb*{r}_{i+1}-\vb*{r}_{i}=\vb*{a}_{0}=a_{0}\vb*{n}_{\mathrm{c}}$ for all $i$, with $k_{\mathrm{F}}a_{0}=1$, so that the nearest-neighbor spins couple ferromagnetically with sufficient strength compared with the longer-range couplings.
In equilibrium ($J_{\mathrm{cc}}=J_{\mathrm{sc}}=0$), the spins therefore align ferromagnetically, in an arbitrary direction because the present model has no spin anisotropy.
When we introduce a sufficiently small amount of spin-polarized charge current ($0<J_{\mathrm{sc}} \le J_{\mathrm{cc}} \ll 1$), the spins align ferromagnetically antiparallel to $\vb*{n}_{\mathrm{s}}$ (Sec.~\ref{sec:sim-ferro-chaos}).
This orientation is consistent with the sign of the canting ratio $\Delta$ of Eq.~(\ref{eq:Delta}), which is negative for $J_{\mathrm{cc}}J_{\mathrm{sc}}>0$ at the wavenumbers $q_{\mathrm{helix}}$ realized in Sec.~\ref{sec:helix-dynamic}; the ferromagnetic state corresponds to the limit $\Delta\to-1$.

The previous study on the two-spin system provides the following insights~\cite{Onoda2006}.
Pure spin current ($J_{\mathrm{cc}}=0$) induces the helimagnetism of a localized spin chain.
This phenomenon comes from the competition between the ferromagnetic RKKY exchange, with $F^{(0)}_{\mathrm{eq}}(\vb*{r})$, and the DM-type coupling, with $\vb*{F}^{(0)}_{\mathrm{sc}}(\vb*{r})$, in Eq.~(\ref{eq:EOM}); the DM-type coupling is oriented along $\vb*{n}_{\mathrm{s}}$ and makes $\vb*{n}_{\mathrm{s}}$ an effective hard axis, i.e., the plane perpendicular to $\vb*{n}_{\mathrm{s}}$ an effective easy plane.
In the nearest-neighbor approximation of Sec.~\ref{sec:helix-static}, the pitch angle $q_{\mathrm{helix}}a_{0}$ of the helix satisfies $\tan(q_{\mathrm{helix}}a_{0})\propto J_{\mathrm{sc}}$ [Eq.~(\ref{eq:pitch-NN})], so that its magnitude increases with $J_{\mathrm{sc}}$. 
This helical order is a steady state under the pure spin current.
When a small amount of $J_{\mathrm{cc}}$ is introduced to the pure spin current, the helical spins rise from the easy plane and precess around $\vb*{n}_{\mathrm{s}}$, as in region II of Fig.~\ref{fig:phase-diagram}.

In the case of a small amount of the weakly spin-polarized current, which belongs to region I, the spins align ferromagnetically as in equilibrium,
but along $\vb*{n}_{\mathrm{s}}$: the spin-polarized current makes $\vb*{n}_{\mathrm{s}}$ an effective easy axis, in contrast to the effective hard axis of the helical states.
In the two-spin system, the spins align parallel to $\vb*{n}_{\mathrm{s}}$ in this region, and so do three spins; for four or more spins, however, the alignment is antiparallel to $\vb*{n}_{\mathrm{s}}$, as noted above. A linear stability analysis of the uniform states $\vb*{\Omega}_{i}=\pm\vb*{n}_{\mathrm{s}}$ of Eq.~(\ref{eq:EOM}) reproduces this reversal, which disappears when $\vb*{F}^{(0)}_{\mathrm{sc}}(\vb*{r})$ is restricted to nearest neighbors (see Supplemental Material~\footnote{See Supplemental Material at [URL will be inserted by publisher] for a linear stability analysis of the ferromagnetic states in short chains; animated visualizations of the spin-chain dynamics ($N_{\mathrm{S}}=60$, open boundary conditions) for six representative parameter sets spanning the static helimagnetic, dynamic helimagnetic, ferromagnetic, and chaotic regimes discussed in Sec.~\ref{sec:simulation}, each shown as a bird's-eye view and three orthogonal projections for normalized times $\tilde{t}=0$ to $2500$, together with the parameters and main-text correspondence of each movie; and a comparison of the numerical results for $N_{\mathrm{S}}=10$--$60$ with the infinite-chain predictions (Fig.~S1).}). The reversal can thus be traced to the sign change of $f^{(0)}_{\mathrm{sc}}(2k_{\mathrm{F}}r)$ beyond the nearest-neighbor distance (Fig.~\ref{fig:RKKY}), that is, to the same long-range coupling that reverses the chirality of the static helix in Sec.~\ref{sec:helix-static}.
However, for a large amount of the strongly spin-polarized current belonging to region IV of Fig.~\ref{fig:phase-diagram}, ferromagnetic order is expected to become unstable.

\subsection{Effective Hamiltonian for static helimagnetism}
\label{sec:helix-static}
Before discussing spin-current-induced helimagnetism, we note that, for a pure spin current ($J_{\mathrm{cc}}=0$), an effective Hamiltonian $H_{\mathrm{eff}}\qty[\vb*{S}]$ for the localized spins can be constructed even in the nonequilibrium steady state:
\begin{align}
H_{\mathrm{eff}}\qty[\vb*{S}]
&= 
\gamma_{\mathrm{ex}}^{2}\epsilon_{\mathrm{F}}
\sum_{i<j}
\Bigl[
F^{(0)}_{\mathrm{eq}}(\vb*{r}_{i}-\vb*{r}_{j})\qty(S^{2}-\vb*{S}_{i}\vdot\vb*{S}_{j})
\Bigr.
\nonumber\\
&\qquad\qquad\qquad
\Bigl.
-\vb*{F}^{(0)}_{\mathrm{sc}}(\vb*{r}_{i}-\vb*{r}_{j})\vdot\qty(\vb*{S}_{i}\times\vb*{S}_{j})
\Bigr].
\label{eq:Heff}
\end{align}
This $H_{\mathrm{eff}}\qty[\vb*{S}]$ generates the dynamics through the following Poisson bracket for classical spin variables:
\begin{align}
&\qty{S_{i}^{\mu}, S_{j}^{\nu}}_{\mathrm{P}} = \delta_{ij}\sum_{\lambda}\epsilon_{\mu\nu\lambda}S_{i}^{\lambda}
.
\label{eq:commutation-relation}
\end{align}
We can derive the approximate equations of motion as follows:
\begin{widetext}
\begin{align}
&
\pdv{\vb*{S}_{i}}{t} \simeq \qty{\vb*{S}_{i},H_{\mathrm{eff}}\qty[\vb*{S}]}_{\mathrm{P}}
\Longrightarrow
\pdv{\vb*{S}_{i}}{t}
\simeq
\gamma_{\mathrm{ex}}^{2}\epsilon_{\mathrm{F}}
\sum_{j\neq i}
\qty[
F^{(0)}_{\mathrm{eq}}(\vb*{r}_{i}-\vb*{r}_{j})
\vb*{S}_{i}\times\vb*{S}_{j}
-\vb*{S}_{i}\times\qty{\vb*{F}^{(0)}_{\mathrm{sc}}(\vb*{r}_{i}-\vb*{r}_{j})\times\vb*{S}_{j}}
]
.
\label{eq:EOM_pure-spin-current}
\end{align}
\end{widetext}
The above equations approximate the equations of motion in Eq.~(\ref{eq:EOM-S}) well under the condition:
\begin{align}
&J_{\mathrm{cc}}=0 \land \qty{\forall i,\abs{\pdv{\vb*{S}_{i}}{t}} \ll \epsilon_{\mathrm{F}}S}
.
\end{align}
Thus, we can interpret $H_{\mathrm{eff}}\qty[\vb*{S}]$ as an effective energy functional whose minima give the static configurations under steady pure spin currents.
The form of $H_{\mathrm{eff}}\qty[\vb*{S}]$ in Eq.~(\ref{eq:Heff}) suggests that dissipative transport spin currents act on localized spins in the same form as equilibrium spin currents~\cite{Katsura2005}, as noted for a two-spin system in Ref.~\cite{Onoda2006}; DM-type couplings induced by spin-polarized nonequilibrium electrons have since been derived for localized spins in tunnel junctions and open quantum systems~\cite{Fransson2014, Liu2018}. This complements the picture of the DM interaction in chiral magnets as a Doppler shift due to an intrinsic spin current induced by spin-orbit coupling~\cite{Kikuchi2016}; in the present model, by contrast, the conduction electrons have no spin-orbit coupling, and the spin current is driven externally and treated perturbatively in the $s$-$d$ exchange. 

To examine spin-current-induced helimagnetism in the chain of Sec.~\ref{sec:four-type}, we introduce the orthonormal basis $\qty{\vb*{e}_{1},\vb*{e}_{2},\vb*{e}_{3}}$ defined for $\vb*{n}_{\mathrm{s}} \nparallel \vb*{n}_{\mathrm{c}}$ by
\begin{align}
&\vb*{e}_{1} = \vb*{e}_{2} \cross \vb*{e}_{3}
,
 \quad 
 \vb*{e}_{2} = 
 \frac{\vb*{n}_{\mathrm{s}} \cross \vb*{n}_{\mathrm{c}}}{\abs{\vb*{n}_{\mathrm{s}} \cross \vb*{n}_{\mathrm{c}}}}
 ,
 \quad
 \vb*{e}_{3} = \vb*{n}_{\mathrm{s}} 
.
\label{eq:coordinate-system}
\end{align}
For the case $\vb*{n}_{\mathrm{s}} \parallel \vb*{n}_{\mathrm{c}}$, we can set $\vb*{e}_{2}$ to an arbitrary unit vector under the condition $ \vb*{e}_{2}\perp \vb*{e}_{3}$.

We consider chains of infinite length for technical simplicity and introduce the total number $N_{\mathrm{S}}$ of localized spins only to define the energy per spin.
As in the two-spin system~\cite{Onoda2006}, the DM-type coupling generated by the spin current and oriented along $\vb*{n}_{\mathrm{s}}$ competes with the ferromagnetic exchange and makes $\vb*{n}_{\mathrm{s}}$ an effective hard axis.
Thus, we examine the following type of helix with a pitch length $2\pi/\abs{q}$:
\begin{align}
\vb*{S}_{i} &= S\qty{\cos(q\vb*{n}_{\mathrm{c}}\vdot\vb*{r}_{i}+\phi_{0})\vb*{e}_{1}+\sin(q\vb*{n}_{\mathrm{c}}\vdot\vb*{r}_{i}+\phi_{0})\vb*{e}_{2}},
\label{eq:static-S_helix}
\end{align}
where $q$ is the signed wavenumber of the helical configuration and $\phi_{0}$ is an arbitrary constant.
For $J_{\mathrm{cc}}=0$, Eq.~(\ref{eq:static-S_helix}) is a static solution of Eq.~(\ref{eq:EOM-S}), with $\pdv*{\vb*{S}_{i}}{t}=0$ for all $i$, for every $q$ and $\phi_{0}$; the equations of motion therefore do not fix $q$, which we determine below by minimizing $H_{\mathrm{eff}}\qty[\vb*{S}]$.
The above configuration gives the following scalar and vector products:
\begin{align}
&
\vb*{S}_{i} \vdot \vb*{S}_{i+m} = S^{2}\cos(mqa_{0}),
\\
&
\vb*{S}_{i} \cross \vb*{S}_{i+m} = S^{2}\sin(mqa_{0})  \vb*{e}_{3},
\end{align}
where we used $\vb*{r}_{i+m}-\vb*{r}_{i} = m \vb*{a}_{0}$ and $\vb*{a}_{0}=a_{0}\vb*{n}_{\mathrm{c}}$.

With the regularization of the system size through the parameter $N_{\mathrm{S}}$, the restriction of $\vb*{S}_{i}$ to the form of Eq.~(\ref{eq:static-S_helix}) reduces the variational functional $H_{\mathrm{eff}}\qty[\vb*{S}]$ to the variational function $H_{\mathrm{eff}}(q)$:
\begin{align}
H_{\mathrm{eff}}(q)
&=
N_{\mathrm{S}}S\tilde{\gamma}^{2}\epsilon_{\mathrm{F}}
\mathcal{F}_{\mathrm{H}}(q)
\label{eq:variational-energy}
,\\
\mathcal{F}_{\mathrm{H}}(q)
&=
\qty{
\tilde{f}^{(0)}_{\mathrm{eq}}\qty(qa_{0}) + J_{\mathrm{sc}}\tilde{f}^{(0)}_{\mathrm{sc}}\qty(qa_{0})
}
\label{eq:F_H}
,
\end{align}
where we used $\tilde{\gamma}^{2} = S\gamma_{\mathrm{ex}}^{2}$, $\vb*{e}_{3} = \vb*{n}_{\mathrm{s}}$, and
\begin{align}
&
F^{(0)}_{\mathrm{eq}}(-\vb*{r})=F^{(0)}_{\mathrm{eq}}(\vb*{r}),
\quad
\vb*{F}^{(0)}_{\mathrm{sc}}(-\vb*{r})=-\vb*{F}^{(0)}_{\mathrm{sc}}(\vb*{r})
.
\end{align}
Here $\tilde{f}^{(0)}_{\mathrm{eq}}\qty(qa_{0})$ and $\tilde{f}^{(0)}_{\mathrm{sc}}\qty(qa_{0})$ are the Fourier-transformed RKKY range functions of Sec.~\ref{sec:F-transformed_RKKY}.

The variational principle provides the variational equation to determine the realizable value of $q$ as
\begin{widetext}
\begin{align}
&
{\dv{H_{\mathrm{eff}}}{q}}(q) =0
\Longleftrightarrow
{\dv{q}}\qty{\tilde{f}^{(0)}_{\mathrm{eq}}\qty(qa_{0}) + J_{\mathrm{sc}}\tilde{f}^{(0)}_{\mathrm{sc}}\qty(qa_{0})} =0
.
\label{eq:variational-principle}
\end{align}
\end{widetext}
Among the roots of this equation, we denote by $q_{\mathrm{helix}}$ the one at which $\mathcal{F}_{\mathrm{H}}(q)$ takes its global minimum; for $k_{\mathrm{F}}a_{0}=1$ and $J_{\mathrm{sc}}=1$, $q_{\mathrm{helix}}a_{0}\approx0.238$.
Since $\mathcal{F}_{\mathrm{H}}(q_{\mathrm{helix}})<0$ and a uniform canting $\Delta$ toward $\vb*{n}_{\mathrm{s}}$ multiplies both $S^{2}-\vb*{S}_{i}\vdot\vb*{S}_{j}$ and $\vb*{n}_{\mathrm{s}}\vdot\qty(\vb*{S}_{i}\times\vb*{S}_{j})$, and hence $H_{\mathrm{eff}}(q)$, by $1-\Delta^{2}$, the planar helix of Eq.~(\ref{eq:static-S_helix}) with $q=q_{\mathrm{helix}}$ has a lower energy than the conical configurations with the same wavenumber. Moreover, because the chain is a Bravais lattice and all the DM-type vectors $\vb*{F}^{(0)}_{\mathrm{sc}}(\vb*{r}_{i}-\vb*{r}_{j})$ are parallel to $\vb*{n}_{\mathrm{s}}$, the Luttinger-Tisza method~\cite{Luttinger1946}, which yields a spiral ground state for any Heisenberg exchange on a Bravais lattice~\cite{Lyons1960}, shows that this planar helix minimizes $H_{\mathrm{eff}}[\vb*{S}]$ among all configurations of the infinite chain.
Although Eq.~(\ref{eq:variational-principle}) is difficult to solve analytically, its closed form in terms of the Clausen functions allows its roots to be found numerically.

For intuition, we first retain only the nearest-neighbor couplings.
In this approximation, the variational function and the variational principle become
\begin{align}
H_{\mathrm{eff}}(q)
&\cong 
N_{\mathrm{S}}S\tilde{\gamma}^{2}\epsilon_{\mathrm{F}}
\Bigl\{
 F^{(0)}_{\mathrm{eq}}(\vb*{a}_{0})\qty{1-\cos(qa_{0})}
 \Bigr.
 \nonumber\\
 \Bigl.
&\qquad\qquad\qquad
+\vb*{n}_{\mathrm{s}} \vdot \vb*{F}^{(0)}_{\mathrm{sc}}(\vb*{a}_{0})\sin(qa_{0})
\Bigr\}
,\\
{\dv{H_{\mathrm{eff}}}{q}}(q) 
&= 0
\Longrightarrow
\tan(qa_{0}) \cong -\vb*{n}_{\mathrm{s}}\vdot\frac{\vb*{F}^{(0)}_{\mathrm{sc}}(\vb*{a}_{0})}{F^{(0)}_{\mathrm{eq}}(\vb*{a}_{0})}
.
\label{eq:pitch-NN}
\end{align}
This result shows that the magnitude of the pitch angle, $\abs{q_{\mathrm{helix}}a_{0}}$, increases with $J_{\mathrm{sc}}$, because $\abs{\vb*{F}^{(0)}_{\mathrm{sc}}(\vb*{a}_{0})}\propto J_{\mathrm{sc}}$.
This trend agrees with the simulations in Sec.~\ref{sec:simulation}.
The nearest-neighbor approximation, however, does not predict the chirality of the helix, i.e., the sign of $q_{\mathrm{helix}}$, correctly, because the original model includes long-range interactions.
For $k_{\mathrm{F}}a_{0}=1$, for example, the nearest-neighbor approximation gives $q_{\mathrm{helix}}<0$ because $F^{(0)}_{\mathrm{eq}}(\vb*{a}_{0})>0$ and $\vb*{n}_{\mathrm{s}}\vdot \vb*{F}^{(0)}_{\mathrm{sc}}(\vb*{a}_{0})>0$. However, virtually all of $\vb*{n}_{\mathrm{s}}\vdot \vb*{F}^{(0)}_{\mathrm{sc}}(m\vb*{a}_{0})$ ($m\ge 2$) have the sign opposite to that of $\vb*{n}_{\mathrm{s}}\vdot \vb*{F}^{(0)}_{\mathrm{sc}}(\vb*{a}_{0})$ and some of $\abs{\vb*{n}_{\mathrm{s}}\vdot \vb*{F}^{(0)}_{\mathrm{sc}}(m\vb*{a}_{0})}$ ($m\ge 2$) are comparable to $\abs{\vb*{n}_{\mathrm{s}}\vdot \vb*{F}^{(0)}_{\mathrm{sc}}(\vb*{a}_{0})}$.
For the DM-type coupling, the contributions beyond the nearest neighbors thus dominate and reverse the sign, giving $q_{\mathrm{helix}}>0$.

To make the above argument on the sign of $q_{\mathrm{helix}}$ more concrete, we return to the original situation including non-nearest-neighbor interactions and adopt another approximation in which the spin current is very weak, i.e., $J_{\mathrm{sc}}\ll 1$.  In this approximation, $\abs{q_{\mathrm{helix}}a_{0}}\ll1$ also holds because $q_{\mathrm{helix}}\to0$ as $J_{\mathrm{sc}}\to0$.
Here, we expand $\sin(mq_{\mathrm{helix}}a_{0})$ and $\cos(mq_{\mathrm{helix}}a_{0})$ to lowest order in $mq_{\mathrm{helix}}a_{0}$, which is valid for $\abs{mq_{\mathrm{helix}}a_{0}}\ll1$, and cut off the $m$-summation at $m_{\Lambda}$; the terms with $\abs{mq_{\mathrm{helix}}a_{0}}\gtrsim1$, where the expansion fails, are suppressed by the oscillation of these factors.
The variational principle gives the following approximate equation for $q_{\mathrm{helix}}$:
\begin{widetext}
\begin{align}
&
\sum_{m=1}^{m_{\Lambda}}
\qty{
m^{2}F^{(0)}_{\mathrm{eq}}(m\vb*{a}_{0})qa_{0}
+m\vb*{n}_{\mathrm{s}} \vdot \vb*{F}^{(0)}_{\mathrm{sc}}(m\vb*{a}_{0})
}
\cong 0
\Longleftrightarrow
qa_{0}
\cong 
-\vb*{n}_{\mathrm{s}} \vdot
\frac{\sum_{m=1}^{m_{\Lambda}}m \vb*{F}^{(0)}_{\mathrm{sc}}(m\vb*{a}_{0})}{\sum_{m=1}^{m_{\Lambda}}m^{2}F^{(0)}_{\mathrm{eq}}(m\vb*{a}_{0})}
.
\end{align}
\end{widetext}
For $k_{\mathrm{F}}a_{0}=1$, $\vb*{n}_{\mathrm{s}} \vdot \sum_{m=2}^{m_{\Lambda}}m \vb*{F}^{(0)}_{\mathrm{sc}}(m\vb*{a}_{0})<0$ outweighs $\vb*{n}_{\mathrm{s}} \vdot \vb*{F}^{(0)}_{\mathrm{sc}}(\vb*{a}_{0})>0$, whereas $F^{(0)}_{\mathrm{eq}}(\vb*{a}_{0})>0$ dominates $\sum_{m=2}^{m_{\Lambda}}m^{2}F^{(0)}_{\mathrm{eq}}(m\vb*{a}_{0})$.
This result supports the argument of the previous paragraph on the sign of $q_{\mathrm{helix}}$.
The origin of this dominance is the nonoscillating $1/r^{2}$ tail of $f^{(0)}_{\mathrm{sc}}(x)\simeq-2\pi(1+\cos x)/x^{2}$: $m\vb*{n}_{\mathrm{s}}\vdot\vb*{F}^{(0)}_{\mathrm{sc}}(m\vb*{a}_{0})$ decays only as $1/m$, so that the numerator grows logarithmically with $m_{\Lambda}$, whereas the denominator converges. Taking $m_{\Lambda}\sim1/\abs{q_{\mathrm{helix}}a_{0}}$ yields, for $k_{\mathrm{F}}a_{0}=1$, $q_{\mathrm{helix}}a_{0}\simeq[\pi/(2D)]\,J_{\mathrm{sc}}\ln[1/(q_{\mathrm{helix}}a_{0})]$ with $D=\sum_{m\ge1}m^{2}F^{(0)}_{\mathrm{eq}}(m\vb*{a}_{0})\approx1.39$; that is, $q_{\mathrm{helix}}$ is linear in $J_{\mathrm{sc}}$ only up to a logarithmic factor, as the exact root of Eq.~(\ref{eq:variational-principle}) shows.

\subsection{Effective Lagrangian for dynamic helimagnetism}
\label{sec:helix-dynamic}
This subsection considers dynamic helimagnetism driven by a charged spin current.
We examine the following form of solutions for the equations of motion, Eq.~(\ref{eq:EOM-S}):
\begin{align}
\vb*{S}_{i}&=
S\sqrt{1-\Delta^{2}(q)}\cos(q\vb*{n}_{\mathrm{c}}\vdot\vb*{r}_{i}-\omega(q) t+\phi_{0})\vb*{e}_{1}
\nonumber\\
&\quad
+S\sqrt{1-\Delta^{2}(q)}\sin(q\vb*{n}_{\mathrm{c}}\vdot\vb*{r}_{i}-\omega(q) t+\phi_{0})\vb*{e}_{2}
\nonumber\\
&\quad
+S\Delta(q) \vb*{e}_{3}
,
\label{eq:dynamic-S_helix}
\end{align}
where $q$, $\omega(q)$, and  $\Delta(q)$ ($-1\le\Delta(q)\le1$) are the signed wavenumber of the helical configuration,  the signed frequency of coherent motion, and the signed canting ratio toward the hard axis $\vb*{e}_{3} = \vb*{n}_{\mathrm{s}}$, respectively.

Inserting Eq.~(\ref{eq:dynamic-S_helix}) into Eq.~(\ref{eq:EOM-S}) gives
\begin{align}
&
\omega(q) \cong \mathcal{G}_{\mathrm{cc}}(q)\epsilon_{\mathrm{F}}
\label{eq:omega}
,\\
&\mathcal{G}_{\mathrm{cc}}(q)
= 
\frac{J_{\mathrm{cc}}\tilde{f}^{(0)}_{\mathrm{cc}}\qty(qa_{0})}
{
\tilde{f}^{(1)}_{\mathrm{eq}}\qty(qa_{0}) + J_{\mathrm{sc}}\tilde{f}^{(1)}_{\mathrm{sc}}\qty(qa_{0})
}
\label{eq:Gcc}
,\\
&
\Delta(q) \cong
\frac{1}{2\tilde{\gamma}^{2}}\cdot
\frac{\mathcal{G}_{\mathrm{cc}}(q)}
{
\mathcal{F}_{\mathrm{H}}(q)
+ J_{\mathrm{cc}}\mathcal{G}_{\mathrm{cc}}(q) \tilde{f}^{(1)}_{\mathrm{cc}}\qty(qa_{0})
}
\label{eq:Delta}
,
\end{align}
where $\mathcal{F}_{\mathrm{H}}(q)$ is the same function as Eq.~(\ref{eq:F_H}) used in the variational energy of Eq.~(\ref{eq:variational-energy}).

Since $\mathcal{G}_{\mathrm{cc}}(q)\propto J_{\mathrm{cc}}$, $\omega(q)$ is proportional to $J_{\mathrm{cc}}$ at a given $q$, and $\Delta(q)$ is likewise proportional to $J_{\mathrm{cc}}$ to leading order for $J_{\mathrm{cc}}\ll1$; the spin-current component $J_{\mathrm{sc}}$ enters both through the denominator of Eq.~(\ref{eq:Gcc}) and through the realized wavenumber $q$, discussed below. Equation~(\ref{eq:Delta}) also shows that $\Delta(q)$ is inversely proportional to $\tilde{\gamma}^{2}$.
Because $\abs{\Delta(q)}\le1$, however, the limit $\tilde{\gamma}^{2}\to0$ cannot be taken at a fixed nonzero $J_{\mathrm{cc}}$: below the value of $\tilde{\gamma}^{2}$ at which $\abs{\Delta(q)}$ reaches $1$, no helix of the form of Eq.~(\ref{eq:dynamic-S_helix}) exists (Sec.~\ref{sec:sim-dynamic}).
For later discussion on numerical simulations, we introduce the dimensionless frequency $\tilde{\omega}(q)$ as follows:
\begin{align}
\tilde{\omega}(q) \tilde{t} = \omega(q) t
\Longrightarrow 
\tilde{\omega}(q) = \frac{\omega(q)}{\tilde{\gamma}^{2}\epsilon_{\mathrm{F}}} =  \frac{1}{\tilde{\gamma}^{2}}\mathcal{G}_{\mathrm{cc}}(q)
\label{eq:omega-tilde}
,
\end{align}
where $\tilde{t}$ is the dimensionless time defined by $\tilde{t}=\tilde{\gamma}^{2}\epsilon_{\mathrm{F}}t$.
Because $\omega(q)$ is independent of $\tilde{\gamma}^{2}$, $\tilde{\omega}(q)$ is inversely proportional to $\tilde{\gamma}^{2}$.
This suggests that, in the numerical simulations with the dimensionless time $\tilde{t}$, the coherent motion of dynamic helimagnetism slows down as $\tilde{\gamma}^{2}$ increases. 

We now turn to the realized value of $q$, the counterpart of $q_{\mathrm{helix}}$ in the static case of Sec.~\ref{sec:helix-static}.
Strictly speaking, we cannot determine such a wavenumber in the dynamic case because we cannot construct an effective Hamiltonian, unlike in the static case.
Indeed, for the infinite chain the ansatz of Eq.~(\ref{eq:dynamic-S_helix}) yields $\omega(q)$ and $\Delta(q)$ for each $q$, so that the equations of motion alone do not fix $q$; its selection is a question of dynamical stability, which we do not address here.

For $J_{\mathrm{cc}}\ll1$, the helix moves slowly, $\omega(q)/\epsilon_{\mathrm{F}}\ll1$, so that one might expect $H_{\mathrm{eff}}\qty[\vb*{S}]$ of Eq.~(\ref{eq:Heff}) to work as an energy functional as in the static case.
Inserting Eq.~(\ref{eq:dynamic-S_helix}) into Eq.~(\ref{eq:Heff}) gives the static variational function of Eq.~(\ref{eq:variational-energy}) rescaled by the factor $1-\Delta^{2}(q)$:
\begin{align}
H_{\mathrm{eff}}(q)
&=
N_{\mathrm{S}}S\tilde{\gamma}^{2}\epsilon_{\mathrm{F}}
\qty{1-\Delta^{2}(q)}
\mathcal{F}_{\mathrm{H}}(q)
\label{eq:variational-energy_dynamic}
.
\end{align} 
However, because $\Delta(q)\propto1/\tilde{\gamma}^{2}$, this variational function makes $q_{\mathrm{helix}}$ depend on $\tilde{\gamma}^{2}$, whereas the simulations of Sec.~\ref{sec:sim-dynamic} show no such dependence.

Another candidate for the variational functional in a coherent motion with period $T$ is the following effective Lagrangian:
\begin{widetext}
\begin{align}
L_{\mathrm{eff}}\qty[\vb*{S}]
&= T_{\mathrm{eff}}\qty[\vb*{S}]-V_{\mathrm{eff}}\qty[\vb*{S}],
\\
T_{\mathrm{eff}}\qty[\vb*{S}]
&=
-\frac{1}{T}\int_{0}^{T} \dd{t}
\frac{1}{S^{2}}\int_{0}^{1} \dd{\tau}\sum_{i}
 \vb*{S}_{i}(\tau, t)\vdot \qty{{\pdv{ \vb*{S}_{i}}{\tau}} (\tau, t) \cross {\pdv{\vb*{S}_{i}}{t}}(\tau, t)}
 ,\\
V_{\mathrm{eff}}\qty[\vb*{S}]
&=
-\frac{1}{T}\int_{0}^{T} \dd{t}
\frac{J_{\mathrm{ex}}v_{\mathrm{uc}}}{2}
\sum_{i}\vb*{S}_{i}(t)\vdot\ev{\hat{\vb*{\sigma}}(t, \vb*{r}_{i})}_{\vb*{S}^{\Delta}=0}
.
\end{align}
\end{widetext}
The first term $T_{\mathrm{eff}}\qty[\vb*{S}]$ is the sum of averaged geometric phases due to the periodic motion of each spin.
More precisely, $-T_{\mathrm{eff}}\qty[\vb*{S}]$ is $S/T$ times the sum of the oriented solid angles surrounded by the trajectories of the spins in the single period $T$.
The second term $-V_{\mathrm{eff}}\qty[\vb*{S}]$ can be roughly interpreted as an effective Hamiltonian multiplied by $(-1)$.
Inserting the helix of Eq.~(\ref{eq:dynamic-S_helix}) into $L_{\mathrm{eff}}\qty[\vb*{S}]$, with $T\to2\pi/\abs{\omega(q)}$, with the signs of $\omega(q)$ and $\Delta(q)$ taken relative to $\vb*{n}_{\mathrm{s}}$ in $T_{\mathrm{eff}}\qty[\vb*{S}]$, and with the regularization $-\vb*{S}_{i}\vdot\vb*{S}_{j}\to S^{2}-\vb*{S}_{i}\vdot\vb*{S}_{j}$ in $V_{\mathrm{eff}}\qty[\vb*{S}]$,
we obtain the following function:
\begin{align}
L_{\mathrm{eff}}(q)
&= T_{\mathrm{eff}}(q)-V_{\mathrm{eff}}(q),
\\
T_{\mathrm{eff}}(q)
&=N_{\mathrm{S}}S\qty{\pm 1-\Delta(q)}\omega(q)
 ,\\
V_{\mathrm{eff}}(q)
&= 2N_{\mathrm{S}}S\tilde{\gamma}^{2}\epsilon_{\mathrm{F}}
\qty{1-\Delta^{2}(q)}
\nonumber\\
&\quad\times
\qty{
\mathcal{F}_{\mathrm{H}}(q)
+ J_{\mathrm{cc}}\mathcal{G}_{\mathrm{cc}}(q) \tilde{f}^{(1)}_{\mathrm{cc}}\qty(qa_{0})
}
.
\end{align}
Note that the solid angle $\Omega(q)$ surrounded by the trajectory of each spin in the single period $2\pi/\abs{\omega(q)}$ is given by
\begin{align}
\Omega(q) & =  -2\pi\qty{\pm 1-\Delta\qty(q)} \mathrm{sgn}\qty(\omega(q))
,
\end{align}
where the upper sign corresponds to the solid angle on the side of $\vb*{n}_{\mathrm{s}}$, whose sign is opposite to that of $\omega(q)$.
The ambiguity $\pm N_{\mathrm{S}}S\omega(q)$ in $T_{\mathrm{eff}}(q)$ comes from the ambiguity $4\pi$ in solid angle, and the factor $2$ in $V_{\mathrm{eff}}(q)$ comes from the double counting of coupling between spins.
For the helix of Eq.~(\ref{eq:dynamic-S_helix}), $\vb*{S}_{i}\vdot\vb*{S}_{i+m}$ and $\vb*{n}_{\mathrm{s}}\vdot\qty(\pdv*{\vb*{S}_{i+m}}{t}\times\vb*{S}_{i})$ are even in $m$, whereas $\vb*{n}_{\mathrm{s}}\vdot\qty(\vb*{S}_{i+m}\times\vb*{S}_{i})$ and $\vb*{S}_{i}\vdot\pdv*{\vb*{S}_{i+m}}{t}$ are odd; because $F^{(n)}_{\mathrm{eq}}(\vb*{r})$ is even and $F^{(n)}_{\mathrm{cc}}(\vb*{r})$ and $\vb*{F}^{(n)}_{\mathrm{sc}}(\vb*{r})$ are odd in $\vb*{r}$, only the terms with $F^{(0)}_{\mathrm{eq}}(\vb*{r})$, $\vb*{F}^{(0)}_{\mathrm{sc}}(\vb*{r})$, and $F^{(1)}_{\mathrm{cc}}(\vb*{r})$ contribute to $V_{\mathrm{eff}}(q)$.
Note that the first term of  $V_{\mathrm{eff}}(q)$ equals $2H_{\mathrm{eff}}(q)$ given in Eq.~(\ref{eq:variational-energy_dynamic}).
Here, we omit the ambiguity in $T_{\mathrm{eff}}(q)$, as $T_{\mathrm{eff}}(q) \to -N_{\mathrm{S}}S\Delta(q)\omega(q)$, to remove the similar ambiguity from the variational equation.
On the other hand, we retain the factor $2$  in $V_{\mathrm{eff}}(q)$ to keep the ratio between the first and second terms.
Finally, we reach the following variational function by using the relation between $\omega(q)$ and $\Delta(q)$ in Eqs.~(\ref{eq:omega}) and (\ref{eq:Delta}):
\begin{align}
L_{\mathrm{eff}}(q)
&= -2N_{\mathrm{S}}S\tilde{\gamma}^{2}\epsilon_{\mathrm{F}}
\qty{
\mathcal{F}_{\mathrm{H}}(q)
+ J_{\mathrm{cc}}\mathcal{G}_{\mathrm{cc}}(q) \tilde{f}^{(1)}_{\mathrm{cc}}\qty(qa_{0})
}
.
\end{align}
Note that the first term of  $L_{\mathrm{eff}}(q)$ equals $-2H_{\mathrm{eff}}(q)$ in the static case of Eq.~(\ref{eq:variational-energy}).
Thus, for $J_{\mathrm{cc}}\ll 1$, we can approximate the solution $q_{\mathrm{helix}}$ based on $L_{\mathrm{eff}}(q)$ well by the solution in the static case.
The second term in $L_{\mathrm{eff}}(q)$ makes $q_{\mathrm{helix}}$ depend on $J_{\mathrm{cc}}$.
Section~\ref{sec:sim-dynamic} compares this dependence with the simulations.

We emphasize that the two prescriptions above---dropping the $4\pi$ ambiguity of the solid angle in $T_{\mathrm{eff}}(q)$ and retaining the factor $2$ in $V_{\mathrm{eff}}(q)$---are not derived from first principles; they are heuristic choices, and $L_{\mathrm{eff}}(q)$ constructed in this way should be regarded as a working hypothesis. It is tested a posteriori in Sec.~\ref{sec:simulation}: the wavenumber $q_{\mathrm{helix}}$ obtained from its variation, together with $\omega$ and $\Delta$ obtained by substituting $q_{\mathrm{helix}}$ into Eqs.~(\ref{eq:omega}) and (\ref{eq:Delta}), agrees with the simulations, typically to within a few percent.
We adopt $L_{\mathrm{eff}}(q)$ as the variational function that determines $q_{\mathrm{helix}}$: the predicted $q_{\mathrm{helix}}$ is the root of the variational equation
\begin{widetext}
\begin{align}
&{\dv{L_{\mathrm{eff}}}{q}}(q) = 0
\Longleftrightarrow
{\dv{}{q}}
\qty{
\tilde{f}^{(0)}_{\mathrm{eq}}\qty(qa_{0}) + J_{\mathrm{sc}}\tilde{f}^{(0)}_{\mathrm{sc}}\qty(qa_{0})
+
J_{\mathrm{cc}}^{2}
\frac{\tilde{f}^{(0)}_{\mathrm{cc}}\qty(qa_{0}) \tilde{f}^{(1)}_{\mathrm{cc}}\qty(qa_{0})}
{
\tilde{f}^{(1)}_{\mathrm{eq}}\qty(qa_{0}) + J_{\mathrm{sc}}\tilde{f}^{(1)}_{\mathrm{sc}}\qty(qa_{0})
}
}
= 0
.
\label{eq:dLeff}
\end{align}
\end{widetext}
Among its roots, we take the one that reduces to the static $q_{\mathrm{helix}}$ as $J_{\mathrm{cc}}\to0$, following it continuously as $J_{\mathrm{cc}}$ increases.
Although this equation is difficult to solve analytically, as in the static case, it shows that the predicted $q_{\mathrm{helix}}$ depends on $J_{\mathrm{cc}}^{2}$ and $J_{\mathrm{sc}}$ but not on $\tilde{\gamma}$.
Because Eq.~(\ref{eq:EOM-S}) retains the response of the conduction electrons only to linear order in the currents, this $J_{\mathrm{cc}}^{2}$ dependence is a property of Eq.~(\ref{eq:EOM-S}) as a model; terms of second order in the currents, omitted from the response, could contribute to $q_{\mathrm{helix}}$ at the same order.
Reversing only the charge-current component, i.e., the sign of $J_{\mathrm{cc}}$, reverses $\omega$ and $\Delta$ [Eqs.~(\ref{eq:omega})--(\ref{eq:Delta})] but not $q_{\mathrm{helix}}$ [Eq.~(\ref{eq:dLeff})]; because $\tilde{f}^{(n)}_{\mathrm{eq}}\qty(qa_{0})$ is even and $\tilde{f}^{(n)}_{\mathrm{cc}}\qty(qa_{0})$ and $\tilde{f}^{(n)}_{\mathrm{sc}}\qty(qa_{0})$ are odd in $q$, reversing only the spin-current component, i.e., the sign of $J_{\mathrm{sc}}$, reverses all three. Hence the spin current alone reverses the chirality, and either current reverses the sense of precession.

For more general cases than those discussed above, we must numerically solve the equations of motion directly.
Our previous analysis focused on the two-spin system and found a region in the $(J_{\mathrm{cc}}, J_{\mathrm{sc}})$ plane where the spin motion can be chaotic. Although various types of motions can occur in this region, we could not identify any sharp boundaries between them, except for a relatively clear change around the dotted line in the phase diagram of the two-spin system of Ref.~\cite{Onoda2006} (Fig.~\ref{fig:phase-diagram}).
Below the dotted line, the upstream spin slowly precesses around $\vb*{n}_{\mathrm{s}}$ with a cycloidal trajectory, and the downstream spin quickly precesses around the upstream spin. Above the dotted line, the motion becomes chaotic. This type of motion is considered to originate from the nonlinearity with the damping and pumping effects, and may be interpreted in a similar manner to forced oscillations~\cite{Thompson2002}. Note that a large amount of strongly spin-polarized charge current leads to this class of motion, where adiabatic spin-transfer theories predict that the uniform ferromagnetic order becomes unstable against spin waves~\cite{Bazaliy1998, Fernandez-Rossier2004} and breaks up into domains above a critical spin current~\cite{Shibata2005}. 

\section{Numerical simulations}
\label{sec:simulation}
Building on the analytical framework of Sec.~\ref{sec:helix-theory}, we now numerically integrate the equations of motion, Eq.~(\ref{eq:EOM-S}), for a finite-size one-dimensional spin chain to test the predicted static and dynamic helimagnetism directly. 
We simulated a chain of $N_{\mathrm{S}}=60$ spins along the $x$ axis. Because the right-hand side of Eq.~(\ref{eq:EOM}) contains the time derivatives $\partial_{\tilde{t}}\vb*{\Omega}_{j}$, we collect all spins into a single $3N_{\mathrm{S}}$-component vector $\vb*{\Omega}$ and write Eq.~(\ref{eq:EOM}) as $(I-M[\vb*{\Omega}])\,\partial_{\tilde{t}}\vb*{\Omega}=\vb*{T}[\vb*{\Omega}]$, where the matrix $M[\vb*{\Omega}]$ collects the coefficients of the time-derivative terms and $\vb*{T}[\vb*{\Omega}]$ the remaining torques. At each evaluation of the right-hand side, $\partial_{\tilde{t}}\vb*{\Omega}=(I-M[\vb*{\Omega}])^{-1}\vb*{T}[\vb*{\Omega}]$ is computed and integrated with the fourth-order Runge-Kutta method, and each spin is renormalized to unit length after every step. This procedure generalizes the standard conversion of the Landau-Lifshitz-Gilbert equation~\cite{Landau1935, Gilbert2004} into the Landau-Lifshitz form to the nonlocal damping terms; the helical solutions of Sec.~\ref{sec:helix-theory} satisfy Eq.~(\ref{eq:EOM}) as it stands.

In all cases below, the electron flow direction is the positive $x$ direction, $\vb*{n}_{\mathrm{c}}=\vb*{e}_{x}$, and the time step is $d\tilde{t}=1/100$. Unless otherwise stated, the initial configuration is a uniform ferromagnetic state along $\vb*{e}_{z}$ for $\vb*{n}_{\mathrm{s}}=\vb*{e}_{x}$ and along $\vb*{e}_{y}$ for $\vb*{n}_{\mathrm{s}}=\vb*{e}_{z}$, i.e., perpendicular to $\vb*{n}_{\mathrm{s}}$, because the uniform states along $\pm\vb*{n}_{\mathrm{s}}$ are stationary. Starting from other uniform directions, twisted configurations, or random configurations leads to the same helical state apart from a global phase, although small residual fluctuations persist for random initial configurations.
We adopt open (free-end) boundary conditions, with no coupling beyond the two termini of the chain; to minimize the resulting edge effects, all quantitative comparisons with the analytical predictions below are restricted to the two central spins of the chain (Sec.~\ref{sec:sim-dynamic}).

In most examples below we take $J_{\mathrm{sc}}=1.000$ for visual clarity; smaller spin-current densities yield longer helical pitches and therefore require longer chains (see Sec.~\ref{sec:sim-dynamic}). As shown in Sec.~\ref{sec:four-type}, this spin-current amplitude corresponds to drift velocities $(J_{\mathrm{cc}}\pm1)(v_{\mathrm{F}}/2)$ of the two spin species, which lie beyond the regime where the linearized distribution of Eq.~(\ref{eq:distribution_function}) is quantitatively controlled; the physically controlled regime is $J_{\mathrm{cc}}, J_{\mathrm{sc}}\ll 1$. Because the simulations integrate Eq.~(\ref{eq:EOM-S}), which is linear in $J_{\mathrm{cc}}$ and $J_{\mathrm{sc}}$ by construction, however, the comparisons below test the analysis of this equation in Sec.~\ref{sec:helix-theory} whether or not the currents lie in this regime. Similarly, faster motion, such as the transient approach to the helix, fluctuating or irregular motion, and the precession at larger $J_{\mathrm{cc}}$, need not satisfy the condition $\omega_{\mathrm{S}}\ll\eta$ of Sec.~\ref{sec:Approximations}; Eq.~(\ref{eq:EOM-S}) is then integrated as a model in its own right. The same applies to the larger values of $\tilde{\gamma}^{2}$ in Sec.~\ref{sec:sim-dynamic} (up to $0.4$), which we use to explore the behavior of the model: at these values, the Born parameter of Sec.~\ref{sec:Model}, of the order of $\tilde{\gamma}\sqrt{S}$, is not small even for $S=1/2$ (about $0.45$), whereas a large spin with a small Born parameter corresponds to a small $\tilde{\gamma}^{2}$ (Sec.~\ref{sec:EOM}). Animated visualizations of the spin dynamics for six representative parameter sets, spanning the static helimagnetic, dynamic helimagnetic, ferromagnetic, and chaotic regimes discussed below, are provided as Supplemental Material~\cite{Note1}. 

\subsection{Static helimagnetism by pure spin current}
\label{sec:sim-static}
\begin{figure}[htb]
  \centering
  \begin{overpic}[width=0.48\linewidth]{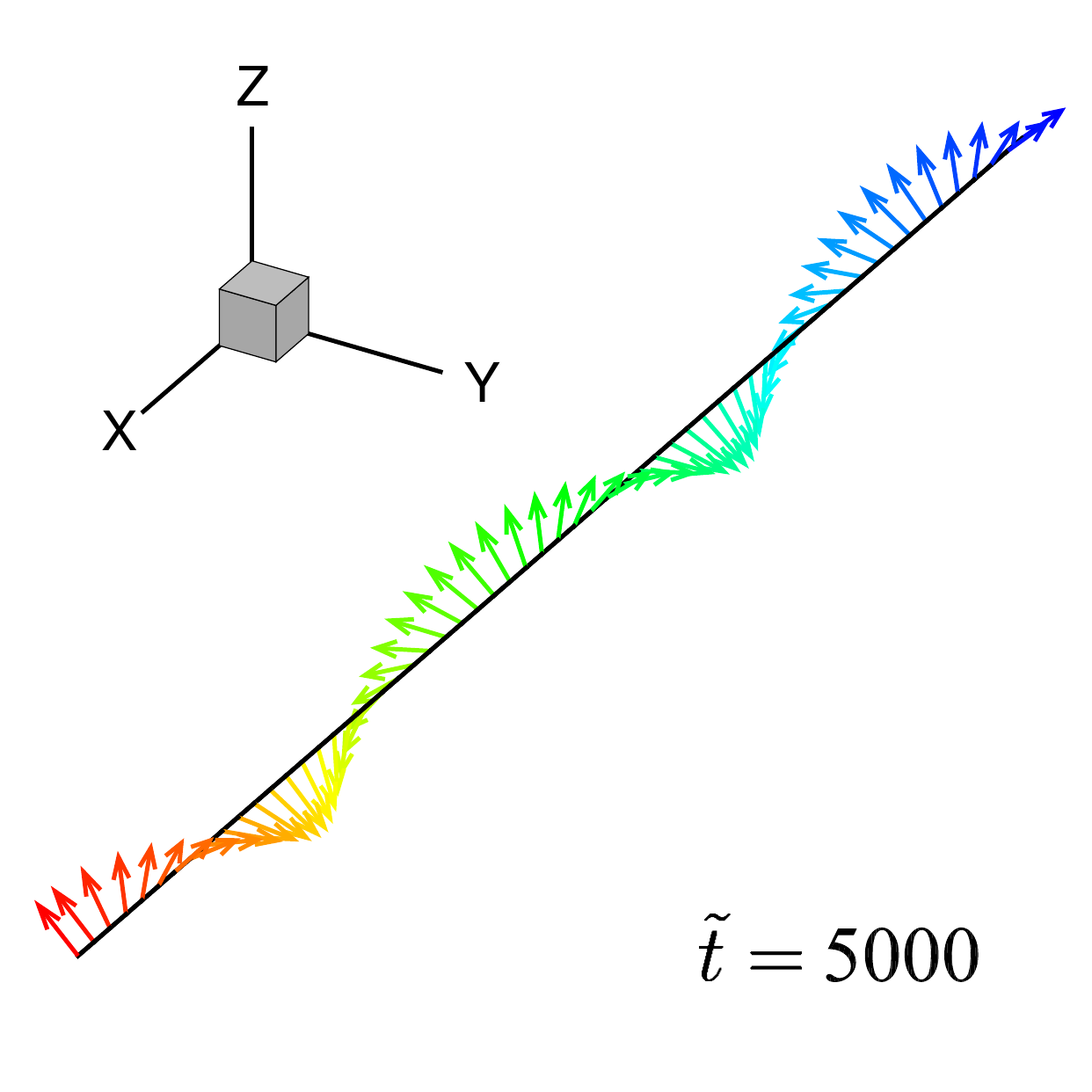}
  \put(0,95){(a)}
  \end{overpic}
  \hfill
  \begin{overpic}[width=0.48\linewidth]{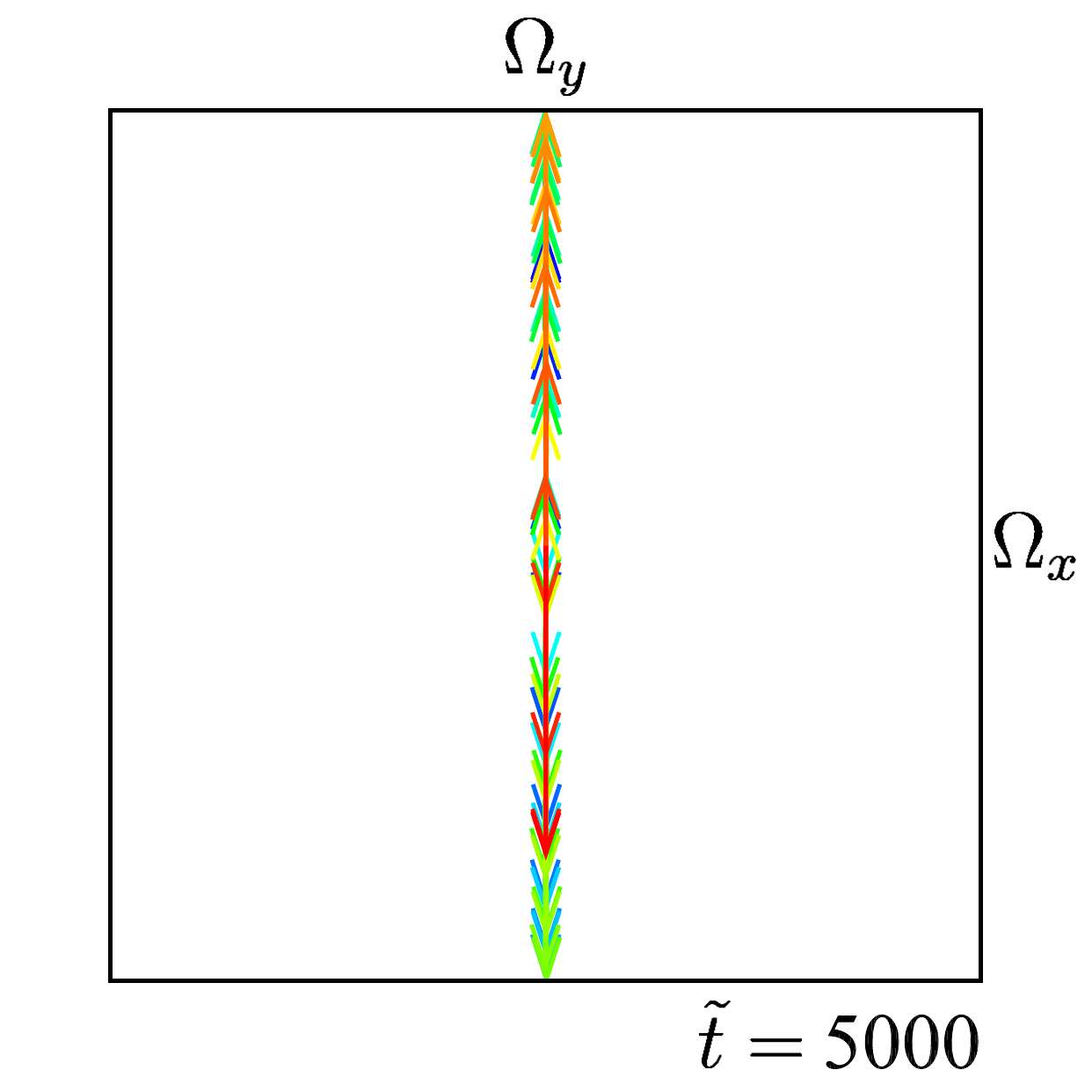}
  \put(10,95){(b)}
  \end{overpic}
  \hfill
  \begin{overpic}[width=0.48\linewidth]{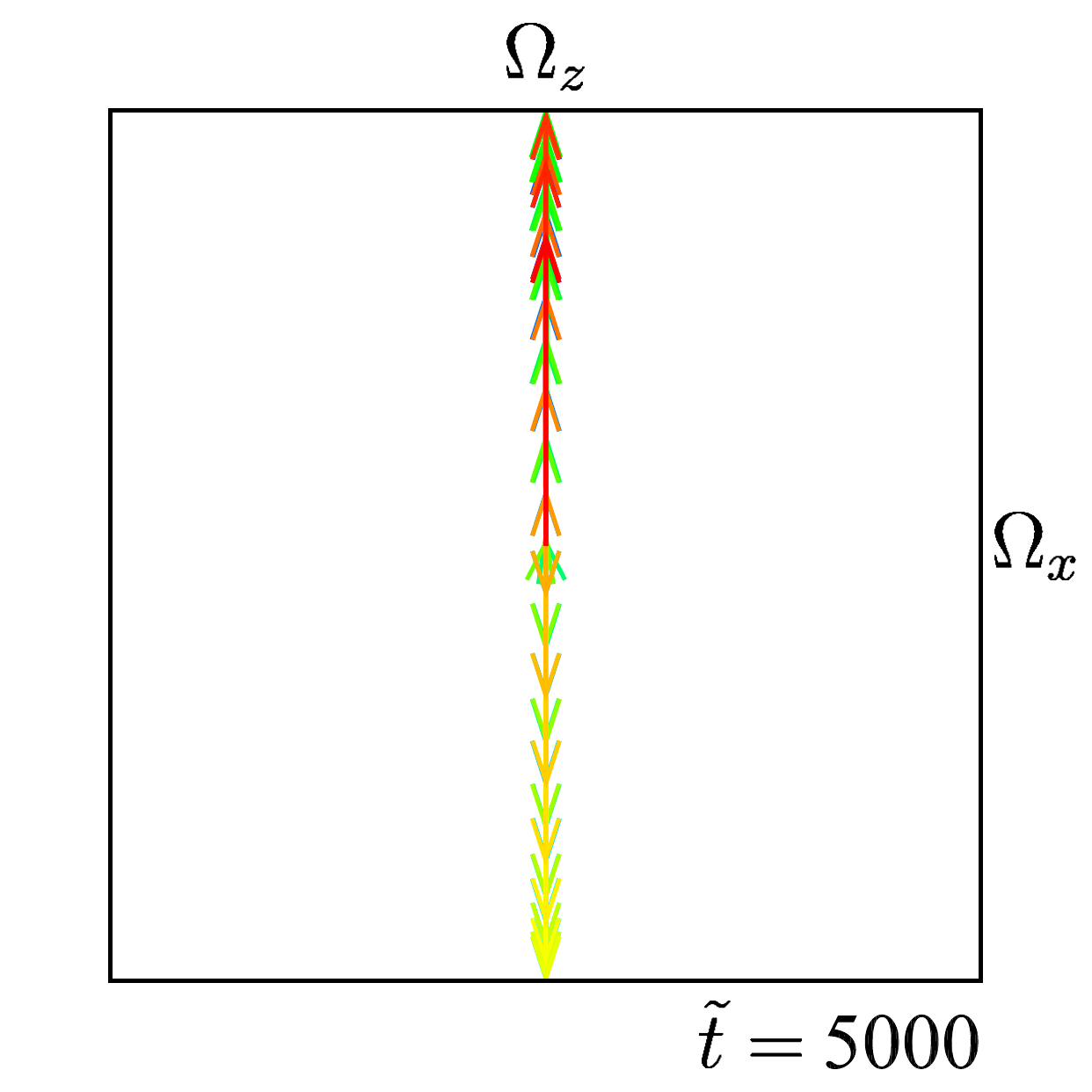}
  \put(10,95){(c)}
  \end{overpic}
   \hfill
  \begin{overpic}[width=0.48\linewidth]{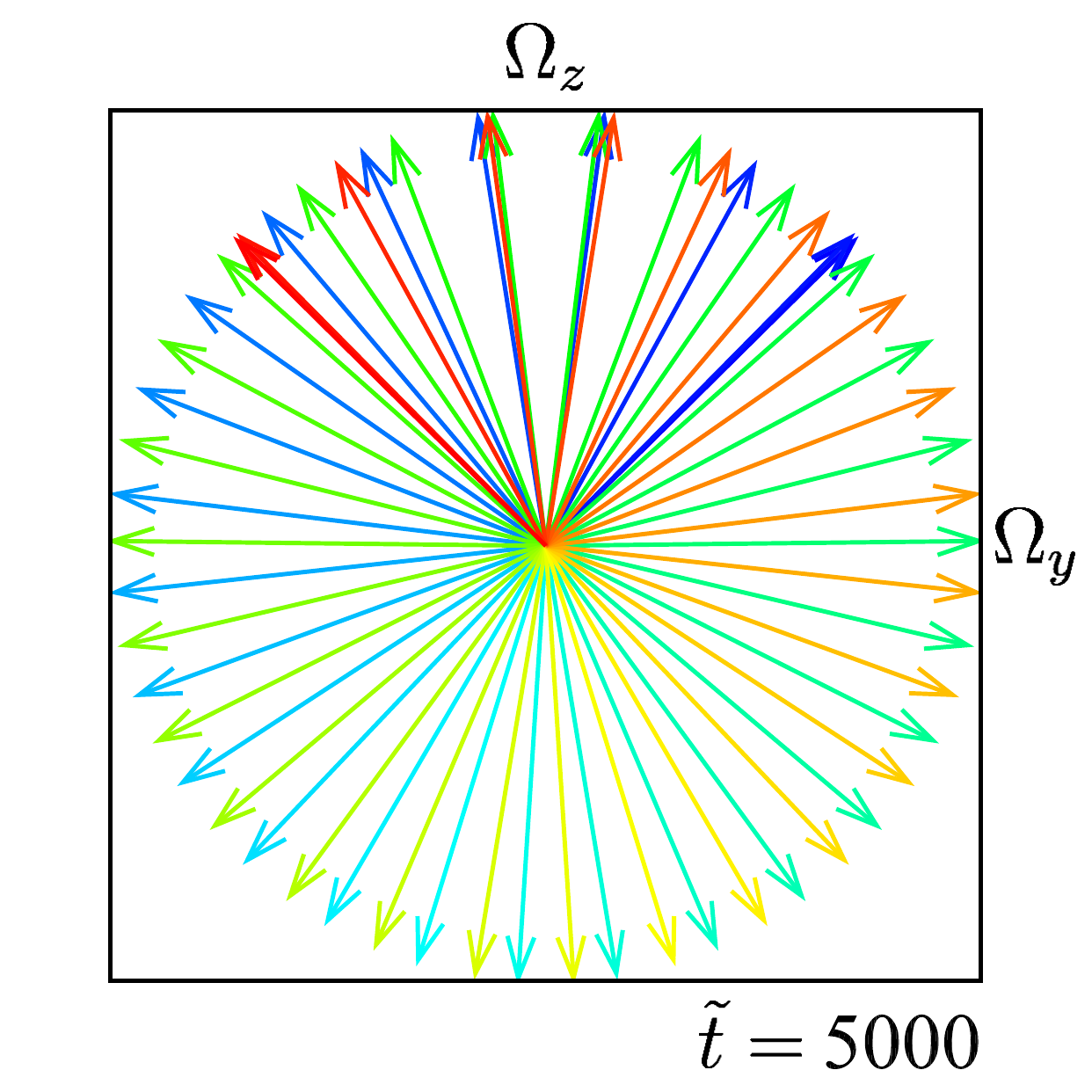}
  \put(10,95){(d)}
  \end{overpic}
\caption{
Snapshots at $\tilde{t} = 5000$ of a spin chain consisting of 60 spins equally spaced along the $x$~direction, 
for $\tilde{\gamma}^2=0.100$, $J_{\mathrm{cc}} = 0.000$, $J_{\mathrm{sc}} = 1.000$, $\vb*{n}_{\mathrm{c}} = \vb*{e}_{x}$, and $\vb*{n}_{\mathrm{s}} = \vb*{e}_{x}$.
(a)~Bird's-eye view of the entire spin chain.
(b)~Normalized $x$ and $y$ components of the spins.
(c)~Normalized $x$ and $z$ components of the spins.
(d)~Normalized $y$ and $z$ components of the spins.
}
\label{fig:snapshots_Jcc0.000_Jsc1.000_nc100_ns100}
\end{figure}

\begin{figure}[htb]
  \centering
  \begin{overpic}[width=0.48\linewidth]{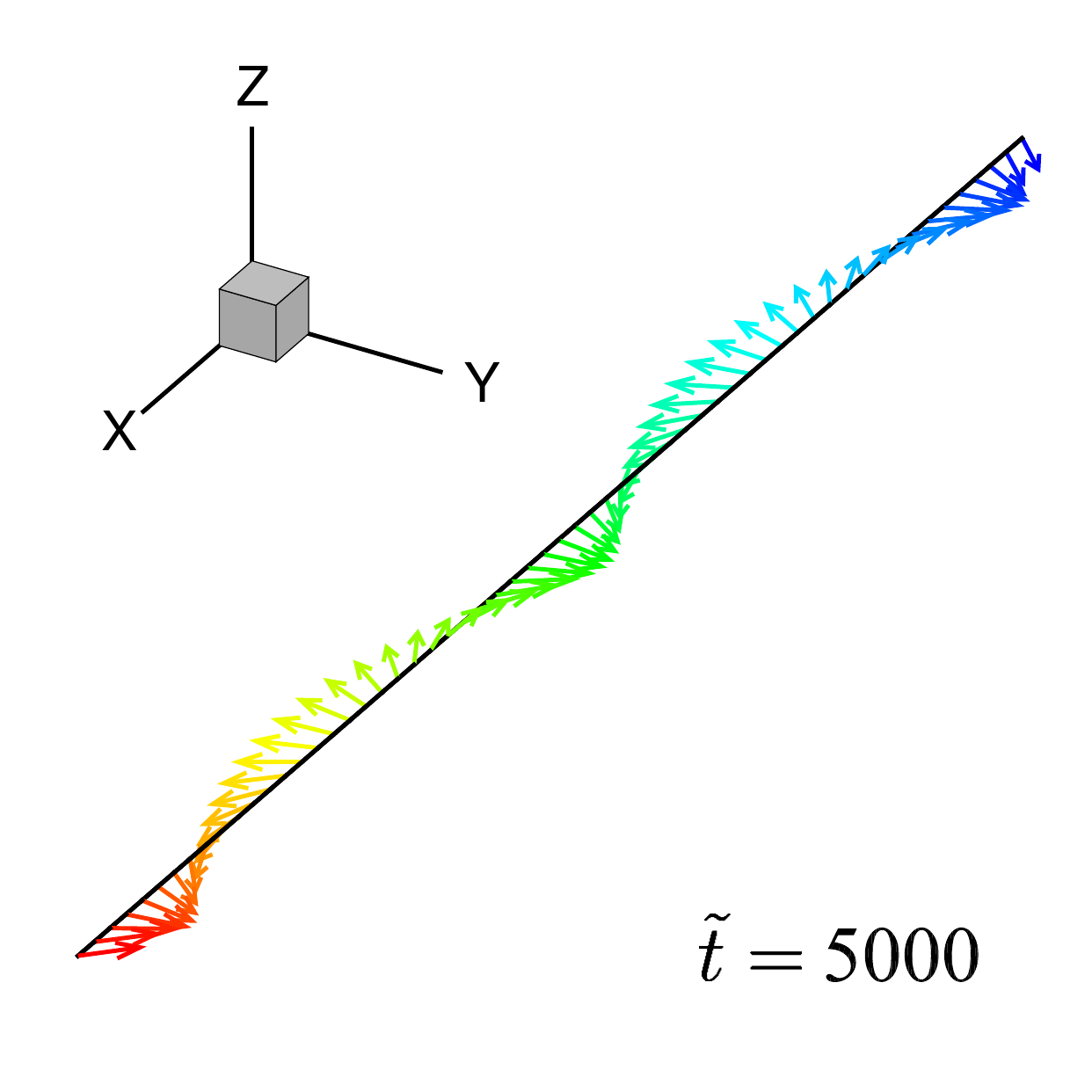}
  \put(0,95){(a)}
  \end{overpic}
  \hfill
  \begin{overpic}[width=0.48\linewidth]{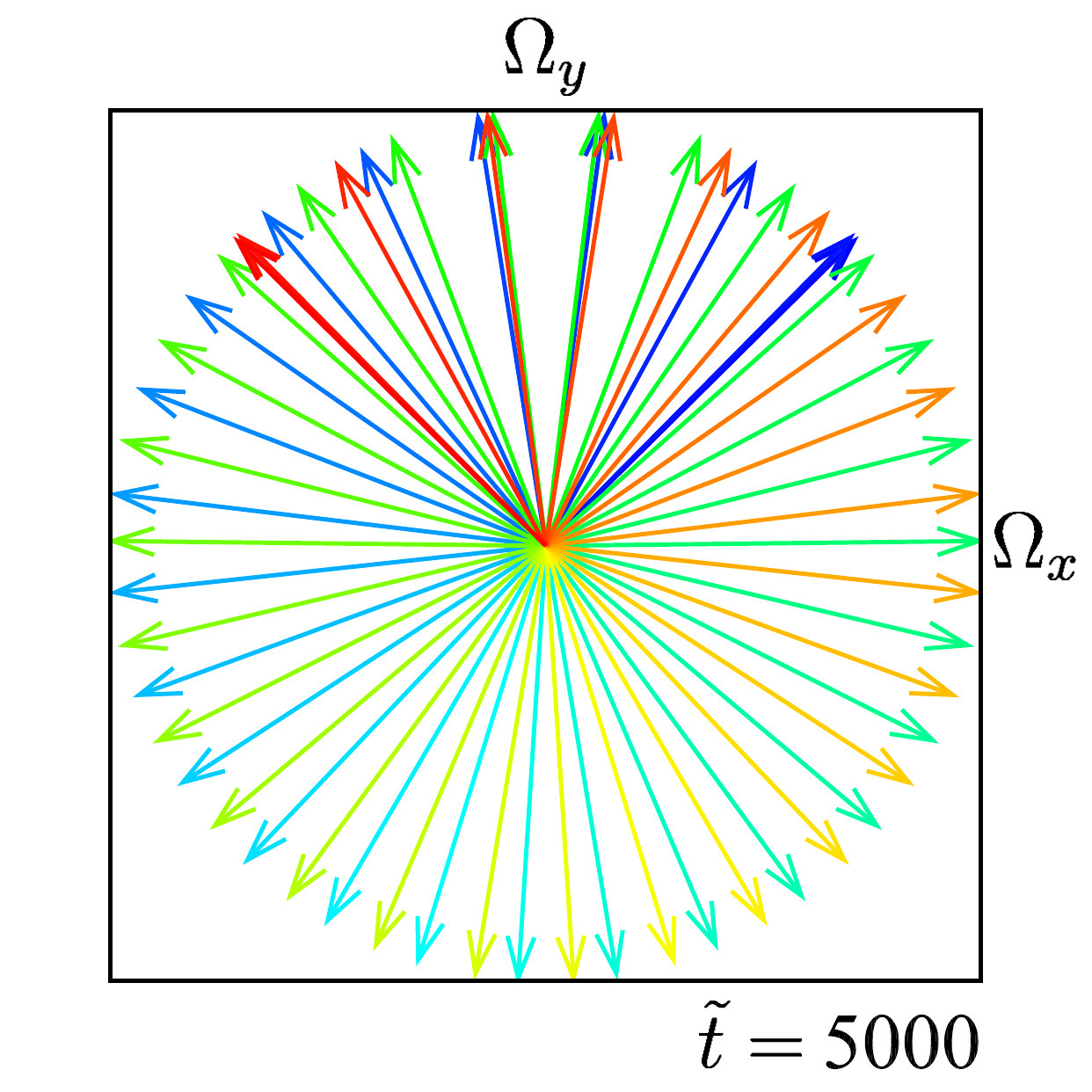}
  \put(10,95){(b)}
  \end{overpic}
  \hfill
  \begin{overpic}[width=0.48\linewidth]{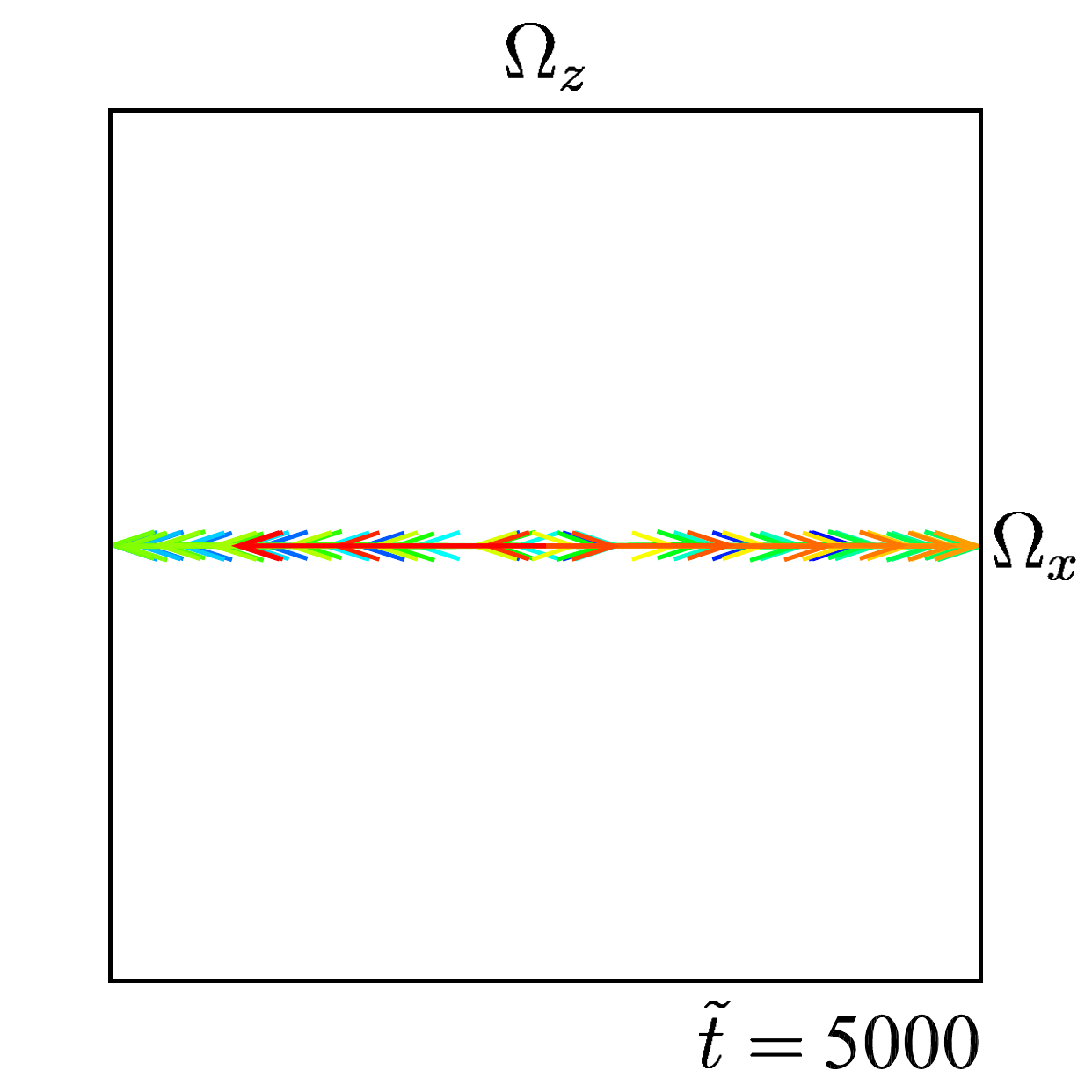}
  \put(10,95){(c)}
  \end{overpic}
   \hfill
  \begin{overpic}[width=0.48\linewidth]{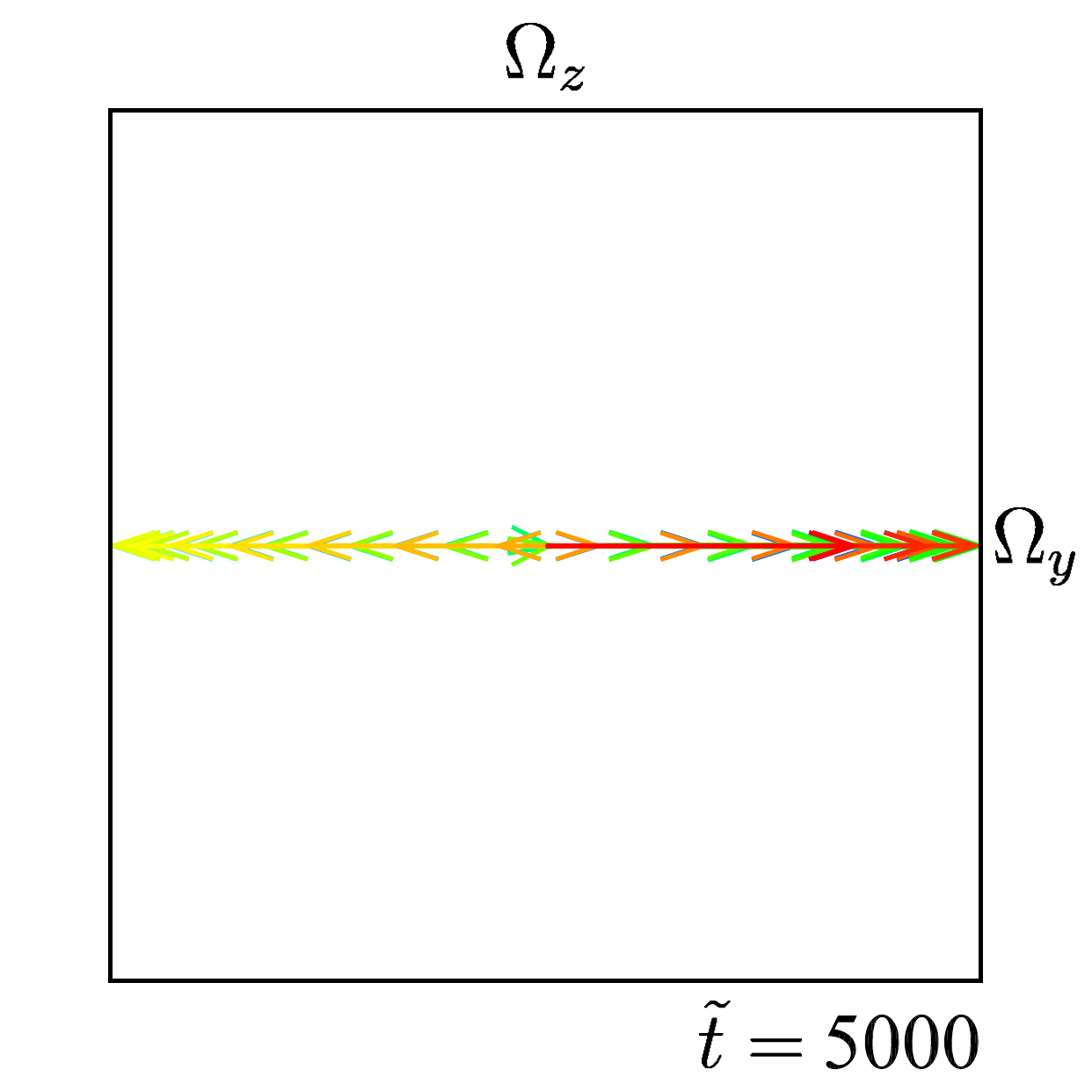}
  \put(10,95){(d)}
  \end{overpic}
\caption{
Snapshots at $\tilde{t} = 5000$ of a spin chain consisting of 60 spins equally spaced along the $x$~direction, 
for $\tilde{\gamma}^2=0.100$, $J_{\mathrm{cc}} = 0.000$, $J_{\mathrm{sc}} = 1.000$, $\vb*{n}_{\mathrm{c}} = \vb*{e}_{x}$, and $\vb*{n}_{\mathrm{s}} = \vb*{e}_{z}$.
(a)~Bird's-eye view of the entire spin chain.
(b)~Normalized $x$ and $y$ components of the spins.
(c)~Normalized $x$ and $z$ components of the spins.
(d)~Normalized $y$ and $z$ components of the spins.
}
\label{fig:snapshots_Jcc0.000_Jsc1.000_nc100_ns001}
\end{figure}

Before turning to dynamic helimagnetism, we test the predicted static helimagnetism and its dependence on the spin polarization $\vb*{n}_{\mathrm{s}}$ of the spin current. Figures~\ref{fig:snapshots_Jcc0.000_Jsc1.000_nc100_ns100} and \ref{fig:snapshots_Jcc0.000_Jsc1.000_nc100_ns001} show static helimagnetism for $\vb*{n}_{\mathrm{s}}= \vb*{e}_{x}$ and $\vb*{n}_{\mathrm{s}}= \vb*{e}_{z}$, respectively. In both cases, the spins rotate in the plane perpendicular to $\vb*{n}_{\mathrm{s}}$, so that $\vb*{n}_{\mathrm{s}}$ is the effective hard axis of the helix (Sec.~\ref{sec:four-type}).
Relative to the chain, however, the two helices differ: the plane of rotation is perpendicular to the chain for $\vb*{n}_{\mathrm{s}}=\vb*{e}_{x}$, as in a proper-screw helix, and contains the chain for $\vb*{n}_{\mathrm{s}}=\vb*{e}_{z}$, as in a cycloid. Which of the two forms appears is set here by the spin polarization of the current, whereas in chiral magnets it is set by the crystal symmetry through the direction of the DM vector relative to the bonds~\cite{Moriya1960}.
Because Eq.~(\ref{eq:EOM-S}) contains no spin-orbit coupling, it is invariant under a global rotation of all spins together with $\vb*{n}_{\mathrm{s}}$; the rotation that takes $\vb*{e}_{x}$ to $\vb*{e}_{z}$ and the initial direction $\vb*{e}_{z}$ to $\vb*{e}_{y}$ maps the simulation of Fig.~\ref{fig:snapshots_Jcc0.000_Jsc1.000_nc100_ns100} onto that of Fig.~\ref{fig:snapshots_Jcc0.000_Jsc1.000_nc100_ns001}. We therefore consider only $\vb*{n}_{\mathrm{s}}=\vb*{e}_{x}$ hereafter.

The static helix also agrees quantitatively with the effective-Hamiltonian analysis of Sec.~\ref{sec:helix-static}. For $J_{\mathrm{sc}}=1.000$, the simulated wavenumber is $q_{\mathrm{helix}}a_{0}\approx0.241$, close to the value $q_{\mathrm{helix}}a_{0}\approx0.238$ that minimizes $H_{\mathrm{eff}}(q)$; the difference of about $1\%$ is comparable to the finite-size deviations discussed in Sec.~\ref{sec:sim-dynamic}. The chain of 60 spins accommodates about two and a quarter turns of the helix [Fig.~\ref{fig:snapshots_Jcc0.000_Jsc1.000_nc100_ns100}(a)]. Although Eq.~(\ref{eq:EOM-S}) admits a static helix of the form of Eq.~(\ref{eq:static-S_helix}) for every $q$ (Sec.~\ref{sec:helix-static}), the chain selects the one that minimizes $H_{\mathrm{eff}}(q)$.

The sign of $q_{\mathrm{helix}}$ gives the chirality of the helix. For $q_{\mathrm{helix}}>0$, the vector spin chirality $\vb*{S}_{i}\times\vb*{S}_{i+1}=S^{2}\sin(q_{\mathrm{helix}}a_{0})\vb*{n}_{\mathrm{s}}$, with $\vb*{S}_{i+1}$ downstream of $\vb*{S}_{i}$ along $\vb*{n}_{\mathrm{c}}$, is parallel to $\vb*{n}_{\mathrm{s}}$; in Fig.~\ref{fig:snapshots_Jcc0.000_Jsc1.000_nc100_ns100}, where $\vb*{n}_{\mathrm{s}}=\vb*{n}_{\mathrm{c}}$, the spins thus form a right-handed screw along the electron flow. The nearest-neighbor approximation, Eq.~(\ref{eq:pitch-NN}), predicts the opposite chirality, $q_{\mathrm{helix}}a_{0}\approx-0.36$ for $J_{\mathrm{sc}}=1.000$. The simulated static helices have $q_{\mathrm{helix}}>0$ for all values of $J_{\mathrm{sc}}$ studied between $0.025$ and $1$ (not shown), as the full theory predicts: the long-range RKKY-type couplings are needed even to select the chirality. Because $\tilde{f}^{(0)}_{\mathrm{sc}}\qty(qa_{0})$ is odd in $q$, reversing the spin current, i.e., the sign of $J_{\mathrm{sc}}$, reverses the chirality [Eq.~(\ref{eq:F_H})].

\subsection{Dynamic helimagnetism by charged spin current}
\label{sec:sim-dynamic}
Figures~\ref{fig:snapshots_Jcc0.025_Jsc1.000_nc100_ns100_initial} and \ref{fig:snapshots_Jcc0.025_Jsc1.000_nc100_ns100_late} show representative real-space snapshots of the spin configuration: driven by the charged spin current, the chain evolves from the uniform initial state into a coherently precessing helix of the form of Eq.~(\ref{eq:dynamic-S_helix}). Figures~\ref{fig:q_all}--\ref{fig:Delta_all} then provide a quantitative test of the theory by comparing the numerically extracted helical wavenumber $q_{\mathrm{helix}}$, precession frequency $\tilde{\omega}$, and canting ratio $\Delta$ against the analytical predictions in Sec.~\ref{sec:helix-theory}, as functions of $\tilde{\gamma}^{2}$, $J_{\mathrm{cc}}$, and $J_{\mathrm{sc}}$. Because these three observables depend on the model parameters in qualitatively different ways---for instance, $\tilde\omega$ and $\Delta$ are both predicted to scale as $\tilde\gamma^{-2}$, while $q_{\mathrm{helix}}$ is predicted to be independent of $\tilde\gamma^{2}$ altogether---the comparison below checks several distinct parameter dependences of the analysis of the equations of motion rather than a single curve.

To minimize edge effects, we extracted the signed wavenumber of the helical configuration, the signed frequency of coherent motion, and the signed canting ratio from data sampled at intervals of $\Delta\tilde{t}=1$ from $\tilde{t}=4000$ to $\tilde{t}=5000$ for the two central spins, i.e., the $(N_{\mathrm{S}}/2)$th and $(N_{\mathrm{S}}/2+1)$th spins (the 30th and 31st spins for $N_{\mathrm{S}}=60$). Denoting by $\phi_{i}(\tilde{t})$ the azimuthal angle of $\vb*{\Omega}_{i}$ about $\vb*{n}_{\mathrm{s}}$, we took $q_{\mathrm{helix}}a_{0}=\phi_{N_{\mathrm{S}}/2+1}-\phi_{N_{\mathrm{S}}/2}$, $\tilde{\omega}=-[\phi_{i}(\tilde{t}+\Delta\tilde{t})-\phi_{i}(\tilde{t})]/\Delta\tilde{t}$, and $\Delta=\vb*{n}_{\mathrm{s}}\vdot\vb*{\Omega}_{i}$, with the angle differences reduced to $(-\pi,\pi]$ and the last two quantities averaged over the two central spins. For a coherent helix, these local values represent the whole chain: in the simulation of Figs.~\ref{fig:snapshots_Jcc0.025_Jsc1.000_nc100_ns100_initial} and \ref{fig:snapshots_Jcc0.025_Jsc1.000_nc100_ns100_late}, for example, all spins precess at the same $\tilde{\omega}$, and except for about ten spins at each end the azimuthal angles follow a linear function of position to within $0.04$~rad, while the local wavenumber and canting ratio vary by a few percent along the chain owing to its finite length. The symbols in the figures below represent the time averages of these quantities, and the error bars represent the standard deviations of their temporal fluctuations. The theoretical curves are evaluated from the expressions of Sec.~\ref{sec:helix-theory} at the same parameter values as the simulations, without any fitting to the numerical data. The residual discrepancy between the two is typically a few percent---for instance, below $2\%$ for $q_{\mathrm{helix}}$ in the close-up view of Fig.~\ref{fig:q_Jcc} for $J_{\mathrm{cc}}\leq 0.175$.

To assess how this discrepancy depends on the chain length, we repeated the simulations for $N_{\mathrm{S}}=10, 20, \ldots, 60$ (Fig.~S1 of the Supplemental Material~\cite{Note1}). For $N_{\mathrm{S}}\geq 40$, the deviations of $q_{\mathrm{helix}}$, $\tilde{\omega}$, and $\Delta$ from the infinite-chain predictions at $\tilde{\gamma}^{2}=0.100$, $J_{\mathrm{cc}}=0.025$, and $J_{\mathrm{sc}}=1.000$ remain below $2\%$ and show no systematic trend with $N_{\mathrm{S}}$.
More generally, the deviation of $q_{\mathrm{helix}}$ is governed mainly by the ratio of the helical pitch length $2\pi/\abs{q_{\mathrm{helix}}}$ to the chain length $N_{\mathrm{S}}a_{0}$: it remains below $2\%$ as long as this ratio is smaller than about $0.7$, i.e., as long as the central spins are separated from the free ends by more than about $0.7$ pitch lengths, and becomes appreciable only when the pitch length becomes comparable to or exceeds the chain length. The latter situation occurs at the smallest values of $J_{\mathrm{sc}}$ in Figs.~\ref{fig:q_all}(c), \ref{fig:omega_all}(c), and \ref{fig:Delta_all}(c) and accounts for the larger deviations there.

Parameter points with $J_{\mathrm{sc}}\leq J_{\mathrm{cc}}$, where the chain orders ferromagnetically (see Sec.~\ref{sec:sim-ferro-chaos}), and those with $\tilde{\gamma}^{2}<0.010$, where the coherent dynamic helix becomes unstable, are omitted from Figs.~\ref{fig:q_all}--\ref{fig:Delta_all}; the ferromagnetic order and the loss of coherence at these points are found for all $N_{\mathrm{S}}$ studied and are therefore not finite-size artifacts. For $J_{\mathrm{cc}}=0.025$ and $J_{\mathrm{sc}}=1.000$, Eq.~(\ref{eq:Delta}) gives $\abs{\Delta}=1$ at $\tilde{\gamma}^{2}\approx0.0051$, below which no helix of the form of Eq.~(\ref{eq:dynamic-S_helix}) exists; the loss of coherence observed at $\tilde{\gamma}^{2}=0.005$ is consistent with this bound. The points at $\tilde{\gamma}^{2}=0.010$, where the deviations and fluctuations are already enhanced, are retained to indicate the approach to this instability. Similarly, the points at $J_{\mathrm{cc}}=0.200$ and $0.225$ in Figs.~\ref{fig:q_all}(b), \ref{fig:q_Jcc}(b), \ref{fig:omega_all}(b), and \ref{fig:Delta_all}(b) lie close to the onset of the chaotic regime (Sec.~\ref{sec:sim-ferro-chaos}): the chain still forms a recognizable dynamic helix, but fluctuations are superimposed on it, so that $q_{\mathrm{helix}}$, $\tilde{\omega}$, and $\Delta$ can still be extracted, albeit with larger deviations and error bars.

\newcommand{\FigLabelPutSim}[1]{\put(0,95){\textbf{#1}}}

\begin{figure*}[htb]
\centering
\newcommand{\FigDirSnap}{Figs/Jcc0025Jsc1000nc100ns100_0000-0020}
\newcommand{\FigWSnap}{0.195\textwidth}
\newcommand{\FigGapSnap}{1mm}

\begin{overpic}[width=\FigWSnap]{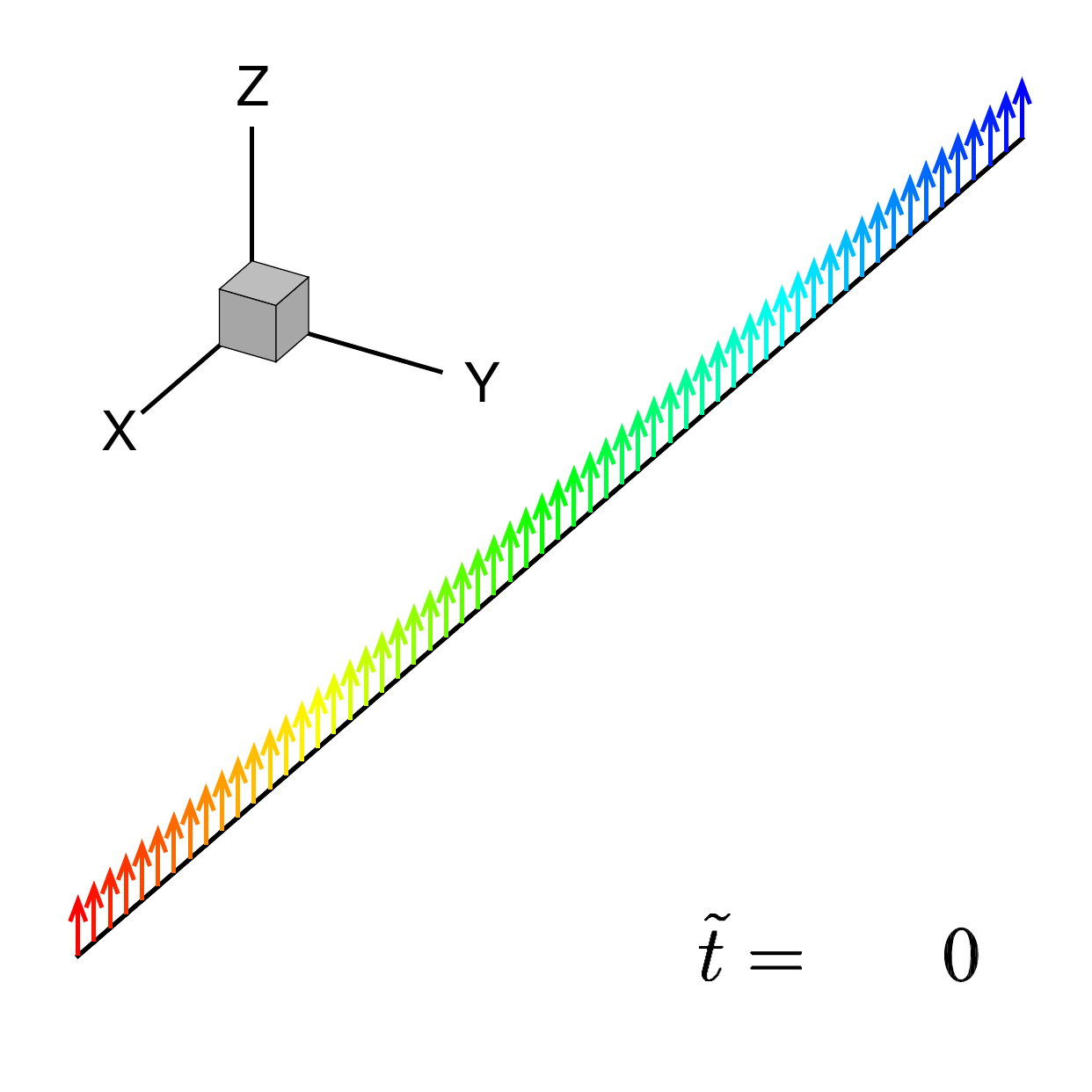}
\FigLabelPutSim{(a)}
\end{overpic}\hspace{\FigGapSnap}%
\includegraphics[width=\FigWSnap]{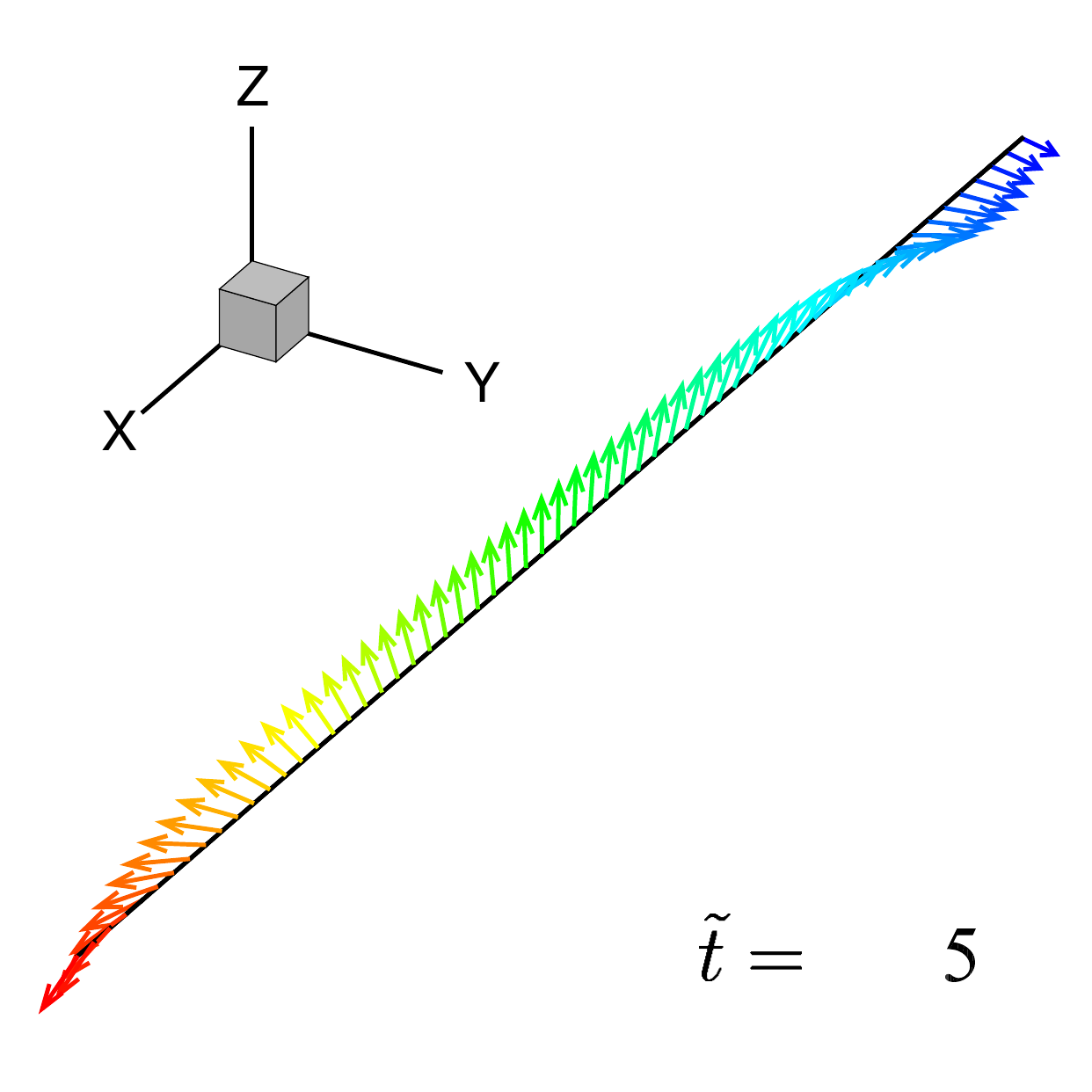}\hspace{\FigGapSnap}%
\includegraphics[width=\FigWSnap]{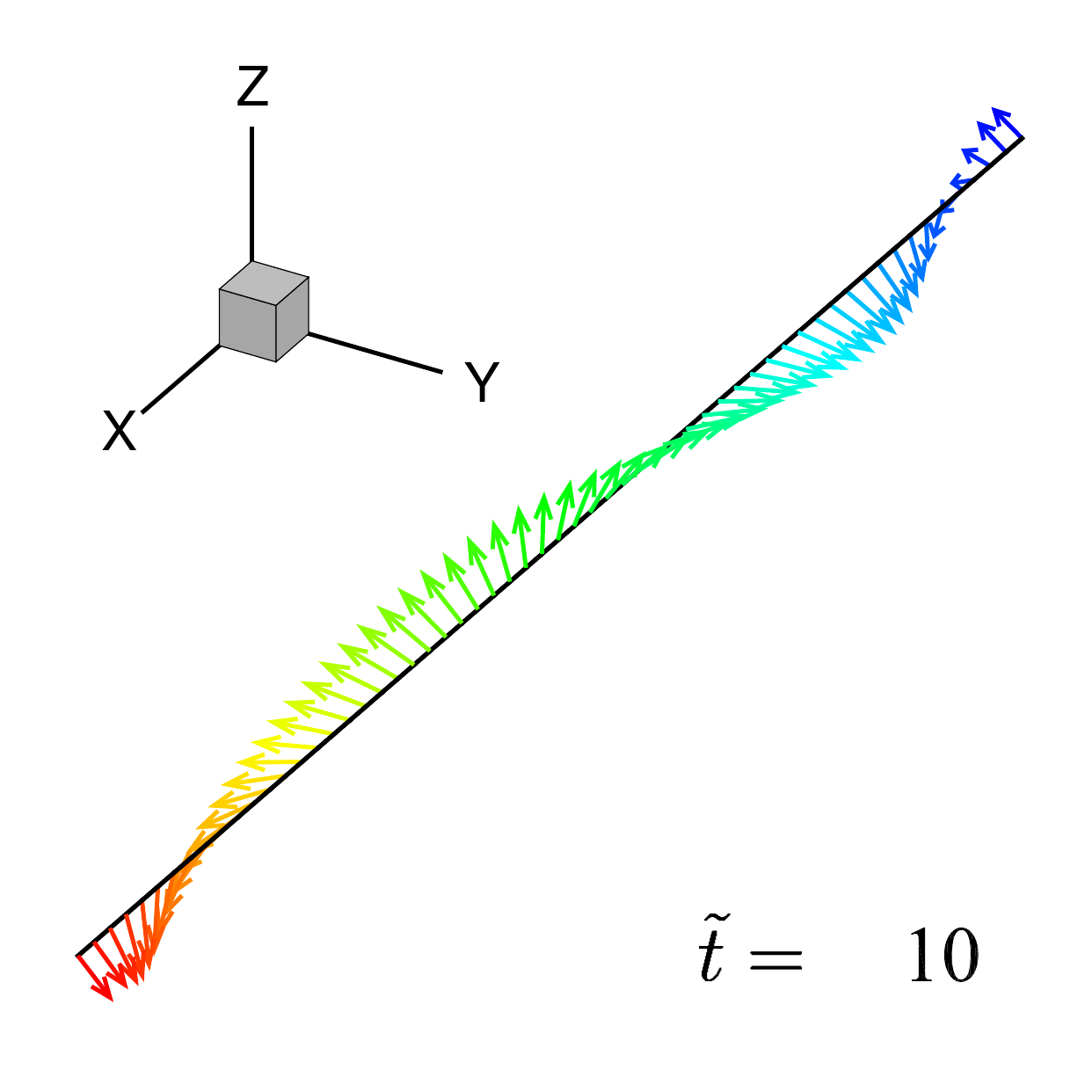}\hspace{\FigGapSnap}%
\includegraphics[width=\FigWSnap]{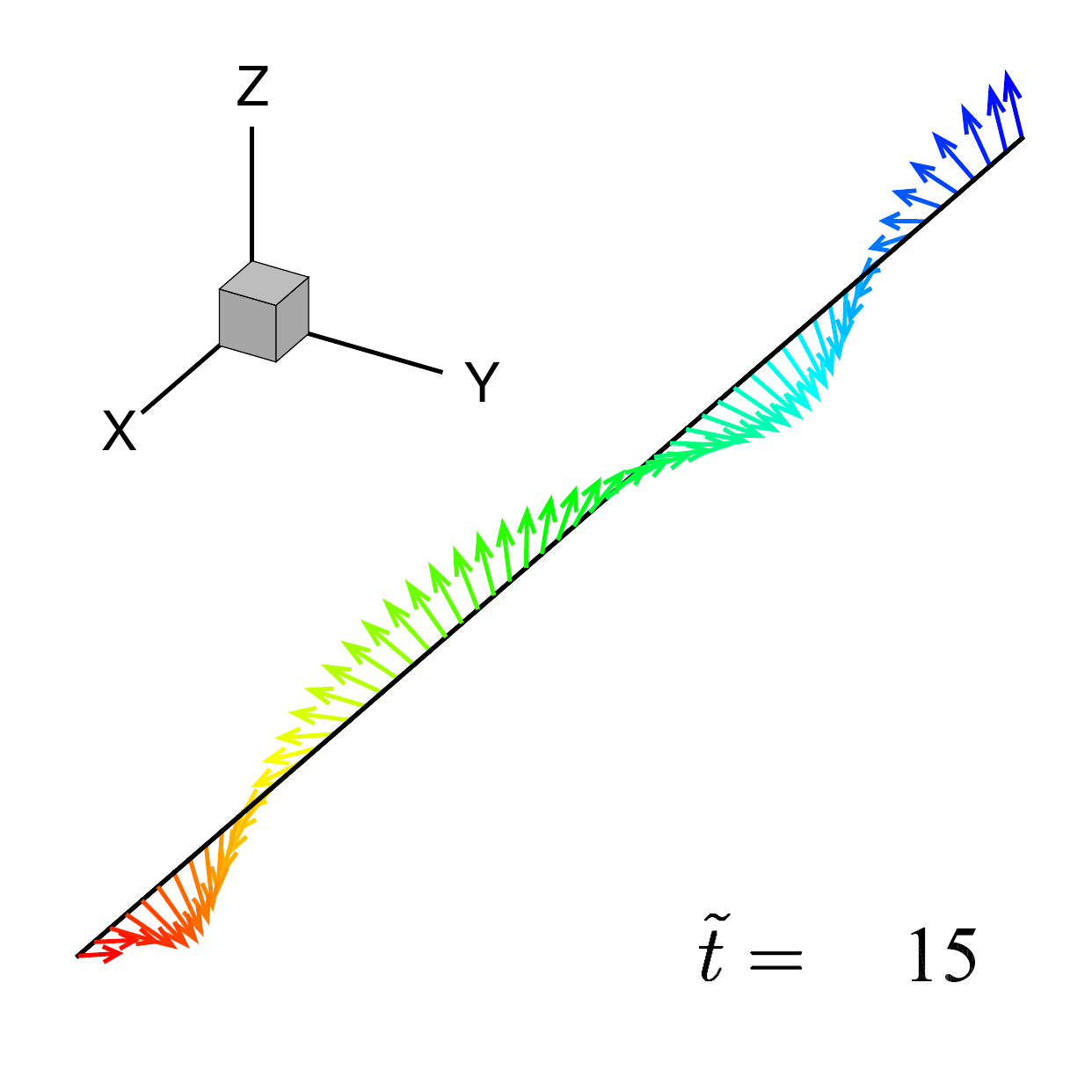}\hspace{\FigGapSnap}%
\includegraphics[width=\FigWSnap]{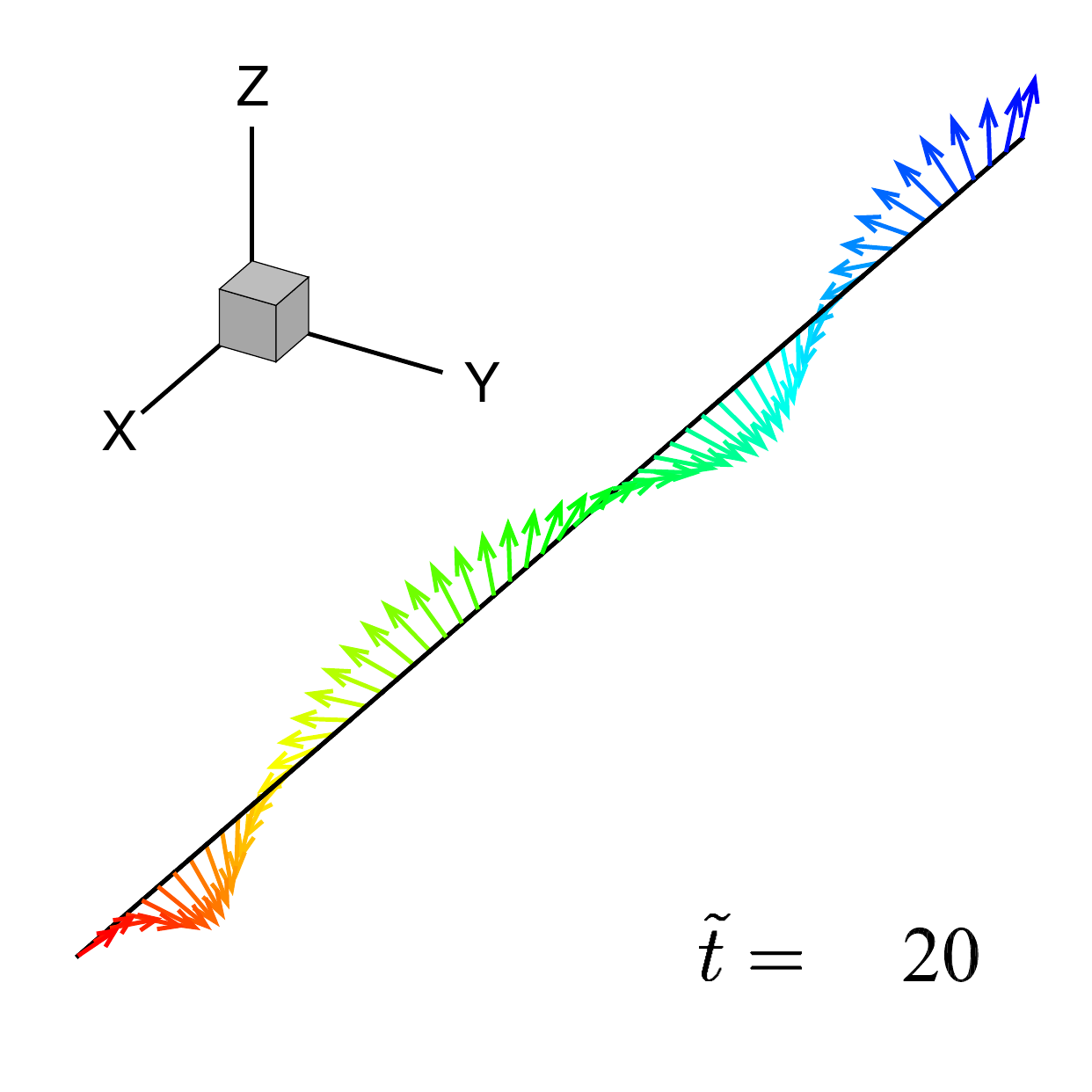}
\\[2mm]

\begin{overpic}[width=\FigWSnap]{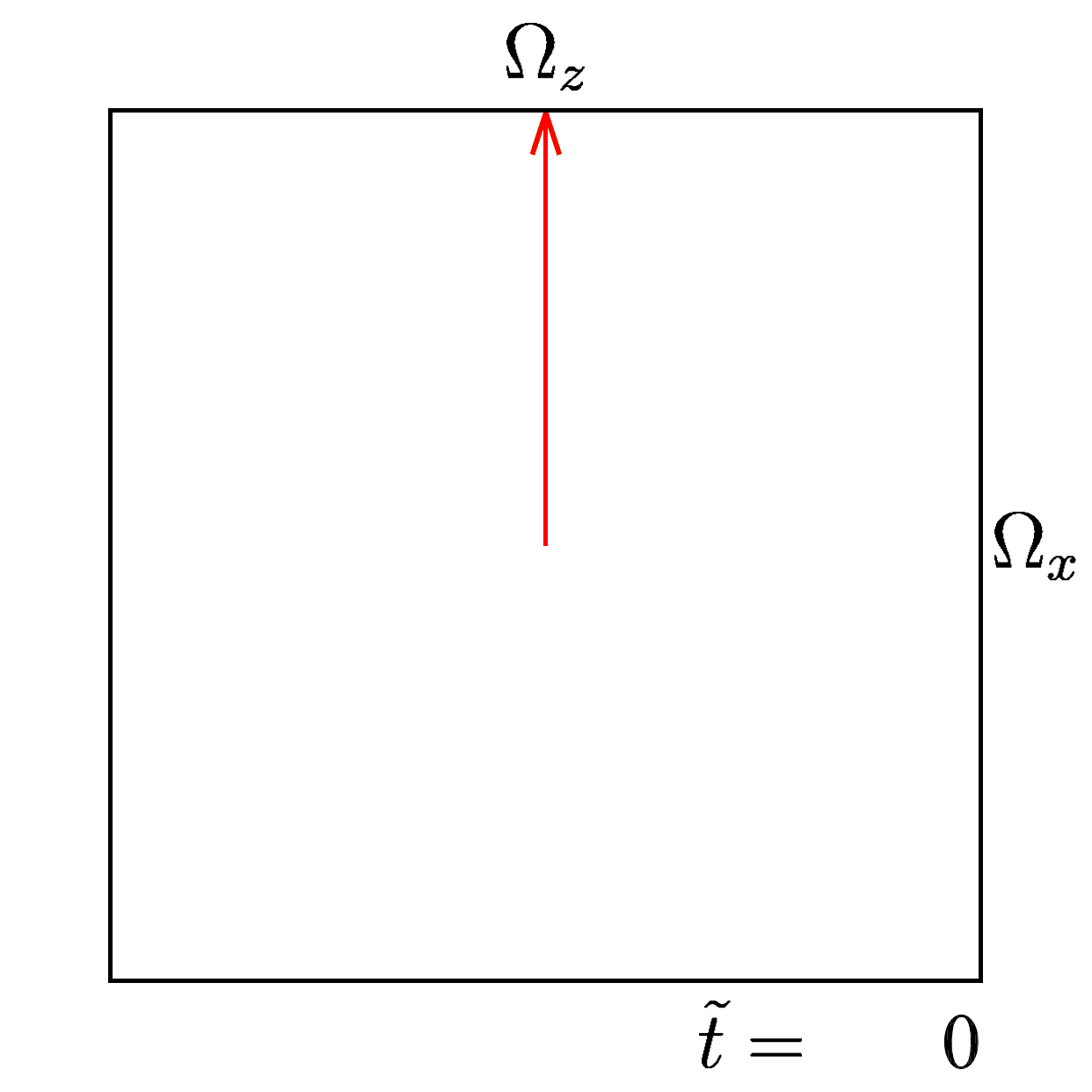}
\FigLabelPutSim{(b)}
\end{overpic}\hspace{\FigGapSnap}%
\includegraphics[width=\FigWSnap]{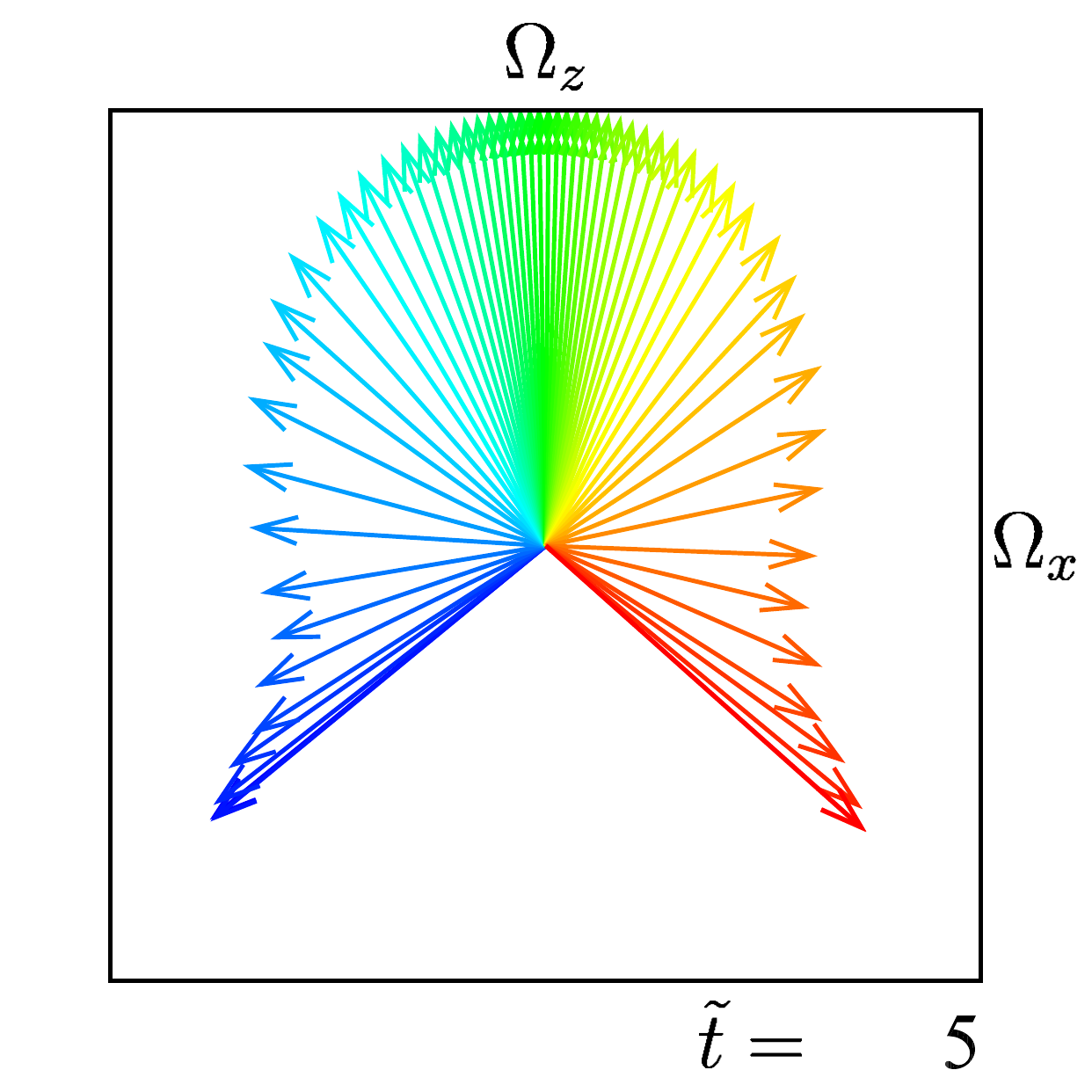}\hspace{\FigGapSnap}%
\includegraphics[width=\FigWSnap]{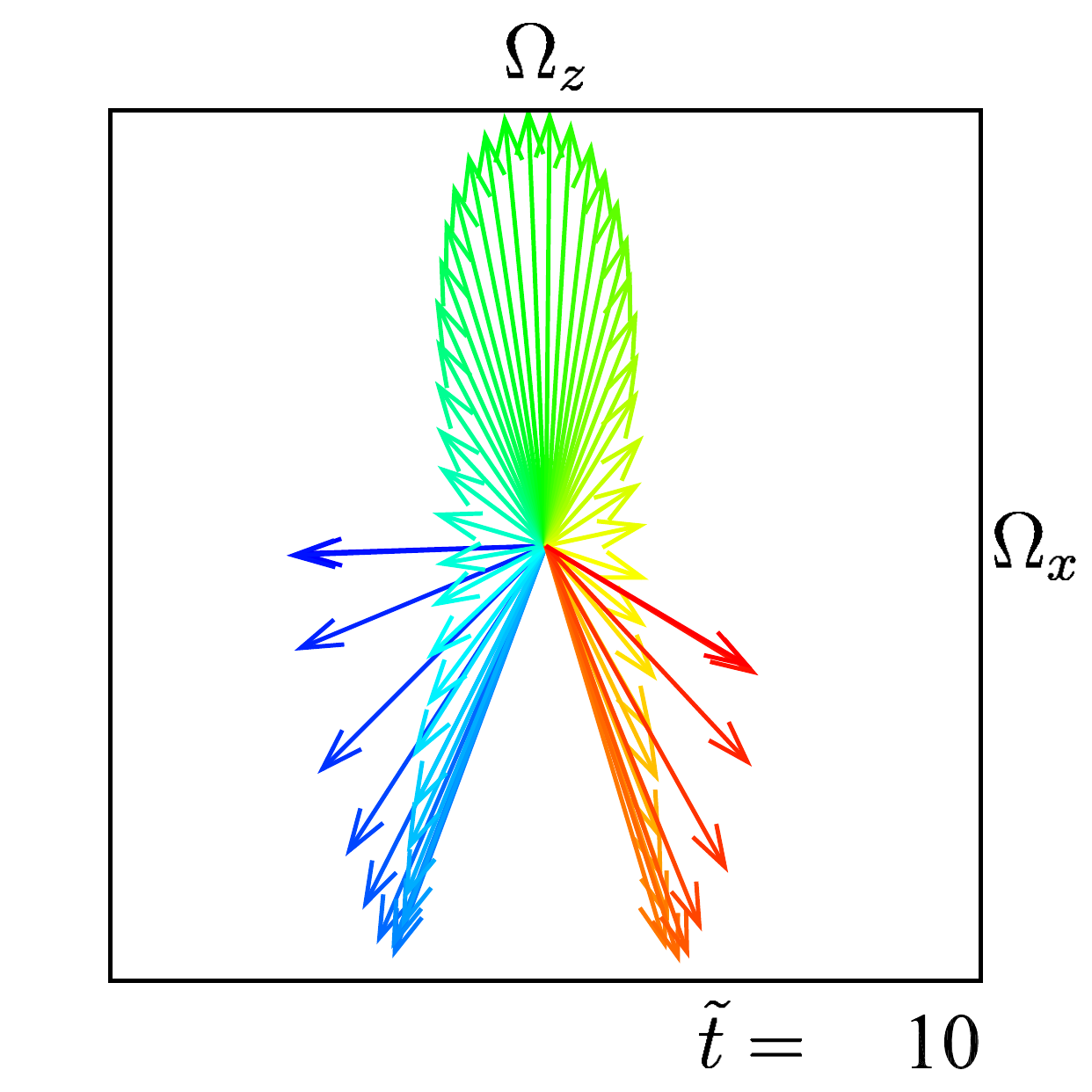}\hspace{\FigGapSnap}%
\includegraphics[width=\FigWSnap]{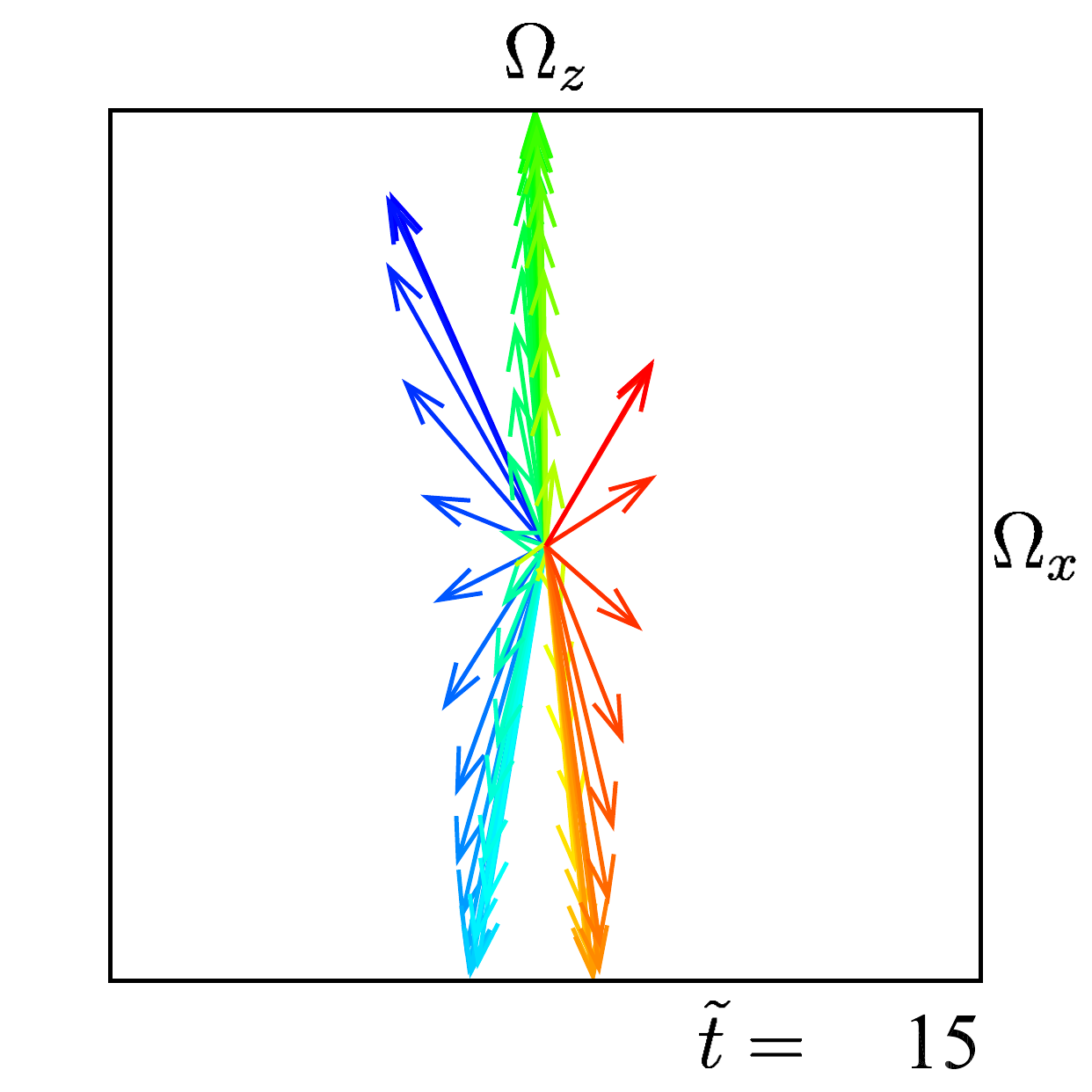}\hspace{\FigGapSnap}%
\includegraphics[width=\FigWSnap]{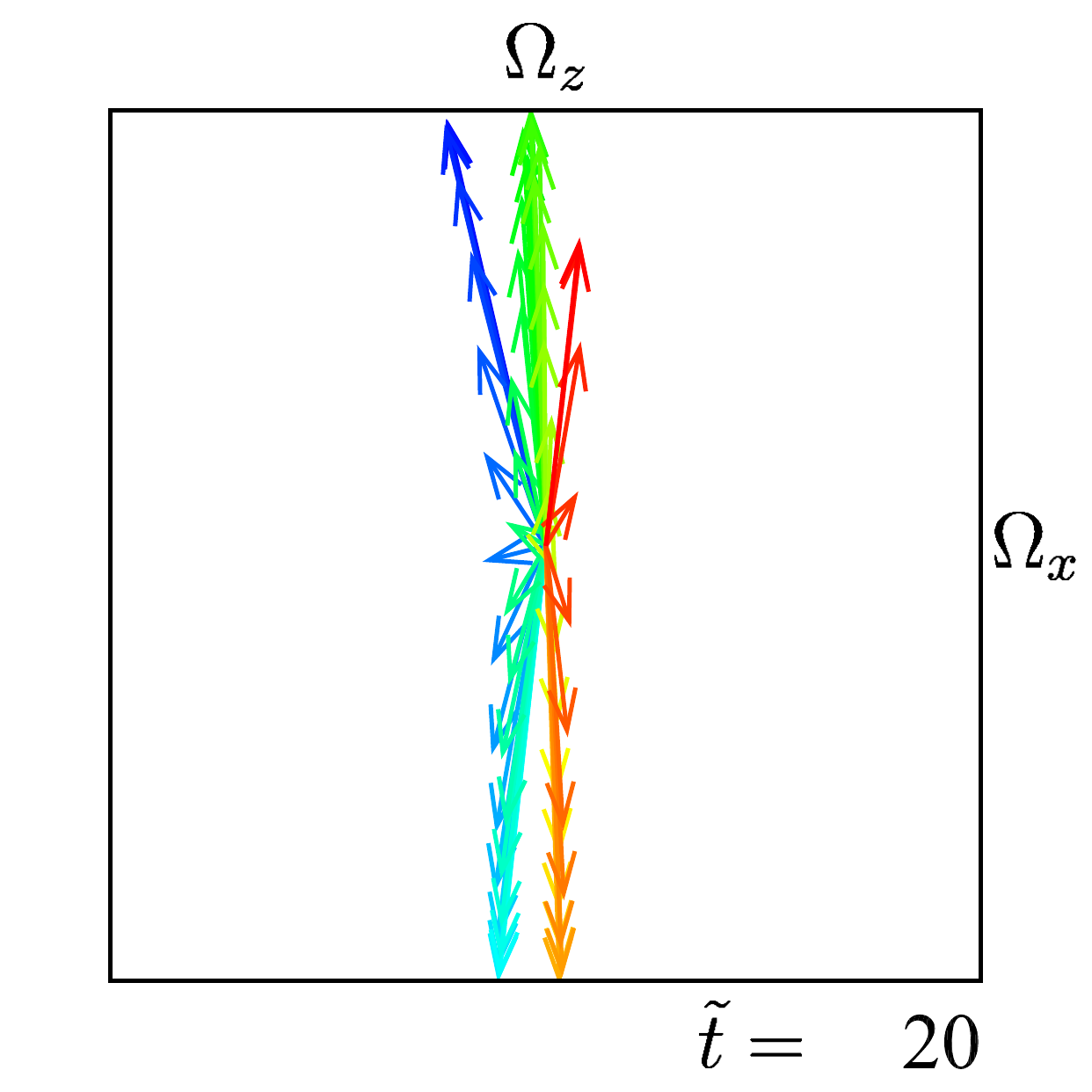}
\\[2mm]

\begin{overpic}[width=\FigWSnap]{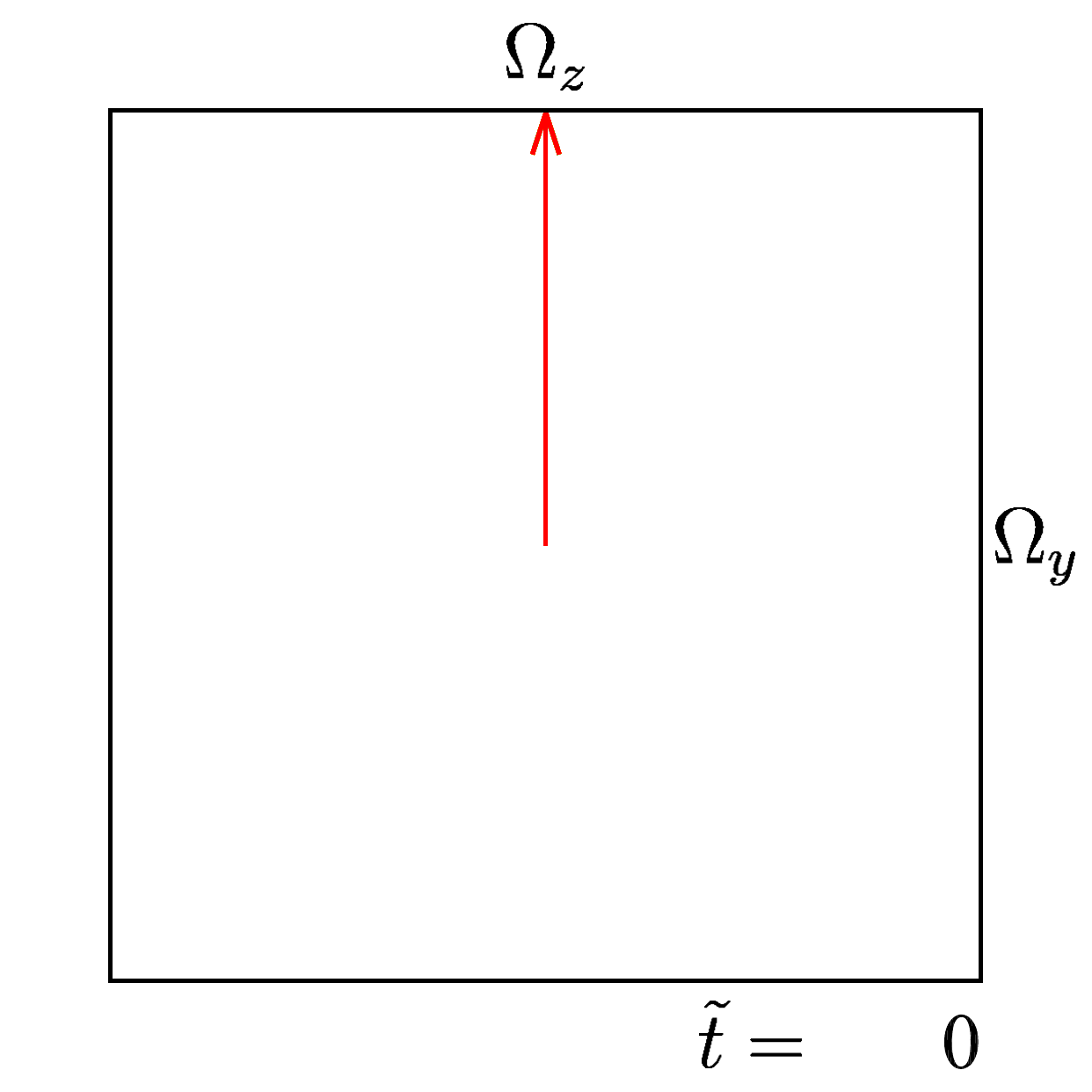}
\FigLabelPutSim{(c)}
\end{overpic}\hspace{\FigGapSnap}%
\includegraphics[width=\FigWSnap]{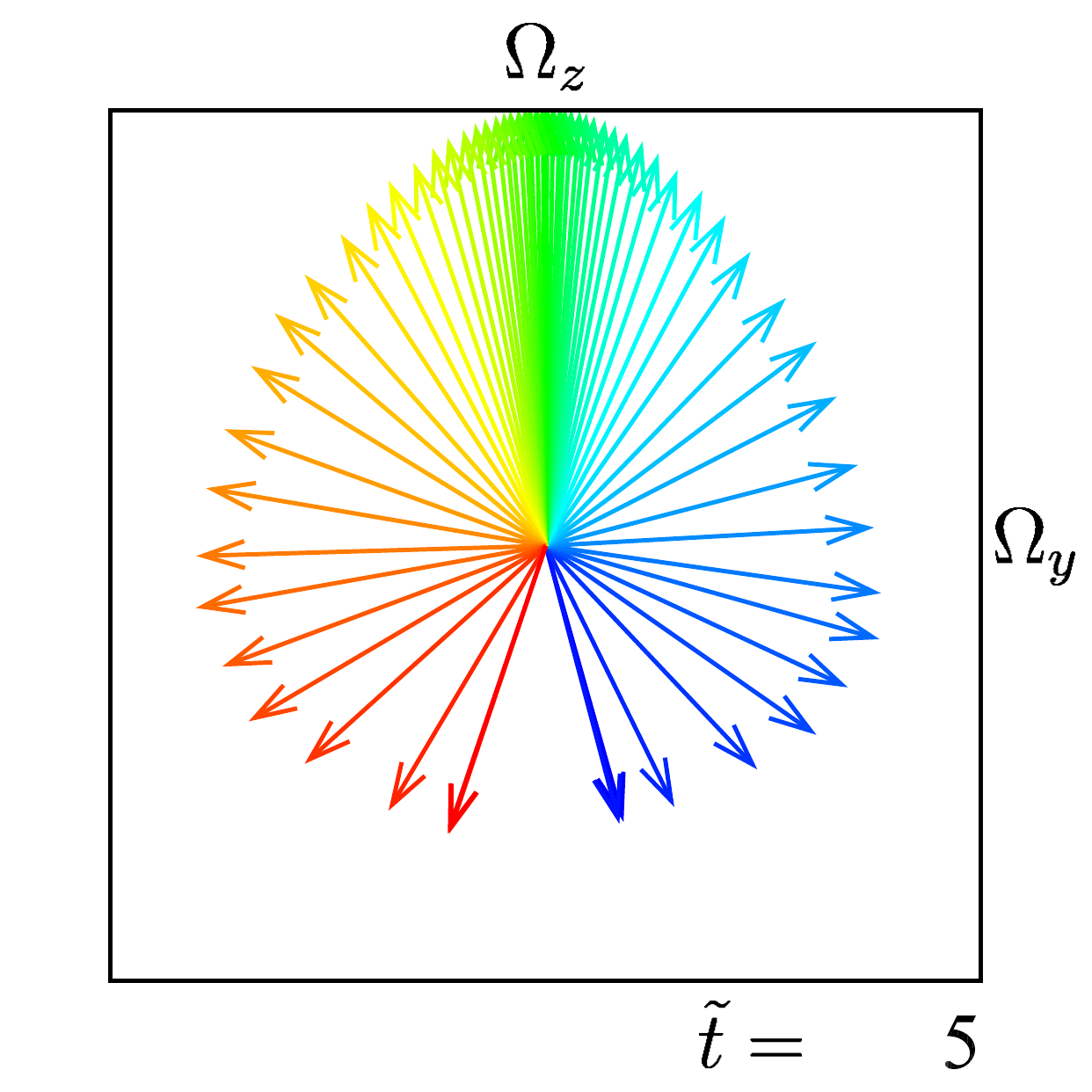}\hspace{\FigGapSnap}%
\includegraphics[width=\FigWSnap]{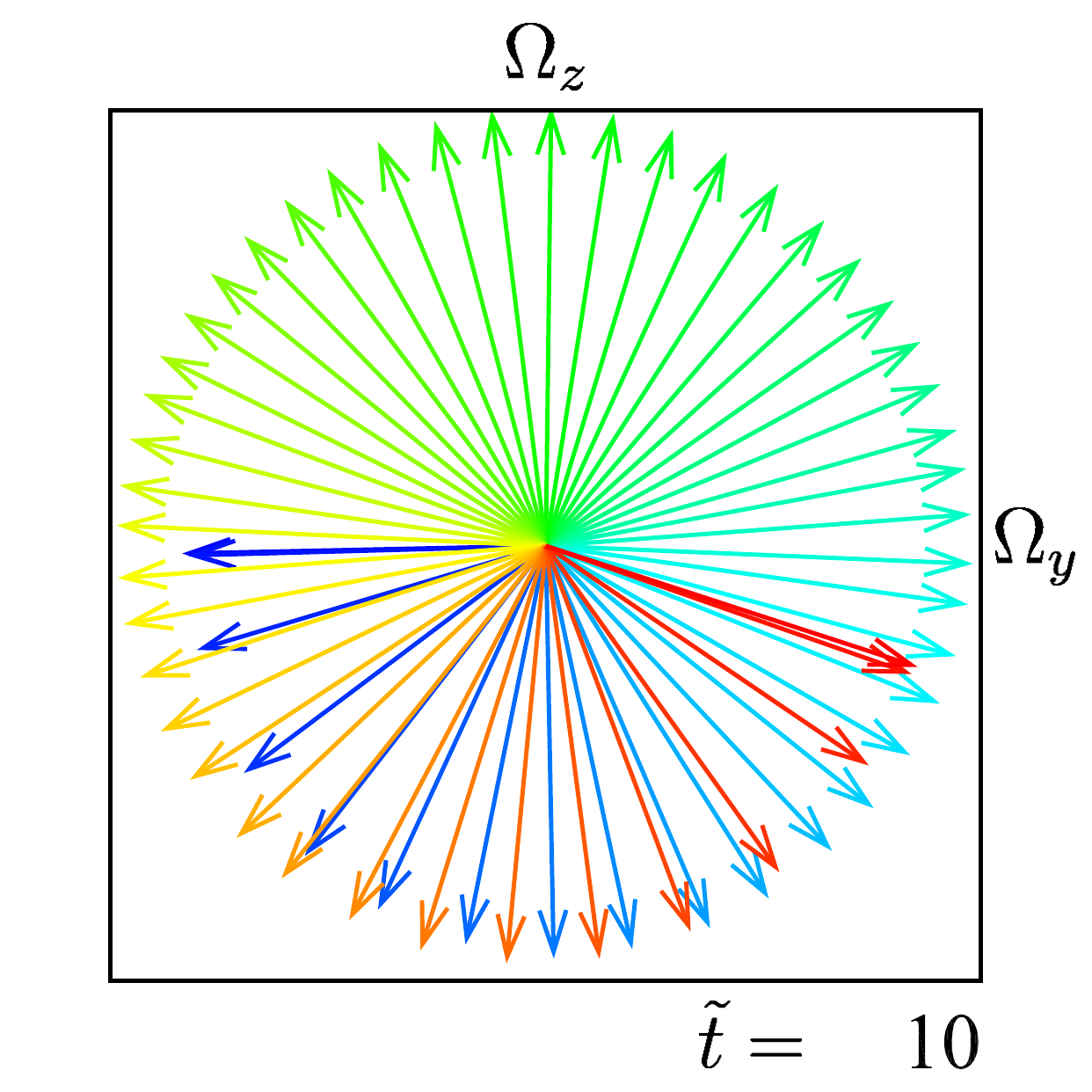}\hspace{\FigGapSnap}%
\includegraphics[width=\FigWSnap]{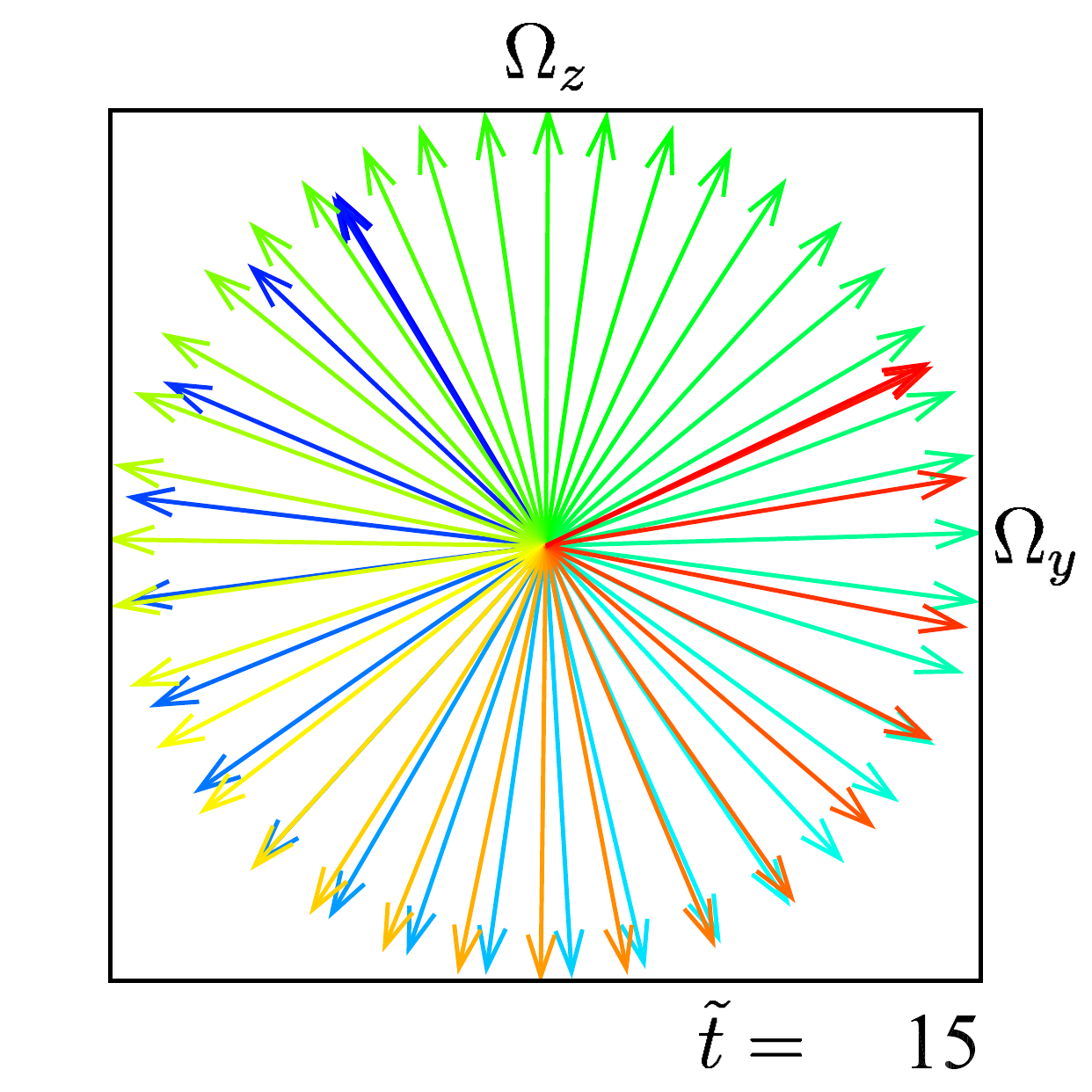}\hspace{\FigGapSnap}%
\includegraphics[width=\FigWSnap]{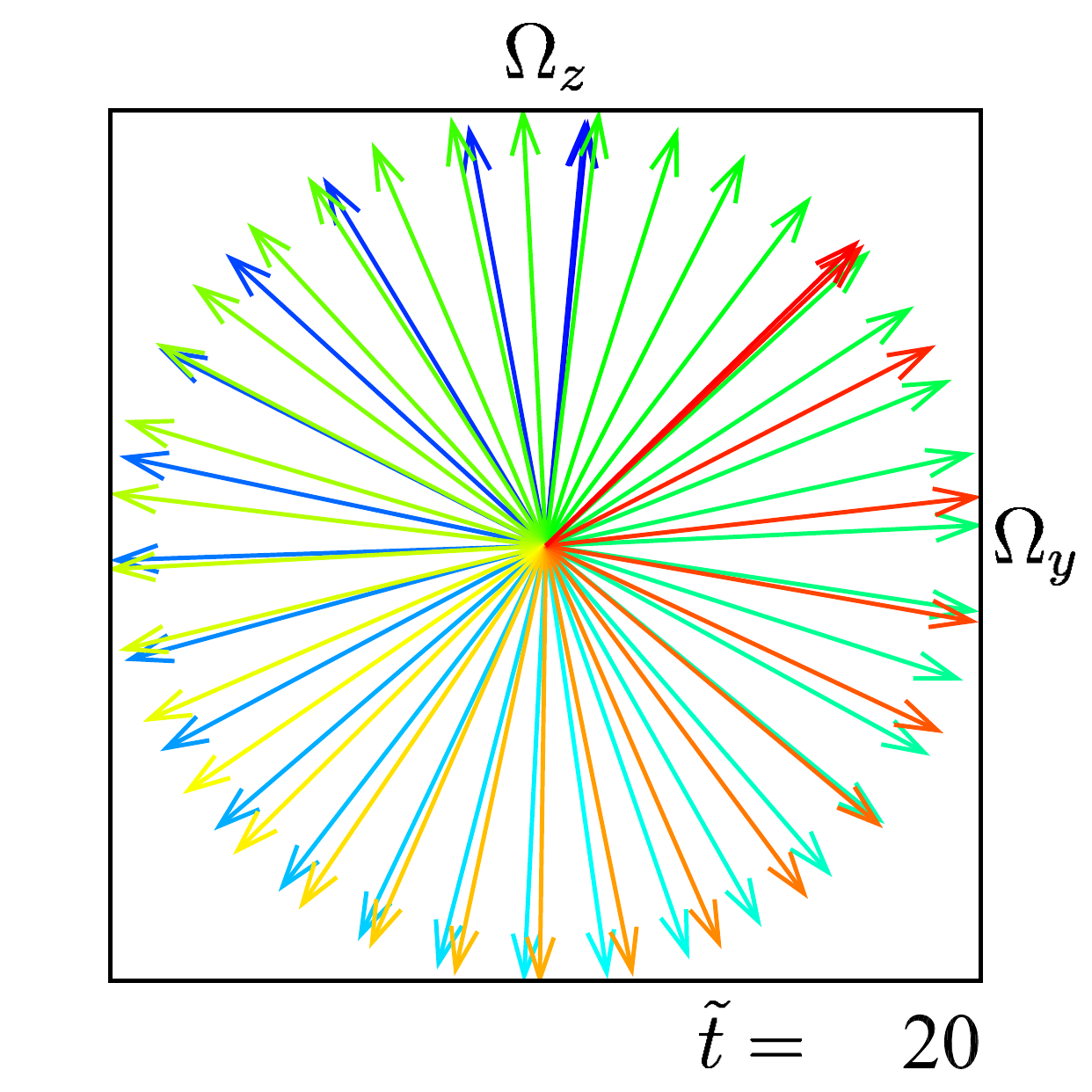}

\caption{
Initial time evolution of a spin chain consisting of 60 spins equally spaced along the $x$~direction, 
for $\tilde{\gamma}^2=0.100$, $J_{\mathrm{cc}} = 0.025$, $J_{\mathrm{sc}} = 1.000$, $\vb*{n}_{\mathrm{c}} = \vb*{e}_{x}$, and $\vb*{n}_{\mathrm{s}} = \vb*{e}_{x}$. 
Columns show snapshots at dimensionless times $\tilde{t} = 0, 5, 10, 15, 20$ (left to
right).
(a)~Bird's-eye view of the entire spin chain.
(b)~Normalized $x$ and $z$ components of the spins.
(c)~Normalized $y$ and $z$ components of the spins.
}
\label{fig:snapshots_Jcc0.025_Jsc1.000_nc100_ns100_initial}
\end{figure*}

\begin{figure*}[htb]
\centering
\newcommand{\FigDirSnapB}{Figs/Jcc0025Jsc1000nc100ns100_4900-5000}
\newcommand{\FigWSnapB}{0.195\textwidth}
\newcommand{\FigGapSnapB}{1mm}

\begin{overpic}[width=\FigWSnapB]{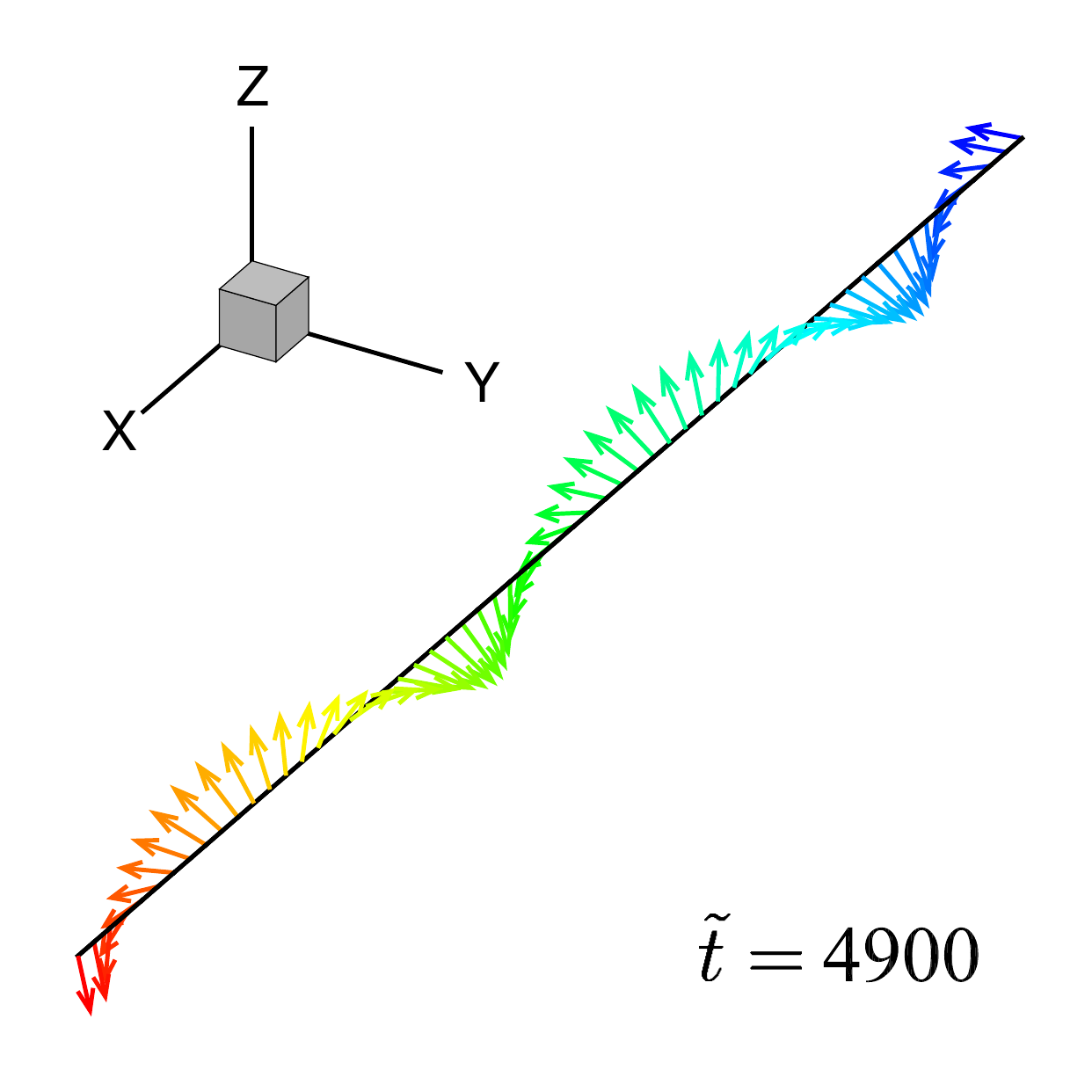}
\FigLabelPutSim{(a)}
\end{overpic}\hspace{\FigGapSnapB}%
\includegraphics[width=\FigWSnapB]{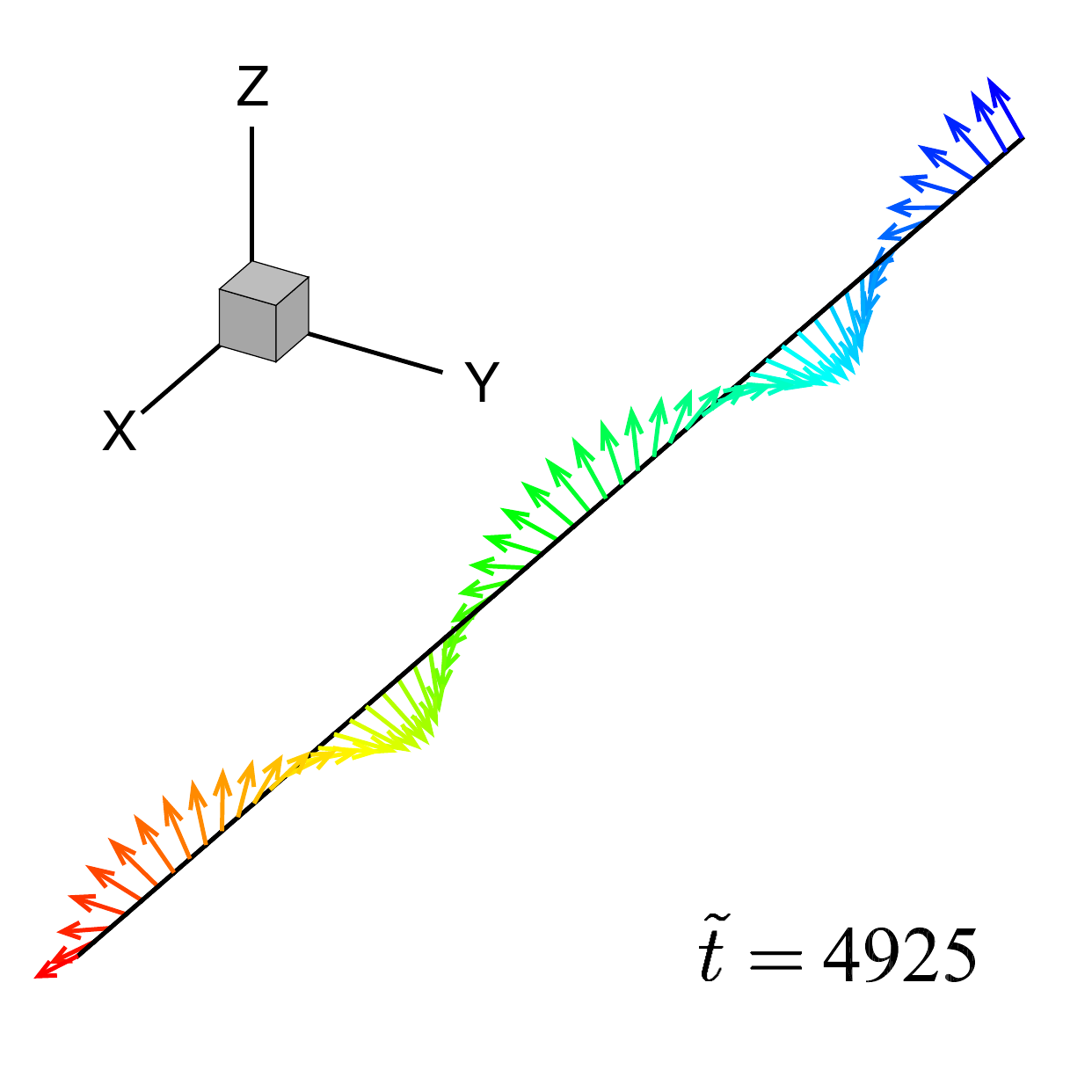}\hspace{\FigGapSnapB}%
\includegraphics[width=\FigWSnapB]{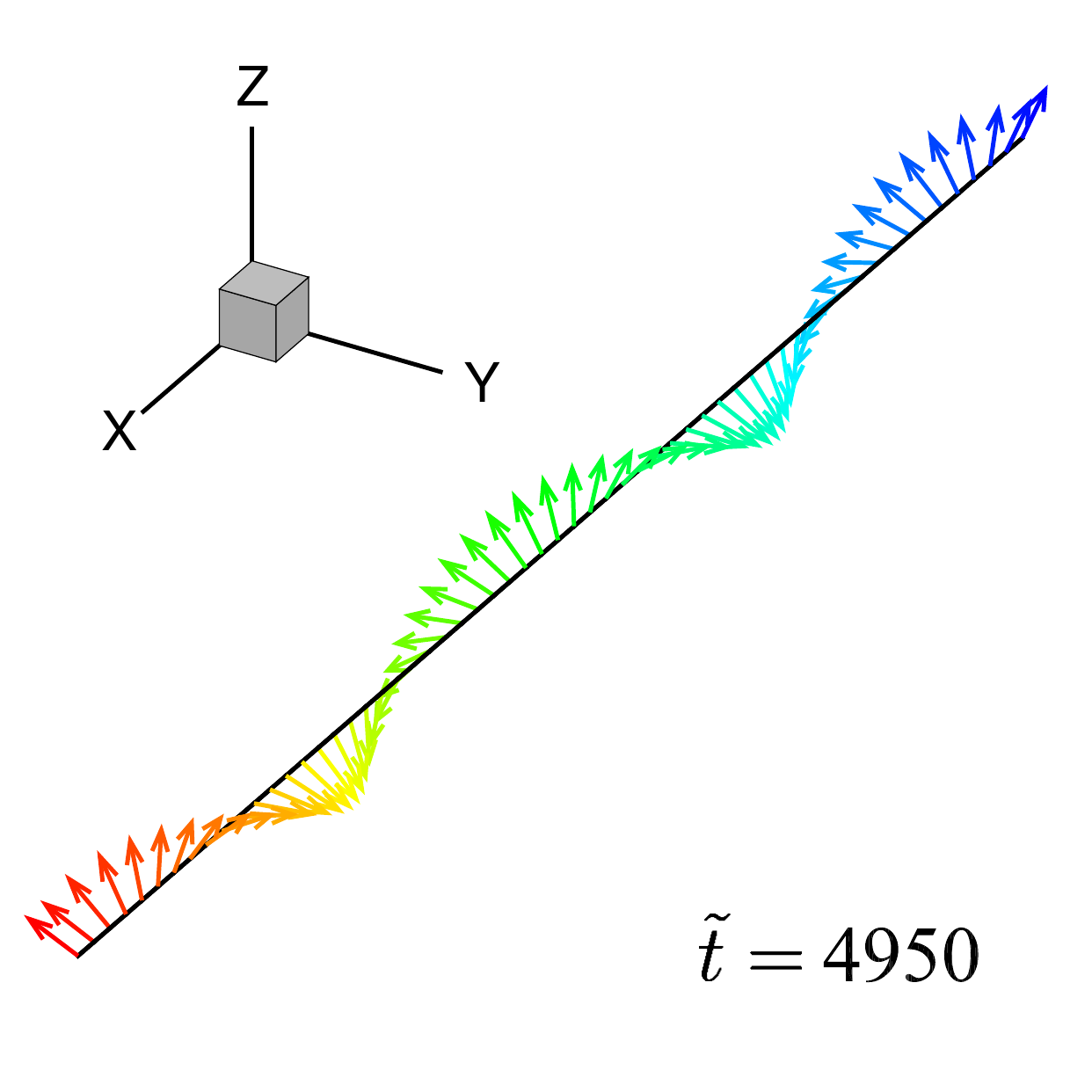}\hspace{\FigGapSnapB}%
\includegraphics[width=\FigWSnapB]{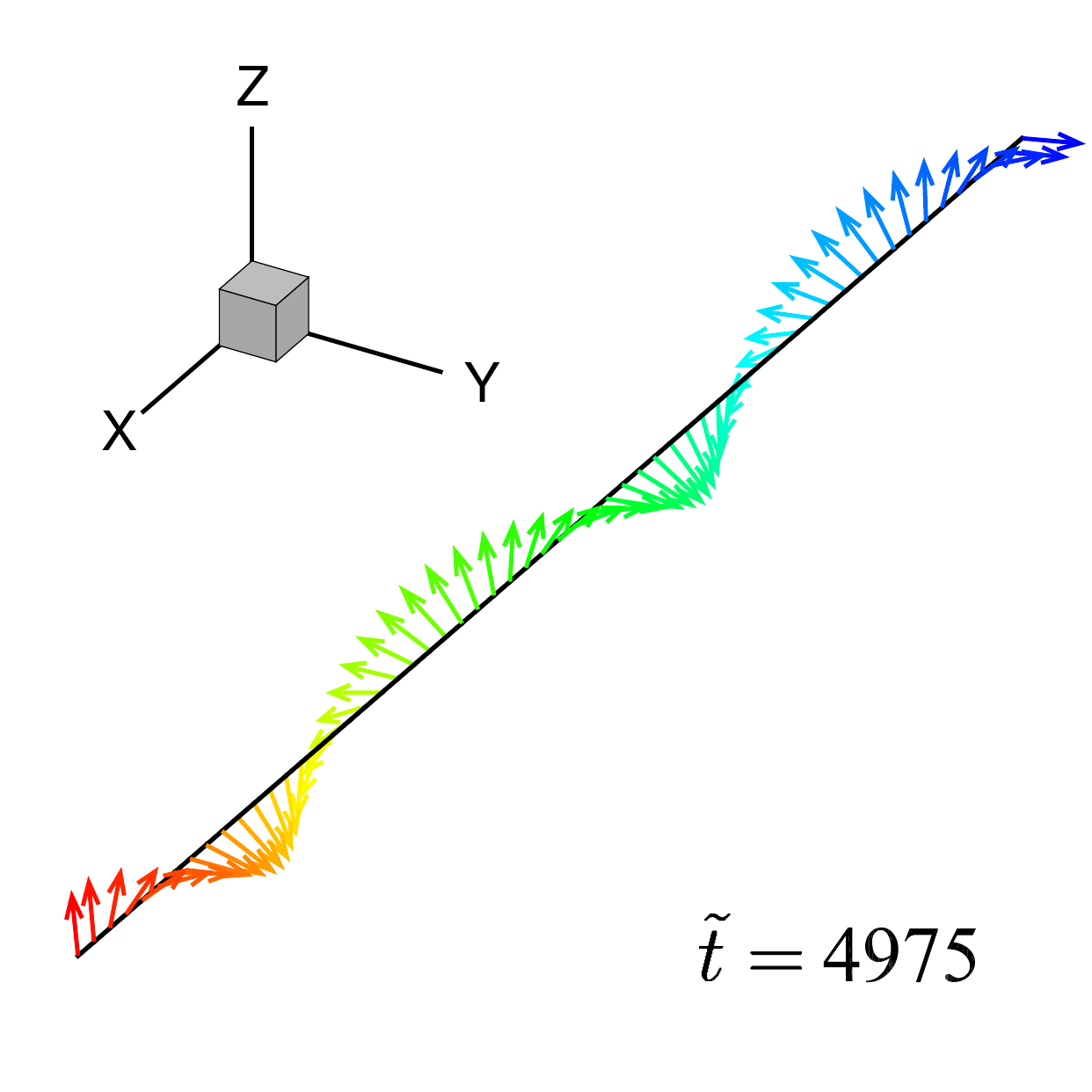}\hspace{\FigGapSnapB}%
\includegraphics[width=\FigWSnapB]{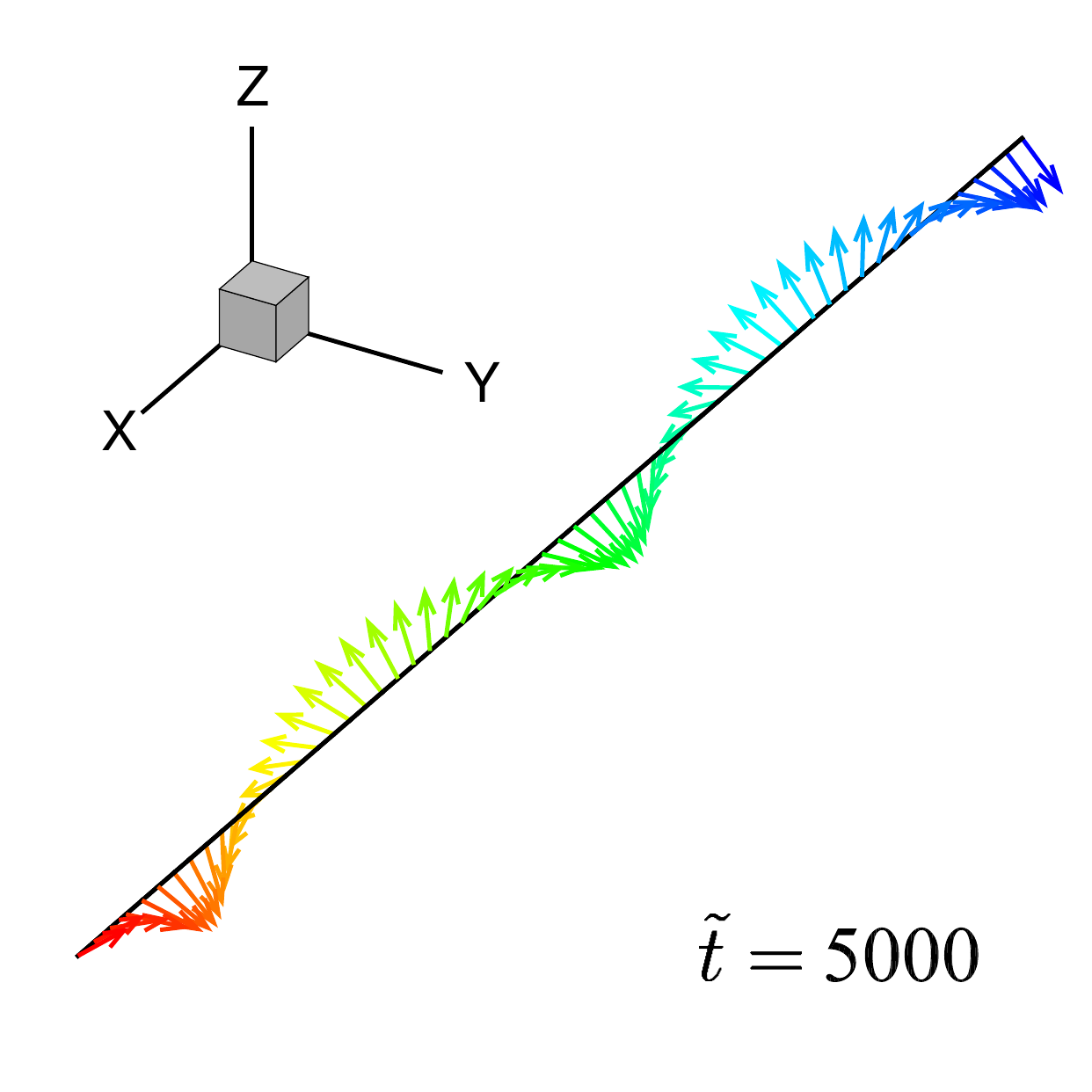}
\\[2mm]

\begin{overpic}[width=\FigWSnapB]{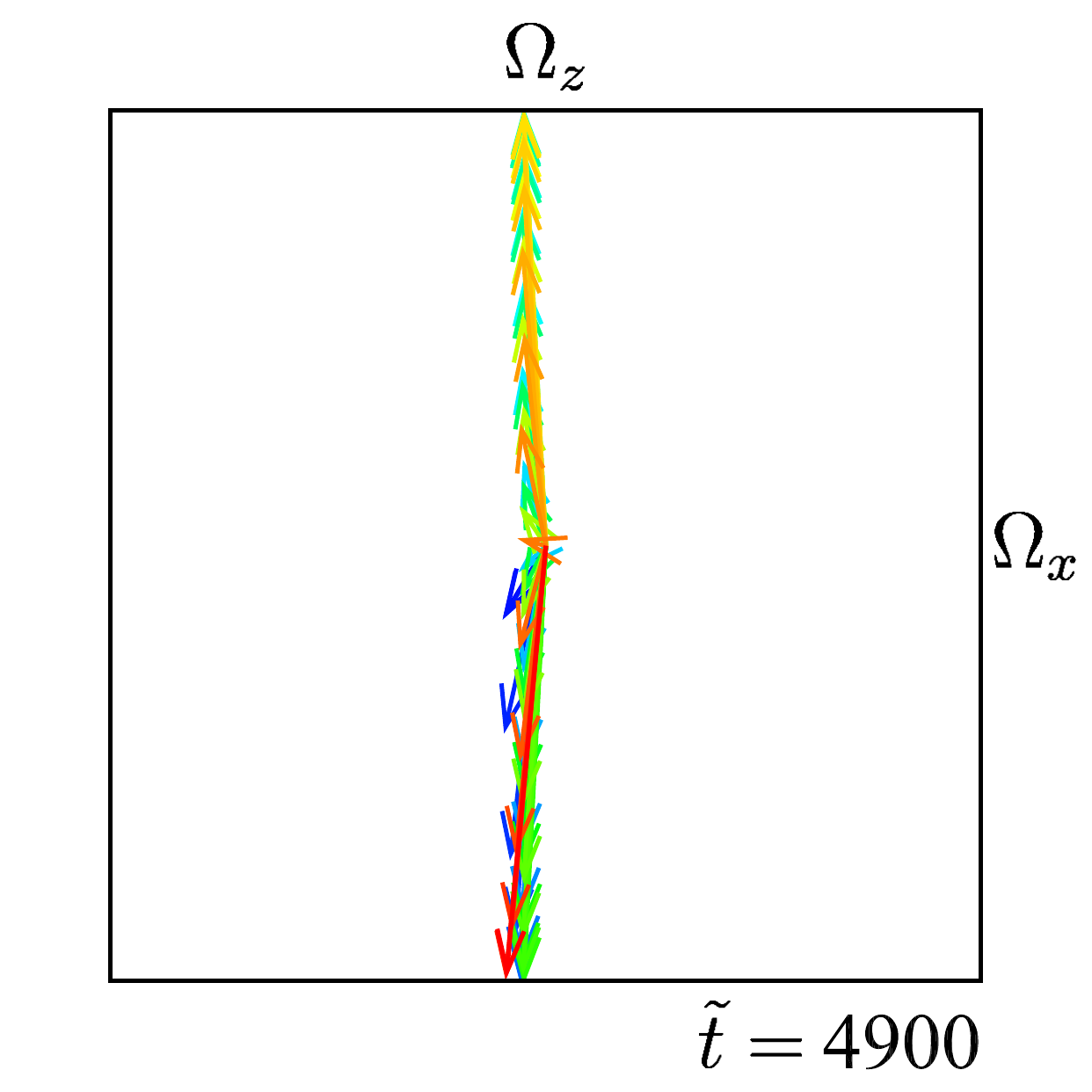}
\FigLabelPutSim{(b)}
\end{overpic}\hspace{\FigGapSnapB}%
\includegraphics[width=\FigWSnapB]{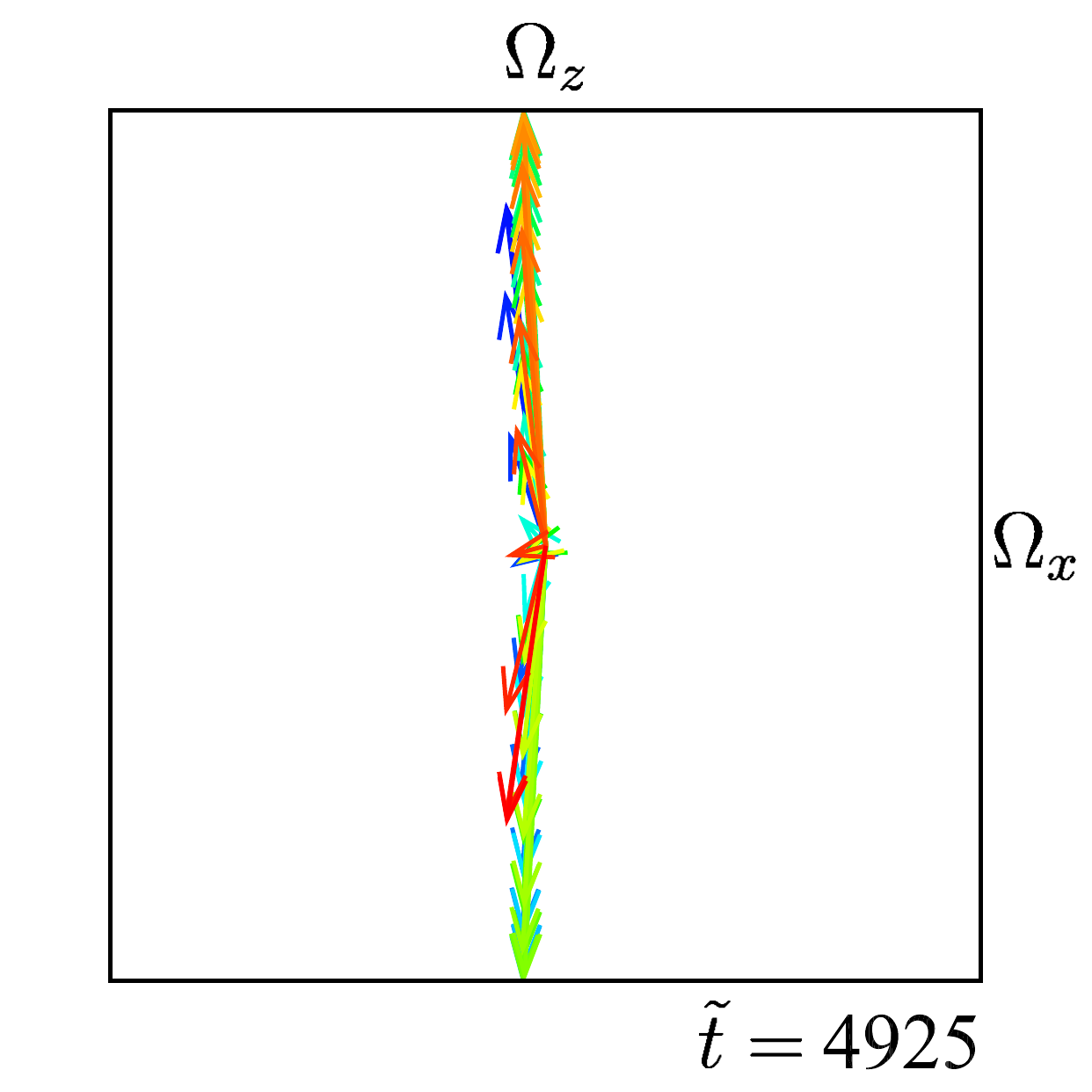}\hspace{\FigGapSnapB}%
\includegraphics[width=\FigWSnapB]{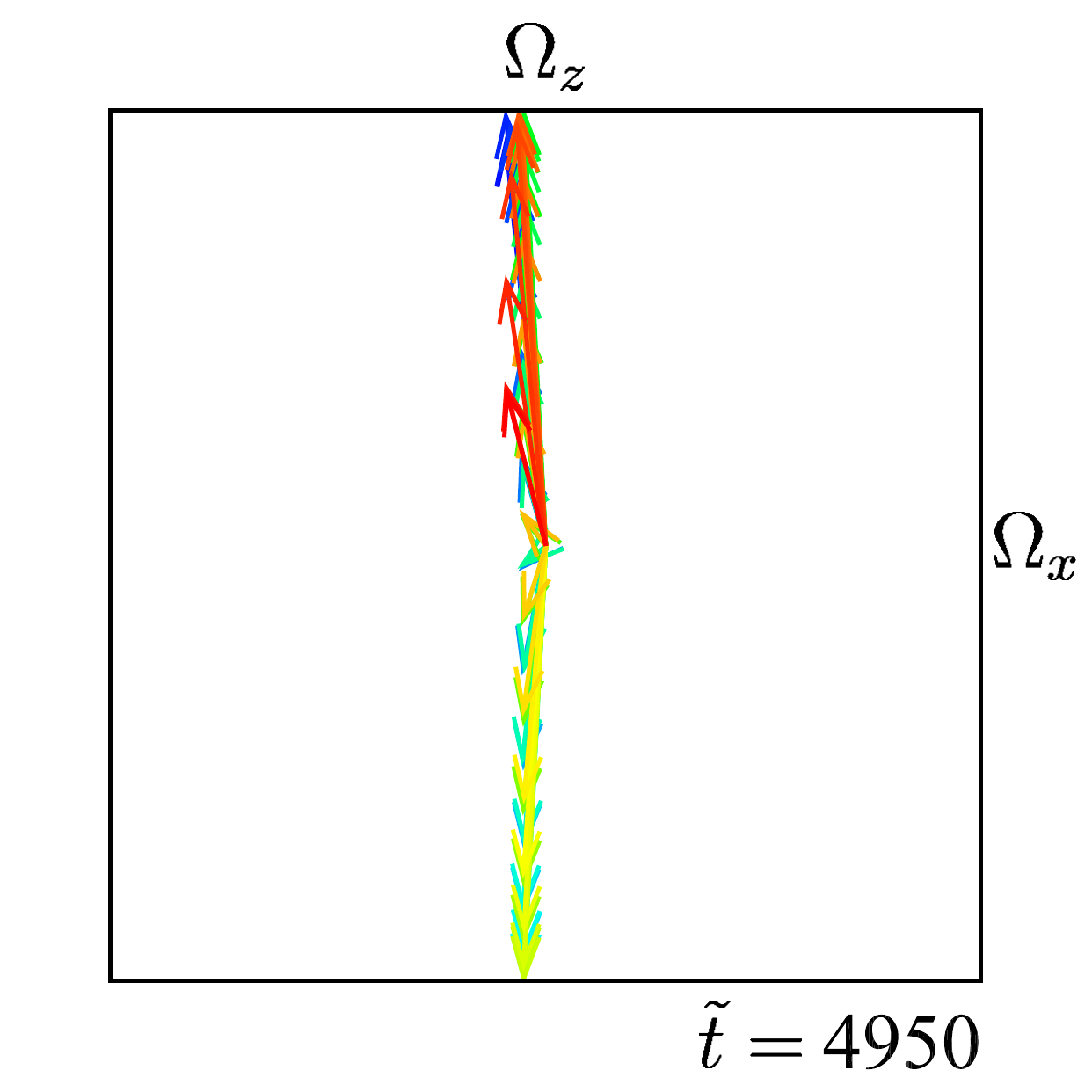}\hspace{\FigGapSnapB}%
\includegraphics[width=\FigWSnapB]{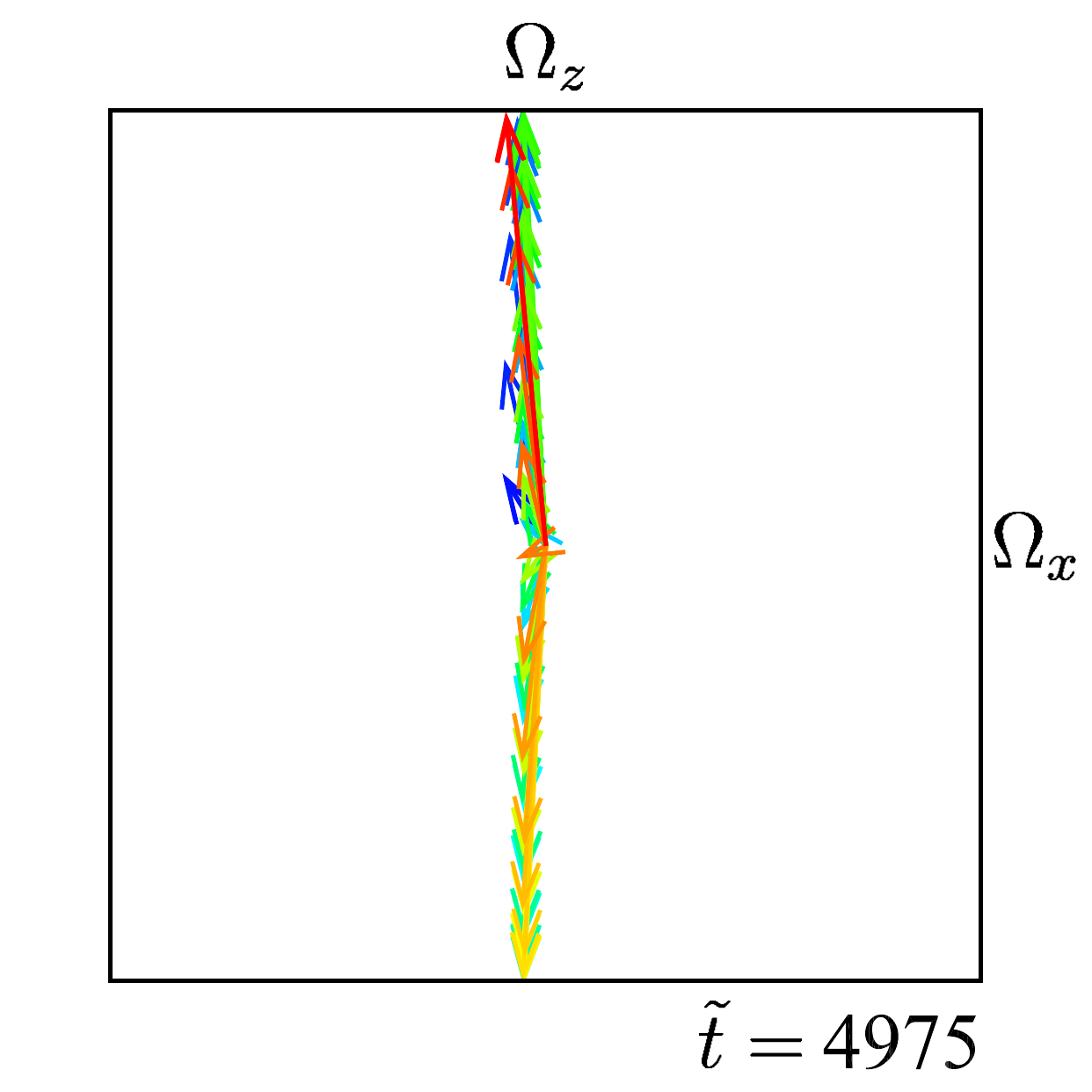}\hspace{\FigGapSnapB}%
\includegraphics[width=\FigWSnapB]{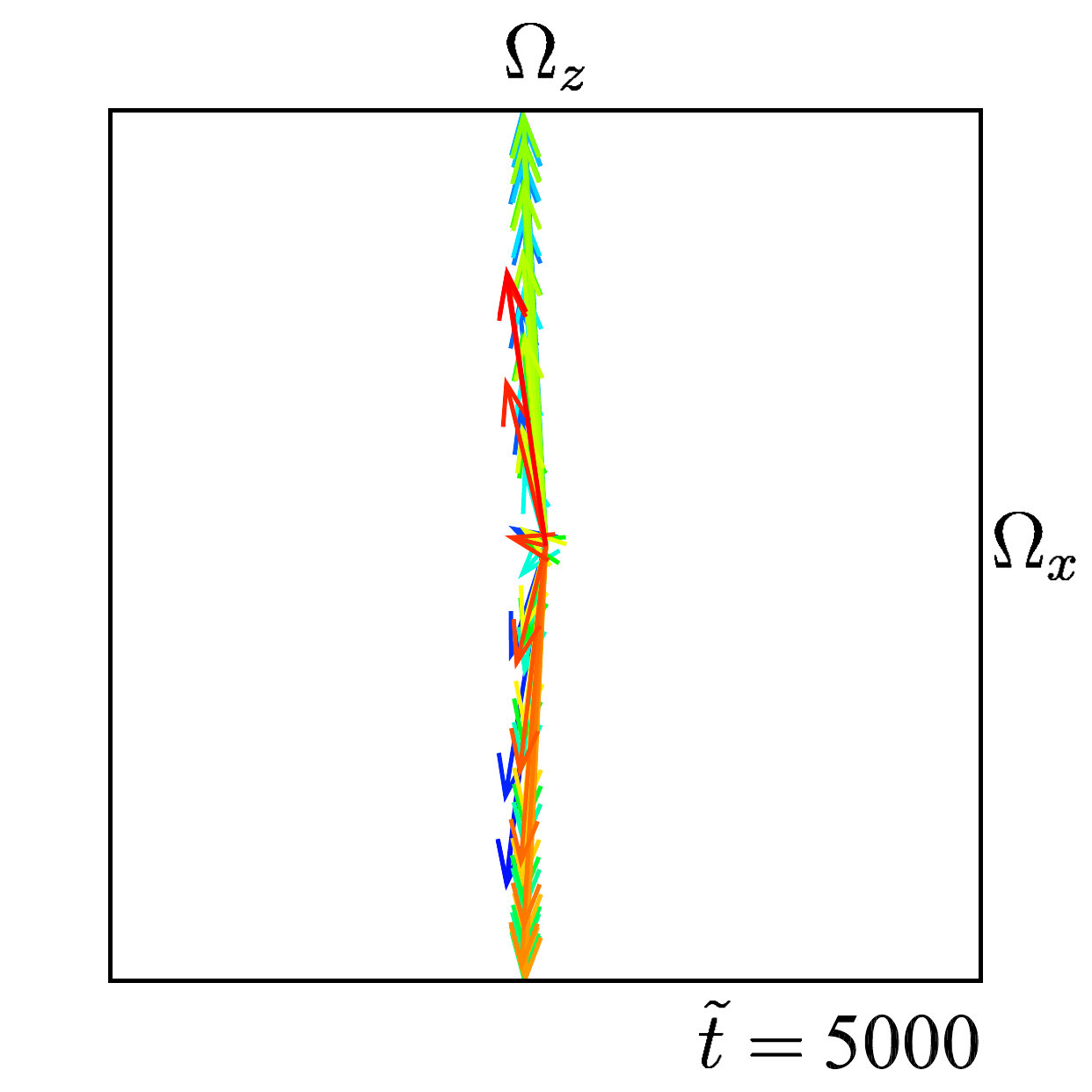}
\\[2mm]

\begin{overpic}[width=\FigWSnapB]{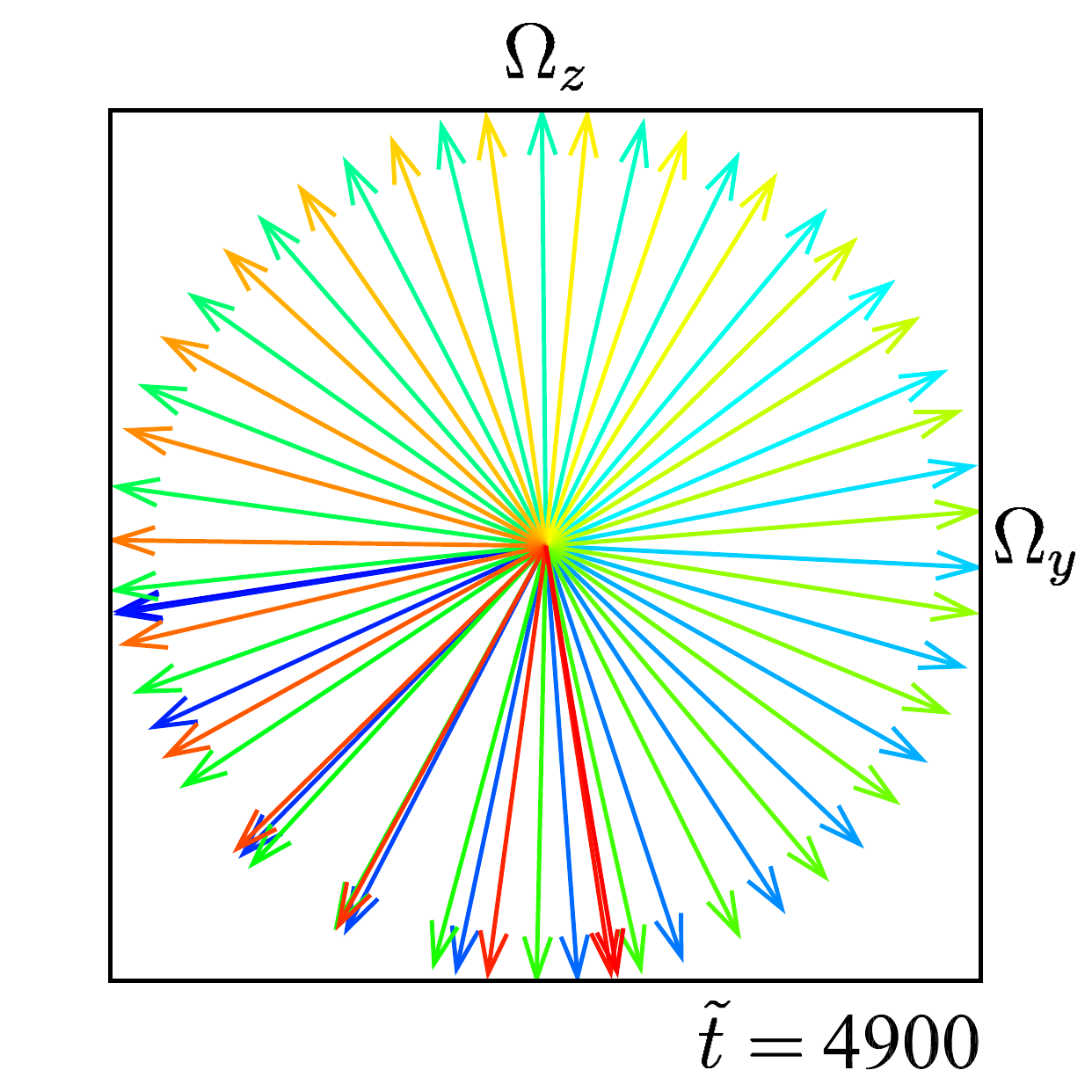}
\FigLabelPutSim{(c)}
\end{overpic}\hspace{\FigGapSnapB}%
\includegraphics[width=\FigWSnapB]{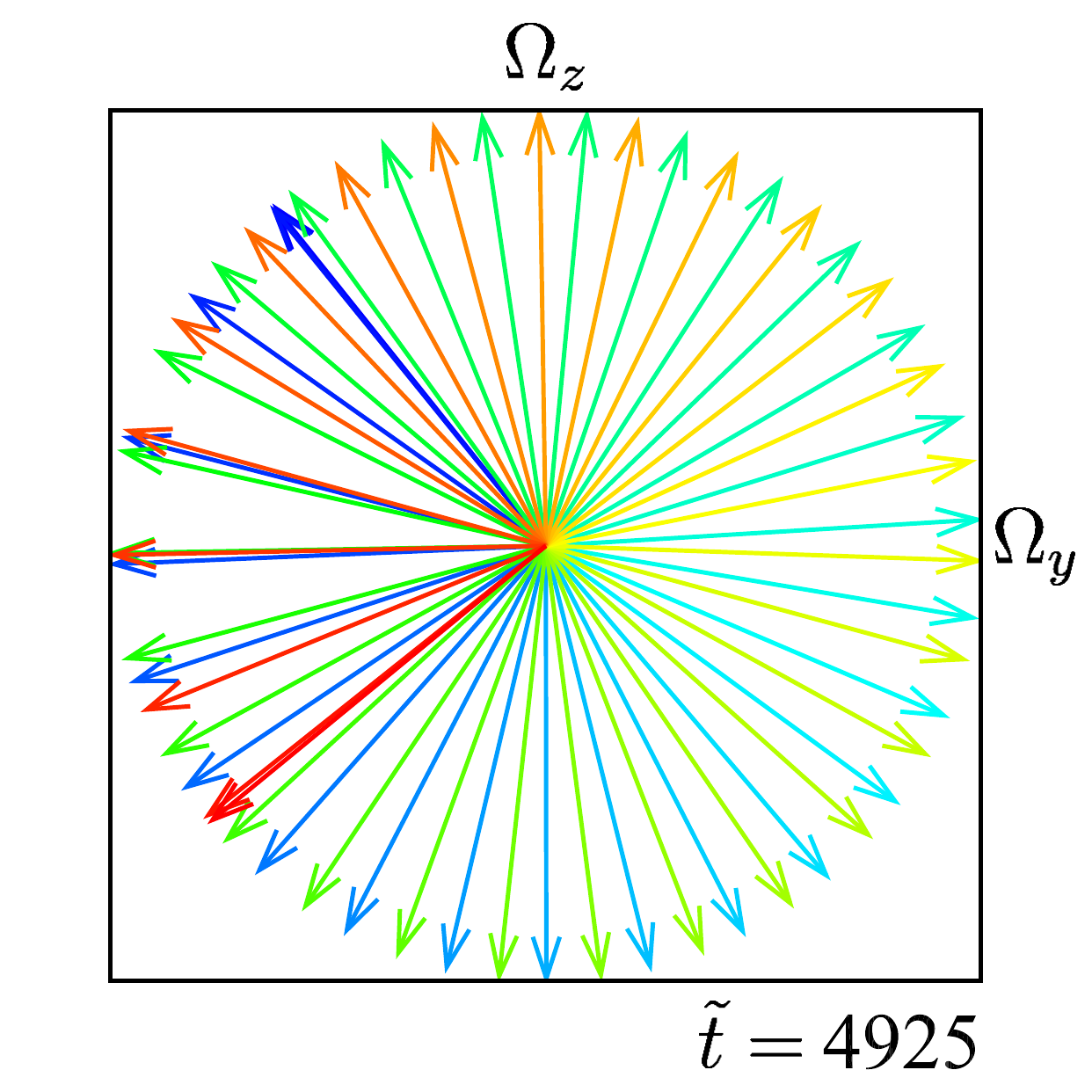}\hspace{\FigGapSnapB}%
\includegraphics[width=\FigWSnapB]{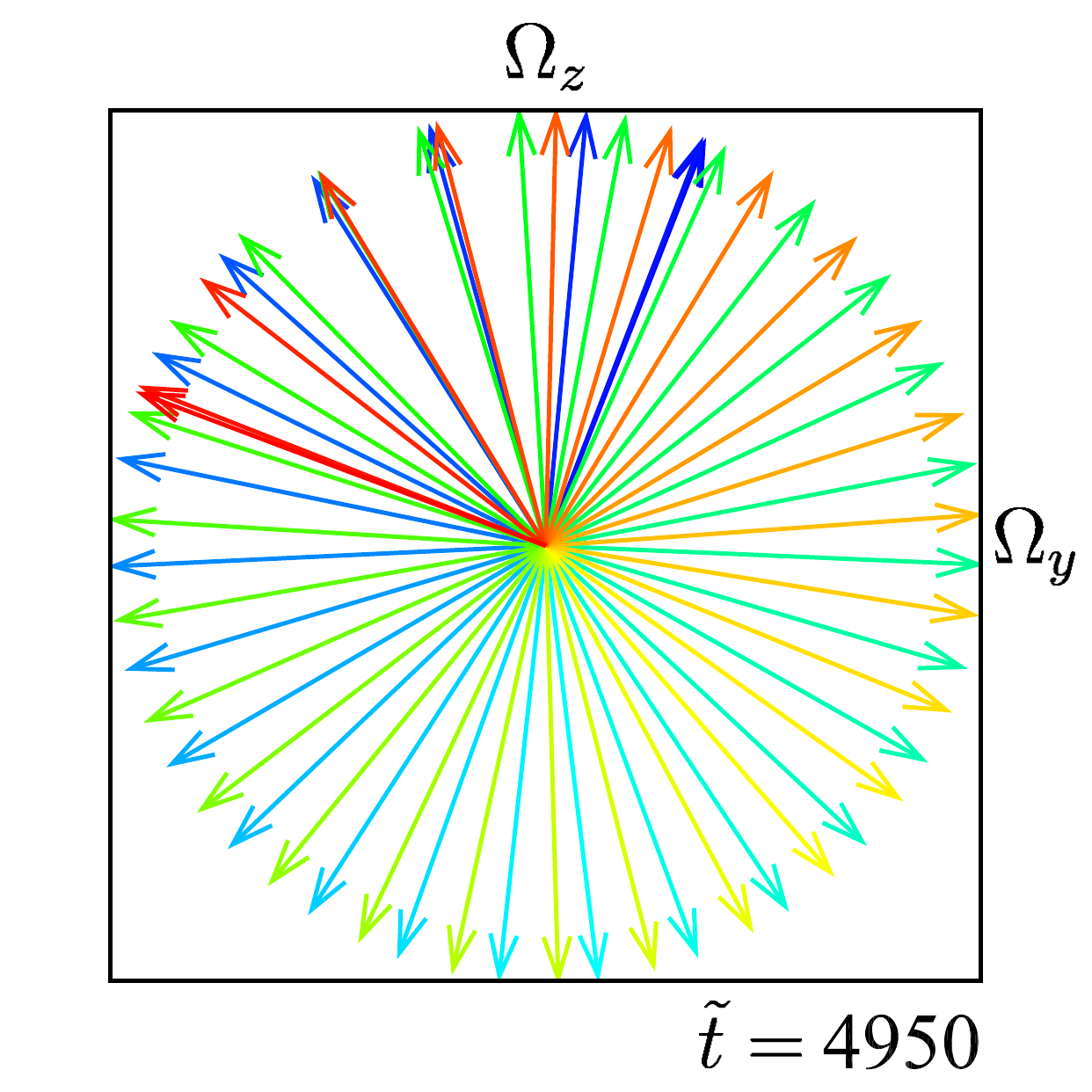}\hspace{\FigGapSnapB}%
\includegraphics[width=\FigWSnapB]{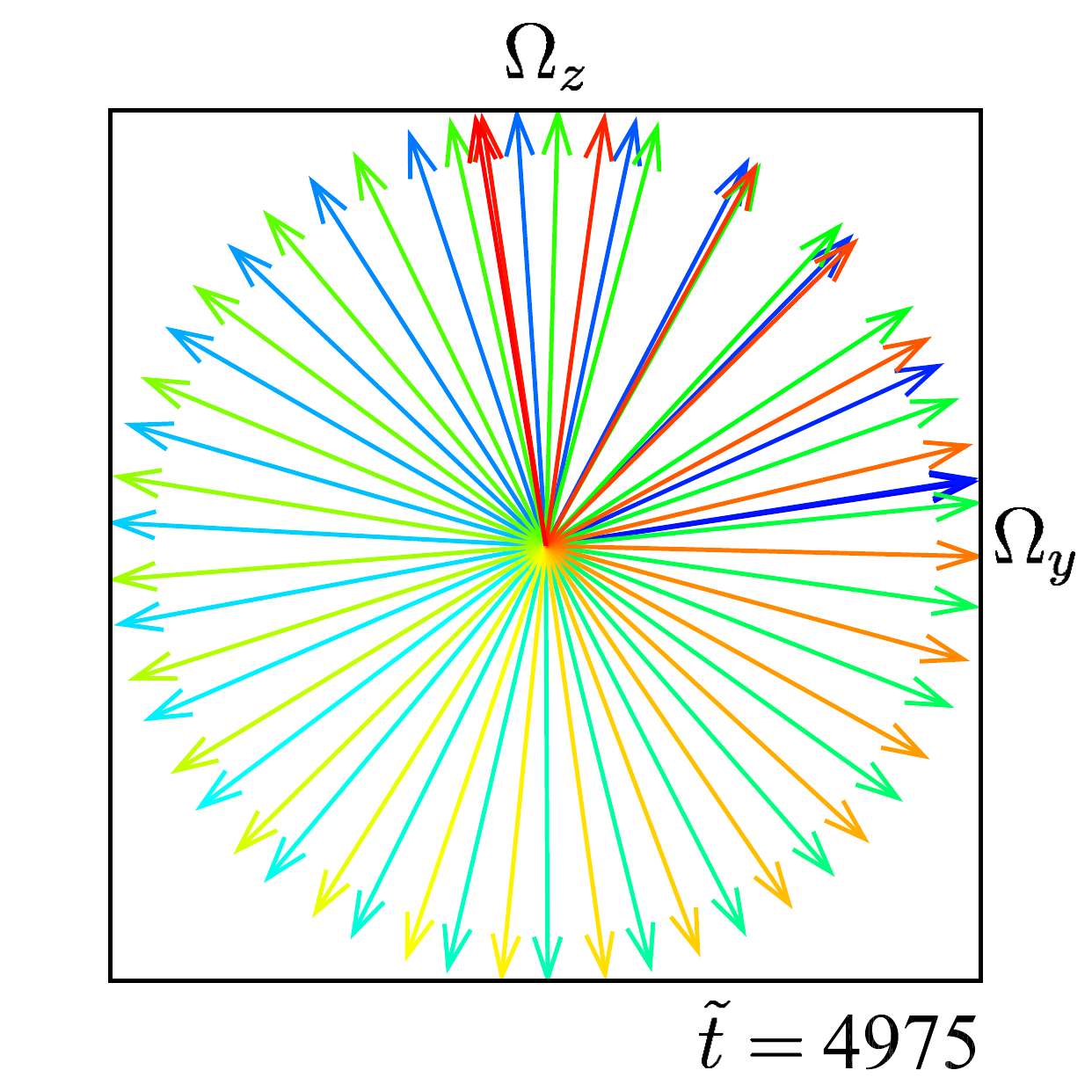}\hspace{\FigGapSnapB}%
\includegraphics[width=\FigWSnapB]{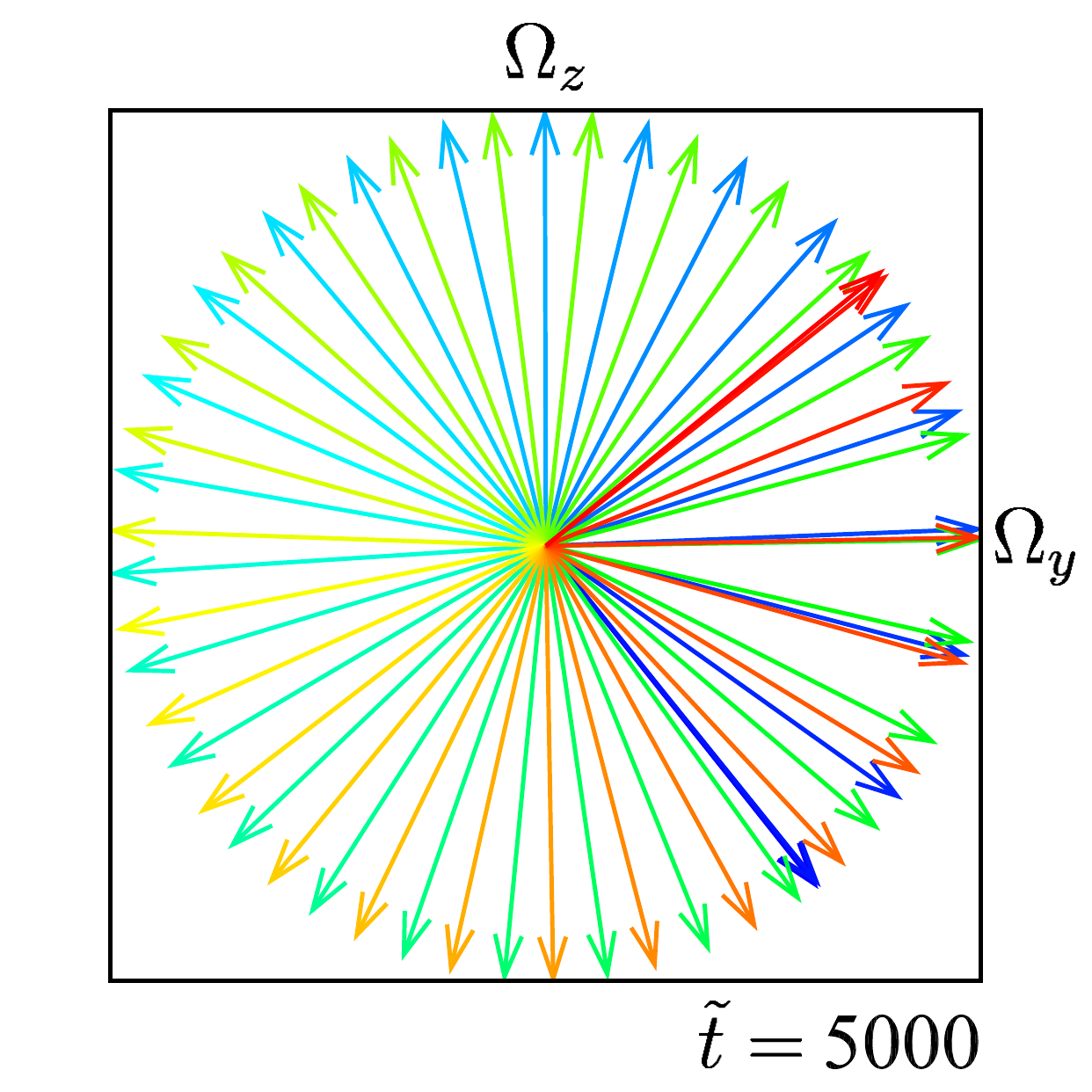}

\caption{
Late-time evolution of a spin chain consisting of 60 spins equally spaced along the $x$~direction, 
for $\tilde{\gamma}^2=0.100$, $J_{\mathrm{cc}} = 0.025$, $J_{\mathrm{sc}} = 1.000$, $\vb*{n}_{\mathrm{c}} = \vb*{e}_{x}$, and $\vb*{n}_{\mathrm{s}} = \vb*{e}_{x}$. 
Columns show snapshots at dimensionless times $\tilde{t} = 4900, 4925, 4950, 4975, 5000$
(left to right).
(a)~Bird's-eye view of the entire spin chain.
(b)~Normalized $x$ and $z$ components of the spins.
(c)~Normalized $y$ and $z$ components of the spins.
}
\label{fig:snapshots_Jcc0.025_Jsc1.000_nc100_ns100_late}
\end{figure*}

Figures~\ref{fig:q_all} and \ref{fig:q_Jcc}, the latter a close-up of Figs.~\ref{fig:q_all}(a) and \ref{fig:q_all}(b), test two aspects of the theory for $q_{\mathrm{helix}}$. First, the precessing helices keep the chirality of the static helix (Sec.~\ref{sec:sim-static}): the time-averaged $q_{\mathrm{helix}}$ is positive at all parameter points, as the full theory predicts. Second, the close-up views in Fig.~\ref{fig:q_Jcc} show that $q_{\mathrm{helix}}$ hardly depends on $\tilde{\gamma}^{2}$ and depends only weakly on $J_{\mathrm{cc}}$; these dependences distinguish the full effective Lagrangian $L_{\mathrm{eff}}(q)$ of Sec.~\ref{sec:helix-dynamic} from simpler alternatives.
Because the geometric-phase term $T_{\mathrm{eff}}(q)$ removes the factor $1-\Delta^{2}(q)$, $L_{\mathrm{eff}}(q)$ predicts that $q_{\mathrm{helix}}$ is independent of $\tilde{\gamma}^{2}$ and increases slightly with $J_{\mathrm{cc}}$ through the $J_{\mathrm{cc}}$-dependent term of $V_{\mathrm{eff}}(q)$, by about $0.3\%$ at $J_{\mathrm{cc}}=0.100$. The naive rescaling of the static effective Hamiltonian $H_{\mathrm{eff}}[\vb*{S}]$, Eq.~(\ref{eq:variational-energy_dynamic}), and $V_{\mathrm{eff}}(q)$ alone, by contrast, retain this factor and predict that $q_{\mathrm{helix}}$ decreases as $\abs{\Delta}$ grows: by about $7\%$ at $J_{\mathrm{cc}}=0.100$ for $\tilde{\gamma}^{2}=0.100$ and by about $10\%$ at $\tilde{\gamma}^{2}=0.020$ for $J_{\mathrm{cc}}=0.025$. The simulations clearly exclude these decreases. For $J_{\mathrm{cc}}=0.025$, the simulated $q_{\mathrm{helix}}$ varies by less than $0.1\%$ between $\tilde{\gamma}^{2}=0.020$ and $0.4$ [Fig.~\ref{fig:q_Jcc}(a)], and it increases slightly with $J_{\mathrm{cc}}$ [Fig.~\ref{fig:q_Jcc}(b)]; the size of this increase, below $1\%$ for $J_{\mathrm{cc}}\le0.100$, is comparable to the finite-size deviations discussed above, so that the simulations test its sign but not its magnitude.
\newcommand{\FigLabelPutGraph}[1]{\put(15,70){\textbf{#1}}}
\begin{figure*}[htb]
  \centering
  \begin{overpic}[width=0.32\textwidth]{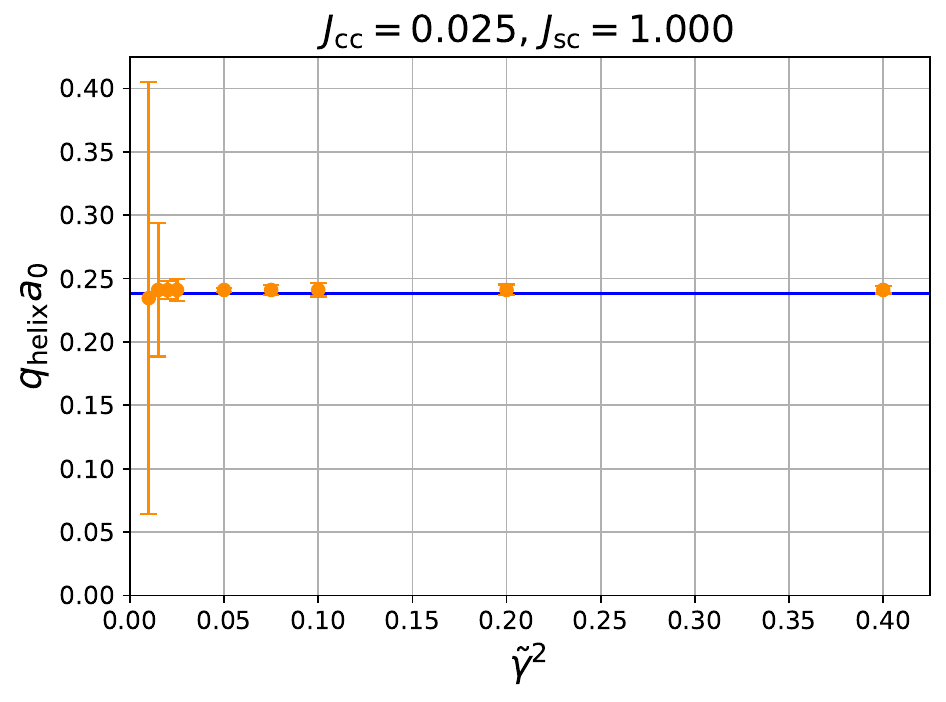}
  \FigLabelPutGraph{(a)}
  \end{overpic}
  \hfill
  \begin{overpic}[width=0.32\textwidth]{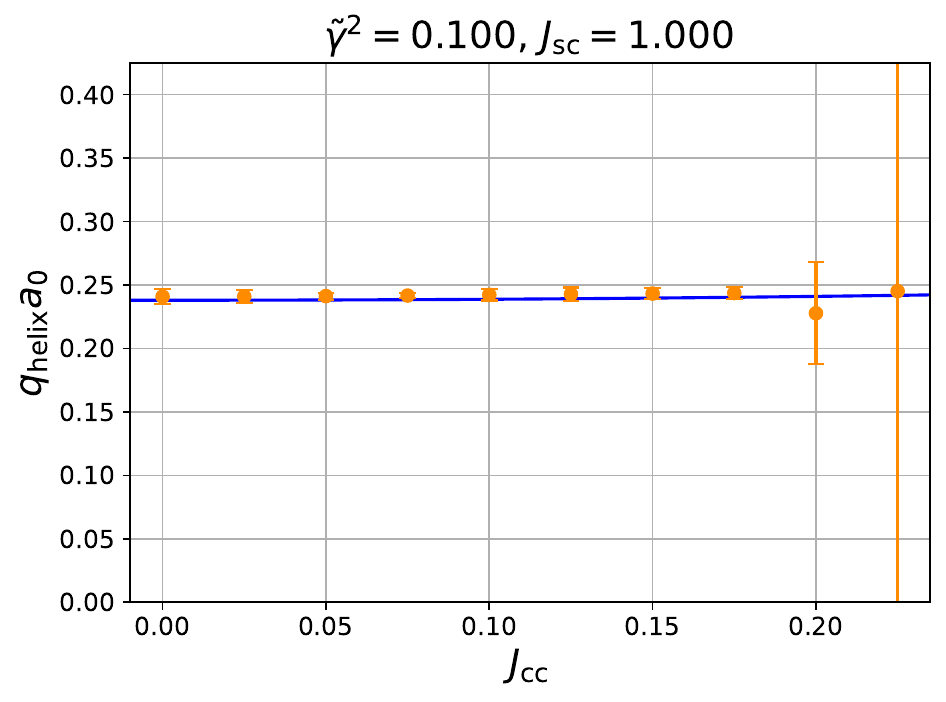}
  \FigLabelPutGraph{(b)}
  \end{overpic}
  \hfill
  \begin{overpic}[width=0.32\textwidth]{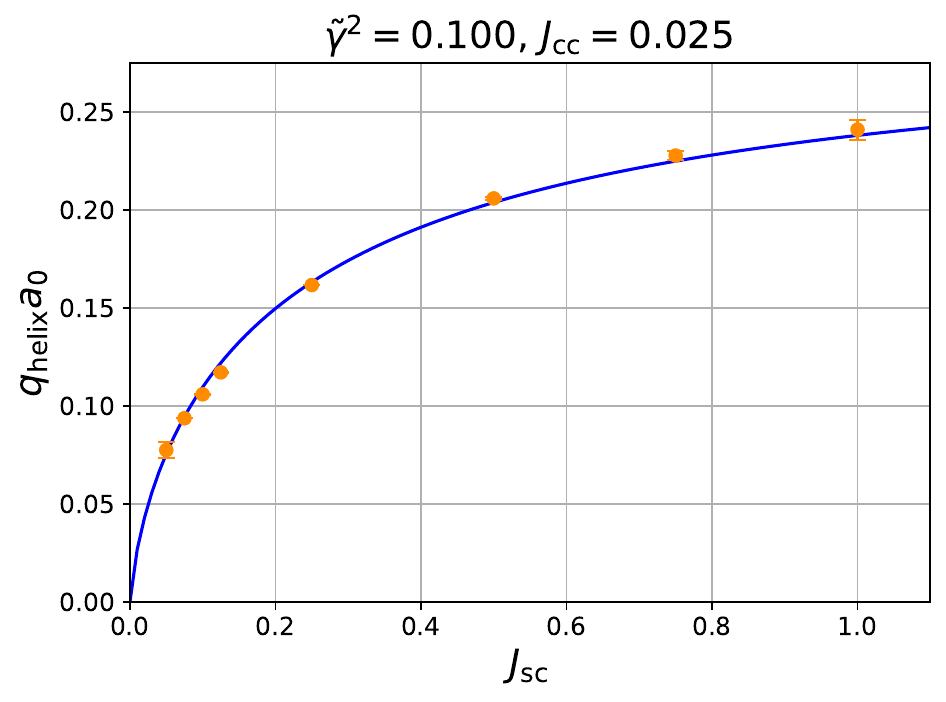}
  \FigLabelPutGraph{(c)}
  \end{overpic}
   \caption{
   Dependence of the helical wavenumber $q_{\mathrm{helix}}$ on the model parameters. 
  (a) $q_{\mathrm{helix}}$ as a function of $\tilde{\gamma}^2$ for fixed $J_{\mathrm{cc}}=0.025$ and $J_{\mathrm{sc}}=1.000$.
  (b) $q_{\mathrm{helix}}$ as a function of $J_{\mathrm{cc}}$ for fixed $\tilde{\gamma}^2=0.100$ and $J_{\mathrm{sc}}=1.000$.
  (c) $q_{\mathrm{helix}}$ as a function of $J_{\mathrm{sc}}$ for fixed $\tilde{\gamma}^2=0.100$ and $J_{\mathrm{cc}}=0.025$.
  In all panels, the solid blue line is the theoretical curve; the orange circles are time averages of the numerical results over $\tilde{t}=4000$--$5000$, and the error bars indicate the standard deviations of their temporal fluctuations.
  }
  \label{fig:q_all}
\end{figure*}

\begin{figure}[htb]
  \centering
  \begin{overpic}[width=0.85\linewidth]{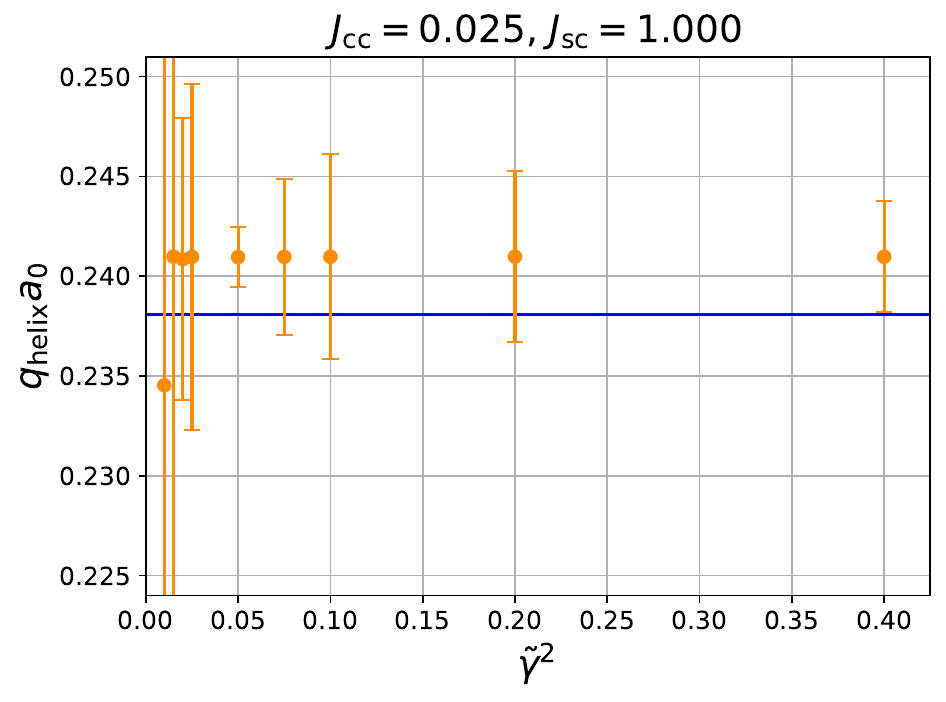}
  \FigLabelPutGraph{(a)}
  \end{overpic}
  \hfill
  \begin{overpic}[width=0.85\linewidth]{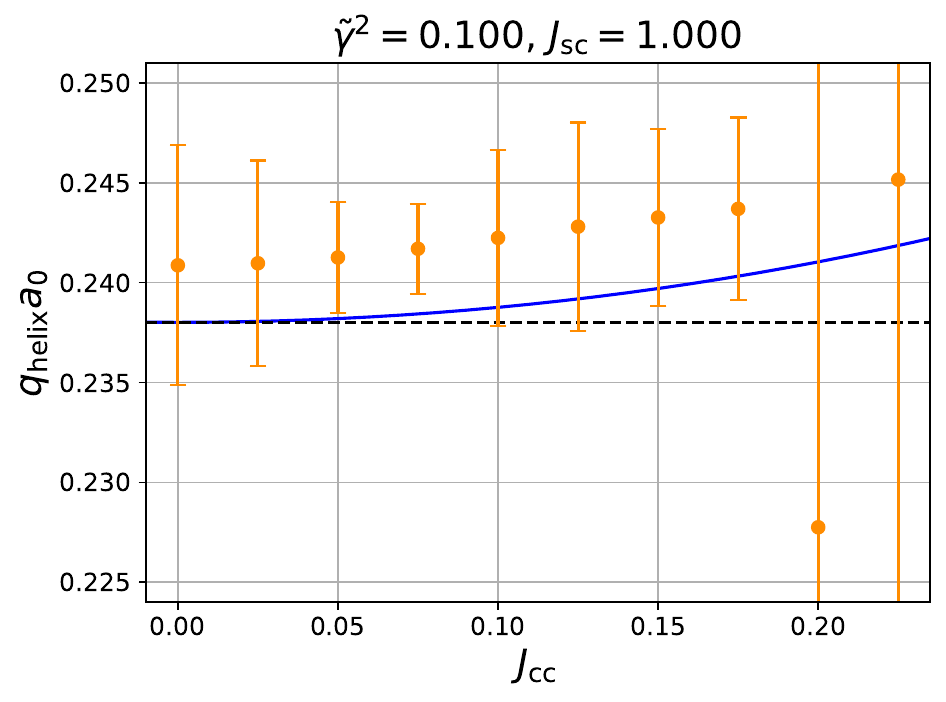}
  \FigLabelPutGraph{(b)}
  \end{overpic}
  \caption{Close-up view of Figs.~\ref{fig:q_all}(a) and \ref{fig:q_all}(b), showing the dependence of the helical wavenumber $q_{\mathrm{helix}}$ on the model parameters. 
  (a) $q_{\mathrm{helix}}$ as a function of $\tilde{\gamma}^2$ for fixed $J_{\mathrm{cc}}=0.025$ and $J_{\mathrm{sc}}=1.000$.
  (b) $q_{\mathrm{helix}}$ as a function of $J_{\mathrm{cc}}$ for fixed $\tilde{\gamma}^2=0.100$ and $J_{\mathrm{sc}}=1.000$.
  Lines and symbols are as in Fig.~\ref{fig:q_all} except for the dashed black line in (b) denoting the theoretical value at $J_{\mathrm{cc}}=0$.
  }
  \label{fig:q_Jcc}
\end{figure}

Figure~\ref{fig:omega_all} tests the prediction that the dimensionless precession frequency
$\tilde\omega(q)=\omega(q)/(\tilde\gamma^{2}\epsilon_{\mathrm{F}})$ is inversely proportional to
$\tilde\gamma^{2}$ [Eqs.~(\ref{eq:omega}) and (\ref{eq:omega-tilde})] because $\omega(q)$ itself is
independent of $\tilde\gamma^{2}$.
Figure~\ref{fig:Delta_all} similarly tests the predicted $\tilde\gamma^{-2}$ scaling of the canting
ratio $\Delta$ toward the hard axis [Eq.~(\ref{eq:Delta})], which vanishes continuously as
$J_{\mathrm{cc}}\to0$, consistent with the recovery of the static helix of Sec.~\ref{sec:helix-static}.

\begin{figure*}[htb]
  \centering
  \begin{overpic}[width=0.32\textwidth]{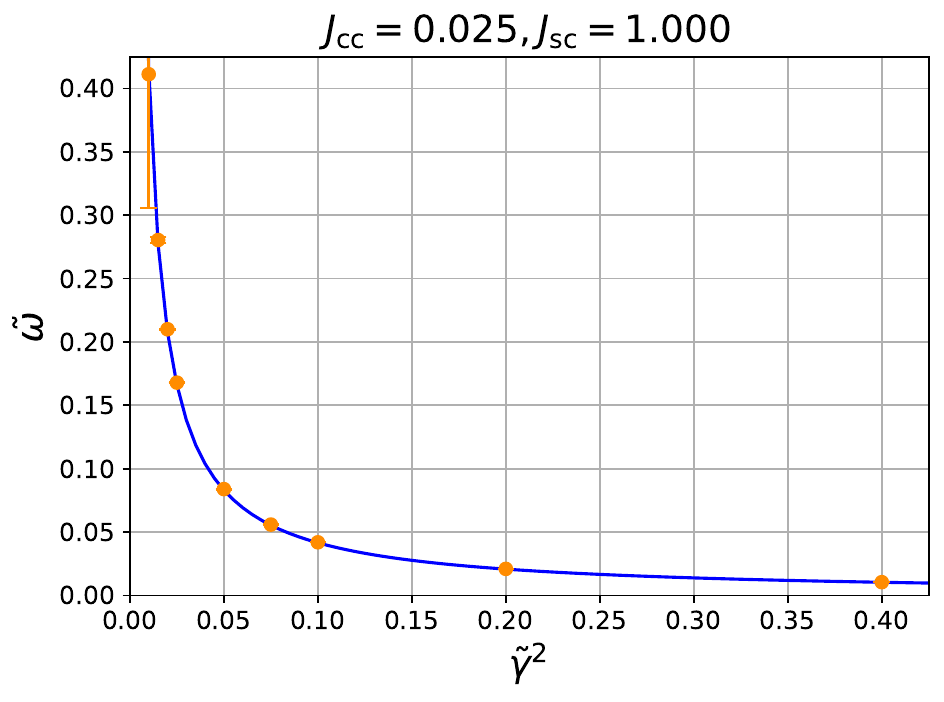}
  \FigLabelPutGraph{(a)}
  \end{overpic}
  \hfill
  \begin{overpic}[width=0.32\textwidth]{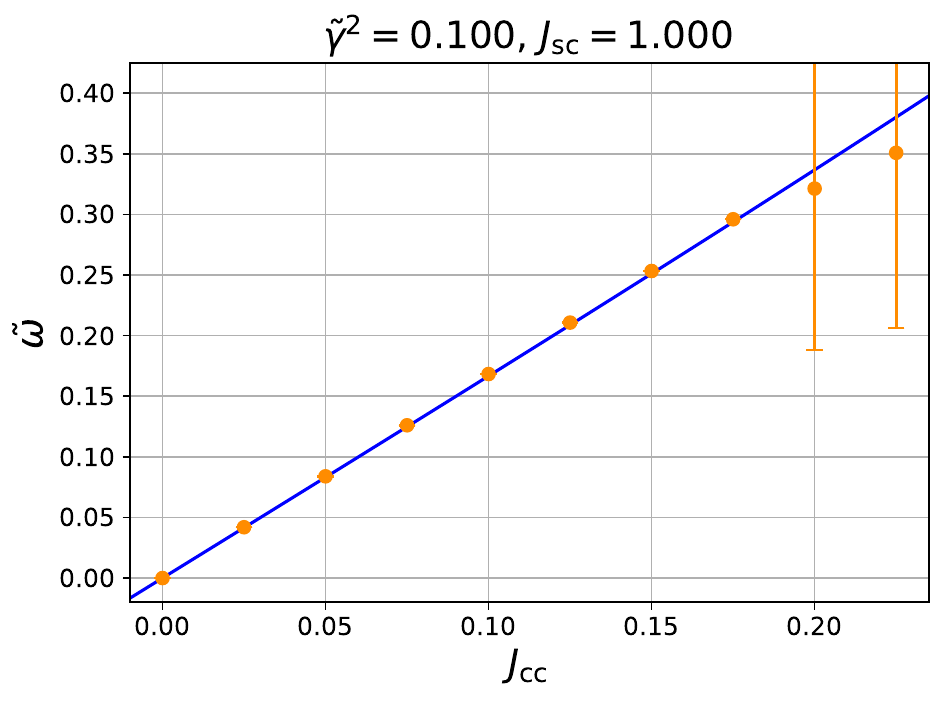}
  \FigLabelPutGraph{(b)}
  \end{overpic}
  \hfill
  \begin{overpic}[width=0.32\textwidth]{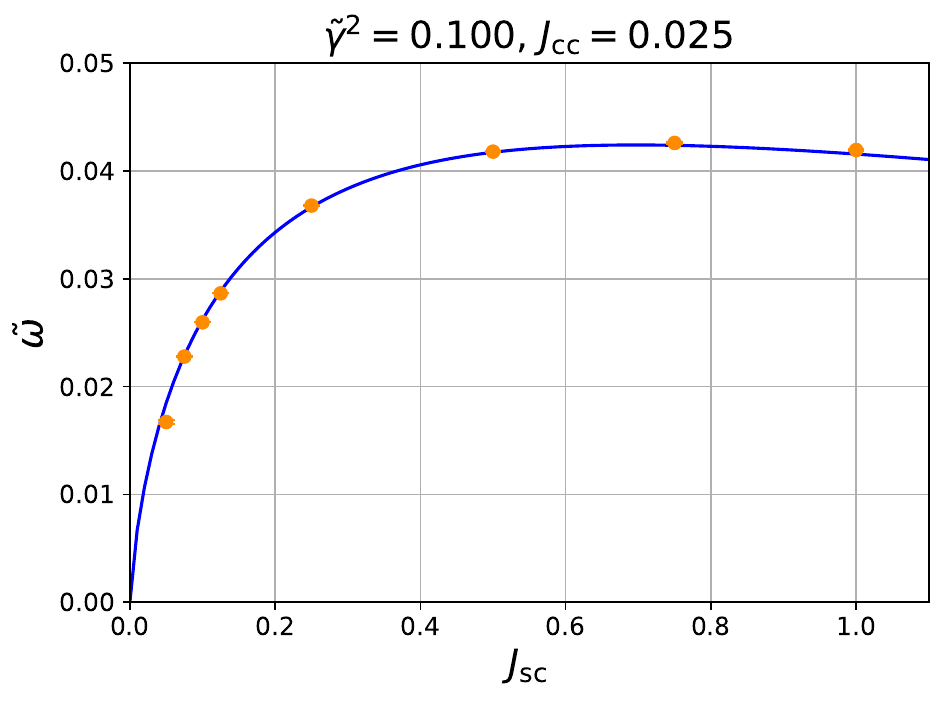}
  \FigLabelPutGraph{(c)}
  \end{overpic}
  \caption{
  Dependence of the dimensionless precession frequency $\tilde{\omega}$ on the model parameters. 
  (a) $\tilde{\omega}$ as a function of $\tilde{\gamma}^2$ for fixed $J_{\mathrm{cc}}=0.025$ and $J_{\mathrm{sc}}=1.000$. 
  (b) $\tilde{\omega}$ as a function of $J_{\mathrm{cc}}$ for fixed $\tilde{\gamma}^2=0.100$ and $J_{\mathrm{sc}}=1.000$. 
  (c) $\tilde{\omega}$ as a function of $J_{\mathrm{sc}}$ for fixed $\tilde{\gamma}^2=0.100$ and $J_{\mathrm{cc}}=0.025$. 
  Lines and symbols are as in Fig.~\ref{fig:q_all}.
  }
  \label{fig:omega_all}
\end{figure*}

\begin{figure*}[htb]
  \centering
  \begin{overpic}[width=0.32\textwidth]{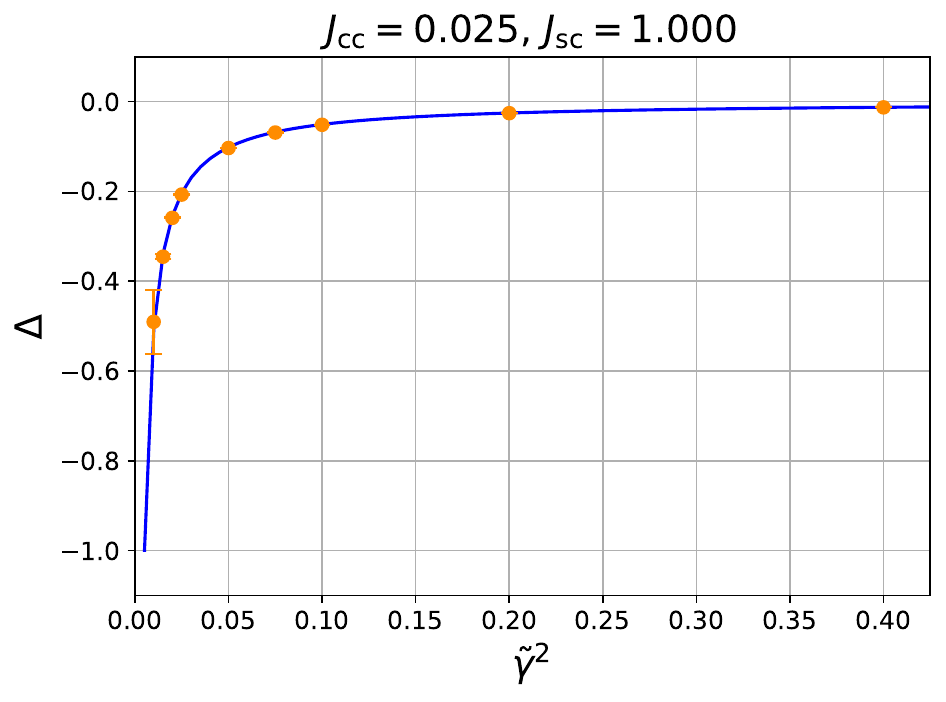}
  \FigLabelPutGraph{(a)}
  \end{overpic}
  \hfill
  \begin{overpic}[width=0.32\textwidth]{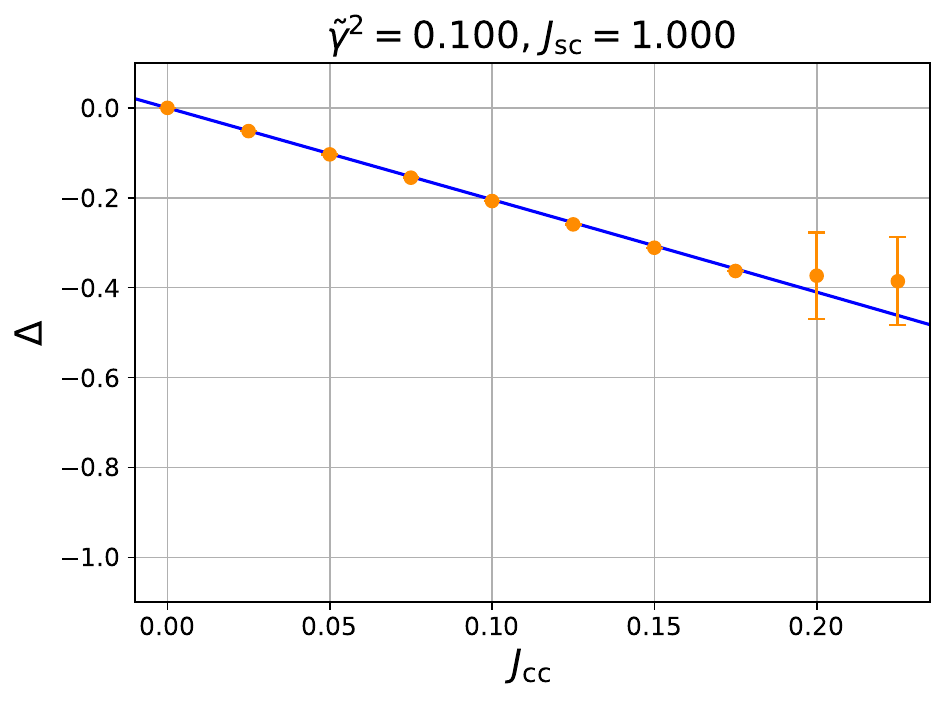}
  \FigLabelPutGraph{(b)}
  \end{overpic}
  \hfill
  \begin{overpic}[width=0.32\textwidth]{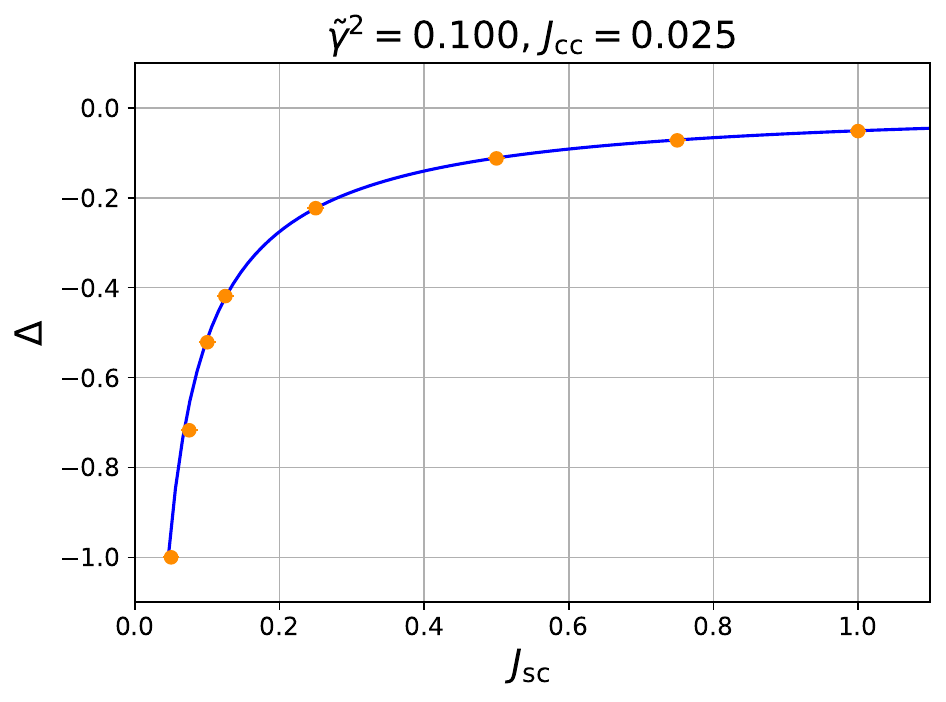}
  \FigLabelPutGraph{(c)}
  \end{overpic}
  \caption{
  Dependence of the canting ratio $\Delta$ toward the hard axis on the model parameters. 
  (a) $\Delta$ as a function of $\tilde{\gamma}^2$ for fixed $J_{\mathrm{cc}}=0.025$ and $J_{\mathrm{sc}}=1.000$.
  (b) $\Delta$ as a function of $J_{\mathrm{cc}}$ for fixed $\tilde{\gamma}^2=0.100$ and $J_{\mathrm{sc}}=1.000$.
  (c) $\Delta$ as a function of $J_{\mathrm{sc}}$ for fixed $\tilde{\gamma}^2=0.100$ and $J_{\mathrm{cc}}=0.025$.
  Lines and symbols are as in Fig.~\ref{fig:q_all}.
  }
  \label{fig:Delta_all}
\end{figure*}

\subsection{Ferromagnetic and chaotic regimes}
\label{sec:sim-ferro-chaos}
\begin{figure}[htb]
  \centering
  \begin{overpic}[width=0.48\linewidth]{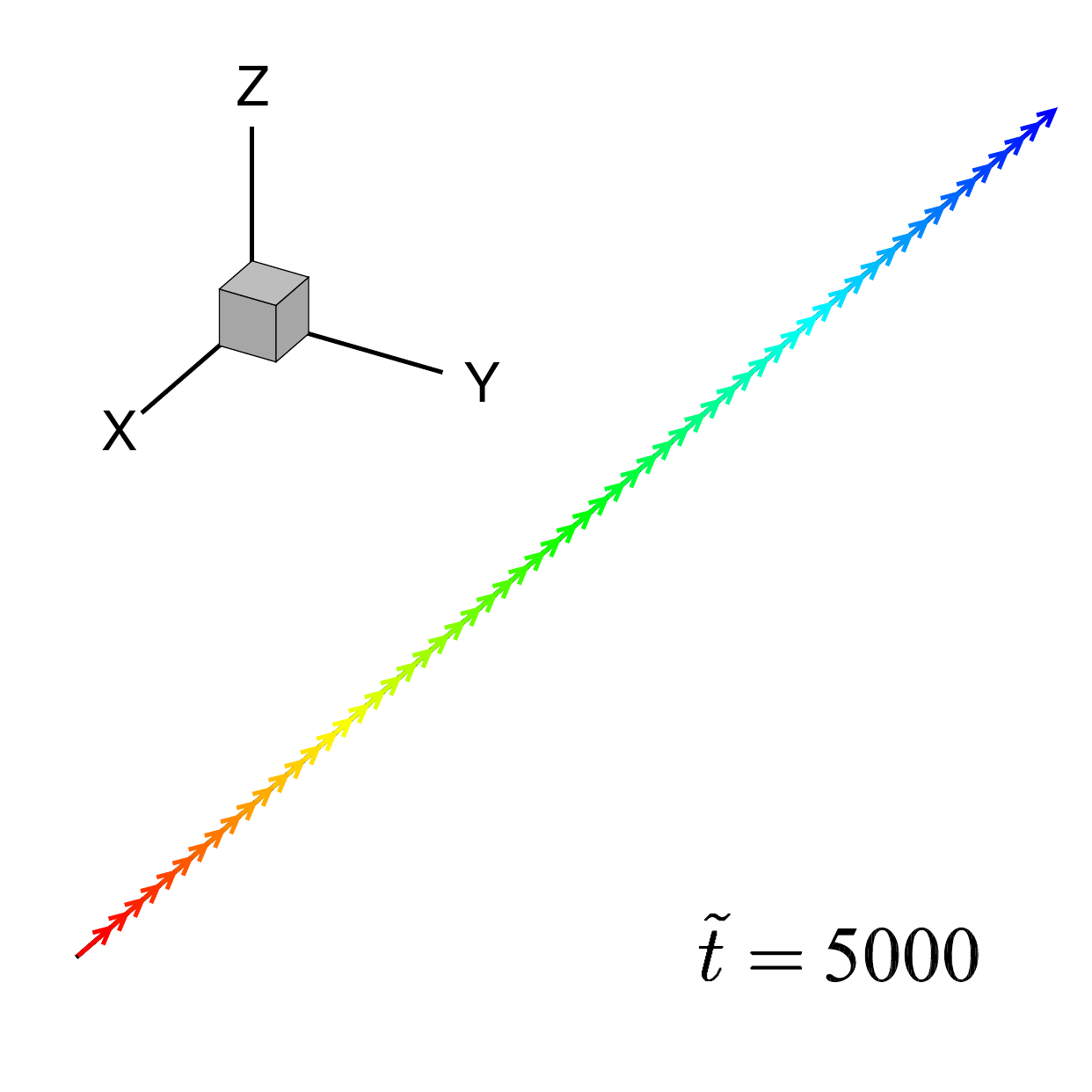}
  \put(0,95){(a)}
  \end{overpic}
  \hfill
  \begin{overpic}[width=0.48\linewidth]{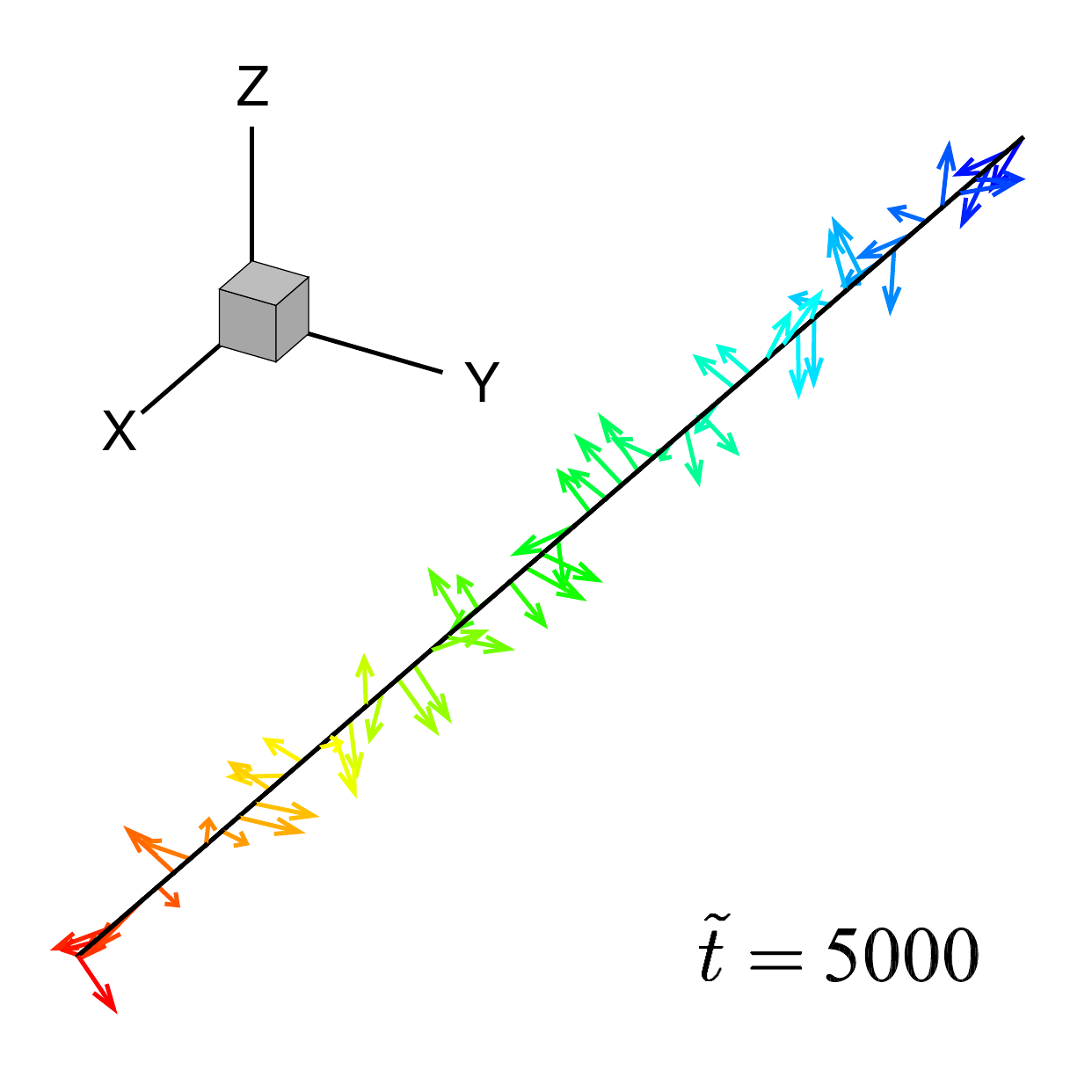}
  \put(0,95){(b)}
  \end{overpic}
\caption{
Snapshots at $\tilde{t} = 5000$ of a spin chain consisting of $N_{\mathrm{S}}=60$ spins equally spaced along the $x$~direction for $\tilde{\gamma}^2=0.100$, $\vb*{n}_{\mathrm{c}} = \vb*{e}_{x}$, and $\vb*{n}_{\mathrm{s}} = \vb*{e}_{x}$.
(a)~Bird's-eye view of the entire spin chain for $J_{\mathrm{cc}} = 0.025$ and $J_{\mathrm{sc}} = 0.025$.
(b)~Bird's-eye view of the entire spin chain for $J_{\mathrm{cc}} = 1.000$ and $J_{\mathrm{sc}} = 1.000$.
}
\label{fig:snapshots_ferro_chaos}
\end{figure}

Beyond the pure- and weakly charged spin-current regimes shown in Figs.~\ref{fig:snapshots_Jcc0.000_Jsc1.000_nc100_ns100}--\ref{fig:Delta_all}, we also examined, without presenting detailed figures here, the response of the chain to small and large spin-polarized charge currents (including the fully spin-polarized limit $J_{\mathrm{cc}}=J_{\mathrm{sc}}$). 
For small spin-polarized currents, the chain settles into ferromagnetic order aligned opposite to the spin-polarization direction $\vb*{n}_{\mathrm{s}}$, so that the current makes $\vb*{n}_{\mathrm{s}}$ an effective easy axis (Sec.~\ref{sec:four-type}) and controls the direction of the ferromagnetic order electrically.
For sufficiently large spin-polarized currents, the dynamics instead becomes irregular and apparently chaotic: the motion is not fully random, but the coherence of the dynamic helix is lost and the spins undergo complex, irregular oscillations, consistent with the chaotic regime identified for the two-spin system in Ref.~\cite{Onoda2006} [see also point (i) in Sec.~\ref{sec:conclusion}]. Along the fully spin-polarized diagonal $J_{\mathrm{cc}}=J_{\mathrm{sc}}$, this onset appears to be gradual rather than abrupt: small-amplitude fluctuations first become visible around $J_{\mathrm{cc}}=J_{\mathrm{sc}}\sim0.1$ and grow steadily as $J_{\mathrm{cc}}=J_{\mathrm{sc}}$ increases further; the onset of these fluctuations is also sensitive to the chain length $N_{\mathrm{S}}$, which we have not characterized systematically. The present simulations therefore do not determine the currents at which the irregular dynamics sets in; doing so requires a dedicated bifurcation and Lyapunov-exponent analysis beyond the comparison pursued here. 
Figures~\ref{fig:snapshots_ferro_chaos}(a) and \ref{fig:snapshots_ferro_chaos}(b) are examples of ferromagnetic ordering and irregular, apparently chaotic motion for small and large fully spin-polarized currents, respectively.

\section{Discussion}
\label{sec:discussion}

The central result of this work is that a single set of microscopically derived equations of motion, Eq.~(\ref{eq:EOM-S}), formulated for an arbitrary current direction $\vb*{n}_{\mathrm{c}}$ and spin-current polarization $\vb*{n}_{\mathrm{s}}$, yields two qualitatively distinct nonequilibrium magnetic phases of the periodic spin chain: a static helix stabilized by a pure spin current and a coherently precessing helix stabilized by a charged spin current. Both phases originate from the same mechanism identified in Sec.~\ref{sec:helix-theory}: the DM-type coupling generated by the spin current and oriented along its polarization $\vb*{n}_{\mathrm{s}}$ competes with the ferromagnetic exchange encoded in $F^{(0)}_{\mathrm{eq}}(\vb*{r})$ and makes $\vb*{n}_{\mathrm{s}}$ an effective hard axis. What distinguishes the static and dynamic cases is not a different mechanism but a different variational principle: an effective Hamiltonian $H_{\mathrm{eff}}[\vb*{S}]$ for $J_{\mathrm{cc}}=0$ versus an effective Lagrangian $L_{\mathrm{eff}}(q)$ that retains the geometric (Berry-phase) contribution for $J_{\mathrm{cc}}\neq0$.

The agreement in Figs.~\ref{fig:q_all}--\ref{fig:Delta_all} shows that the analytical treatment of Sec.~\ref{sec:helix-theory}---the helical ansatz and the two variational principles---captures the helical dynamics generated by Eq.~(\ref{eq:EOM-S}) quantitatively wherever the simulations yield a coherent helix. Because the three observables $q_{\mathrm{helix}}$, $\tilde\omega$, and $\Delta$ depend on $\tilde\gamma^{2}$, $J_{\mathrm{cc}}$, and $J_{\mathrm{sc}}$ in qualitatively different ways, this agreement is unlikely to be accidental, and it lends a posteriori support to the heuristic prescriptions used to construct $L_{\mathrm{eff}}(q)$. Determining $q$ from $V_{\mathrm{eff}}(q)$ alone would not reproduce the simulations; the geometric-phase term $T_{\mathrm{eff}}(q)$, which removes the factor $1-\Delta^{2}(q)$ from the effective Lagrangian, must be retained. A first-principles justification of the two prescriptions remains an open problem. The stability of the helix lies outside the present analysis, which does not predict the irregular, apparently chaotic dynamics that set in at larger $J_{\mathrm{cc}}$ (Secs.~\ref{sec:sim-dynamic} and \ref{sec:sim-ferro-chaos}) and anticipates the loss of coherence at small $\tilde\gamma^{2}$ only through the bound $\abs{\Delta}\le1$ of Eq.~(\ref{eq:Delta}). Near these limits, and where the helical pitch approaches the chain length, the deviations between theory and simulations grow (Sec.~\ref{sec:sim-dynamic}). Because the simulations integrate Eq.~(\ref{eq:EOM-S}) itself, they test this analysis of the equations of motion rather than the approximations made in deriving them (Secs.~\ref{sec:Model}--\ref{sec:EOM}), whose limitations are discussed below. The comparison thus leaves out, for instance, the corrections from a finite decoherence rate: keeping $\eta$ finite in the kernels with $n\le1$ (Sec.~\ref{sec:Approximations}) increases the predicted $q_{\mathrm{helix}}$, $\omega$, and $\abs{\Delta}$ of the representative helix with $J_{\mathrm{cc}}=0.025$ and $J_{\mathrm{sc}}=1.000$ by about $1\%$, $2\%$, and $3\%$ at $\eta/\epsilon_{\mathrm{F}}=10^{-2}$ and by about $4\%$, $5\%$, and $10\%$ at $\eta/\epsilon_{\mathrm{F}}=3\times10^{-2}$.

The chirality of the helix, i.e., the sign of $q_{\mathrm{helix}}$, is not captured correctly by a nearest-neighbor truncation of the RKKY-type couplings: only when longer-range terms are retained does the theory predict the handedness found in the simulations (Secs.~\ref{sec:helix-theory} and \ref{sec:simulation}). Unlike in conventional chiral magnets, where the DM interaction is fixed by spin-orbit coupling and the crystal structure~\cite{Dzyaloshinsky1958, Moriya1960}, the DM-type coupling here is generated by an externally driven, dissipative spin current (Sec.~\ref{sec:helix-theory}), and its sign is fixed by long-range RKKY-type couplings rather than by the lattice. Materials lacking bulk or interfacial DM interactions could therefore also host current-controlled helical order.

Because the helical pitch, precession frequency, and canting ratio are continuously tunable by $J_{\mathrm{cc}}$ and $J_{\mathrm{sc}}$, the helical pattern could in principle serve as a probe of the spin current: the spins rotate in the plane perpendicular to the spin polarization $\vb*{n}_{\mathrm{s}}$, and the pitch and chirality of the helix reflect the spin-current component along the chain. Both the chirality and the sense of precession are also set by the current and can be reversed electrically, in analogy with the electric-current control of spin helicity demonstrated in the itinerant helimagnet MnP~\cite{Jiang2020}. Because the helical state here is sustained by a dissipative spin current, however, using it for nonvolatile information storage, as in domain-wall racetrack memory~\cite{Parkin2008} or skyrmion-based devices~\cite{Fert2017}, would require an additional mechanism.

The present analysis is restricted to a strictly one-dimensional, translationally invariant chain, but the framework of Sec.~\ref{sec:helix-theory} can be extended in several directions. In finite or defective chains, or where the spin-current polarization varies in space, regions of opposite chirality could be separated by chiral domain walls, whose current-driven dynamics would connect to established domain-wall theory~\cite{Tatara2004,Thiaville2005}. Because the
dynamic helix is itself a coherent, current-driven oscillator, arrays of coupled chains may exhibit
synchronization phenomena analogous to those studied for spin-torque nano-oscillators~\cite{Locatelli2014}, potentially enhancing coherence and output power. The periodic helical modulation found here also forms a current-tunable artificial lattice for spin waves; computing its magnonic band structure, including possible nonreciprocity due to the chirality of the texture, would connect this work to magnonics~\cite{Chumak2015,Barman2021,Chumak2022}. Extending the analysis to two-dimensional nanomagnet arrays such as artificial spin ices~\cite{Skjaervo2020} is a further direction. There, interference between helical modulations along different directions might nucleate topological textures such as skyrmions, which spin-polarized currents can create in confined nanostructures~\cite{Iwasaki2013, Sampaio2013}; the present simulations do not address whether such nucleation occurs.

Beyond these spatial extensions, the classical Landau-Lifshitz-Gilbert framework~\cite{Landau1935, Gilbert2004} itself can be refined in at least two respects.
First, Eq.~(\ref{eq:EOM-S}) retains the nonlocal damping only to second order in the $s$-$d$ exchange and first order in time derivatives, whereas nonperturbative quantum-classical simulations reveal additional geometric torques~\cite{Bajpai2020} and ultrabroadband spin pumping~\cite{Petrovic2021}.
Second, spin-orbit coupling renders the nonlocal damping anisotropic~\cite{ReyesOsorio2024}, and electron correlations can strongly enhance its nonlocality~\cite{Lenzing2025}.
Incorporating these effects into the present framework is left for future work.

The theory also has a simple consequence for experiments: the spatial and temporal scales of the dynamic helimagnetic state can be tuned largely independently. For $J_{\mathrm{cc}}\ll1$, the helical wavenumber is well approximated by its static value $q_{\mathrm{helix}}$, which is set by the spin-current component (linear in $J_{\mathrm{sc}}$ up to a logarithmic factor for $J_{\mathrm{sc}}\ll1$; Sec.~\ref{sec:helix-theory}), whereas $\omega$ and $\Delta$ are, to leading order, proportional to the charge-current component $J_{\mathrm{cc}}$ at fixed $J_{\mathrm{sc}}$ [Eqs.~(\ref{eq:omega})--(\ref{eq:Delta})]. For orientation, with $k_{\mathrm{F}}a_{0}=1$ the representative pitch $2\pi/\abs{q_{\mathrm{helix}}}\approx26a_{0}$ corresponds to about $80$~nm for $n_{\mathrm{e}}\sim10^{18}$~cm$^{-3}$ and $8$~nm for $n_{\mathrm{e}}\sim10^{21}$~cm$^{-3}$, and $\omega\approx4\times10^{-3}\epsilon_{\mathrm{F}}$ corresponds to $\omega/(2\pi)\sim1$~GHz and $\sim1$~THz for $\epsilon_{\mathrm{F}}\sim1$~meV and $1$~eV, respectively. Equation~(\ref{eq:EOM-S}) describes deterministic dynamics: it retains the effective action only to first order in $\vb*{S}^{\Delta}$ (Sec.~\ref{sec:Model}) and therefore omits the fluctuations of the localized spins, including the thermal noise that accompanies the damping terms at finite temperature. The helices can be expected to remain coherent only for $k_{\mathrm{B}}T$ well below the energy scale $S\tilde{\gamma}^{2}\epsilon_{\mathrm{F}}$ of the couplings in Eq.~(\ref{eq:EOM-S}); in one dimension, such noise limits their coherence length and lifetime at any finite temperature, which we have not evaluated. Time-resolved magnetic imaging techniques, such as X-ray holography~\cite{Eisebitt2004} and scanning transmission X-ray microscopy~\cite{Baumgartner2017}, could probe such states once suitable material platforms are identified.

\section{Conclusions}
\label{sec:conclusion}

We have presented a theoretical and numerical study of the dynamics of a classical, localized
spin system coupled to conduction electrons carrying both charge and spin currents. Building on the
two-spin phase diagram of our previous letter~\cite{Onoda2006} (Fig.~\ref{fig:phase-diagram}), we derived the effective equations of motion for the localized spins, Eq.~(\ref{eq:EOM-S}), systematically from the $s$-$d$ exchange Hamiltonian (Secs.~\ref{sec:Model}--\ref{sec:EOM}). Applying these equations to a periodically arranged one-dimensional spin chain (Sec.~\ref{sec:helix-theory}), we analytically predicted static helical magnetic order under a pure spin current and dynamic helical order under a charged spin current, together with the helical wavenumber $q_{\mathrm{helix}}$ and, for the precessing helix, the precession frequency $\omega$ and canting ratio $\Delta$. Direct numerical integration of Eq.~(\ref{eq:EOM-S}) agrees quantitatively with these predictions wherever it yields a coherent helix (Sec.~\ref{sec:simulation}, Figs.~\ref{fig:q_all}--\ref{fig:Delta_all}). The helix chirality is set by the long-range RKKY-type couplings and is opposite to that obtained in the nearest-neighbor approximation. For the precessing helix, the selection of $q_{\mathrm{helix}}$ rests on the heuristic effective Lagrangian of Sec.~\ref{sec:helix-theory}, which has yet to be justified from first principles. The present analysis does not address the stability of this helix, which loses coherence at small $\tilde\gamma^{2}$ and at larger $J_{\mathrm{cc}}$ (Sec.~\ref{sec:simulation}).

Section~\ref{sec:discussion} discusses the implications of these results for experiments and devices. Here we instead outline four directions that would extend the theoretical framework itself rather than the specific phenomenon studied above.
(i) Our previous two-spin analysis found a regime of chaotic spin motion for strongly spin-polarized charge currents~\cite{Onoda2006}. In the extended chain considered here, sufficiently large charge-current components likewise produce irregular, apparently chaotic dynamics (Secs.~\ref{sec:sim-dynamic} and \ref{sec:sim-ferro-chaos}); determining their nature, for example whether they constitute spatiotemporal chaos, and the currents at which they set in requires a dedicated bifurcation and Lyapunov-exponent analysis. This analysis is of practical interest because the charge-current component is readily varied in experiments. (ii) Throughout this work, the localized spins were treated classically. In the presence of conduction electrons, this approximation can fail even for large spin length, as well as for large Heisenberg or $s$-$d$ exchange, owing to the dynamical buildup of entanglement~\cite{Mondal2021}. Quantizing the effective theory itself, e.g., via a spin-coherent-state path integral or a Holstein-Primakoff treatment of $H_{\mathrm{eff}}[\vb*{S}]$ and $L_{\mathrm{eff}}(q)$ (Sec.~\ref{sec:helix-theory}), would allow the quantum corrections to $q_{\mathrm{helix}}$, $\omega$, and $\Delta$, and their dependence on the spin quantum number $S$, to be assessed systematically. (iii) The conduction electrons enter the derivation of Secs.~\ref{sec:Approximations} and \ref{sec:EOM} only through the response functions $F^{(n)}_{\mathrm{eq}}(\vb*{r})$, $F^{(n)}_{\mathrm{cc}}(\vb*{r})$, and $\vb*{F}^{(n)}_{\mathrm{sc}}(\vb*{r})$, so the same
methodology applies directly once these functions are recomputed for conduction electrons confined
to one or two dimensions, where the modified Fermi-surface geometry is expected to alter the helical
wavenumber selection of Sec.~\ref{sec:helix-theory}. (iv) The methodology also applies to conduction electrons with a
Dirac-type dispersion, such as the helical surface states of a topological insulator, but there
spin-momentum locking makes the electron propagator nondiagonal in spin, so that the full spin-spin response tensor, not only $\vb*{F}^{(n)}_{\mathrm{sc}}(\vb*{r})$, must be recomputed, together with the spin density induced by the currents themselves. Because the nonequilibrium distribution enters the current-induced part of this response only through the electron states at the Fermi surface (Sec.~\ref{sec:Approximations}), this recomputation requires the dispersion and
spin texture at a generic Fermi wavevector, not only near the Dirac point where a linear continuum
approximation would suffice; lattice models of topological insulators with closed-form surface states valid throughout the surface Brillouin zone~\cite{Onoda2022} therefore provide a convenient starting point. Such an extension could yield helical textures and current-driven
torques beyond those captured by the free-electron treatment adopted here.

\begin{acknowledgments}
The author used Claude (Anthropic, Claude Opus 5.5) to propose revisions to the text and to assist with literature organization and consistency checks, and ChatGPT (OpenAI, GPT-5.6 Sol and GPT-6.1 Sol) to obtain critical pre-submission readings of the manuscript and to check the equations for errors. All comments and suggested corrections were assessed and verified by the author before adoption, and the author takes full responsibility for the content of this article.
\end{acknowledgments}

\appendix
\section{Three pictures on the closed time path}
\label{sec:Three-pictures_in_quantum-theory}
This Appendix briefly reviews the relations among the Schr\"{o}dinger, Heisenberg, and interaction pictures in the CTPGF formalism.
For an arbitrary operator $\hat{\mathcal{O}}$ in the Heisenberg picture, the subscripts ``$\mathrm{S}$'' and ``$\mathrm{I}$'' stand for the corresponding operators in the Schr\"{o}dinger and interaction pictures as $\hat{\mathcal{O}}_{\mathrm{S}}$ and $\hat{\mathcal{O}}_{\mathrm{I}}$, respectively.
As a theoretical expedient, we tentatively set the initial time to a finite value $t_{0}$ and introduce two values $t$ and $t'$ with the condition $t_{0} \le t' \le t$ for later convenience.
With $t_{\pm}$ as the time variables for the integration on the forward and backward paths, respectively, the interaction parts $\hat{V}_{\mathrm{S}}(t_{+})$ and $\hat{V}_{\mathrm{S}}(t_{-})$ of the Hamiltonian can differ from each other even when $t_{+}$ and $t_{-}$ take the same value, i.e., $\hat{V}_{\mathrm{S}}(t_{+})\vert_{t_{+}=t} \neq \hat{V}_{\mathrm{S}}(t_{-})\vert_{t_{-}=t}$.

First, we define the operators for the time evolution from $t'$ to $t$ on the forward path and for the reverse evolution from $t$ to $t'$ on the backward path, as follows:
\begin{align}
&
\hat{U}_{\mathrm{S}+}\qty(t,t') = \mathcal{T}_{\mathrm{ctp}}e^{-i\int_{t'}^{t}\dd{t_{+}}\hat{H}_{\mathrm{S}}(t_{+})}
,\\
&
\hat{U}_{\mathrm{S}-}\qty(t',t) = \mathcal{T}_{\mathrm{ctp}}e^{-i\int_{t}^{t'}\dd{t_{-}}\hat{H}_{\mathrm{S}}(t_{-})}
,\\
&
\hat{U}_{\mathrm{I}+}\qty(t,t') = \hat{U}_{0}\qty(t_{0},t)\hat{U}_{\mathrm{S}+}\qty(t,t')\hat{U}_{0}\qty(t',t_{0})
,\\
&
\hat{U}_{\mathrm{I}-}\qty(t',t) = \hat{U}_{0}\qty(t_{0},t')\hat{U}_{\mathrm{S}-}\qty(t',t)\hat{U}_{0}\qty(t, t_{0})
,
\end{align}
where $\hat{U}_{0}\qty(t,t') = e^{-i\hat{H}_{0}\qty(t-t')}$ and $\mathcal{T}_{\mathrm{ctp}}$ represents the path-ordering of operators from right to left according to the closed time path, i.e., the time-ordering on the forward path and the anti-time-ordering on the backward path.
Direct differentiation shows that $\hat{U}_{\mathrm{S}+}\qty(t,t')$ and $\hat{U}_{\mathrm{S}-}\qty(t',t)$ are the formal solutions of the following differential equations:
\begin{align}
& 
\dv{\hat{U}_{\mathrm{S}+}}{t}\qty(t,t') = -i\hat{H}_{\mathrm{S}}\qty(t) \hat{U}_{\mathrm{S}+}\qty(t,t')
,\\
&
\dv{\hat{U}_{\mathrm{S}-}}{t}\qty(t',t) = \hspace{2.5mm} i\hat{U}_{\mathrm{S}-}\qty(t',t) \hat{H}_{\mathrm{S}}\qty(t)
.
\end{align}
In the above equations, $\hat{H}_{\mathrm{S}}(t)$ stands for $\hat{H}_{\mathrm{S}}(t_{+})\vert_{t_{+}=t}$ in the first one and for $\hat{H}_{\mathrm{S}}(t_{-})\vert_{t_{-}=t}$ in the second one.
Note that neither $\hat{U}_{\mathrm{S}-}\qty(t',t) \hat{U}_{\mathrm{S}+}\qty(t,t')$ nor $\hat{U}_{\mathrm{S}+}\qty(t,t')\hat{U}_{\mathrm{S}-}\qty(t',t)$ is necessarily the identity operator $\hat{I}$ as long as $\hat{V}_{\mathrm{S}}(t_{+})\vert_{t_{+}=t} \neq \hat{V}_{\mathrm{S}}(t_{-})\vert_{t_{-}=t}$, while $\hat{U}_{0}\qty(t',t)\hat{U}_{0}\qty(t,t')=\hat{U}_{0}\qty(t,t')\hat{U}_{0}\qty(t',t)=\hat{I}$ always holds.

Taking an operator $\hat{\mathcal{O}}_{\mathrm{S}}\qty(t)$ in the Schr\"{o}dinger picture, we define the corresponding operators in other pictures as follows:
\begin{align}
& 
\hat{\mathcal{O}}\qty(t) 
= \hat{U}_{\mathrm{S}-}\qty(t_{0},t) \hat{\mathcal{O}}_{\mathrm{S}}\qty(t) \hat{U}_{\mathrm{S}+}\qty(t,t_{0})
,\\
& 
\hat{\mathcal{O}}_{\mathrm{I}}\qty(t) 
= \hat{U}_{0}\qty(t_{0},t) \hat{\mathcal{O}}_{\mathrm{S}}\qty(t) \hat{U}_{0}\qty(t,t_{0})
.
\end{align}
The second of the above definitions and the definitions of $\hat{U}_{\mathrm{I}+}\qty(t,t')$ and $\hat{U}_{\mathrm{I}-}\qty(t', t)$ show that the operators satisfy the following differential equations, respectively:
\begin{align}
& 
\dv{\hat{U}_{\mathrm{I}+}}{t}\qty(t,t') = -i\hat{V}_{\mathrm{I}}\qty(t) \hat{U}_{\mathrm{I}+}\qty(t,t')
,\\
&
\dv{\hat{U}_{\mathrm{I}-}}{t}\qty(t',t) = \hspace{2.5mm} i\hat{U}_{\mathrm{I}-}\qty(t',t) \hat{V}_{\mathrm{I}}\qty(t)
.
\end{align}
Similarly, $\hat{V}_{\mathrm{I}}(t)$ stands for $\hat{V}_{\mathrm{I}}(t_{+})\vert_{t_{+}=t}$ in the first and for $\hat{V}_{\mathrm{I}}(t_{-})\vert_{t_{-}=t}$ in the second of the above equations.
Thus, we can also express $\hat{U}_{\mathrm{I}+}\qty(t,t')$ and $\hat{U}_{\mathrm{I}-}\qty(t', t)$ as the formal solutions of these equations as follows:
\begin{align}
& 
\hat{U}_{\mathrm{I}+}\qty(t,t')
=
\mathcal{T}_{\mathrm{ctp}}e^{-i\int_{t'}^{t}\dd{t_{+}}\hat{V}_{\mathrm{I}}(t_{+})}
,\\
&
\hat{U}_{\mathrm{I}-}\qty(t',t)
=
\mathcal{T}_{\mathrm{ctp}}e^{-i\int_{t}^{t'}\dd{t_{-}}\hat{V}_{\mathrm{I}}(t_{-})}
 .
\end{align}

To close this Appendix, we mention the relation between the partition function $\mathcal{Z}_{\mathrm{Q}}[\vb*{S}]$ of Eq.~(\ref{eq:partition-function}) and the density matrix operators in the Schr\"{o}dinger and interaction pictures, $\hat{\rho}_{\mathrm{S}}\qty(t)$ and $\hat{\rho}_{\mathrm{I}}\qty(t)$.
The time-evolution operators give the following representations of the density matrix operator in the two pictures:
\begin{align}
\hat{\rho}_{\mathrm{S}}\qty(t)
& = \hat{U}_{\mathrm{S}+}\qty(t,t_{0}) \hat{\rho}_{0} \hat{U}_{\mathrm{S}-}\qty(t_{0},t)
,\\
\hat{\rho}_{\mathrm{I}}\qty(t) 
& = \hat{U}_{0}\qty(t_{0},t) \hat{\rho}_{\mathrm{S}}\qty(t) \hat{U}_{0}\qty(t,t_{0})
\nonumber\\
& = \hat{U}_{\mathrm{I}+}\qty(t,t_{0}) \hat{\rho}_{0}\hat{U}_{\mathrm{I}-}\qty(t_{0},t)
.
\end{align}
The relation $\Tr\hat{\rho}_{\mathrm{S}}\qty(t) = \Tr\hat{\rho}_{\mathrm{I}}\qty(t)$ follows from $\hat{U}_{0}\qty(t_{0},t)\hat{U}_{0}\qty(t,t_{0})=\hat{U}_{0}\qty(t,t_{0})\hat{U}_{0}\qty(t_{0},t)=\hat{I}$, whereas these traces are not necessarily equal to $\Tr\hat{\rho}_{0}$ as long as $\hat{U}_{\mathrm{S}-}\qty(t_{0},t) \hat{U}_{\mathrm{S}+}\qty(t,t_{0}) \neq \hat{I}$, i.e., as long as $\hat{V}_{\mathrm{S}}(t_{+})\vert_{t_{+}=t} \neq \hat{V}_{\mathrm{S}}(t_{-})\vert_{t_{-}=t}$.
Finally, taking the limits $t_{0}\to -\infty$ and $t \to \infty$ for the initial and final times, we obtain the following relation:
\begin{align}
\mathcal{Z}_{\mathrm{Q}}[\vb*{S}] & =  \Tr\hat{\rho}_{\mathrm{S}}\qty(\infty) = \Tr\hat{\rho}_{\mathrm{I}}\qty(\infty)
.
\end{align}
The expressions of $\mathcal{Z}_{\mathrm{Q}}[\vb*{S}]$ in Eq.~(\ref{eq:partition-function}) come from the representations of the time-evolution operators as the formal solutions of the differential equations.

\section{Variational procedure for the action}
\label{sec:Variational-procedure_for_the-action}
This Appendix derives the equations of motion for $\qty{\vb*{S}_{\vb*{r}}(t)}$, Eq.~(\ref{eq:EOM-formal}), from the effective action $\mathcal{A}_{\mathrm{eff}}[\vb*{S}]$ given in Eq.~(\ref{eq:effective-action}) through the variational principle.
After expanding $\mathcal{A}_{\mathrm{eff}}[\vb*{S}]$ to first order in $\qty{\vb*{S}^{\Delta}_{\vb*{r}}(t)}$, we apply the variational principle under the constraint $\abs{\vb*{S}_{\vb*{r}}(\tau, t)}=S$ in two ways: the method of Lagrange multipliers and the representation in spherical coordinates.
Both approaches yield these equations of motion, and their agreement provides a check of the derivation.
\subsection{First-order expansion}
We first recall the extension of localized spin variables $\qty{\vb*{S}_{\vb*{r}}(t)}$ to $\qty{\vb*{S}_{\vb*{r}}(\tau, t)}$ in Eq.~(\ref{eq:extended_variables}), which imposes $\vb*{S}_{\vb*{r}}(0, t)=\vb*{S}_{0}$ and $\vb*{S}_{\vb*{r}}(1, t)=\vb*{S}_{\vb*{r}}(t)$.
Because $\vb*{S}_{0}$ is a constant vector, the first condition gives
\begin{align}
&{\pdv{\vb*{S}_{\vb*{r}}}{t}} (0, t) = 0.
\label{eq:time-derivative-at-tau=0}
\end{align}
Next, we recall the subscripts ``$\pm$'' and the superscripts ``$\mathrm{c}$'' and ``$\Delta$'' of Eqs.~(\ref{eq:superscripts_pm})--(\ref{eq:superscript_Delta}): for a variable $f(t)$ on the closed time path, $f^{\mathrm{c}}(t)$ and $f^{\Delta}(t)$ are the average and the difference of its values $f_{\pm}(t)$ on the forward and backward paths, and $f^{\Delta}(\pm\infty)=0$ because the time path is closed.
Hence $f_{\pm}(t)=f^{\mathrm{c}}(t)\pm f^{\Delta}(t)/2$.

Then, $\mathcal{A}_{\mathrm{WZWN}}[\vb*{S}]$ defined in Eq.~(\ref{eq:WZWN}) takes the following expression using $\vb*{S}^{\mathrm{c}}_{\vb*{r}}(\tau, t)$ and $\vb*{S}^{\Delta}_{\vb*{r}}(\tau, t)$:
\begin{widetext}
\begin{align}
\mathcal{A}_{\mathrm{WZWN}}[\vb*{S}] 
&=\mathcal{A}^{\text{1st-order}}_{\mathrm{WZWN}}[\vb*{S}] + \mathcal{A}^{\text{3rd-order}}_{\mathrm{WZWN}}[\vb*{S}]
,\\
\mathcal{A}^{\text{1st-order}}_{\mathrm{WZWN}}[\vb*{S}] 
&=
-\frac{1}{S^{2}}\int_{0}^{1} \dd{\tau}\int_{-\infty}^{\infty} \dd{t}\sum_{i}
\bigg[
 \vb*{S}^{\Delta}_{\vb*{r}_{i}}(\tau, t) \vdot \qty{{\pdv{ \vb*{S}^{\mathrm{c}}_{\vb*{r}_{i}}}{\tau}} (\tau, t) \cross {\pdv{\vb*{S}^{\mathrm{c}}_{\vb*{r}_{i}}}{t}}(\tau, t)}
\nonumber\\
&\qquad+
 \vb*{S}^{\mathrm{c}}_{\vb*{r}_{i}}(\tau, t) \vdot \qty{{\pdv{ \vb*{S}^{\Delta}_{\vb*{r}_{i}}}{\tau}} (\tau, t) \cross {\pdv{\vb*{S}^{\mathrm{c}}_{\vb*{r}_{i}}}{t}}(\tau, t)}
 +
 \vb*{S}^{\mathrm{c}}_{\vb*{r}_{i}}(\tau, t) \vdot \qty{{\pdv{ \vb*{S}^{\mathrm{c}}_{\vb*{r}_{i}}}{\tau}} (\tau, t) \cross {\pdv{\vb*{S}^{\Delta}_{\vb*{r}_{i}}}{t}}(\tau, t)}
\bigg]
,\\
\mathcal{A}^{\text{3rd-order}}_{\mathrm{WZWN}}[\vb*{S}] 
&=
-\frac{1}{4S^{2}}\int_{0}^{1} \dd{\tau}\int_{-\infty}^{\infty} \dd{t}\sum_{i}
 \vb*{S}^{\Delta}_{\vb*{r}_{i}}(\tau, t) \vdot \qty{{\pdv{ \vb*{S}^{\Delta}_{\vb*{r}_{i}}}{\tau}} (\tau, t) \cross {\pdv{\vb*{S}^{\Delta}_{\vb*{r}_{i}}}{t}}(\tau, t)}
.
\end{align}
\end{widetext}
Every term of the variation of $\mathcal{A}^{\text{3rd-order}}_{\mathrm{WZWN}}[\vb*{S}]$ at $\vb*{S}^{\Delta}=0$ contains either of the following derivatives, which vanish at $\vb*{S}^{\Delta}=0$:
\begin{align}
& \left.{\pdv{\vb*{S}^{\Delta}_{\vb*{r}_{i}}}{\tau}} (\tau, t)\right\vert_{\vb*{S}^{\Delta}=0}=0
,\quad
\left.{\pdv{\vb*{S}^{\Delta}_{\vb*{r}_{i}}}{t}} (\tau, t)\right\vert_{\vb*{S}^{\Delta}=0}=0
.
\end{align}
Hence, $\mathcal{A}^{\text{3rd-order}}_{\mathrm{WZWN}}[\vb*{S}]$ does not contribute to the equations of motion for $\qty{\vb*{S}^{\mathrm{c}}_{\vb*{r}}(t)}$ at $\vb*{S}^{\Delta}=0$, and we focus on $\mathcal{A}^{\text{1st-order}}_{\mathrm{WZWN}}[\vb*{S}]$ hereafter.
For later convenience, we rewrite $\mathcal{A}^{\text{1st-order}}_{\mathrm{WZWN}}[\vb*{S}]$ by partial integration as follows:
\begin{widetext}
\begin{align}
\mathcal{A}^{\text{1st-order}}_{\mathrm{WZWN}}[\vb*{S}] 
&=
-\frac{1}{S^{2}}\int_{-\infty}^{\infty} \dd{t}\sum_{i}
\bigg[
 \vb*{S}^{\Delta}_{\vb*{r}_{i}}(1, t) \vdot \qty{{\pdv{\vb*{S}^{\mathrm{c}}_{\vb*{r}_{i}}}{t}}(1, t) \cross \vb*{S}^{\mathrm{c}}_{\vb*{r}_{i}}(1, t)}
\nonumber\\
&\qquad
+3\int_{0}^{1} \dd{\tau}
\vb*{S}^{\Delta}_{\vb*{r}_{i}}(\tau, t) \vdot \qty{{\pdv{ \vb*{S}^{\mathrm{c}}_{\vb*{r}_{i}}}{\tau}}(\tau, t) \cross {\pdv{\vb*{S}^{\mathrm{c}}_{\vb*{r}_{i}}}{t}}(\tau, t)}
\bigg],
\end{align}
\end{widetext}
where we used $\partial\vb*{S}^{\mathrm{c}}_{\vb*{r}_{i}}(0, t)/\partial t =0$ derived from Eq.~(\ref{eq:time-derivative-at-tau=0})
and $\vb*{S}^{\Delta}_{\vb*{r}_{i}}(\tau, \pm\infty) =0$ derived from the closedness of the time path.

All the above arguments lead to the following expression for the effective action $\mathcal{A}_{\mathrm{eff}}[\vb*{S}]$ in the first-order approximation with respect to $\qty{\vb*{S}^{\Delta}_{\vb*{r}}(t)}$:
\begin{widetext}
\begin{align}
&\mathcal{A}_{\mathrm{eff}}[\vb*{S}]
\cong
\mathcal{A}^{\text{1st-order}}_{\mathrm{WZWN}}[\vb*{S}] 
+g\mu_{\mathrm{B}} \int_{-\infty}^{\infty} \dd{t} \sum_{i}
\vb*{S}^{\Delta}_{\vb*{r}_{i}}(t)\vdot\vb*{B}(t, \vb*{r}_{i})
+\frac{J_{\mathrm{ex}}v_{\mathrm{uc}}}{2}
\int_{-\infty}^{\infty} \dd{t} \sum_{i}
\vb*{S}^{\Delta}_{\vb*{r}_{i}}(t) \vdot \ev{\hat{\vb*{\sigma}}(t, \vb*{r}_{i})}_{\vb*{S}^{\Delta}=0}
.
\end{align}
\end{widetext}

\subsection{Method of Lagrange multipliers}
This subsection discusses the variational procedure under the constraint $\abs{\vb*{S}_{\vb*{r}}(\tau, t)}=S$ by adopting the method of Lagrange multipliers.
To this end, we extend the effective action $\mathcal{A}_{\mathrm{eff}}[\vb*{S}]$ to the following one:
\begin{widetext}
\begin{align}
\mathcal{A}_{\text{ex-eff}}[\vb*{S}]
&=
\mathcal{A}_{\mathrm{eff}}[\vb*{S}]
+
\mathcal{A}_{\lambda}[\vb*{S}]
,\\
\mathcal{A}_{\lambda}[\vb*{S}]
&=
-\frac{1}{2}\int_{0}^{1} \dd{\tau}\oint \dd{t}\sum_{i}
\lambda_{\vb*{r}_{i}}(\tau, t)\qty{\vb*{S}_{\vb*{r}_{i}}(\tau, t) \vdot \vb*{S}_{\vb*{r}_{i}}(\tau, t)-S^{2}}
 \nonumber\\
&=
-\frac{1}{2}\int_{0}^{1} \dd{\tau} \int_{-\infty}^{\infty}\dd{t} \sum_{i}
\nonumber\\
&\qquad\times
\bigg[
\lambda^{\Delta}_{\vb*{r}_{i}}(\tau, t)
\qty{
\vb*{S}^{\mathrm{c}}_{\vb*{r}_{i}}(\tau, t) \vdot \vb*{S}^{\mathrm{c}}_{\vb*{r}_{i}}(\tau, t)
+\frac{1}{4}\vb*{S}^{\Delta}_{\vb*{r}_{i}}(\tau, t) \vdot \vb*{S}^{\Delta}_{\vb*{r}_{i}}(\tau, t)-S^{2}
}
+2\lambda^{\mathrm{c}}_{\vb*{r}_{i}}(\tau, t)
\vb*{S}^{\Delta}_{\vb*{r}_{i}}(\tau, t) \vdot \vb*{S}^{\mathrm{c}}_{\vb*{r}_{i}}(\tau, t)
\bigg]
.
\end{align}
\end{widetext}
The equations of motion for $\qty{\vb*{S}^{\mathrm{c}}_{\vb*{r}}(\tau, t)}_{\vb*{S}^{\Delta}=0}$ come from the variational principle for $\mathcal{A}_{\text{ex-eff}}[\vb*{S}]$ with respect to $\qty{\vb*{S}^{\Delta}_{\vb*{r}}(\tau, t)}$ and $\qty{\lambda^{\Delta}_{\vb*{r}}(\tau, t)}$, i.e.,
\begin{align}
\left.\fdv{\mathcal{A}_{\text{ex-eff}}[\vb*{S}]}{\vb*{S}^{\Delta}_{\vb*{r}_{i}}(\tau, t)}\right\vert_{\vb*{S}^{\Delta}=0}=0,
\quad
\left.\fdv{\mathcal{A}_{\text{ex-eff}}[\vb*{S}]}{\lambda^{\Delta}_{\vb*{r}_{i}}(\tau, t)}\right\vert_{\vb*{S}^{\Delta}=0}=0.
\label{eq:EOM-spherical}
\end{align}
In the rest of this subsection, we drop the superscript ``c'' of $\vb*{S}^{\mathrm{c}}_{\vb*{r}_{i}}(t)$ and $\lambda^{\mathrm{c}}_{\vb*{r}_{i}}(t)$, with the conditions $\vb*{S}^{\Delta}=0$ and $\lambda^{\Delta}=0$ understood.

The first one in Eq.~(\ref{eq:EOM-spherical}) is equivalent to 
\begin{widetext}
\begin{align}
&
\delta(\tau-1)\qty{
-\frac{1}{S^{2}}{\pdv{\vb*{S}_{\vb*{r}_{i}}}{t}}(1, t) \cross \vb*{S}_{\vb*{r}_{i}}(1, t)
+g\mu_{\mathrm{B}} \vb*{B}(t, \vb*{r}_{i})+\frac{J_{\mathrm{ex}}v_{\mathrm{uc}}}{2}\ev{\hat{\vb*{\sigma}}(t, \vb*{r}_{i})}_{\vb*{S}^{\Delta}=0}
}
\nonumber\\
&\hspace{14mm}
-\frac{3}{S^{2}}{\pdv{\vb*{S}_{\vb*{r}_{i}}}{\tau}}(\tau, t) \cross {\pdv{\vb*{S}_{\vb*{r}_{i}}}{t}}(\tau, t)
-\lambda_{\vb*{r}_{i}}(\tau, t) \vb*{S}_{\vb*{r}_{i}}(\tau, t)
=0,
\label{eq:EOM_with_lambda}
\end{align}
\end{widetext}
where we used the relation $\vb*{S}^{\Delta}_{\vb*{r}_{i}}(t)= \vb*{S}^{\Delta}_{\vb*{r}_{i}}(1, t)$ and 
\begin{align}
&\fdv{\vb*{S}^{\Delta}_{\vb*{r}_{i}}(\tau, t)}{\vb*{S}^{\Delta}_{\vb*{r}_{j}}(\tau', t')}
=\delta_{ij}\delta(\tau-\tau')\delta(t-t').
\end{align}
The second one in Eq.~(\ref{eq:EOM-spherical}) means the constraint:
\begin{align}
&\abs{\vb*{S}_{\vb*{r}_{i}}(\tau, t)}^{2} -S^{2} = 0
.
\label{eq:constraint}
\end{align}
This constraint implies 
\begin{align}
& \vb*{S}_{\vb*{r}_{i}}(\tau, t)\vdot{\pdv{\vb*{S}_{\vb*{r}_{i}}}{\tau}} (\tau, t)=0
,
\quad
\vb*{S}_{\vb*{r}_{i}}(\tau, t)\vdot{\pdv{\vb*{S}_{\vb*{r}_{i}}}{t}} (\tau, t)=0
.
\label{eq:orthogonality}
\end{align}

Then, the inner product between $\vb*{S}_{\vb*{r}_{i}}(\tau, t)$ and Eq.~(\ref{eq:EOM_with_lambda}) gives the result:
\begin{widetext}
\begin{align}
\lambda_{\vb*{r}_{i}}(\tau, t)
&=
\frac{1}{S^{2}}\vb*{S}_{\vb*{r}_{i}}(\tau, t) \vdot
\bigg[
\delta(\tau-1)
\qty{
-\frac{1}{S^{2}}{\pdv{\vb*{S}_{\vb*{r}_{i}}}{t}}(1, t) \cross \vb*{S}_{\vb*{r}_{i}}(1, t)
+g\mu_{\mathrm{B}} \vb*{B}(t, \vb*{r}_{i})+\frac{J_{\mathrm{ex}}v_{\mathrm{uc}}}{2}\ev{\hat{\vb*{\sigma}}(t, \vb*{r}_{i})}_{\vb*{S}^{\Delta}=0}
}
\nonumber\\
&\hspace{41mm}
-\frac{3}{S^{2}}{\pdv{\vb*{S}_{\vb*{r}_{i}}}{\tau}}(\tau, t) \cross {\pdv{\vb*{S}_{\vb*{r}_{i}}}{t}}(\tau, t)
\bigg]
.
\end{align}
\end{widetext}
Inserting this result into Eq.~(\ref{eq:EOM_with_lambda}) projects Eq.~(\ref{eq:EOM_with_lambda}) onto the orthogonal complement of $\vb*{S}_{\vb*{r}_{i}}(\tau, t)$. The projected equation is equivalent to the cross product of $\vb*{S}_{\vb*{r}_{i}}(\tau, t)$ with Eq.~(\ref{eq:EOM_with_lambda}),
\begin{widetext}
\begin{align}
&\delta(\tau-1)\qty[
-{\pdv{\vb*{S}_{\vb*{r}_{i}}}{t}}(1, t)
+\vb*{S}_{\vb*{r}_{i}}(1, t) \cross 
\qty{g\mu_{\mathrm{B}} \vb*{B}(t, \vb*{r}_{i})+\frac{J_{\mathrm{ex}}v_{\mathrm{uc}}}{2}\ev{\hat{\vb*{\sigma}}(t, \vb*{r}_{i})}_{\vb*{S}^{\Delta}=0}}
]
=0,
\label{eq:EOM_without_lambda}
\end{align}
\end{widetext}
where we used Eqs.~(\ref{eq:constraint}) and (\ref{eq:orthogonality}).
Integrating over $\tau$, or simply removing the factor $\delta(\tau-1)$, and using $\vb*{S}_{\vb*{r}_{i}}(1, t) = \vb*{S}_{\vb*{r}_{i}}(t)$ from Eq.~(\ref{eq:extended_variables}), we find that Eq.~(\ref{eq:EOM_without_lambda}) is equivalent to Eq.~(\ref{eq:EOM-formal}) in the main text.

\subsection{Representation in spherical coordinates}
This subsection shows that the variational procedure using the spherical-coordinate representation provides the same equations of motion as that using the method of Lagrange multipliers in the previous subsection.
The spherical-coordinate representation satisfies the constraint $\abs{\vb*{S}_{\vb*{r}}(\tau, t)}=S$ identically:
\begin{align}
\vb*{S}_{\vb*{r}}(\tau, t)
&=
S
\mqty(
\sin\theta_{\vb*{r}}(\tau,t)\cos\phi_{\vb*{r}}(\tau,t) \\
\sin\theta_{\vb*{r}}(\tau,t)\sin\phi_{\vb*{r}}(\tau,t) \\
\cos\theta_{\vb*{r}}(\tau,t)
).
\end{align}
This definition yields the following relations:
\begin{widetext}
\begin{align}
\vb*{S}^{\mathrm{c}}_{\vb*{r}}(\tau, t)
&=
S
\mqty(
\sin\theta^{\mathrm{c}}_{\vb*{r}}(\tau,t)\cos\phi^{\mathrm{c}}_{\vb*{r}}(\tau,t) \cos\frac{\theta^{\Delta}_{\vb*{r}}(\tau,t)}{2}\cos\frac{\phi^{\Delta}_{\vb*{r}}(\tau,t)}{2}
-\cos\theta^{\mathrm{c}}_{\vb*{r}}(\tau,t)\sin\phi^{\mathrm{c}}_{\vb*{r}}(\tau,t) \sin\frac{\theta^{\Delta}_{\vb*{r}}(\tau,t)}{2}\sin\frac{\phi^{\Delta}_{\vb*{r}}(\tau,t)}{2}
\\
\sin\theta^{\mathrm{c}}_{\vb*{r}}(\tau,t)\sin\phi^{\mathrm{c}}_{\vb*{r}}(\tau,t) \cos\frac{\theta^{\Delta}_{\vb*{r}}(\tau,t)}{2}\cos\frac{\phi^{\Delta}_{\vb*{r}}(\tau,t)}{2}
+\cos\theta^{\mathrm{c}}_{\vb*{r}}(\tau,t)\cos\phi^{\mathrm{c}}_{\vb*{r}}(\tau,t) \sin\frac{\theta^{\Delta}_{\vb*{r}}(\tau,t)}{2}\sin\frac{\phi^{\Delta}_{\vb*{r}}(\tau,t)}{2}
\\
\cos\theta^{\mathrm{c}}_{\vb*{r}}(\tau,t)\cos\frac{\theta^{\Delta}_{\vb*{r}}(\tau,t)}{2}
)
,\\
\vb*{S}^{\Delta}_{\vb*{r}}(\tau, t)
&=
2S
\mqty(
\cos\theta^{\mathrm{c}}_{\vb*{r}}(\tau,t)\cos\phi^{\mathrm{c}}_{\vb*{r}}(\tau,t) \sin\frac{\theta^{\Delta}_{\vb*{r}}(\tau,t)}{2}\cos\frac{\phi^{\Delta}_{\vb*{r}}(\tau,t)}{2}
-\sin\theta^{\mathrm{c}}_{\vb*{r}}(\tau,t)\sin\phi^{\mathrm{c}}_{\vb*{r}}(\tau,t) \cos\frac{\theta^{\Delta}_{\vb*{r}}(\tau,t)}{2}\sin\frac{\phi^{\Delta}_{\vb*{r}}(\tau,t)}{2}
\\
\cos\theta^{\mathrm{c}}_{\vb*{r}}(\tau,t)\sin\phi^{\mathrm{c}}_{\vb*{r}}(\tau,t) \sin\frac{\theta^{\Delta}_{\vb*{r}}(\tau,t)}{2}\cos\frac{\phi^{\Delta}_{\vb*{r}}(\tau,t)}{2}
+\sin\theta^{\mathrm{c}}_{\vb*{r}}(\tau,t)\cos\phi^{\mathrm{c}}_{\vb*{r}}(\tau,t) \cos\frac{\theta^{\Delta}_{\vb*{r}}(\tau,t)}{2}\sin\frac{\phi^{\Delta}_{\vb*{r}}(\tau,t)}{2}
\\
-\sin\theta^{\mathrm{c}}_{\vb*{r}}(\tau,t)\sin\frac{\theta^{\Delta}_{\vb*{r}}(\tau,t)}{2}
)
.
\end{align}
\end{widetext}
The physically appropriate condition is the following relation:
\begin{align}
&\forall \qty{\theta^{\mathrm{c}}_{\vb*{r}}(\tau, t), \phi^{\mathrm{c}}_{\vb*{r}}(\tau, t)}, \vb*{S}^{\Delta}_{\vb*{r}}(\tau, t) = 0
\nonumber\\
&\Longleftrightarrow
\theta^{\Delta}_{\vb*{r}}(\tau, t)=0 \land \phi^{\Delta}_{\vb*{r}}(\tau, t) = 0
.
\end{align}
Here, since the angles are defined modulo $2\pi$, the above equivalence is understood in the neighborhood of $\theta^{\Delta}=\phi^{\Delta}=0$, where the variation is taken.
Thus, we abbreviate the conditions $\theta^{\Delta}_{\vb*{r}_{i}}(\tau, t)=0$ and $\phi^{\Delta}_{\vb*{r}_{i}}(\tau, t) = 0$ for all $(i,\tau, t)$ to $\vb*{S}^{\Delta}=0$, as before.

Before taking the variation of $\mathcal{A}_{\mathrm{eff}}[\vb*{S}]$ with respect to the spherical coordinates $\qty{\theta^{\Delta}_{\vb*{r}}(\tau,t)}$ and $\qty{\phi^{\Delta}_{\vb*{r}}(\tau,t)}$, we provide the variation of $\qty{\vb*{S}^{\mathrm{c}}_{\vb*{r}}(\tau, t)}$ and $\qty{\vb*{S}^{\Delta}_{\vb*{r}}(\tau, t)}$ with respect to these variables:
\begin{align}
&\left.\fdv{\vb*{S}^{\mathrm{c}}_{\vb*{r}_{i}}(\tau, t)}{\theta^{\Delta}_{\vb*{r}_{j}}(\tau',t')}\right\vert_{\vb*{S}^{\Delta}=0}
=0,
\quad
\left.\fdv{\vb*{S}^{\mathrm{c}}_{\vb*{r}_{i}}(\tau, t)}{\phi^{\Delta}_{\vb*{r}_{j}}(\tau',t')}\right\vert_{\vb*{S}^{\Delta}=0}
=0,
\\
&
\left.
\fdv{\vb*{S}^{\Delta}_{\vb*{r}_{i}}(\tau, t)}{\theta^{\Delta}_{\vb*{r}_{j}}(\tau',t')}
\right\vert_{\vb*{S}^{\Delta}=0}
=
\delta_{ij}\delta(\tau-\tau')\delta(t-t')
{\pdv{\vb*{S}_{\vb*{r}_{i}}(\tau, t)}{\theta_{\vb*{r}_{i}}(\tau,t)}}
,\\
&
\left.
\fdv{\vb*{S}^{\Delta}_{\vb*{r}_{i}}(\tau, t)}{\phi^{\Delta}_{\vb*{r}_{j}}(\tau',t')}
\right\vert_{\vb*{S}^{\Delta}=0}
=
\delta_{ij}\delta(\tau-\tau')\delta(t-t')
{\pdv{\vb*{S}_{\vb*{r}_{i}}(\tau, t)}{\phi_{\vb*{r}_{i}}(\tau,t)}}
,
\end{align}
where
\begin{align}
&
{\pdv{\vb*{S}_{\vb*{r}_{i}}(\tau, t)}{\theta_{\vb*{r}_{i}}(\tau,t)}}
=
S
\mqty(
\cos\theta_{\vb*{r}_{i}}(\tau,t)\cos\phi_{\vb*{r}_{i}}(\tau,t)
\\
\cos\theta_{\vb*{r}_{i}}(\tau,t)\sin\phi_{\vb*{r}_{i}}(\tau,t)
\\
-\sin\theta_{\vb*{r}_{i}}(\tau,t)
)
, \label{eq:basis_of_orthocomplement-space_theta}
\\
&
{\pdv{\vb*{S}_{\vb*{r}_{i}}(\tau, t)}{\phi_{\vb*{r}_{i}}(\tau,t)}}
=
S
\mqty(
-\sin\theta_{\vb*{r}_{i}}(\tau,t)\sin\phi_{\vb*{r}_{i}}(\tau,t)
\\
\sin\theta_{\vb*{r}_{i}}(\tau,t)\cos\phi_{\vb*{r}_{i}}(\tau,t)
\\
0
)
.\label{eq:basis_of_orthocomplement-space_phi}
\end{align}
In the above expressions and in the rest of this subsection, we drop the superscript ``c'', with the condition $\vb*{S}^{\Delta}=0$ understood.
Note that, for $\sin\theta_{\vb*{r}_{i}}(\tau,t)\neq0$, the vectors of Eqs.~(\ref{eq:basis_of_orthocomplement-space_theta}) and (\ref{eq:basis_of_orthocomplement-space_phi}) form a basis of the orthogonal complement of $\vb*{S}_{\vb*{r}_{i}}(\tau, t)$; at the poles, the same argument holds in a rotated coordinate system.

The equations of motion for $\qty{\vb*{S}_{\vb*{r}}(\tau, t)}$ come from the variational principle for $\mathcal{A}_{\mathrm{eff}}[\vb*{S}]$ with respect to $\qty{\theta^{\Delta}_{\vb*{r}}(\tau, t)}$ and $\qty{\phi^{\Delta}_{\vb*{r}}(\tau, t)}$ at $\vb*{S}^{\Delta}=0$, i.e.,
\begin{align}
\left.\fdv{\mathcal{A}_{\mathrm{eff}}[\vb*{S}]}{\theta^{\Delta}_{\vb*{r}_{i}}(\tau, t)}\right\vert_{\vb*{S}^{\Delta}=0}=0,
\quad
\left.\fdv{\mathcal{A}_{\mathrm{eff}}[\vb*{S}]}{\phi^{\Delta}_{\vb*{r}_{i}}(\tau, t)}\right\vert_{\vb*{S}^{\Delta}=0}=0.
\label{eq:EOM_spherical-coordinates}
\end{align}
The left-hand sides of the above equations are, respectively, the projections of the following vector variable onto the two vectors of Eqs.~(\ref{eq:basis_of_orthocomplement-space_theta}) and (\ref{eq:basis_of_orthocomplement-space_phi}):
\begin{widetext}
\begin{align}
&
\delta(\tau-1)\qty{
-\frac{1}{S^{2}}{\pdv{\vb*{S}_{\vb*{r}_{i}}}{t}}(1, t) \cross \vb*{S}_{\vb*{r}_{i}}(1, t)
+g\mu_{\mathrm{B}} \vb*{B}(t, \vb*{r}_{i})+\frac{J_{\mathrm{ex}}v_{\mathrm{uc}}}{2}\ev{\hat{\vb*{\sigma}}(t, \vb*{r}_{i})}_{\vb*{S}^{\Delta}=0}
}
-\frac{3}{S^{2}}{\pdv{\vb*{S}_{\vb*{r}_{i}}}{\tau}}(\tau, t) \cross {\pdv{\vb*{S}_{\vb*{r}_{i}}}{t}}(\tau, t)
.
\end{align}
\end{widetext}
Since the two vectors of Eqs.~(\ref{eq:basis_of_orthocomplement-space_theta}) and (\ref{eq:basis_of_orthocomplement-space_phi}) form a basis of the orthogonal complement of $\vb*{S}_{\vb*{r}_{i}}(\tau, t)$, Eq.~(\ref{eq:EOM_spherical-coordinates}) means that the cross product between $\vb*{S}_{\vb*{r}_{i}}(\tau, t)$ and the above vector variable is zero.
The result is again equivalent to Eq.~(\ref{eq:EOM-formal}) in the main text.

\section{Spin-spin retarded Green function}
\label{sec:Spin-spin_retarded-Green-function}
This Appendix evaluates the spin-spin retarded Green function of the conduction-electron system $\Pi^{\mathrm{R}\mu\nu}(t,\vb*{r}; t', \vb*{r}')$ defined by Eq.~(\ref{eq:spin-spin_RGF}).
Section~\ref{sec:Time-derivative_expansion} sets up the time-derivative expansion of $\Pi^{\mathrm{R}\mu\nu}_{\mathrm{eff}}(t,\vb*{r})$ and derives the general $n$th-order response functions $\Pi^{\mathrm{R}(n)}_{\mathrm{eq}}(\vb*{r})$, $\Pi^{\mathrm{R}(n)}_{\mathrm{cc}}(\vb*{r})$, and $\vb*{\Pi}^{\mathrm{R}(n)}_{\mathrm{sc}}(\vb*{r})$; Sec.~\ref{sec:evaluation_low-temp} evaluates these functions explicitly in the low-temperature regime $k_{\mathrm{B}}T\ll\epsilon_{\mathrm{F}}$, which gives the dimensionless functions $F^{(n)}_{\mathrm{eq}}(\vb*{r})$, $F^{(n)}_{\mathrm{cc}}(\vb*{r})$, and $\vb*{F}^{(n)}_{\mathrm{sc}}(\vb*{r})$ of Sec.~\ref{sec:Approximations}; and Sec.~\ref{sec:clean-limit-eta0} takes the limit $\eta\to0$, which gives the RKKY range functions of Sec.~\ref{sec:EOM}.
\subsection{Time-derivative expansion}
\label{sec:Time-derivative_expansion}
As mentioned in the main text, we assume the following isotropic dispersion for free electrons with effective mass $m^{*}$:
\begin{align}
&
\epsilon_{\vb*{k}} = \frac{k^{2}}{2m^{*}} \quad(k=\abs{\vb*{k}}) ,
\quad
\vb*{v}_{\vb*{k}} = \grad_{\vb*{k}} \epsilon_{\vb*{k}}= \frac{\vb*{k}}{m^{*}}
.
\label{eq:isotropic_dispersion}
\end{align}

We begin with the approximation of Eq.~(\ref{eq:approximation_of_spin-spin_RGF}) in Sec.~\ref{sec:Approximations}, which replaces $\Pi^{\mathrm{R}\mu\nu}(t, \vb*{r}; t', \vb*{r}')$ with $\Pi^{\mathrm{R}\mu\nu}_{\mathrm{eff}}(t- t', \vb*{r}-\vb*{r}')$.
Inserting the explicit form of $\overleftrightarrow{g}(\vb*{k})$, Eq.~(\ref{eq:distribution_function}), into $\Pi^{\mathrm{R}\mu\nu}_{\mathrm{eff}}(t, \vb*{r})$ according to Eq.~(\ref{eq:spin-spin_RGF_eff}), we obtain the following expressions:
\begin{widetext}
\begin{align}
\Pi^{\mathrm{R}\mu\nu}_{\mathrm{eff}}(t, \vb*{r})
&=
\delta_{\mu\nu} \qty{\Pi^{\mathrm{R}}_{\mathrm{eq}}(t, \vb*{r})+\Pi^{\mathrm{R}}_{\mathrm{cc}}(t, \vb*{r})}
+\sum_{\lambda}\epsilon_{\mu\nu\lambda}\Pi^{\mathrm{R}\lambda}_{\mathrm{sc}}(t, \vb*{r})
,\label{eq:approx_spin-spin_RGF}
\\
\Pi^{\mathrm{R}}_{\mathrm{eq}}(t, \vb*{r})
&=
\int\frac{\dd{\vb*{k}}}{(2\pi)^{3}}
f_{\mathrm{F}}(\epsilon_{\vb*{k}})
\int\frac{\dd{\vb*{k}'}}{(2\pi)^{3}}
\Phi^{[\vb*{k};\vb*{k}']}_{-}(t, \vb*{r})
,\\
\Pi^{\mathrm{R}}_{\mathrm{cc}}(t, \vb*{r})
&=
\int\frac{\dd{\vb*{k}}}{(2\pi)^{3}}
\qty{-{\dv{f_{\mathrm{F}}}{\epsilon}} (\epsilon_{\vb*{k}})}
\qty(
\tau_{\mathrm{e}}\vb*{v}_{\vb*{k}}\vdot e\vb*{E}
)
\int\frac{\dd{\vb*{k}'}}{(2\pi)^{3}}
\Phi^{[\vb*{k};\vb*{k}']}_{-}(t, \vb*{r})
,\\
\Pi^{\mathrm{R}\mu}_{\mathrm{sc}}(t, \vb*{r})
&=
\int\frac{\dd{\vb*{k}}}{(2\pi)^{3}}
\qty{-{\dv{f_{\mathrm{F}}}{\epsilon}} (\epsilon_{\vb*{k}})}
\qty(
\tau_{\mathrm{e}}\vb*{v}_{\vb*{k}}\vdot e_{\mathrm{s}}\vb*{E}_{\mathrm{s}}^{\mu}
)
\int\frac{\dd{\vb*{k}'}}{(2\pi)^{3}}
\Phi^{[\vb*{k};\vb*{k}']}_{+}(t, \vb*{r})
,\\
\Phi^{[\vb*{k};\vb*{k}']}_{\pm}(t, \vb*{r})
&=2 \qty(-i)^{\frac{1\mp1}{2}} \theta(t)
\qty{
e^{-i(\vb*{k}-\vb*{k}')\vdot\vb*{r}+i(\epsilon_{\vb*{k}}-\epsilon_{\vb*{k}'})t-\eta t} 
\pm e^{i(\vb*{k}-\vb*{k}')\vdot\vb*{r}-i(\epsilon_{\vb*{k}}-\epsilon_{\vb*{k}'})t-\eta t}
}
.
\end{align}
\end{widetext}
The subscript ``eq'' in $\Pi^{\mathrm{R}}_{\mathrm{eq}}(t, \vb*{r})$ indicates that this part exists even in equilibrium, without the driving forces for the charge and spin currents.
On the other hand, $\Pi^{\mathrm{R}}_{\mathrm{cc}}(t, \vb*{r})$ and $\vb*{\Pi}^{\mathrm{R}}_{\mathrm{sc}}(t, \vb*{r})$ depend on the driving forces for charge and spin currents, respectively.
In the subsequent part of this Appendix, we call them the equilibrium part, the charge-current part, and the spin-current part of $\Pi^{\mathrm{R}\mu\nu}_{\mathrm{eff}}(t, \vb*{r})$, respectively.

As in Sec.~\ref{sec:Approximations}, we make the time-derivative expansion of $\qty{S^{\mu}_{\vb*{r}}(t)}$ in $\ev*{\hat{\sigma}^{\mu}(t, \vb*{r})}_{\vb*{S}^{\Delta}=0}$, taking advantage of the homogeneity in time of the right-hand side of Eq.~(\ref{eq:approx_spin-spin_RGF}), and obtain the following response functions for the $n$th-order derivatives:
\begin{widetext}
\begin{align}
\Pi^{\mathrm{R}\mu\nu(n)}_{\mathrm{eff}}(\vb*{r})
& = \int_{-\infty}^{\infty} \dd{t}\frac{(-t)^n}{n!}\Pi^{\mathrm{R}\mu\nu}_{\mathrm{eff}}\qty(t,\vb*{r})
 =
\delta_{\mu\nu} \qty{\Pi^{\mathrm{R}(n)}_{\mathrm{eq}}(\vb*{r})+\Pi^{\mathrm{R}(n)}_{\mathrm{cc}}(\vb*{r})}
+\sum_{\lambda}\epsilon_{\mu\nu\lambda}\Pi^{\mathrm{R}\lambda(n)}_{\mathrm{sc}}(\vb*{r}),
\\
\Pi^{\mathrm{R}(n)}_{\mathrm{eq}}(\vb*{r}) 
&=
\int\frac{\dd{\vb*{k}}}{(2\pi)^{3}}
f_{\mathrm{F}}(\epsilon_{\vb*{k}})
\int\frac{\dd{\vb*{k}'}}{(2\pi)^{3}}
\Phi^{[\vb*{k};\vb*{k}'](n)}_{-}(\vb*{r})
,\\
\Pi^{\mathrm{R}(n)}_{\mathrm{cc}}(\vb*{r})
&=
\int\frac{\dd{\vb*{k}}}{(2\pi)^{3}}
\qty{-{\dv{f_{\mathrm{F}}}{\epsilon}} (\epsilon_{\vb*{k}})}
\qty(
\tau_{\mathrm{e}}\vb*{v}_{\vb*{k}}\vdot e\vb*{E}
)
\int\frac{\dd{\vb*{k}'}}{(2\pi)^{3}}
\Phi^{[\vb*{k};\vb*{k}'](n)}_{-}(\vb*{r})
,\\
\Pi^{\mathrm{R}\mu(n)}_{\mathrm{sc}}(\vb*{r})
&=
\int\frac{\dd{\vb*{k}}}{(2\pi)^{3}}
\qty{-{\dv{f_{\mathrm{F}}}{\epsilon}} (\epsilon_{\vb*{k}})}
\qty( 
\tau_{\mathrm{e}}\vb*{v}_{\vb*{k}}\vdot e_{\mathrm{s}}\vb*{E}_{\mathrm{s}}^{\mu}
)
\int\frac{\dd{\vb*{k}'}}{(2\pi)^{3}}
\Phi^{[\vb*{k};\vb*{k}'](n)}_{+}(\vb*{r})
,\\
\Phi^{[\vb*{k};\vb*{k}'](n)}_{\pm}(\vb*{r})
& =  \int_{-\infty}^{\infty} \dd{t} \frac{(-t)^{n}}{n!} 
\Phi^{[\vb*{k};\vb*{k}']}_{\pm}(t, \vb*{r})
= -2 \qty(-i)^{\frac{1\mp1}{2}} 
\qty{
\frac{i^{n+1}e^{-i(\vb*{k}-\vb*{k}')\vdot\vb*{r}}}{\qty(\epsilon_{\vb*{k}'}-\epsilon_{\vb*{k}}-i\eta)^{n+1}}
\pm 
\frac{\qty(-i)^{n+1}e^{i(\vb*{k}-\vb*{k}')\vdot\vb*{r}}}{\qty(\epsilon_{\vb*{k}'}-\epsilon_{\vb*{k}}+i\eta)^{n+1}}
}
.
\end{align}
\end{widetext}
In the limit $\eta\to 0$, the terms with $n\ge 2$ can diverge in the long-distance limit $k_{\mathrm{F}}r \to \infty$ (Secs.~\ref{sec:evaluation_low-temp} and \ref{sec:clean-limit-eta0}); a finite $\eta$ is therefore essential to justify the time-derivative expansion, even in the weak-coupling regime.

\subsection{Evaluation in the low-temperature regime}
\label{sec:evaluation_low-temp}
This subsection presents an approximate but explicit form of the spin-spin retarded Green function $\Pi^{\mathrm{R}\mu\nu}(t,\vb*{r}; t', \vb*{r}')$ through the functions $\Pi^{\mathrm{R}(n)}_{\mathrm{eq}}(\vb*{r})$, $\Pi^{\mathrm{R}(n)}_{\mathrm{cc}}(\vb*{r})$, and $\vb*{\Pi}^{\mathrm{R}(n)}_{\mathrm{sc}}(\vb*{r})$ introduced in the previous subsection.
As in Sec.~\ref{sec:Approximations}, we focus on the low-temperature regime $k_{\mathrm{B}}T \ll \epsilon_{\mathrm{F}}$ and adopt the approximation:
\begin{align}
& 
f_{\mathrm{F}}(\epsilon) \cong \theta(\epsilon_{\mathrm{F}}-\epsilon)
,\quad
-{\dv{f_{\mathrm{F}}}{\epsilon}} (\epsilon) \cong \delta(\epsilon-\epsilon_{\mathrm{F}})
.
\end{align}
Before going further, we recall the relations among the Fermi energy $\epsilon_{\mathrm{F}}$,
the Fermi wavenumber $k_{\mathrm{F}}$, the Fermi wavelength $\lambda_{\mathrm{F}}$, and the number density $n_{\mathrm{e}}$ of the conduction-electron system:
\begin{align}
& 
k_{\mathrm{F}} =\sqrt{2m^{*}\epsilon_{\mathrm{F}}}
,\quad
\lambda_{\mathrm{F}} = \frac{2\pi}{k_{\mathrm{F}}}
,\quad
n_{\mathrm{e}} = \frac{k_{\mathrm{F}}^{3}}{3\pi^{2}}
.
\end{align}

For later convenience, we introduce some formulae to calculate the integration with respect to $\vb*{k}$ and $\vb*{k}'$.
We begin with the following two formulae for the integration over the angular variables $\vb*{\Omega}_{\vb*{k}}=\vb*{k}/k$ of $\vb*{k}$:
\begin{align}
&\int\dd{\Omega_{\vb*{k}}} e^{\pm i \vb*{k}\vdot\vb*{r}} 
= 4\pi\frac{\sin\qty(kr)}{kr},
\label{eq:Omega-integration_plain}
\\
&\int\dd{\Omega_{\vb*{k}}}
\qty(
\vb*{\Omega}_{\vb*{k}}\vdot \vb*{g}
)
e^{\pm i \vb*{k}\vdot\vb*{r}}
\nonumber\\
&\qquad
=
\pm 4\pi i 
\qty(
\frac{\vb*{r}}{r}\vdot\vb*{g}
)
\frac{\sin\qty(kr)-\qty(kr)\cos\qty(kr)}{\qty(kr)^{2}},
\label{eq:Omega-integration_inner-product}
\end{align}
where $\vb*{g}$ is an arbitrary constant vector, and the polar angles in the measure $\dd{\Omega_{\vb*{k}}}=\sin\theta\dd{\theta}\dd{\phi}$ are defined by
\begin{align}
&
\frac{\vb*{r}}{r}\vdot \vb*{\Omega}_{\vb*{k}} = \cos\theta,
\quad
\frac{\vb*{r}\cross\qty(\vb*{g}\cross\vb*{r})}{\abs{\vb*{r}\cross\qty(\vb*{g}\cross\vb*{r})}}\vdot \vb*{\Omega}_{\vb*{k}} = \sin\theta \cos\phi
.
\end{align}
If $\vb*{g}$ is parallel to $\vb*{r}$, the azimuthal angle $\phi$ is measured from an arbitrary direction perpendicular to $\vb*{r}$.
We also use the following formula to change the integration range:
\begin{align}
&\int_{0}^{\infty}\dd{k}k \sin\qty(kr)g(k^{2}) 
= \frac{1}{2i}\int_{-\infty}^{\infty}\dd{k}k e^{ikr} g(k^{2}),
\end{align}
where $g(k^{2})$ is a function of $k^{2}$.
For $r=0$, the integrals on both sides of the above formula can be ill-defined if $g(k^{2})$ does not decay sufficiently fast as $k^{2}$ increases.
This happens in the calculations related to $\Pi^{\mathrm{R}(0)}_{\mathrm{eq}}(\vb*{r})$ and $\Pi^{\mathrm{R}\lambda(0)}_{\mathrm{sc}}(\vb*{r})$.

For the isotropic dispersion $\epsilon_{\vb*{k}} = k^{2}/\qty(2m^{*})$, 
the above formulae enable us to integrate $\Phi^{[\vb*{k};\vb*{k}'](n)}_{\pm}(\vb*{r})$ with respect to the angular variables of momenta $\vb*{k}$ and $\vb*{k}'$ as follows:
\begin{widetext}
\begin{align}
&\int\frac{\dd{\vb*{k}'}}{(2\pi)^{3}}
\Phi^{[\vb*{k};\vb*{k}'](n)}_{\pm}(\vb*{r})
 = \frac{i^{\frac{1\pm1}{2}}}{2\pi^{2}r}
\int_{-\infty}^{\infty}\dd{k'}k'
e^{i k'r}
\qty{
\frac{i^{n+1}e^{-i\vb*{k}\vdot\vb*{r}}}{\qty(\epsilon_{\vb*{k}'}-\epsilon_{\vb*{k}}-i\eta)^{n+1}}
\pm
\frac{\qty(-i)^{n+1}e^{i\vb*{k}\vdot\vb*{r}}}{\qty(\epsilon_{\vb*{k}'}-\epsilon_{\vb*{k}}+i\eta)^{n+1}}
},
\\
&\int\frac{\dd{\vb*{k}}}{(2\pi)^{3}}
f_{\mathrm{F}}(\epsilon_{\vb*{k}})
\int\frac{\dd{\vb*{k}'}}{(2\pi)^{3}}
\Phi^{[\vb*{k};\vb*{k}'](n)}_{\pm}(\vb*{r})
\nonumber\\
&\qquad
= \frac{i^{\frac{1\pm1}{2}}}{4\pi^{4}r^{2}}
\int_{0}^{\infty}\dd{k}k
f_{\mathrm{F}}(\epsilon_{\vb*{k}})
\sin\qty(kr)
\int_{-\infty}^{\infty}\dd{k'}k'
e^{i k'r}
\qty{
\frac{i^{n+1}}{\qty(\epsilon_{\vb*{k}'}-\epsilon_{\vb*{k}}-i\eta)^{n+1}}
\pm
\frac{\qty(-i)^{n+1}}{\qty(\epsilon_{\vb*{k}'}-\epsilon_{\vb*{k}}+i\eta)^{n+1}}
},
\\
&\int\frac{\dd{\vb*{k}}}{(2\pi)^{3}}
\qty{-{\dv{f_{\mathrm{F}}}{\epsilon}} (\epsilon_{\vb*{k}})}
\qty(
\vb*{k}\vdot \vb*{g}
)
\int\frac{\dd{\vb*{k}'}}{(2\pi)^{3}}
\Phi^{[\vb*{k};\vb*{k}'](n)}_{\pm}(\vb*{r})
\nonumber\\
&\qquad
= \frac{\qty(-i)^{\frac{1\mp1}{2}}}{4\pi^{4}r^{3}}
\qty(
\frac{\vb*{r}}{r}\vdot \vb*{g}
)
\int_{0}^{\infty}\dd{k}k
\qty{-{\dv{f_{\mathrm{F}}}{\epsilon}} (\epsilon_{\vb*{k}})}
\qty{\sin\qty(kr)-\qty(kr)\cos\qty(kr)}
\nonumber\\
&\qquad\quad\times
\int_{-\infty}^{\infty}\dd{k'}k'
e^{i k'r}
\qty{
\frac{i^{n+1}}{\qty(\epsilon_{\vb*{k}'}-\epsilon_{\vb*{k}}-i\eta)^{n+1}}
\mp
\frac{\qty(-i)^{n+1}}{\qty(\epsilon_{\vb*{k}'}-\epsilon_{\vb*{k}}+i\eta)^{n+1}}
}.
\end{align}
\end{widetext}
Among these integrals, the $k'$ integrals for $\Pi^{\mathrm{R}(n)}_{\mathrm{eq}}(\vb*{r})$ and $\Pi^{\mathrm{R}\lambda(n)}_{\mathrm{sc}}(\vb*{r})$ are not well-defined for $n=0$ and $r=0$.

With the low-temperature approximation and the variable transformations $k \to u = k/k_{\mathrm{F}}$ and $k' \to v = k'/k_{\mathrm{F}}$, we obtain the following interim expressions:
\begin{widetext}
\begin{align}
&
\Pi^{\mathrm{R}(n)}_{\mathrm{eq}}(\vb*{r}) 
\cong
\frac{1}{\epsilon_{\mathrm{F}}^{n+1}}\qty(\frac{3n_{\mathrm{e}}}{2})^{2}
\frac{1}{\qty(k_{\mathrm{F}}r)^{2}}
\int_{0}^{1}\dd{u}u\sin\qty(k_{\mathrm{F}}ru) 
H^{(n)}_{-}\qty(u^{2};\frac{\eta}{\epsilon_{\mathrm{F}}};k_{\mathrm{F}}r)
,\label{eq:Pi(n)eq_integral-rep}
\\
&
\Pi^{\mathrm{R}(n)}_{\mathrm{cc}}(\vb*{r})
\cong
\frac{1}{\epsilon_{\mathrm{F}}^{n+1}}\qty(\frac{3n_{\mathrm{e}}}{2})^{2}
\qty(\frac{3\lambda_{\mathrm{F}}^{2}}{4e\epsilon_{\mathrm{F}}})
\qty(
\frac{\vb*{r}}{r}\vdot\frac{e^{2}n_{\mathrm{e}}\tau_{\mathrm{e}}}{m^{*}}\vb*{E}
)
\frac{1}{2}
\cdot
\frac{\sin\qty(k_{\mathrm{F}}r)-\qty(k_{\mathrm{F}}r)\cos\qty(k_{\mathrm{F}}r)}{\qty(k_{\mathrm{F}}r)^{3}}
H^{(n)}_{+}\qty(1;\frac{\eta}{\epsilon_{\mathrm{F}}};k_{\mathrm{F}}r)
,\label{eq:Pi(n)cc_integral-rep}
\\
&
\Pi^{\mathrm{R}\mu(n)}_{\mathrm{sc}}(\vb*{r})
\cong
\frac{1}{\epsilon_{\mathrm{F}}^{n+1}}\qty(\frac{3n_{\mathrm{e}}}{2})^{2}
\qty(\frac{3\lambda_{\mathrm{F}}^{2}}{4e_{\mathrm{s}}\epsilon_{\mathrm{F}}})
\qty(
\frac{\vb*{r}}{r}\vdot\frac{e_{\mathrm{s}}^{2}n_{\mathrm{e}}\tau_{\mathrm{e}}}{m^{*}}\vb*{E}_{\mathrm{s}}^{\mu}
)
\frac{1}{2}
\cdot
\frac{\sin\qty(k_{\mathrm{F}}r)-\qty(k_{\mathrm{F}}r)\cos\qty(k_{\mathrm{F}}r)}{\qty(k_{\mathrm{F}}r)^{3}}
H^{(n)}_{-}\qty(1;\frac{\eta}{\epsilon_{\mathrm{F}}};k_{\mathrm{F}}r)
,\label{eq:Pi(n)sc_integral-rep}
\\
&
H^{(n)}_{\pm}\qty(u^{2};\frac{\eta}{\epsilon_{\mathrm{F}}};k_{\mathrm{F}}r)
=
\qty(-i)^{\frac{1\pm1}{2}}
\int_{-\infty}^{\infty} \dd{v}v e^{i k_{\mathrm{F}}rv}
\qty[
\frac{i^{n+1}}{\qty{v^{2}-w^{2}(u^{2})}^{n+1}}
\pm\frac{\qty(-i)^{n+1}}{\qty{v^{2}-\overline{w}^{2}(u^{2})}^{n+1}}
]
,\label{eq:H(n)pm_integral-rep}
\end{align}
\end{widetext}
where
\begin{align}
&
w(u^{2})
= \sqrt{u^{2}+i\frac{\eta}{\epsilon_{\mathrm{F}}}}
,\quad
\overline{w}(u^{2})
= \sqrt{u^{2}-i\frac{\eta}{\epsilon_{\mathrm{F}}}}
.
\end{align}
Again, the $v$ integral of $H^{(n)}_{-}\qty(u^{2};\eta/\epsilon_{\mathrm{F}};k_{\mathrm{F}}r)$ is not well-defined for $n=0$ and $r=0$, so that $\Pi^{\mathrm{R}(0)}_{\mathrm{eq}}(\vb*{r})$ and $\Pi^{\mathrm{R}\lambda(0)}_{\mathrm{sc}}(\vb*{r})$ are not well-defined at $\vb*{r}=0$.

Next, the residue theorem enables us to evaluate $H^{(n)}_{\pm}\qty(u^{2};\eta/\epsilon_{\mathrm{F}};k_{\mathrm{F}}r)$ after the Taylor expansion of $e^{i k_{\mathrm{F}}rv}$ as
\begin{align}
e^{i k_{\mathrm{F}}rv}
&=
e^{i k_{\mathrm{F}}r w(u^{2})}\sum_{m=0}^{\infty}\frac{1}{m!}\qty[i k_{\mathrm{F}}r\qty{v-w(u^{2})}]^{m}
.
\end{align}
The resultant expressions of $H^{(n)}_{\pm}\qty(u^{2};\eta/\epsilon_{\mathrm{F}};k_{\mathrm{F}}r)$ are
\begin{widetext}
\begin{align}
&
H^{(n)}_{\pm}\qty(u^{2};\frac{\eta}{\epsilon_{\mathrm{F}}};k_{\mathrm{F}}r)
=
\frac{2\pi i^{\frac{1\mp1}{2}}}{n!}
\sum_{m=0}^{n}
\qty{
D^{(n)}_{m}\qty(w(u^{2}); k_{\mathrm{F}}r)
\pm
\overline{D}^{(n)}_{m}\qty(\overline{w}(u^{2}); k_{\mathrm{F}}r)
}
,\label{eq:I(n)pm_D-rep}
\\
&
D^{(n)}_{m}\qty(w; k_{\mathrm{F}}r)
=
\frac{n!}{m!\qty(n-m)!} \qty(i k_{\mathrm{F}}r)^{m} i^{n+1} e^{i k_{\mathrm{F}}rw}
\left.
\dv[n-m]{v}
\qty{
\frac{1}{\qty(v+w)^{n}}-\frac{w}{\qty(v+w)^{n+1}}
}
\right\vert_{v=w}
,\\
&
\overline{D}^{(n)}_{m}\qty(\overline{w}; k_{\mathrm{F}}r)
=
\frac{n!}{m!\qty(n-m)!} \qty(i k_{\mathrm{F}}r)^{m} \qty(-i)^{n+1} e^{-i k_{\mathrm{F}}r\overline{w}}
\left.
\dv[n-m]{v}
\qty{
\frac{1}{\qty(v-\overline{w})^{n}}+\frac{\overline{w}}{\qty(v-\overline{w})^{n+1}}
}
\right\vert_{v=-\overline{w}}
,
\end{align}
\end{widetext}
where $0\le m \le n$. 
The change of the dummy variable $v\to -v$ in $\overline{D}^{(n)}_{m}\qty(\overline{w}; k_{\mathrm{F}}r)$ shows 
\begin{align}
\overline{D}^{(n)}_{m}\qty(\overline{w}; k_{\mathrm{F}}r)
&= \overline{D^{(n)}_{m}\qty(w; k_{\mathrm{F}}r)}
.\label{eq:D-conjugation}
\end{align}
It therefore suffices to derive the explicit expression of $D^{(n)}_{m}\qty(w; k_{\mathrm{F}}r)$:
\begin{align}
D^{(n)}_{m}\qty(w; k_{\mathrm{F}}r)
&=
\frac{i}{2^{2n+1}}
C^{(n)}_{m}
\qty(2k_{\mathrm{F}}r)^{m}\qty(-i)^{n+m}\frac{e^{ik_{\mathrm{F}}rw}}{w^{2n-m}}
,\label{eq:D(n)m}
\end{align}
where $C^{(n)}_{m}$ is the coefficient defined in Eqs.~(\ref{eq:C(n)m}) and (\ref{eq:C(n)0}).
Applying Eqs.~(\ref{eq:D-conjugation}) and (\ref{eq:D(n)m}) to Eq.~(\ref{eq:I(n)pm_D-rep}), we obtain the explicit forms of $H^{(n)}_{\pm}\qty(u^{2};\eta/\epsilon_{\mathrm{F}};k_{\mathrm{F}}r)$:
\begin{widetext}
\begin{align}
H^{(n)}_{\pm}\qty(u^{2};\frac{\eta}{\epsilon_{\mathrm{F}}};k_{\mathrm{F}}r)
&=
\frac{\pi i^{\frac{3\mp1}{2}}}{2^{2n}n!}
\sum_{m=0}^{n}
C^{(n)}_{m}
\qty(2k_{\mathrm{F}}r)^{m}
\qty{
\frac{(-i)^{n+m}e^{i k_{\mathrm{F}}r\sqrt{u^{2}+i\frac{\eta}{\epsilon_{\mathrm{F}}}}}}{\qty(u^{2}+i\frac{\eta}{\epsilon_{\mathrm{F}}})^{\frac{2n-m}{2}}}
\mp
\frac{i^{n+m}e^{-i k_{\mathrm{F}}r\sqrt{u^{2}-i\frac{\eta}{\epsilon_{\mathrm{F}}}}}}{\qty(u^{2}-i\frac{\eta}{\epsilon_{\mathrm{F}}})^{\frac{2n-m}{2}}}
}
.
\label{eq:H(n)pm_final}
\end{align}
\end{widetext}
This completes the evaluation of $\Pi^{\mathrm{R}(n)}_{\mathrm{cc}}(\vb*{r})$ and $\vb*{\Pi}^{\mathrm{R}(n)}_{\mathrm{sc}}(\vb*{r})$, which contain $H^{(n)}_{\pm}\qty(u^{2};\eta/\epsilon_{\mathrm{F}};k_{\mathrm{F}}r)$ only at $u=1$ [Eqs.~(\ref{eq:Pi(n)cc_integral-rep}) and (\ref{eq:Pi(n)sc_integral-rep})].

The evaluation of $\Pi^{\mathrm{R}(n)}_{\mathrm{eq}}(\vb*{r})$, in contrast, further requires the integration of $H^{(n)}_{-}\qty(u^{2};\eta/\epsilon_{\mathrm{F}};k_{\mathrm{F}}r)$ over $u$ in Eq.~(\ref{eq:Pi(n)eq_integral-rep}).
To complete this integration, we first transform the interim expression of Eq.~(\ref{eq:Pi(n)eq_integral-rep}) with the explicit forms of $H^{(n)}_{-}\qty(u^{2};\eta/\epsilon_{\mathrm{F}};k_{\mathrm{F}}r)$ in Eq.~(\ref{eq:H(n)pm_final}) and introduce the integral $I^{(n)}_{m}\qty(\eta/\epsilon_{\mathrm{F}};k_{\mathrm{F}}r)$ as follows:
\begin{widetext}
\begin{align}
&
\frac{1}{\qty(k_{\mathrm{F}}r)^{2}}
\int_{0}^{1}\dd{u}u\sin\qty(k_{\mathrm{F}}ru) 
H^{(n)}_{-}\qty(u^{2};\frac{\eta}{\epsilon_{\mathrm{F}}};k_{\mathrm{F}}r)
=
-
\frac{\pi}{2^{2n}n!}
\sum_{m=0}^{n} C^{(n)}_{m}
(2k_{\mathrm{F}}r)^{m}
I^{(n)}_{m}\qty(\frac{\eta}{\epsilon_{\mathrm{F}}};k_{\mathrm{F}}r)
.\label{eq:H(n)pm-I(n)m_relation}
\end{align}
\end{widetext}

At this stage, through the Drude-type approximation in Eq.~(\ref{eq:Drude-type_approximation}), we replace the driving forces $\vb*{E}$ and $\vb*{E}_{\mathrm{s}}^{\mu}$ for charge and spin currents with the expectation values of these currents, $\ev*{\hat{\vb*{j}}_{\mathrm{cc}}}_{J_{\mathrm{ex}}=0}$ and $\ev*{\hat{\vb*{j}}_{\mathrm{sc}}^{\mu}}_{J_{\mathrm{ex}}=0}$, respectively.
As in Eq.~(\ref{eq:nondimensionalization}), we nondimensionalize the currents into the corresponding dimensionless ones, $\vb*{J}_{\mathrm{cc}}$ and $\vb*{J}_{\mathrm{sc}}^{\mu} $, by defining the units of charge and spin currents as $4e\epsilon_{\mathrm{F}}/(3\lambda_{\mathrm{F}}^{2})$ and $4e_{\mathrm{s}}\epsilon_{\mathrm{F}}/(3\lambda_{\mathrm{F}}^{2})$, respectively.
In addition, we rewrite $\Pi^{\mathrm{R}(n)}_{\mathrm{eq}}(\vb*{r})$, $\Pi^{\mathrm{R}(n)}_{\mathrm{cc}}(\vb*{r})$, and $\vb*{\Pi}^{\mathrm{R}(n)}_{\mathrm{sc}}(\vb*{r})$ with the dimensionless functions $F^{(n)}_{\mathrm{eq}}(\vb*{r})$, $F^{(n)}_{\mathrm{cc}}(\vb*{r})$, and $\vb*{F}^{(n)}_{\mathrm{sc}}(\vb*{r})$ of Eqs.~(\ref{eq:Pi(n)eq_F(n)eq})--(\ref{eq:Pi(n)sc_F(n)sc}) in Sec.~\ref{sec:Approximations}.

Collecting the results above, we obtain $F^{(n)}_{\mathrm{eq}}(\vb*{r})$, $F^{(n)}_{\mathrm{cc}}(\vb*{r})$, and $\vb*{F}^{(n)}_{\mathrm{sc}}(\vb*{r})$ in the forms of Eqs.~(\ref{eq:F(n)eq_I(n)m-rep})--(\ref{eq:Theta(n)m}) in Sec.~\ref{sec:Approximations}, where the first line of Eq.~(\ref{eq:I(n)m}) gives $I^{(n)}_{m}\qty(\eta/\epsilon_{\mathrm{F}};k_{\mathrm{F}}r)$ in integral form.
It remains to perform this integration.
We change the integration variable from $u$ to $s$ by $u^{2}=1-s$
and integrate over $s$ after the double Taylor expansions of the factors $\sin(k_{\mathrm{F}}r\sqrt{1-s})$ and 
$e^{\pm i k_{\mathrm{F}}r\sqrt{1-s\pm i\eta/\epsilon_{\mathrm{F}}}}$.
Then, every integral takes the form of Euler's integral representation of the hypergeometric function ${}_{2}F_{1}\qty(\alpha, \beta; \gamma; z)$, where $0 < \Re \beta < \Re \gamma$, $\abs{z} <1$, and
\begin{align}
{}_{2}F_{1}\qty(\alpha, \beta; \gamma; z)
&=
\frac{\Gamma\qty(\gamma)}{\Gamma\qty(\beta)\Gamma\qty(\gamma-\beta)}
\int_{0}^{1}\dd{s} \frac{s^{\beta-1}\qty(1-s)^{\gamma-\beta-1}}{\qty(1-zs)^{\alpha}}
\nonumber\\
&= \sum_{j=0}^{\infty}\frac{(\alpha)_{j}(\beta)_{j}}{(\gamma)_{j}} \cdot \frac{z^{j}}{j!}
,\\
(\delta)_{j} 
&= \delta(\delta+1)\cdots(\delta+j-1) =\frac{\Gamma\qty(\delta+j)}{\Gamma\qty(\delta)}
.
\end{align}
We provide below the first step of the above procedure and the final form of $I^{(n)}_{m}\qty(\eta/\epsilon_{\mathrm{F}};k_{\mathrm{F}}r)$ in terms of ${}_{2}F_{1}\qty(\alpha, \beta; \gamma; z)$:
\begin{widetext}
\begin{align}
I^{(n)}_{m}\qty(\frac{\eta}{\epsilon_{\mathrm{F}}};k_{\mathrm{F}}r)
&=
\frac{1}{2\qty(k_{\mathrm{F}}r)^{2}}
\int_{0}^{1}\dd{s}\,\sin(k_{\mathrm{F}}r\sqrt{1-s})
\qty{
\frac{(-i)^{n+m}e^{i k_{\mathrm{F}}r\sqrt{1-s+i\frac{\eta}{\epsilon_{\mathrm{F}}}}}}{\qty(1-s+i\frac{\eta}{\epsilon_{\mathrm{F}}})^{\frac{2n-m}{2}}}
+\frac{i^{n+m}e^{-i k_{\mathrm{F}}r\sqrt{1-s-i\frac{\eta}{\epsilon_{\mathrm{F}}}}}}{\qty(1-s-i\frac{\eta}{\epsilon_{\mathrm{F}}})^{\frac{2n-m}{2}}}
}
\nonumber\\
&= 
\frac{1}{\qty(k_{\mathrm{F}}r)^{2}}
\sum_{p,q=0}^{\infty} \frac{(-1)^{p}\qty(k_{\mathrm{F}}r)^{2p+q+1}}{(2p+1)! q!} 
\nonumber\\
&\quad\times\frac{1}{2p+3}
\Biggl\{
(-i)^{n+m-q}\qty(1+i\frac{\eta}{\epsilon_{\mathrm{F}}})^{\frac{q-2n+m}{2}} {}_{2}F_{1}\qty(\frac{2n-m-q}{2}, 1; \frac{2p+5}{2}; \frac{1}{1+i\frac{\eta}{\epsilon_{\mathrm{F}}}})
\Biggr.
\nonumber\\
&\qquad\qquad\qquad\quad
\Biggl.
+
i^{n+m-q}\qty(1-i\frac{\eta}{\epsilon_{\mathrm{F}}})^{\frac{q-2n+m}{2}} {}_{2}F_{1}\qty(\frac{2n-m-q}{2}, 1; \frac{2p+5}{2}; \frac{1}{1-i\frac{\eta}{\epsilon_{\mathrm{F}}}})
\Biggr\}
.
\label{eq:I(n)m_final}
\end{align}
\end{widetext}
Applying Eq.~(\ref{eq:I(n)m_final}) to Eq.~(\ref{eq:F(n)eq_I(n)m-rep}) and using Eq.~(\ref{eq:Pi(n)eq_F(n)eq}), we obtain the explicit expression of $\Pi^{\mathrm{R}(n)}_{\mathrm{eq}}(\vb*{r})$.

To close this subsection, we comment on the role of the damping factor $\eta$.
Note that Eq.~(\ref{eq:I(n)m}) suggests $\eta$ also works as a regularization parameter for the integration with respect to the variable $u$.
The damping also leads to the factor $e^{-r/\ell}$ in the final expressions of $\Pi^{\mathrm{R}(n)}_{\mathrm{cc}}(\vb*{r})$ and $\vb*{\Pi}^{\mathrm{R}(n)}_{\mathrm{sc}}(\vb*{r})$, while the damping of $\Pi^{\mathrm{R}(n)}_{\mathrm{eq}}(\vb*{r})$ appears implicitly in the integrand of Eq.~(\ref{eq:I(n)m}).
Sections~\ref{sec:Approximations} and \ref{sec:EOM} discuss how this damping, together with the slow motion of the localized spins, justifies retaining only the terms with $n\le1$ and setting $\eta\to0$ in them.

\subsection{Response functions in the limit \texorpdfstring{$\eta\to0$}{\unichar{"03B7}\textrightarrow 0}}
\label{sec:clean-limit-eta0}
This subsection takes the limit $\eta\to0$ of $\Pi^{\mathrm{R}(n\leq 3)}_{\mathrm{eq}}(\vb*{r})$, $\Pi^{\mathrm{R}(n)}_{\mathrm{cc}}(\vb*{r})$, and $\vb*{\Pi}^{\mathrm{R}(n)}_{\mathrm{sc}}(\vb*{r})$ through $F^{(n\leq 3)}_{\mathrm{eq}}(\vb*{r})$, $F^{(n)}_{\mathrm{cc}}(\vb*{r})$, and $\vb*{F}^{(n)}_{\mathrm{sc}}(\vb*{r})$.
Throughout this subsection, $\vert_{\eta\to0}$ denotes the limit $\eta\to0^{+}$ taken after all the integrations.
We first present the general forms of these functions in this limit, except for $\Pi^{\mathrm{R}(n)}_{\mathrm{eq}}(\vb*{r})$ with $n\ge2$, for which the limit $\eta \to 0$ cannot be taken term by term:
\begin{widetext}
\begin{align}
\left.F^{(n\leq 1)}_{\mathrm{eq}}(\vb*{r})\right\vert_{\eta\to0}
&=
\frac{\pi}{2^{2n}n!}
\sum_{m=0}^{n} C^{(n)}_{m}
(2k_{\mathrm{F}}r)^{m}
\frac{1}{\qty(k_{\mathrm{F}}r)^{2}}
\sum_{p,q=0}^{\infty} \frac{(-1)^{p}\qty(k_{\mathrm{F}}r)^{2p+q+1}}{(2p+1)! q!} 
\cdot \frac{1}{2p+3}
\nonumber\\
&\qquad\times
\qty{
(-i)^{n+m-q}{}_{2}F_{1}\qty(\frac{2n-m-q}{2}, 1; \frac{2p+5}{2}; 1)
+
i^{n+m-q} {}_{2}F_{1}\qty(\frac{2n-m-q}{2}, 1; \frac{2p+5}{2}; 1)
}
,\label{eq:F(n)eq_limit}
\\
\left.F^{(n)}_{\mathrm{cc}}(\vb*{r})\right\vert_{\eta\to0}
&=
\qty(\frac{\vb*{r}}{r}\vdot\vb*{J}_{\mathrm{cc}})
\frac{\pi}{2^{2n}n!}
\sum_{m=0}^{n} C^{(n)}_{m}
\qty(2k_{\mathrm{F}}r)^{m}
\frac{\sin\qty(k_{\mathrm{F}}r)-\qty(k_{\mathrm{F}}r)\cos\qty(k_{\mathrm{F}}r)}{\qty(k_{\mathrm{F}}r)^{3}}
\sin
\qty(
-\frac{\pi}{2}(n+m)
+k_{\mathrm{F}}r
)
,\label{eq:F(n)cc_limit}
\\
\left.F^{\mu(n)}_{\mathrm{sc}}(\vb*{r})\right\vert_{\eta\to0}
&=
\qty(\frac{\vb*{r}}{r}\vdot\vb*{J}_{\mathrm{sc}}^{\mu})
\frac{\pi}{2^{2n}n!}
\sum_{m=0}^{n} C^{(n)}_{m}
\qty(2k_{\mathrm{F}}r)^{m}
\frac{\sin\qty(k_{\mathrm{F}}r)-\qty(k_{\mathrm{F}}r)\cos\qty(k_{\mathrm{F}}r)}{\qty(k_{\mathrm{F}}r)^{3}}
\cos
\qty(
-\frac{\pi}{2}(n+m)
+k_{\mathrm{F}}r
)
.\label{eq:F(n)sc_limit}
\end{align}
\end{widetext}
As for $F^{(n\leq 1)}_{\mathrm{eq}}(\vb*{r})\vert_{\eta\to0}$, the above expression is based on the final form of $I^{(n)}_{m}\qty(\eta/\epsilon_{\mathrm{F}};k_{\mathrm{F}}r)$ in Eq.~(\ref{eq:I(n)m_final}) through Eq.~(\ref{eq:F(n)eq_I(n)m-rep}).
On the other hand, the original form of $I^{(n)}_{m}\qty(\eta/\epsilon_{\mathrm{F}};k_{\mathrm{F}}r)$ in Eq.~(\ref{eq:I(n)m}) is convenient for directly deriving a closed-form expression for each case of $F^{(n)}_{\mathrm{eq}}(\vb*{r})\vert_{\eta\to0}$ and thus for each case of $\Pi^{\mathrm{R}(n)}_{\mathrm{eq}}(\vb*{r})\vert_{\eta\to0}$ with a specific $n$.
We can show the equivalence between them by Gauss's hypergeometric theorem for ${}_{2}F_{1}\qty(\alpha, \beta; \gamma; 1)$, which holds under the condition $\Re\gamma >0$ and $\Re\qty(\gamma-\alpha-\beta)>0$:
\begin{align}
{}_{2}F_{1}\qty(\alpha, \beta; \gamma; 1)
&= \frac{\Gamma\qty(\gamma)\Gamma\qty(\gamma-\alpha-\beta)}{\Gamma\qty(\gamma-\alpha)\Gamma\qty(\gamma-\beta)}
.
\end{align}

The closed-form expressions for $F^{(n\le 1)}_{\mathrm{eq}}(\vb*{r})\vert_{\eta\to0}$ are those of Eqs.~(\ref{eq:Feq_limit_f}), (\ref{eq:f0eq}), and (\ref{eq:f1eq}) in Sec.~\ref{sec:EOM}.
As for $F^{(n\ge 2)}_{\mathrm{eq}}(\vb*{r})\vert_{\eta\to0}$, we can no longer use the expression of Eq.~(\ref{eq:F(n)eq_limit}),
because the condition needed by Gauss's hypergeometric theorem for ${}_{2}F_{1}\qty(\alpha, \beta; \gamma; 1)$ does not always hold;
for $n=2$, the term with $m=1$ and $p=q=0$ has $\gamma-\alpha-\beta=0$, so that ${}_{2}F_{1}(3/2,1;5/2;1)$ diverges while its phase factor $(-i)^{3}+i^{3}$ vanishes.
In addition, for $n\ge 2$, we cannot take the limit $\eta\to0$ before the integration in Eq.~(\ref{eq:I(n)m}),
because each of the two terms in the integrand of $I^{(n)}_{m}\qty(\eta/\epsilon_{\mathrm{F}};k_{\mathrm{F}}r)$ with $2n-m\ge3$ is singular at $u\to0$ for $\eta\to0$; even when these singularities cancel in the sum, as for $n=2$ and $m=1$, the limit is nonuniform near $u=0$,
and the region $u\sim\sqrt{\eta/\epsilon_{\mathrm{F}}}$ leaves a contribution that does not vanish as $\eta\to0$.
For $n=2$, such a contribution arises only from the term with $m=1$, and keeping $\eta$ finite during the integration yields
\begin{align}
&\lim_{\eta\to0^{+}}I^{(2)}_{1}\qty(\frac{\eta}{\epsilon_{\mathrm{F}}};k_{\mathrm{F}}r)
\nonumber\\
&\quad=-\frac{2}{\qty(k_{\mathrm{F}}r)^{2}}\int_{0}^{1}\dd{u}\frac{\sin^{2}\qty(k_{\mathrm{F}}ru)}{u^{2}}
+\frac{\pi}{2k_{\mathrm{F}}r}
,
\end{align}
where the last term comes from the region $u\sim\sqrt{\eta/\epsilon_{\mathrm{F}}}$.
Consequently, we obtain
\begin{align}
&\left.F^{(2)}_{\mathrm{eq}}(\vb*{r})\right\vert_{\eta\to0}
= f^{(2)}_{\mathrm{eq}}(2k_{\mathrm{F}}r)
,\\
&f^{(2)}_{\mathrm{eq}}(x)
=
-
\frac{\pi}{2^{2}2!}
\cdot
\frac{x\qty{\mathrm{Si}\qty(x)-\frac{\pi}{2}}-2\qty(1-\cos x)}{x}
,
\end{align}
where $\mathrm{Si}\qty(z)$ is the sine integral function defined by
\begin{align}
\mathrm{Si}\qty(z)
&= \int_{0}^{z}\dd{t}\frac{\sin t}{t}
.
\end{align}
The constant term $\pi^{2}/2^{4}$ in $f^{(2)}_{\mathrm{eq}}(x)$, i.e., the term with $\pi/2$ in the numerator, would be missed if the limit $\eta\to0$ were taken before the integration.
For $F^{(3)}_{\mathrm{eq}}(\vb*{r})$, we can carry out the integration with $\eta$ kept finite and obtain the following result:
\begin{widetext}
\begin{align}
&F^{(3)}_{\mathrm{eq}}(\vb*{r})
=
\frac{\pi}{2^{4}3!}
\cdot
\frac{1}{\qty(2k_{\mathrm{F}}r)}
\qty{
\mathcal{P}^{\mathrm{R}(3)}_{\mathrm{eq}}\qty(1;\frac{\eta}{\epsilon_{\mathrm{F}}};k_{\mathrm{F}}r)
-\mathcal{P}^{\mathrm{R}(3)}_{\mathrm{eq}}\qty(0;\frac{\eta}{\epsilon_{\mathrm{F}}};k_{\mathrm{F}}r)
}
\nonumber\\
&\hspace{13mm}
=
\frac{\pi}{2^{4}3!}
\cdot
\frac{1}{\qty(2k_{\mathrm{F}}r)}
\mathcal{P}^{\mathrm{R}(3)}_{\mathrm{eq}}\qty(1;\frac{\eta}{\epsilon_{\mathrm{F}}};k_{\mathrm{F}}r)
,\quad
\text{since }
\mathcal{P}^{\mathrm{R}(3)}_{\mathrm{eq}}\qty(0;\frac{\eta}{\epsilon_{\mathrm{F}}};k_{\mathrm{F}}r) =0
\text{ for }\eta>0
,\\
&\mathcal{P}^{\mathrm{R}(3)}_{\mathrm{eq}}\qty(u;\frac{\eta}{\epsilon_{\mathrm{F}}};k_{\mathrm{F}}r)
\nonumber\\
&=
2\sin(k_{\mathrm{F}}ru)
\qty[
-2\qty{
\frac{e^{ik_{\mathrm{F}}r\sqrt{u^{2}+i\frac{\eta}{\epsilon_{\mathrm{F}}}}}}{\qty(u^{2}+i\frac{\eta}{\epsilon_{\mathrm{F}}})^{\frac{3}{2}}}
+
\frac{e^{-ik_{\mathrm{F}}r\sqrt{u^{2}-i\frac{\eta}{\epsilon_{\mathrm{F}}}}}}{\qty(u^{2}-i\frac{\eta}{\epsilon_{\mathrm{F}}})^{\frac{3}{2}}}
}
+i(2k_{\mathrm{F}}r)
\qty{
\frac{e^{ik_{\mathrm{F}}r\sqrt{u^{2}+i\frac{\eta}{\epsilon_{\mathrm{F}}}}}}{u^{2}+i\frac{\eta}{\epsilon_{\mathrm{F}}}}
-
\frac{e^{-ik_{\mathrm{F}}r\sqrt{u^{2}-i\frac{\eta}{\epsilon_{\mathrm{F}}}}}}{u^{2}-i\frac{\eta}{\epsilon_{\mathrm{F}}}}
}
]
\nonumber\\
&\quad
+\frac{2}{i\frac{\eta}{\epsilon_{\mathrm{F}}}}
(2k_{\mathrm{F}}r) 
\qty{
u \cos(k_{\mathrm{F}}ru)
\qty(
\frac{e^{ik_{\mathrm{F}}r\sqrt{u^{2}+i\frac{\eta}{\epsilon_{\mathrm{F}}}}}}{\sqrt{u^{2}+i\frac{\eta}{\epsilon_{\mathrm{F}}}}}
-
\frac{e^{-ik_{\mathrm{F}}r\sqrt{u^{2}-i\frac{\eta}{\epsilon_{\mathrm{F}}}}}}{\sqrt{u^{2}-i\frac{\eta}{\epsilon_{\mathrm{F}}}}}
)
-i\sin(k_{\mathrm{F}}ru)
\qty(
e^{ik_{\mathrm{F}}r\sqrt{u^{2}+i\frac{\eta}{\epsilon_{\mathrm{F}}}}}
+
e^{-ik_{\mathrm{F}}r\sqrt{u^{2}-i\frac{\eta}{\epsilon_{\mathrm{F}}}}}
)
}
.
\end{align}
\end{widetext}
For $\eta>0$, $\mathcal{P}^{\mathrm{R}(3)}_{\mathrm{eq}}\qty(u;\eta/\epsilon_{\mathrm{F}};k_{\mathrm{F}}r)$ vanishes at $u=0$, whereas its limit $\eta\to0$ behaves as $-6(2k_{\mathrm{F}}r)/u^{2}$ for $u\to0$.
The two limits $u\to0$ and $\eta\to0$ therefore do not commute: if the limit $\eta\to0$ were taken before the integration over $u$, the integral would diverge at $u=0$.
A finite $\eta$ thus regularizes the integration near $u=0$, and we can take the limit $\eta\to 0$ in $\mathcal{P}^{\mathrm{R}(3)}_{\mathrm{eq}}\qty(1;\eta/\epsilon_{\mathrm{F}};k_{\mathrm{F}}r)$
and obtain $F^{(3)}_{\mathrm{eq}}(\vb*{r})\vert_{\eta\to0}$ as follows:
\begin{align}
&\left.F^{(3)}_{\mathrm{eq}}(\vb*{r})\right\vert_{\eta\to0}
=f^{(3)}_{\mathrm{eq}}(2k_{\mathrm{F}}r)
,\\
&f^{(3)}_{\mathrm{eq}}(x)
=-
\frac{\pi}{2^{4}3!}
\cdot
\frac{4\sin x+ x\qty(3-\cos x)}{x}
.
\end{align}

We can see that $\left.F^{(3)}_{\mathrm{eq}}(\vb*{r})\right\vert_{\eta\to0}$ does not decay at long distances and that $\left.F^{(2)}_{\mathrm{eq}}(\vb*{r})\right\vert_{\eta\to0}$ decays only as $1/(k_{\mathrm{F}}r)$, much more slowly than $F^{(0)}_{\mathrm{eq}}(\vb*{r})$ and $F^{(1)}_{\mathrm{eq}}(\vb*{r})$.
This fact suggests that the limit $\eta\to0$ is no longer physically adequate for $F^{(n\ge 2)}_{\mathrm{eq}}(\vb*{r})$.

As for $F^{(n)}_{\mathrm{cc}}(\vb*{r})\vert_{\eta\to0}$ and $\vb*{F}^{(n)}_{\mathrm{sc}}(\vb*{r})\vert_{\eta\to0}$, Eqs.~(\ref{eq:F(n)cc_limit}) and (\ref{eq:F(n)sc_limit}) are already in closed form with elementary functions.
However, unlike the specific expressions of $F^{(n\leq 3)}_{\mathrm{eq}}(\vb*{r})\vert_{\eta\to0}$ discussed above, their dependence on $2k_{\mathrm{F}}r$ is not explicit in these expressions.
We make this dependence explicit by transforming Eqs.~(\ref{eq:F(n)cc_limit}) and (\ref{eq:F(n)sc_limit}) with the sum and double-angle formulae of trigonometric functions:
\begin{align}
&\left.F^{(n)}_{\mathrm{cc}}(\vb*{r})\right\vert_{\eta\to0}
= \qty(\frac{\vb*{r}}{r}\vdot\vb*{J}_{\mathrm{cc}}) f^{(n)}_{\mathrm{cc}}(2k_{\mathrm{F}}r)
\label{eq:F(n)cc_limit_2kFr}
,\\
&\left.F^{\mu(n)}_{\mathrm{sc}}(\vb*{r})\right\vert_{\eta\to0}
= \qty(\frac{\vb*{r}}{r}\vdot\vb*{J}_{\mathrm{sc}}^{\mu})  f^{(n)}_{\mathrm{sc}}(2k_{\mathrm{F}}r)
\label{eq:F(n)sc_limit_2kFr}
,
\end{align}
where
\begin{widetext}
\begin{align}
f^{(n)}_{\mathrm{cc}}(x)
&=
\frac{2\pi}{2^{2n}n!}
\sum_{m=0}^{n} C^{(n)}_{m}
x^{m}
\qty{
-\sin
\qty(
\frac{\pi}{2}(n+m)
)
\frac{2\sin x - x \qty(1+\cos x)}{x^{3}}
+\cos
\qty(
\frac{\pi}{2}(n+m)
)
\frac{2\qty(1-\cos x)-x \sin x}{x^{3}}
}
,
\\
f^{(n)}_{\mathrm{sc}}(x)
&=
\frac{2\pi}{2^{2n}n!}
\sum_{m=0}^{n} C^{(n)}_{m}
x^{m}
\qty{
\quad
\cos
\qty(
\frac{\pi}{2}(n+m)
)
\frac{2\sin x - x \qty(1+\cos x)}{x^{3}}
+\sin
\qty(
\frac{\pi}{2}(n+m)
)
\frac{2\qty(1-\cos x) - x \sin x}{x^{3}}
}
.
\end{align}
\end{widetext}

We close this Appendix with the explicit expressions of $f^{(n\le 3)}_{\mathrm{cc}}(x)$ and $f^{(n\le 3)}_{\mathrm{sc}}(x)$; those with $n\le1$ are the RKKY range functions $f^{(n)}_{\mathrm{cc}}(x)$ and $f^{(n)}_{\mathrm{sc}}(x)$ of Sec.~\ref{sec:EOM}:
\begin{widetext}
\begin{align}
f^{(0)}_{\mathrm{cc}}(x)
&=
\quad
2\pi
\cdot
\frac{2\qty(1-\cos x)-x\sin x}{x^{3}}
,\\
f^{(1)}_{\mathrm{cc}}(x)
&=
-\frac{\pi}{2}
\cdot
\frac{2\qty(1-\cos x)-x\sin x}{x^{2}}
,\\
f^{(2)}_{\mathrm{cc}}(x)
&=
\quad
\frac{\pi}{2^{3}2!}
\cdot
\frac{\qty(4-x^{2})\sin x-4x\cos x}{x^{2}}
,\\
f^{(3)}_{\mathrm{cc}}(x)
&=
\quad
\frac{\pi}{2^{5}3!}
\cdot
\frac{
2\qty(12-x^{2})\qty{1-\cos x}
+6x^{2}\qty(1+\cos x)
-\qty(24-x^{2})x\sin x
}
{x^{2}}
,\\
f^{(0)}_{\mathrm{sc}}(x)
&=
\quad
2\pi
\cdot
\frac{2\sin x-x\qty(1+\cos x)}{x^{3}}
,\\
f^{(1)}_{\mathrm{sc}}(x)
&=
-\frac{\pi}{2}
\cdot
\frac{2\sin x- x\qty(1+\cos x)}{x^{2}}
,\\
f^{(2)}_{\mathrm{sc}}(x)
&=
-\frac{\pi}{2^{3}2!}
\cdot
\frac{
4\qty(1-\cos x)+x^{2}\qty(1+\cos x)-4x\sin x
}
{x^{2}}
,\\
f^{(3)}_{\mathrm{sc}}(x)
&=
\quad
\frac{\pi}{2^{5}3!}
\cdot
\frac{
8\qty(3-x^{2})\sin x
+12x\qty(1-\cos x)
-\qty(12-x^{2})x\qty(1+\cos x)
}
{x^{2}}
.
\end{align}
\end{widetext}
We can see that  $\left.F^{(n\ge 2)}_{\mathrm{cc}}(\vb*{r})\right\vert_{\eta\to0}$ and $\left.\vb*{F}^{(n\ge 2)}_{\mathrm{sc}}(\vb*{r})\right\vert_{\eta\to0}$ do not decay at long distances.
The limit $\eta\to0$ is no longer physically adequate for them.

 \bibliography{DHMCSC_References}

\end{document}